\documentclass{article}
\usepackage{epubtk}
\usepackage{amssymb}
\usepackage{booktabs}
\usepackage{graphicx}
\usepackage{textcomp}
\usepackage{xspace}
\usepackage{changes}

\def\M{\mathcal{M}}
\def\R{\mathcal{R}}
\def\D{\mathcal{D}}

\newcommand{\kms}{km~s\super{-1}\xspace}

\newcommand{\ms}{\,M_{\odot}}
\newcommand{\md}{\mbox {$\dot{M}_{\mathrm{d}}$}} 
\newcommand{\rs}{\,R_{\odot}}
\newcommand{\ls}{\mbox {$L_{\odot}$}}
\newcommand{\porb}{\mbox {$P_{\mathrm{orb}}$}}
\newcommand{\ace}{\mbox {$\alpha_{\mathrm{CE}}$}}
\newcommand{\al}{\mbox {$\alpha_{\mathrm{CE}}\lambda$}}
\newcommand{\pyr}{\mathrm{\ yr}^{-1}}
\newcommand{\myr}{M_{\odot}\mathrm{\ yr}^{-1}}

\newcommand{\mdot}{\mbox {$\dot{M}$}}
\newcommand{\dotm}{\mbox {$\dot{M}$}}
\newcommand{\medd}{\mbox {$\dot{M}_{\mathrm{Edd}}$}}

\newcommand{\mch}{\mbox {$M_{\mathrm{Ch}}$}}
\newcommand{\mwd}{\mbox {$M_{\mathrm{WD}}$}}
\newcommand{\rescale}[2]{$#1 \times (\rho_\mathrm{eff}/5)^{-#2}$} 
\newcommand\apgt{\ {\raise-.5ex\hbox{$\buildrel>\over\sim$}}\ }
\newcommand\aplt{\ {\raise-.5ex\hbox{$\buildrel<\over\sim$}}\ }

\newcommand{\am}{AM~CVn\xspace}

\newcommand\sna{SN~Ia\xspace}
\newcommand\sne{SNe~Ia\xspace}

\begin{document}

\title{The Evolution of Compact Binary Star Systems}
\author{\epubtkAuthorData{Konstantin A.\ Postnov}
        {Moscow M.V.Lomonosov State University\\
         Sternberg Astronomical Institute \\
         13 Universitetskij Pr. \\
         119992 Moscow \\
         Russia}
        {kpostnov@gmail.com}
        {http://www.sai.msu.ru}
        \and
        \epubtkAuthorData{Lev R.\ Yungelson}
        {Institute of Astronomy of Russian Academy of Sciences \\
         48 Pyatnitskaya Str. \\
         119017 Moscow \\
         Russia}
        {lry@inasan.ru}
        {http://www.inasan.ru}}

\date{}
\maketitle


\begin{abstract}
We review the formation and evolution of compact binary stars
consisting of white dwarfs (WDs), neutron stars (NSs), and black holes
(BHs). Mergings of compact binary stars are expected to be the most
important sources for the forthcoming gravitational-wave (GW)
astronomy. In the first part of the review, we discuss 
observational manifestations of close binary stars with NS and/or
black components and their merger rate,
crucial points
in the formation and evolution of compact stars in binary systems,
including the treatment of the natal kicks which 
NSs and BHs
acquire during the core collapse of massive stars and the
common envelope phase of binary evolution, which are most relevant to
the merging rates of NS-NS, NS-BH and BH-BH binaries. The second part of
the review is devoted 
mainly to formation and evolution of binary WDs and their
observational manifestations, including their role as progenitors of 
cosmologically important thermonuclear SN~Ia. 
We also consider AM~CVn-stars which are thought to be 
the best verification binary GW sources for
future low-frequency GW space interferometers.
\end{abstract}

\epubtkKeywords{astrophysics, binary systems, gravitational wave
  sources, supernovae, neutron stars, black holes, white dwarfs, AM
  CVn stars}

\epubtkUpdate
    [Id=A,
     ApprovedBy=subjecteditor,
     AcceptDate={11 October 2013},
     PublishDate={11 October 2013}]{%
Major revision, updated and expanded.

References are updated by 2007\,--\,early-2014 publications, some outdated
items removed.  The reference list contains now 881 items
instead of 475. The new version of the review has 39 figures.

In Section~\ref{section:introduction}, stellar evolution is outlined in more
detail, special attention is paid to electron-capture supernovae. 
More attention than before is paid to the uncertainties in the estimates
of the ranges of precursors of white dwarfs, neutron stars and black holes and their 
masses.

In Section~\ref{section:observational_situation}, a subsection
(\ref{sec:BH_inbinaries}) on black holes in binaries is added and an
independent estimate of double black hole/neutron star coalescence
rate is given.

In Section~\ref{section:physical_principles}, a subsection on mass
transfer between binary components is added, discussion of common
envelopes is expanded.

In Section~\ref{section:wd_formation}, the discussion of the scenario
for formation of short-period binaries is appended by a more detailed
description of ``intermediate'' stages, like post-common envelope
binaries, the discussion of properties of binary WD is expanded,
the comparison of results of population synthesis calculations for close
WD with observations is updated, thanks to significant increase of the
number of observed objects. A much more extended comparison of
expectations from scenarios for single-degenerate and
double-degenerate SNe~Ia (including violent mergers) is presented.

In Section~\ref{section:observations} all published information on
double-degenerates with measured periods and estimates of masses of
components and some subgiant+[white~dwarf] systems is summarised (up
to February 2014).

In Section~\ref{section:am-evol} more attention than previously is
paid to initial stages of mass exchange in the ``white dwarf'' channel
of formation of AM~CVn stars and to comparison of ``white dwarf'' and
``helium star'' channels. Final stages of evolution of AM~CVn stars
and SN~.Ia are discussed.

In Section~\ref{section:waves} the GW-signals of detached and
interacting double-degenerates are compared in more detail, as well as
their detection probabilities with planned ``short-arm'' (eLISA)
interferometer.

The former section on overlap of EM- and GW-signals from AM~CVn stars is
deleted because of (temporary?) cancellation of LISA-mission.
}

\newpage


\section{Introduction}
\label{section:introduction}

Close binary stars consisting of two compact stellar remnants -- white dwarfs
(WDs), neutron stars (NSs) or black holes (BHs) are considered as primary
targets of the forthcoming field of gravitational wave (GW) astronomy
(see, for a review, \cite{Thorne87, Grishchuk_al01, Sathyaprakash_Schutz09, 2013CRPhy..14..272S}),
 since
their orbital evolution is entirely controlled by emission of gravitational
waves and leads to ultimate coalescence (merger) 
and possible explosive disruption of the components. Emission of
gravitational waves accompanies the latest stages of evolution of stars and
manifests instabilities in relativistic objects
\cite{2012CQGra..29l4016A,2013CQGra..30s3002A}. Close compact binaries can thus
serve as testbeds for theories of gravity \cite{2013LRR....16....7G}.
The double NS(BH) mergers which release  $\sim$~10\super{52}~erg  as
  GWs~\cite{Clark_Eardley77, Clark_al79} should be
the brightest GW events in the 10\,--\,1000~Hz frequency band of the existing or
future ground-based GW detectors like LIGO~\cite{Barish_Weiss99},
VIRGO~\cite{Acernese_al05}, GEO600~\cite{Ricci_Brillet97},
KAGRA(LCGT)~\cite{2012CQGra..29l4007S} (see also \cite{2013PrPNP..68....1R} for
review of the current state of existing and 2nd- and 3rd-generation ground-based
detectors).
Mergers of double NS(BH) can be accompanied by the release of a
huge amount of electromagnetic energy in a burst and manifest themselves as
short gamma-ray bursts (GRBs). A lot of observational support to NS-NS/NS-BH 
mergers as sources of short GRBs have been obtained (see, e.g.,  studies of 
short GRB locations in the host galaxies \cite{FongBerger13, Tanvir_al13} and references therein). As well, relativistic jets, associated with GRB
of any nature may be sources of GW in the ground-based detectors range
\cite{2013PhRvD..87l3007B}.

Double WDs, especially interacting binary WDs observed as AM~CVn-stars
and ultracompact X-ray binaries (UCXB), are potential GW sources
within the frequency band (10\super{-4}\,--\,1)~Hz of the space GW
interferometers like {\bf currently cancelled} LISA~\cite{fbhhs89}
\epubtkFootnote{ 
\url{http://lisa.gsfc.nasa.gov/cosmic_vision_changes.html},
\url{http://sci.esa.int/science-e/www/object/index.cfm?fobjectid=48661}.},
NGO (eLISA)~\cite{2012CQGra..29l4016A,2013GWN.....6....4A},
DEGIGO~\cite{2013IJMPD..2241013Y}
and other proposed or planned
low-frequency GW
detectors~\cite{Crowder_Cornish05,bender_post_lisa13}. 
At the moment, eLISA is selected for the third Large-class mission in ESA’s Cosmic 
Vision science 
program (L3). Its first step should be the launch of ESA's LISA Pathfinder (LPF) 
mission in 2015; the launch of eLISA itself is currently planned for 2034\\
(\url{https://www.elisascience.org/news/top-news/gravitationaluniverseselectedasl3}).

The double WD
mergers are among the primary candidate mechanisms for type~Ia
supernovae (SNe~Ia) explosions. The NIR magnitudes of the latter are
considered as the best ``standard candles'' \cite{2012MNRAS.425.1007B}
and, in this guise, are crucial in modern cosmological
studies~\cite{1998AJ....116.1009R, perlmutter_cosm99}. Further
improvements of precision of standardisation of \sne fluxes is
possible, e.g., by account of their environmental
dependence~\cite{2013A&A...560A..66R,2013arXiv1309.1182R, 2013Sci...340..170W}. On the
other hand, usually, as an event beneath the ``standard candle'', an
explosion of a non-rotating WD with the Chandrasekhar mass ($\simeq 1.38\ms$) is
considered. Rotation of progenitors may increase the 
critical mass and  makes SNe~Ia less reliable for cosmological
use~\cite{2006ApJ...644...21D, 2007MmSAI..78..549D}.

\sne are suggested to be responsible for production of about 50\% of heavy
elements in the Universe \cite{timmes95}. Mergers of binary WD are, probably,
one of the main-mechanisms of formation of massive ($>0.8\ms$) WD and WD with
strong magnetic fields (see, e.g., \cite{ji_mgn_wd13,2013MNRAS.431.2778K} for
the recent studies and review of earlier work).    

A comparison of SN~Ia rates (for the different models of their progenitors) with
observations may, in principle, shed light on both the star formation history
and on the nature of the progenitors (see, e.g.,
\cite{ruiz_canal_cosm98,yl98,madau_prog98,yl00,2010AstL...36..780Y,2011MNRAS.417..916G,2012PASA...29..447M,
2013MNRAS.430.1746G,mmn_snia_rev13}).



\vspace{0.3cm}
\noindent
Compact binaries are the end products of the evolution of binary stars, and the main purpose of the present review is to describe the
astrophysical knowledge on their formation and evolution. We shall
discuss the present situation with the main parameters determining
their evolution and the rates of coalescence of double NSs/BHs and WDs.

The problem is to evaluate as accurately as possible (i)~the
  physical parameters of the coalescing binaries (masses of the
  components and, if possible, their spins, magnetic fields, etc.) and
  (ii)~the occurrence rate of mergers in the Galaxy and in the local
  Universe. Masses of NSs in binaries are known with a rather good
  accuracy of 10\% or better from, e.g., pulsar
  studies~\cite{Thorsett_Chakrabarty99, 2013arXiv1309.6635K}; see
  also \cite{2010NewAR..54..101L,Ozel_al12}
 for  recent summaries of NS mass  measurements.

The case is not so good with the rate of coalescence of relativistic
  binary stars. Unfortunately, there is no way to derive it from
  the first principles -- neither the formation rate of the progenitor
  binaries for compact double stars nor stellar evolution are known
  well enough. However, the situation is not completely hopeless,
  especially in the case of double NS systems. Natural appearance of
  rotating NSs with magnetic fields as radio pulsars allows searching
  for binary pulsars with secondary compact companion using powerful
  methods of modern radio astronomy (for example, in dedicated pulsar
  surveys such as the Parkes multi-beam pulsar
  survey~\cite{Manchester_al01, Faulkner_al04}).

Based on the observational statistics of the Galactic binary pulsars
  with  NS companions, one can evaluate the Galactic rate of
  binary NS formation and merging~\cite{Phinney91, nps91}.
  On the other hand, a direct simulation of binary star
  evolution in the Galaxy (the \emph{population synthesis} method) can
  also predict the formation and merger rates of close compact
  binaries as a function of (numerous) parameters of binary star
  formation and evolution.
  Both kinds of estimates are plagued by badly constrained parameters or selection
effects, but it is, nevertheless,
  encouraging that most likely Galactic rates of events obtained in two ways
  currently differ by a factor of  $\approx$~3 only: 80/Myr from observations and 30/Myr 
  from population synthesis; 
  see \cite{2010CQGra..27k4007M} for a recent review
  of observational and theoretical estimates and Section~\ref{sec:secI:DetectRate}.
 It is important and encouraging that both
  estimates (observational, as inferred from recent measurements of
  binary pulsars~\cite{Burgay_al03, Kalogera_al04}, and theoretical
  from the population synthesis; see
  Section~\ref{sec:secI:DetectRate}) now give very close estimates for
  the double NS star merger rate in the Galaxy of about one event per
  10\,000 years. 

   No binary BH or NS\,+\,BH systems have been found so far,
  so merger rates of compact binaries with BHs have been evaluated as
  yet only from population synthesis studies.

\subsection{Formation of stars and end products of their evolution}

\vspace{0.3cm}
\noindent
Let us briefly remind the key facts about star formation and evolution.
Around 6\% of the baryonic matter in the Universe is confined in
stars~\cite{Fukugita_Peebles04}. Recent observational data suggests that, first,
long thin filaments form inside molecular clouds and, next, these filaments
fragment into protostellar cores due to gravitational instability, if their mass
per unit length exceeds certain threshold~\cite{2010AA...518L.102A}. An object
may be called a ``star'' if it is able to generate energy by nuclear fusion at a
level sufficient to halt the contraction \cite{Kumar63}. For solar chemical
composition, $M_{\min}\approx 0.075\,M_{\odot}$ and $M_{\min}$ increases if
stellar metallicity is lower than solar \cite{1993ApJ...406..158B}. Currently,
among observed stars the lowest mass value derived from photometric and
spectroscopic observations has AB~Dor~C: $(0.090\pm0.005)\ms$
\cite{2007ApJ...665..736C}. 

The maximum mass of the stars is probably set by the proximity of the
luminosity to the Eddington limit and  pulsational instability. For
solar chemical composition, this limit is close to
$1000\ms$~\cite{2007ApJ...659.1576B, 2008AA...477..223Y}.%
\epubtkFootnote{Existence of such a high-mass very short-living
  objects is deemed possible due to runaway stellar collisions in
  dense young compact stellar clusters~\cite{1999AA...348..117P}; for
  the prospects of observing them see, e.g.,
  \cite{2012MNRAS.425L..91P}.}
But conditions of stellar formation, apparently, define a much lower
mass limit. Currently, components of a $\porb=3.77$\,day eclipsing
binary NGC3603-A1 have maximum dynamically measured masses:
$(116\pm31)\ms$ and $(89\pm16)\ms$ \cite{2008MNRAS.389L..38S}. Since both
stars are slightly evolved main-sequence objects (WN6h), subject to
severe stellar wind mass loss, their inferred \emph{initial} masses
could be higher: $148^{+40}_{-27}\ms$ and $106^{+92}_{-15}\ms$
\cite{2010MNRAS.408..731C}. Indirect evidence, based on photometry and
spectral analysis suggests possible existence of
$(200\mbox{\,--\,}300)\ms$ stars, see \cite{2013arXiv1302.2021V} for
references and discussion.

For metal-free stars which formed first in the Universe (the so-called Population III stars), the upper mass limit is rather uncertain. For example, it can be below $100\ms$ \cite{2002Sci...295...93A,susa_masses13}, because
of competition of accretion onto first formed compact core and nuclear burning
and influence of UV-irradiation from nearby protostars. On the other hand, masses of Population III stars could be much higher due to the absence of effective coolants in the
primordial gas. Then, their masses could be limited by pulsations, though, the
upper mass limit remains undefined \cite{2012MNRAS.423.3397S} and masses up to
$\simeq 1000\ms$ are often inferred in theoretical studies.

The initial mass function (IMF) of main-sequence stars can be approximated by a power-law $dN/dM\sim
M^{-\beta}$ \cite{salpeter55} and most simply may be presented taking $\beta
\approx1.3$ for $0.07 \lesssim M/M_\odot \lesssim 0.5$ and $\beta\approx2.3$ for
$M/ M_\odot \gtrsim0.5$ \cite{2011ASPC..440...19W}. Note, there are also claims
that for the most massive stars ($M \gtrsim 7\ms$) IMF is much more steep -- up
to $\beta=3.8\pm0.5$, see \cite{2013ApJ...763..101L} and references therein.
Estimates of current star formation rate (SFR) in the Galaxy differ depending on
the method applied for the estimate, but most of them at present converge to
$\simeq2\myr$ \cite{2011AJ....142..197C,2012ARAA..50..531K,2013ASSP...34...29N}.
This implies that in the past Galactic SFR was much higher.  For the most
recent review of observations and models pertinent to star formation process,
origin of initial mass function, clustering of stars see
\cite{2014arXiv1402.0867K}.   
 
The evolution of a single star and the nature of its compact remnant
are determined by the main-sequence mass $M_0$ and chemical
composition. If $M_0$ is lower than the minimum mass of stars that
ignite carbon in the core, $M_{\mathrm{up}}$, after exhaustion of
hydrogen and helium in the core a carbon-oxygen WD forms. If  $M_0$
exceeds certain limiting $M_{\mathrm{mas}}$, the star proceeds to
formation of an iron core which collapses into a NS or a BH. In the
intermediate range between $M_{\mathrm{up}}$ and $M_{\mathrm{mas}}$
(so-called ``super-AGB stars'') carbon ignites in a partially
degenerate core and converts its matter (completely or partially) into
oxygen-neon mixture. 
As conjectured by Paczynski and Ziolkowski \cite{1968AcA....18..255P}, TP-AGB stage 
of evolution of stars terminates when in the envelope of the star appear
shells with positive specific binding energy 
$\varepsilon_\textrm{bind}=u_\textrm{int}+\varepsilon_\textrm{grav}$.
In the latter expression internal energy term $u_\textrm{int}$
accounts for ideal gas, radiation, ionisation, dissociation and electron
degeneracy, as suggested by Han et al. \cite{1994MNRAS.270..121H}.
The upper limit of masses of model precursors of CO white dwarfs 
depends primarily on stellar metallicity,
on the treatment of mixing in stellar interiors, accepted rates of mass 
loss during the AGB and  on more subtle details of the models,
see, e. g., \cite{2005ARA&A..43..435H,2013pss4.book..397H,2011ASPC..445....3K}.
Modern studies suggest that for solar metallicity models it does not exceed 
$1.2\ms$  \cite{2012AA...542A...1L}. 

As concerns more massive stars, if due to He-burning in the shell mass of the
core reaches $\approx1.375\ms$, electron captures by \super{24}Mg and
\super{20}Ne~ensue and the core collapses to a neutron star producing the so-called ``electron-capture supernova'' (ECSN)
\cite{1980PASJ...32..303M}). Otherwise, an ONe WD forms.%
\epubtkFootnote{Yet another possibility of formation of neutron stars
  exists: if sufficiently cold white dwarf close to critical mass
  accretes matter, the time scale of its gravitational collapse allows neutronisation before
  the onset of pycnonuclear reactions \cite{canal_aic_76}.}
Figure~\ref{f:single_remn} shows, as an example, the endpoints of
stellar evolution for intermediate mass stars, computed by
Siess~\cite{siess_agb06,2007AA...476..893S}.%
\epubtkFootnote{See these papers for detailed references and
  discussion of earlier studies.}
Note, results of computations very strongly depend on still uncertain
rates of nuclear burning, treatment of convection, the rate of assumed
stellar wind mass loss, poorly known from observations for the stars
in this transition mass range, as well as on other subtle details of
stellar models (see discussion in 
\cite{2008ApJ...675..614P,2012IAUS..279..341J,2013arXiv1310.1898C,2013ApJ...771...28T}).
For instance,
ZAMS masses of solar composition
progenitors of ECSN found by Poelarends et al.~\cite{2008ApJ...675..614P} are
$M_0 \approx (9.0\mbox{\,--\,}9.25)\ms$.%
\epubtkFootnote{Results obtained by the same group for lower
  metallicities may be found in Poelarend's PhD thesis at
  \url{http://www.astro.uni-bonn.de/~nlanger/siu_web/pre.html}}
In~\cite{2013ApJ...765L..43I}, the lower mass limit for core-collapse
SN at solar metallicity is found to be equal to $9.5\ms$. According to
Jones et al.~\cite{2013ApJ...772..150J}, $M_{\mathrm{ZAMS}}=8.8\ms$
star experiences ECSN, while $9.5\ms$ star evolves to Fe-core
collapse. Takahashi et al.~\cite{2013ApJ...771...28T}  find that
$M_{\mathrm{ZAMS}} = 10.4\ms$ solar metallicity star explodes as ECSN,
while  $M_{\mathrm{ZAMS}} =11\ms$ star experiences  Fe-core collapse,
consistent with Siess' results.

\epubtkImage{siessIAU252.png}{%
  \begin{figure}[htb]
    \centerline{\includegraphics[width=0.8\textwidth, angle=-90]{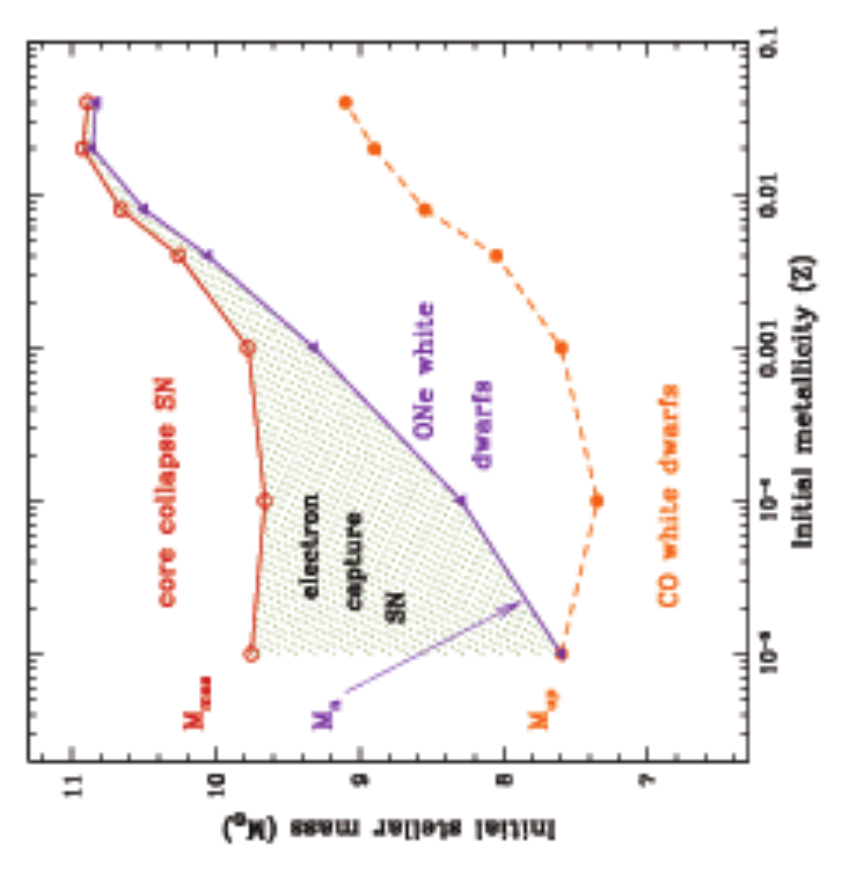}}
    \caption{Endpoints of evolution of moderate mass nonrotating
      single stars depending on initial mass and metallicity
      \cite{2008IAUS..252..297S}.}
    \label{f:single_remn}
  \end{figure}
}

ECSN eject only a small amount of Ni ($<0.015\ms$), 
and it was suggested that they may
be identified with some subluminous type II-P supernovae~\cite{kitaura_one06}.
Tominaga et al. \cite{2013ApJ...771L..12T} were able to simulate ECSN starting
from first principles and to reproduce their multicolour light curves. They have
shown that observed features of SN~1054 (see \cite{2002ISAA....5.....S}) share
the predicted characteristics of ECSN.   

We should enter a caveat that in~\cite{2012AA...542A...1L} it is claimed that
opacity-related instability at the base of convective envelope of
$(7\mbox{\,--\,}10)\ms$ stars may result in ejection of the envelope and,
consequently, prevent ECSN. 

Electron-capture SNe, if really happen, presumably produce NS with short spin periods and, if in binaries, the systems with short orbital periods and
low eccentricities. An observational evidence for ECSN is provided by Be/X-ray
binaries which harbour two subpopulations, typical for post-core-collapse
objects and for post-ECSN ones \cite{2011Natur.479..372K}. Low kicks
implied for NS formed via ECSN \cite{Podsiadlowski_al04} may explain the
dichotomous nature of Be/X-ray binaries. As well, thanks to low kicks, NS
produced via ECSN may be more easily retained in globular clusters (e.g. \cite{2006AstL...32..393K}) 
or in the
regions of recent star formation.

The existence of the ONe-variety of WDs, predicted theoretically 
\cite{brr74,1980PASJ...32..303M,it85}
and later confirmed by observations \cite{truran_livio_one86}, is 
 important in the general context 
of binary compact stars evolution, since their accretion-induced collapse (AIC) 
may result in formation of low-mass neutron stars (almost) without natal kicks
 and may be, thus,
relevant to formation of binary neutron stars and low-mass X-ray binaries
 (see~\cite{1990ApJ...353..159B,dessart_onecoll06,2010NewAR..54..140V} and references therein).   
But since for the
purpose of detection of gravitational waves they are not different
from much more numerous CO WDs, we will, as a rule, not
consider them below as a special class.

If $M_0\gtrsim M_{\mathrm{mas}}$ (or $8\mbox{\,--\,}12\,M_{\odot}$),
thermonuclear evolution proceeds until iron-peak elements are produced
in the core. Iron cores are subject to instabilities (neutronization,
nuclei photo-disintegration),
that lead to gravitational collapse. The core collapse of massive
stars results in the formation of a neutron star or, for very massive
stars, a black hole%
\epubtkFootnote{Pair-creation instability in the oxygen cores of
  slowly rotating $M_0 \approx(140\mbox{\,--\,}260)\ms$ and
  metallicity $Z \lesssim Z_\odot/3$ stars or rapidly rotating stars
  with $M_0 > 80\ms$ and $Z/Z_\odot < 10^{-3}$ may lead to their
  collapse which results in explosive ignition of oxygen which leads
  to the disruption of the star~\cite{2007AA...475L..19L}.}
and is associated with the brightest astronomical phenomena such as
supernova explosions (of type~II, Ib, or~Ib/c, according to the
astronomical classification based on the spectra and light curves
properties). If the pre-collapsing core retains significant rotation,
powerful gamma-ray bursts lasting up to hundreds of seconds may be
produced~\cite{Woosley_Heger06}.

The boundaries between the masses of progenitors of WDs or NSs and NSs or
BHs are fairly uncertain (especially for BHs). Usually accepted mean masses of stellar remnants for single non-rotating  solar chemical composition stars are summarized in Table~\ref{table:ms_evolution}. 

\begin{table}
  \renewcommand{\arraystretch}{1.2}
  \caption{Types of compact 
 remnants of single stars (the ranges of
    progenitor mass are shown for solar composition stars).}
  \label{table:ms_evolution}
  \centering
  {\small
  \begin{tabular}{ccc}
    \toprule
    Initial mass [$M_\odot$] &
    remnant type &
    mean remnant mass [$M_\odot$] \\
    \midrule
    $ \phantom{0\mbox{\,--\,}} 0.95 < M < 8\mbox{\,--\,}12 \phantom{0} $ &
    WD &
    $ \phantom{\sim 0} 0.6 \phantom{0} $ \\
    $ \phantom{0.} 8\mbox{\,--\,}11 < M < 25\mbox{\,--\,}30 $ &
    NS &
    $ \phantom{\sim 0} 1.35 $ \\
$ \phantom{.} {\bf 20} < M < 150 \phantom{\mbox{\,--\,}0} $ &
    BH &
    $ \sim 10 \phantom{.00} $ \\
    \bottomrule
  \end{tabular}}
  \renewcommand{\arraystretch}{1.0}
\end{table}

Note that the above-described mass ranges for different outcomes of stellar
evolution were obtained by computing 1D-models of non-rotating stars.
However, all star are rotating and some of them possess magnetic
fields, thus, the problem is 2D (or 3D, if magnetic field
effects are considered). For a physical description of effects of 
rotation on stellar evolution see~\cite{2012ARAA..50..107L,Palacios13}. 
Realistic models with rotation should account
for deviation from spherical symmetry, modification of gravity due to
centrifugal force, variation of radiative flux with local effective
gravity, transfer of angular momentum and transport of chemical
species. Up to now, as a rule, models with rotation are computed
in 1D-approximation which, typically, makes use of the fact that in 
rotating stars mass is constant within isobar surfaces
\cite{1976ApJ...210..184E}. Regretfully, for low and intermediate mass
stars model sequences of rotating stars covering evolution to advanced phases of
AGB-evolution are 
absent. However, as noted by Dom\'{i}nguez
et~al.~\cite{1996ApJ...472..783D}, rotation must strongly influence
evolution at core helium exhaustion and formation of CO-core stage,
when the radius of the core strongly decreases. With increase of
angular velocity, lower pressure is necessary to balance gravity,
hence the temperature of He-burning shells of rotating stars should be
lower than in non-rotating stars of the same mass. This extends AGB-lifetime and
results in increase of the mass of CO-cores, i.e., C-ignition limit
may be shifted to  lower masses compared to non-rotating stars.  Most
recently, grids of 1-D  models of rotating stars were published, e.g.,
in~\cite{2011AA...530A.115B, 2013AA...553A..24G, 2013A&A...558A.103G}.

The algorithm and first results of self consistent calculations of rapidly rotating 2D
stellar models of stars in early stages of evolution 
are described, e.g., in \cite{2013LNP...865...49R,2013A&A...552A..35E}.  
We may note that these models are very important 
for deriving physical parameters of the stars from astroseismological
data. 
   
For a more detailed introduction into the physics and evolution of stars the
reader is referred to the classical fundamental
textbook~\cite{Cox_Giuli68} and to several more modern
ones~\cite{2000itss.book.....P, 2001spfc.book.....B,
  2011stph.book.....B, 2006epbm.book.....E,2013sepp.book.....I,
2013sepa.book.....I}. Formation and physics of compact objects
is described in more detail in the
monographs~\cite{Shapiro_Teukolsky83, 2011stph.book.....B}.
 For recent studies and reviews of the evolution of massive stars and the
mechanisms of core-collapse supernovae we refer to 
\cite{2012ARAA..50..107L, 2013RvMP...85..245B, 2012ARNPS..62..407J,
  2013MNRAS.433.1114Y,2013CQGra..30x4002F,2014arXiv1402.1237S}.

\subsection{Binary stars}
 
A fundamental property of stars is their multiplicity. Among stars
which complete their nuclear evolution in the Hubble time, the estimated
binary fraction varies from $\sim$~(40\,--\,60)\% for $M\sim \ms$ 
stars~\cite{Duquennoy_Mayor91, 2010ApJS..190....1R} to
almost 100\% for more massive A/B and O-stars, e.g., 
\cite{1998AJ....115..821M, 2007AA...474...77K, 2007ApJ...670..747K,
  2009AJ....137.3358M, 2012MNRAS.424.1925C, 2012Sci...337..444S},
(but, e.g., in~\cite{2013AA...550A..27M,2011MNRAS.416..817S} a
substantially lower binary fraction for massive stars is claimed).

Based on the summary of data on binary fraction $\mathcal{B}(M)$
provided in~\cite{2009AA...493..979K, kraus_binarity09,
  2012Sci...337..444S}, van Haaften et~al.~\cite{2013AA...552A..69V}
suggested an approximate formula 
\begin{equation}
 \label{eq:bin_rate}
  \mathcal{B}(M) \approx \frac{1}{2} + \frac{1}{4}\log(M) \qquad (0.08 \le M/M_{\odot} \le 100)\,,
\end{equation}
considering all multiple systems as binaries.

The most crucial parameters of binary stars include the component separation $a$ and mass ratio $q$, since for close
binaries they define the outcome of the Roche lobe overflow (in fact, the
fate of the system). The most recent estimates for M-dwarfs and solar type stars 
confirm earlier findings that the $q$-distribution does not strongly deviate 
from a flat one: $dN/dq \propto q^{\beta}$, with
$\beta=0.25\pm0.29$ \cite{2013AA...553A.124R}. This distribution is
defined by the star-formation process and dynamical evolution in stellar
clusters, see, e.g., \cite{bate_12,2013MNRAS.432.2378P}. Distribution over $a$ is flat in log between contact and $\simeq 10^6\rs$~\cite{Popova_al82}.

We note, cautionary, that all estimates of the binary fraction, mass-ratios
of components, distributions over separations of components  are
plagued by numerous selection effects (see~\cite{2006PhDT........27K}
for a thorough simulation of observations and modeling observational
bias). A detailed summary of studies of multiplicity rates,
distributions over orbital periods and mass ratios of components for
different groups of stars may be found in~\cite{Duchen_dupl13}.

In the binary stars with sufficiently large orbital separations
(``wide binaries'') the presence of the secondary component does not
influence significantly the evolution of the components. In ``close
binaries'' the evolutionary expansion of stars leads to the overflow
of the critical (Roche) lobe and mass exchange between the components --
the so-called Roche lobe overflow. Consequently, the formation of compact
remnants in binary stars differs from single stars  (see for 
Section~\ref{section:physical_principles} more
details).	

As discussed above, the lower mass limit of NS progenitors is
uncertain by several $M_\odot$. 
This limit is even more uncertain for the BH progenitors. For example, the
presence of a magnetar (neutron star with an extremely large
magnetic field) in the open cluster Westerlund~1 means that it
descends from a star which is more massive than currently observed
most massive main-sequence cluster stars (because for a massive star the duration of the main-sequence stage is inversely proportional to its mass squared). 
The most massive main-sequence stars in this stellar cluster are found to have masses as high as 40~$M_\odot$, suggesting 
$M_{\mathrm{pre-BH}} \gtrsim 40\ms$ \cite{2010AA...520A..48R}.  
On the other hand, it was speculated, based
on the properties of  X-ray sources, that in the initial mass range
$(20\mbox{\,--\,}50)\ms$ the mass of pre-BH may vary depending on such
badly known parameters as rotation or magnetic
fields~\cite{1998AA...331L..29E}.

Binaries with compact remnants are primary potential detectable GW sources (see
Figure~\ref{f:GW_sources}). This figure plots the sensitivity of ground-based
interferometers LIGO, as well as the space laser interferometers LISA and eLISA,
in the terms of dimensionless GW strain $h$ measured over one year. The strongest
Galactic sources at all frequencies are the most compact double NSs, binary WDs,
and (still hypothetical) BHs. Double NS/BH systems are formed from initially massive binaries, while double WDs descend from low-mass binaries. 

\epubtkImage{figure02.png}{
  \begin{figure}[htbp!]
\centerline{\includegraphics[]{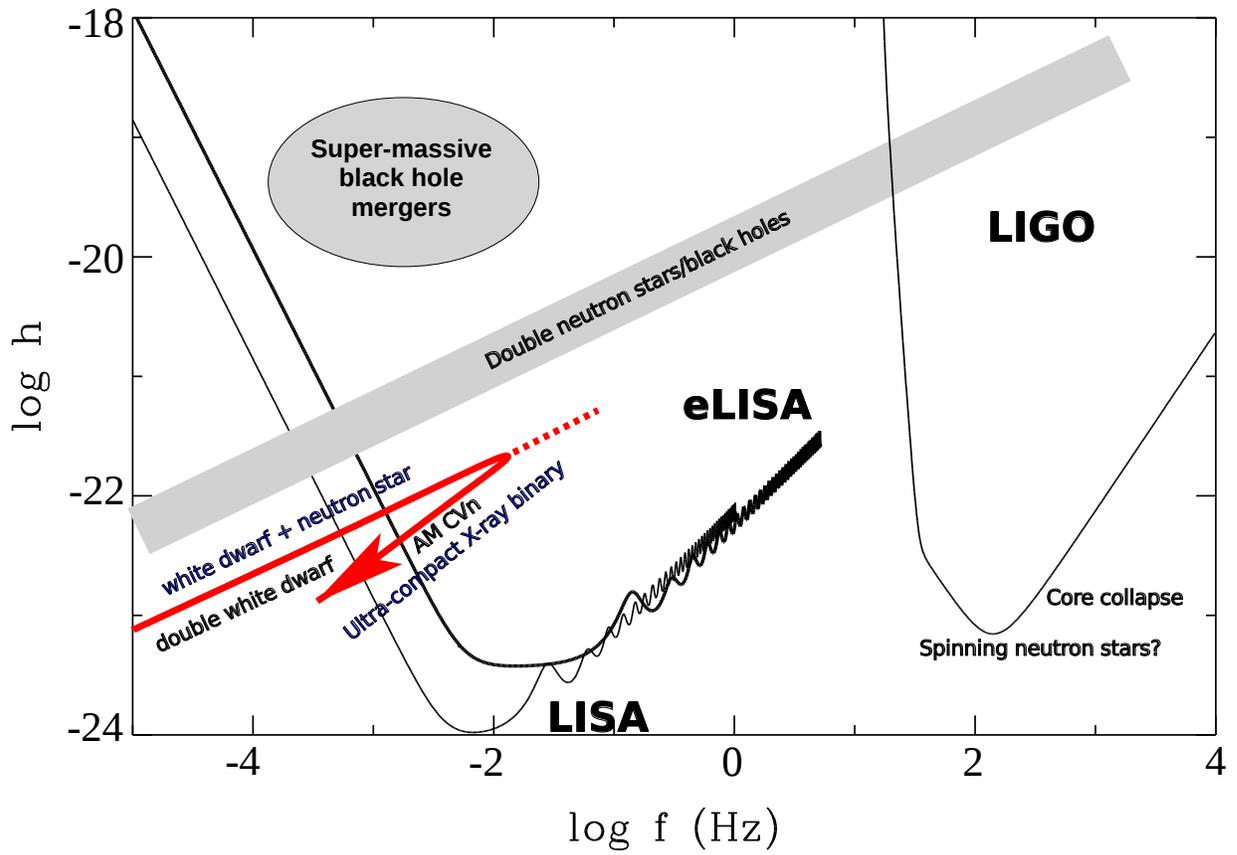}}
    \caption{Sensitivity limits of GW detectors and the regions of
      the $f$--$h$ diagram occupied by some of the potential GW
      sources. (Courtesy G.~Nelemans).}
    \label{f:GW_sources}
\vspace{-0.5cm}
\end{figure}} 

In this review we shall concentrate on the formation and evolution of
binary compact stars most relevant for GW studies. The article is
organized as follows. We start in
Section~\ref{section:observational_situation} with a review of the main
observational data on double NSs, especially measurements of masses of
NSs and BHs, which are most important for the estimate of the amplitude
of the expected GW signal. We briefly discuss the empirical methods to
determine the double NS coalescence rate. The basic principles of binary
stellar evolution are discussed in
Section~\ref{section:physical_principles}. Then, in
Section~\ref{section:evolution_high_mass} we describe the evolution of
massive binary stars. Next, we discuss the Galactic rate of formation of
binaries with NSs and BHs in
Section~\ref{section:binary_NS}. Theoretical estimates of detection
rates for mergers of binary relativistic stars are discussed in
Section~\ref{sec:secI:DetectRate}. Further, we proceed to the analysis
of formation of short-period binaries with WD components in
Section~\ref{section:wd_formation}, and consider observational data on
binary white dwarfs in Section~\ref{section:observations}. A model for
the evolution of interacting double-degenerate systems is presented in
Section~\ref{section:am-evol}. In Section~\ref{section:waves} we
describe gravitational waves from compact binaries with white dwarf
components. Section~\ref{section:opt+x} 
is devoted
to the model of optical and X-ray emission of
AM~CVn-stars.
Our conclusions follow in Section~\ref{sec:concl}.


\newpage
\section{Observations of Double Compact Stars}   
\label{section:observational_situation}


\subsection{Compact binaries with neutron stars}

Double NSs have been discovered because one of the components of the
binary is observed as a radio pulsar~\cite{Hulse_Taylor75}. The
precise pulsar timing allows one to search for a periodic variation
due to the binary motion. This technique is reviewed in detail by
Lorimer~\cite{Lorimer_LRR01}; applications of pulsar timing for
general relativity tests are reviewed by
Stairs~\cite{Stairs_LRR03}. Techniques and results of measurements of
NS mass and radii in different types of binaries are summarized by
Lattimer~\cite{2012ARNPS..62..485L}.

Basically, pulsar timing provides the following Keplerian orbital
parameters of the binary system: the binary orbital period $P_\mathrm{b}$ as
measured from periodic Doppler variations of the pulsar spin, the
projected semimajor axis $x=a\sin i$ as measured from the semi-amplitude
of the pulsar radial velocity curve ($i$ is the binary inclination angle
defined such that $i=0$ for face-on systems), the orbital eccentricity
$e$ as measured from the shape of the pulsar radial velocity curve,
and the longitude of periastron $\omega$ at a particular epoch
$T_0$. The first two parameters allow one to construct the \emph{mass
  function} of the secondary companion,
\begin{equation}
  f(M_\mathrm{p},M_\mathrm{c}) =
  \frac{4\pi^2x^3}{P_\mathrm{b}^2 T_\odot} =
  \frac{(M_\mathrm{c}\sin i)^3}{(M_\mathrm{c}+M_\mathrm{p})^2}.
  \label{eq:mass_function}
\end{equation}
In this expression, $x$ is measured in light-seconds, $T_\odot\equiv
GM_\odot/c^3=4.925490947 \mathrm{\ \mu s}$, and $M_\mathrm{p}$ and
$M_\mathrm{c}$ denote masses of the
pulsar and its companion, respectively. This function gives the strict
lower limit on the mass of the unseen companion. However, assuming the
pulsar mass to have the typical value of a NS mass (for example,
confined between the lowest measured NS mass $1.25\,M_\odot$ for
\epubtkSIMBAD{PSR~J0737--3039B}~\cite{Lyne_al04} and the maximum measured NS mass of
$2.1\,M_\odot$ in the NS--WD binary \epubtkSIMBAD{PSR~J0751+1807}~\cite{Nice_al04}), one
can estimate the mass of the secondary star even without knowing the
binary inclination angle $i$.

Long-term pulsar timing allows measurements of several relativistic 
phenomena: the advance of periastron $\dot{\omega}$, the redshift parameter
$\gamma$, the Shapiro delay within the binary system qualified through
post-Keplerian parameters $r$, $s$, and the binary orbit decay $\dot
P_\mathrm{b}$. From the post-Keplerian parameters the individual masses $M_\mathrm{p}$,
$M_\mathrm{c}$ and the binary inclination angle $i$ can be
calculated~\cite{Brumberg_al75}.

Of the post-Keplerian parameters of binary pulsars, the periastron advance rate
is usually measured most readily. Assuming it to be entirely due to general
relativity, the total mass of the system can be evaluated:
\begin{equation}
  \dot{\omega} = 3 \left( \frac{2\pi}{P_\mathrm{b}} \right)^{5/3}
  \frac{T_\odot^{2/3}(M_\mathrm{c}+M_\mathrm{p})^{2/3}}{(1-e^2)}.
  \label{eq:periastron_adv}
\end{equation}
High value of the derived total mass of a system 
($\gtrsim 2.5\,M_\odot$) suggests the presence of another NS or even
BH\epubtkFootnote{Unless the companion is directly observable and its
  mass can be estimated by other means.}.

If the  masses of components, binary period, and eccentricity of a compact
binary system are known, it is easy to calculate the time it takes for
the binary companions to coalesce due to GW emission using the
quadrupole formula for GW emission~\cite{Peters64} (see
Section~\ref{A:GW_evol} for more detail):
\begin{equation}
  \tau_\mathrm{GW} \approx 4.8\times 10^{10} \mathrm{\ yr}
  \left( \frac{P_\mathrm{b}}{\mathrm{\ d}} \right)^{8/3}
  \left( \frac{\mu}{M_\odot} \right)^{-1}
  \left( \frac{M_\mathrm{c}+M_\mathrm{p}}{M_\odot} \right)^{-2/3}
  (1-e^2)^{7/2}.
  \label{eq:tau_GW}
\end{equation}
Here $\mu=M_\mathrm{p}M_\mathrm{c}/(M_\mathrm{p}+M_\mathrm{c})$ is the
reduced mass of the binary. Some
observed and derived parameters of known compact binaries with NSs are
collected in Tables~\ref{table:doubleCB_o} and~\ref{table:doubleCB_d}.

\begin{table}
  \renewcommand{\arraystretch}{1.2}
  \caption{Observed parameters of double neutron star binaries.}
  \label{table:doubleCB_o}
  \centering
  {\small
  \begin{tabular}{lccccccc}
    \toprule
    PSR & $P$ & $P_\mathrm{b}$ & $a_1 \sin i$ & $e$ &
    $\dot{\omega}$ & $\dot{P}_{\mathrm{b}}$ & Ref. \\
    & [ms] & [d] & [lt-s] & &
    [deg~yr\super{-1}] & [\texttimes~10\super{-12}] \\
    \midrule
    \epubtkSIMBAD[PSR~J0737--3039A]{J0737--3039A} & 22.70 & 0.102 & 1.42 & 0.088 & 16.88 & --1.24 &
    \cite{Burgay_al03} \\
    \epubtkSIMBAD[PSR~J0737--3039B]{J0737--3039B} & 2773 & --- & --- & --- & --- & --- &
    \cite{Lyne_al04} \\
    \epubtkSIMBAD[PSR~J1518+4904]{J1518+4904} & 40.93 & 8.634 & 20.04 & 0.249 & 0.011 & ? &
    \cite{Nice_al96} \\
    \epubtkSIMBAD[PSR~B1534+12]{B1534+12} & 37.90 & 0.421 & 3.73 & 0.274 & 1.756 & --0.138 &
    \cite{Wolszczan90, Stairs_al02} \\
    \epubtkSIMBAD[PSR~J1756--2251]{J1756--2251} & 28.46 & 0.320 & 2.75 & 0.181 & 2.585 & ? &
    \cite{Faulkner_al05} \\
    \epubtkSIMBAD[PSR~J1811--1736]{J1811--1736} & 104.18 & 18.779 & 34.78 & 0.828 & 0.009 & \textless~30 &
    \cite{Lyne_al00} \\
\epubtkSIMBAD[PSR~J1829+2456]{J1829+2456} & 41.00 & 1.176 & 7.236 & 0.139 & 0.28 & ? &
    \cite{Champion_al04} \\
    \epubtkSIMBAD[PSR~J1906+0746]{J1906+0746} & 144.07 & 0.116 & 1.42 & 0.085 & 7.57 & ? &
    \cite{Lorimer_al06} \\
    \epubtkSIMBAD[PSR~B1913+16]{B1913+16} & 59.03 & 0.323 & 2.34 & 0.617 & 4.227 & --2.428 &
    \cite{Hulse_Taylor75} \\
    \epubtkSIMBAD[PSR~B2127+11C]{B2127+11C} & 30.53 & 0.335 & 2.52 & 0.681 & 4.457 & --3.937 &
    \cite{Anderson_al90, Prince_al91} \\
    \bottomrule
  \end{tabular}}
  \renewcommand{\arraystretch}{1.0}
\end{table}

\begin{table}
  \renewcommand{\arraystretch}{1.2}
  \caption{Derived parameters of double neutron star binaries}
  \label{table:doubleCB_d}
  \centering
  {\small
  \begin{tabular}{lcccc}
    \toprule
    PSR & $f(m)$ & $M_{\mathrm{c}}+M_{\mathrm{p}}$ &
    $\tau_c=P/(2\dot{P})$ & $\tau_{\mathrm{GW}}$ \\
    & [$M_{\odot}$] & [$M_{\odot}$] & [Myr] & [Myr] \\
    \midrule
    \epubtkSIMBAD[PSR~J0737--3039A]{J0737--3039A} & 0.29 & 2.58 & 210 & 87 \\
    \epubtkSIMBAD[PSR~J0737--3039B]{J0737--3039B} & --- & --- & 50 & --- \\
    \epubtkSIMBAD[PSR~J1518+4904]{J1518+4904} & 0.12 & 2.62 & & 9.6~\texttimes~10\super{6} \\
    \epubtkSIMBAD[PSR~B1534+12]{B1534+12} & 0.31 & 2.75 & 248 & 2690 \\
    \epubtkSIMBAD[PSR~J1756--2251]{J1756--2251} & 0.22 & 2.57 & 444 & 1690 \\
    \epubtkSIMBAD[PSR~J1811--1736]{J1811--1736} & 0.13 & 2.6 & & 1.7~\texttimes~10\super{6} \\
\epubtkSIMBAD[PSR~J1829+2456]{J1829+2456} & 0.29 & 2.53 & & 60~\texttimes~10\super{3} \\    
\epubtkSIMBAD[PSR~J1906+0746]{J1906+0746} & 0.11 & 2.61 & 0.112 & 300 \\
    \epubtkSIMBAD[PSR~B1913+16]{B1913+16} & 0.13 & 2.83 & 108 & 310 \\
    \epubtkSIMBAD[PSR~B2127+11C]{B2127+11C} & 0.15 & 2.71 & 969 & 220 \\
    \bottomrule
  \end{tabular}}
  \renewcommand{\arraystretch}{1.0}
\end{table}


\subsection{How frequent are double NS coalescences?}
\label{sec:ns_freq}

As it is seen from Table~\ref{table:doubleCB_d}, only six double NS
systems presently known will merge over a time interval shorter
than $\approx$~10~Gyr: 
\epubtkSIMBAD[PSR~J0737--3039A]{J0737--3039A},
\epubtkSIMBAD[PSR~B1534+12]{B1534+12},
\epubtkSIMBAD[PSR~J1756--2251]{J1756--2251},
\epubtkSIMBAD[PSR~J1906+0746]{J1906+0746}, 
\epubtkSIMBAD[PSR~B1913+16]{B1913+16}, and
\epubtkSIMBAD[PSR~B2127+11C]{B2127+11C}. Of these six systems, one
(\epubtkSIMBAD{PSR~B2127+11C}) is located in the globular cluster
\epubtkSIMBAD{M15}. This system may have a different formation history, so usually it
is not included in the analysis of the coalescence rate of the Galactic
double compact binaries. The formation and evolution of relativistic
binaries in dense stellar systems is reviewed
elsewhere~\cite{Benacquista_LRR02, Benacquista_LRR13}. For a general review of
pulsars in globular clusters see also~\cite{Camilo_Rasio05}.

Let us try to estimate the double NS merger rate from binary
pulsar statistics, which is free from many uncertainties of binary star evolution.
Usually, the estimate is based on the following
extrapolation~\cite{nps91, Phinney91}. Suppose, we observe $i$
classes of Galactic binary pulsars. Taking into account various
selection effects of pulsar surveys (see, e.g., \cite{Narayan87, Kim_al03}),
the Galactic number of pulsars $N_i$ in each class can be
evaluated. To compute the Galactic merger rate of double NS binaries,
we need to know the time since the birth of the NS observed as a pulsar
in the given binary system. This time is the sum of the observed
characteristic pulsar age $\tau_\mathrm{c}$ and the time required for the
binary system to merge due to GW orbit decay $\tau_\mathrm{GW}$. With the
exception of \epubtkSIMBAD{PSR~J0737--3039B} and the recently
discovered \epubtkSIMBAD{PSR~J1906+0746}, 
pulsars that we observe in binary NS systems are old recycled pulsars
which were spun-up by accretion from the secondary companion to the
period of several tens of ms (see Table~\ref{table:doubleCB_o}). Thus,
their characteristic ages can be estimated as the time since
termination of spin-up by accretion (for the younger pulsar
\epubtkSIMBAD{PSR~J0737--3039B} this time can be also computed as the dynamical age of
the pulsar, $P/(2\dot{P})$, which gives essentially the same result).

Then the merger rate ${\cal R}_i$ can be calculated as ${\cal R}_i\sim
N_i/(\tau_\mathrm{c}+\tau_\mathrm{GW})$ (summed over all binary pulsars). The
detailed analysis~\cite{Kim_al03} indicates that the Galactic merger rate of
double NSs is mostly determined by pulsars with faint radio luminosity and short
orbital periods. Presently, it is the nearby (600~pc) double-pulsar system
\epubtkSIMBAD{PSR~J0737--3039} with a short orbital period of
2.4~hr~\cite{Burgay_al03} that mostly determines the \emph{empirical} estimate
of the merger rate. According to Kim et~al.~\cite{kim_etal06}, ``the most likely
values of DNS merger rate lie in the range 3\,--\,190 per Myr depending on
different pulsar models''. This estimate has recently been revised
by~\cite{Kim:2013tca} based on the analysis of binary pulsars in the galactic
disk: \epubtkSIMBAD{PSR~1913+16}, \epubtkSIMBAD{PSR~1534+12}, and double pulsar
\epubtkSIMBAD{PSR~J0737--3039A} and
\epubtkSIMBAD[PSR~J0737--3039B]{J0737--3039B}, giving the galactic NS\,+\,NS
coalescence rate $\R_G=21^{+28+40}_{-14-17}$ per Myr (95\% and 90\% confidence
level, respectively). The estimates by \emph{population synthesis codes} are
still plagued by uncertainties in statistics of binaries, in modeling binary
evolution and supernovae. The most optimistic ``theoretical'' predictions amount
to $\simeq$~300~Myr\super{-1}~\cite{ty93b, bel_kal02, Dominik_al12}.

An independent estimate of the binary NS merger rate can also be
obtained using another astrophysical argument, originally suggested by Bailes
\cite{1996IAUS..165..213B}. 1)~Take the formation
rate of single pulsars in the Galaxy ${\cal R}_{\mathrm{PSR}}\sim
1/50\mathrm{\ yr}^{-1}$ (e.g., \cite{FaucherGiguere_Kaspi06}; see also
the discussion on NS formation rate of different types
in~\cite{Keane_Kramer08}), which appears to be correct to within a
factor of two. 2)~Take the fraction of binary NS systems in which one
component is a normal (not recycled) pulsar and which are close enough
to merge within the Hubble time, $f_{\mathrm{DNS}}\sim$ (a few)
$\times$ 1/2000). Assuming a steady state, this fraction yields the
formation rate of coalescing binary NS in the Galaxy ${\cal R}\sim
{\cal R}_{\mathrm{PSR}}f_{\mathrm{DNS}}\sim$ (a few
10s)\texttimes~Myr\super{-1}, in good correspondence with other
empirical estimates  \cite{Kim:2013tca}. Clearly, the Bailes limit
ignores the fact that NS formation rate can be actually higher than
the pulsar formation rate~\cite{Keane_Kramer08} and possible selection
effects related with evolution of stars in binary systems, but the
agreement with other estimates seems to be encouraging.

The extrapolation beyond the Galaxy is usually done by scaling the
Galactic merger rate to the volume over which the merger events can be
detected for given GW detector's sensitivity. The scaling factor
widely used is the ratio between the B-band luminosity density in the
local Universe, correlating with the star-formation rate (SFR), and the
B-band luminosity of the Galaxy~\cite{Phinney91, Kalogera_al01,
  Kopparapu_al08}. One can also use for this purpose the direct ratio
of the Galactic star formation rate SFR$_\mathrm{G}\simeq 2\,M_\odot
\mathrm{\ yr}^{-1}$~\cite{2011AJ....142..197C,2012ARAA..50..531K,2013ASSP...34...29N} to the
(dust-corrected) star formation rate in the local Universe
$\mathrm{SFR}_\mathrm{loc}\simeq 0.03\,M_\odot
\mathrm{\ yr}^{-1}\mathrm{\ Mpc}^{-3}$~\cite{Perez-Gonzalez_al03,
  Schiminovich_al05}. These estimates yield the (Euclidean) relation 
\begin{equation}
  {\cal R}_{\mathrm{V}} \simeq 0.01 {\cal R}_{\mathrm{G}} [\mathrm{Mpc}^{-3}].
  \label{R_V}
\end{equation}
This estimate is very close to the local density of equivalent Milky-Way type 
galaxies found in \cite{Kopparapu_al08}: 0.016 per cubic Mpc.

It should be born in mind that since the double NS coalescences can be strongly delayed compared to the star formation, the estimate of their rate in the nearby Universe based on the present-day star formation density may be underevaluated.
Additionally, these estimates inevitably suffer from many other astrophysical uncertainties. For example, 
a careful study of local SFR from analysis of an
almost complete sample of nearby galaxies within 11~Mpc  
using different SFR indicators and supernova rate measurements during 13 years of observations
\cite{Botticella_al11} results in the local SFR density 
$\mathrm{SFR}_\mathrm{loc}\simeq 
0.008\,M_\odot \mathrm{\ yr}^{-1}\mathrm{\ Mpc}^{-3}$. However, 
Horiuchi et~al.~\cite{Horiuchi_ea13}
argue that estimates of the local SFR can be strongly affected by stellar rotation. Using new 
stellar evolutionary tracks \cite{2012AA...537A.146E}, they derived 
the local SFR density 
$\mathrm{SFR}_\mathrm{loc}\simeq 
0.017\,M_\odot \mathrm{\ yr}^{-1}\mathrm{\ Mpc}^{-3}$ and noted that the estimate of the local SFR from core-collapse supernova counts is higher by factor 2-3. 
 
Therefore, the actual value 
of the scaling factor from the Galactic merger rate is presently 
uncertain to within a factor of at least two.  
The mean SFR density steeply increases with distance (e.g., as $\sim (1+z)^{3.4\pm 0.4}$ 
\cite{Dale_al10}) and can be higher in individual galaxies. 
Thus, for the conservative 
Galactic NS\,+\,NS merger rate ${\cal R}_\mathrm{G}\sim 10^{-5}
\mathrm{\ yr}^{-1}$ using scaling relation (\ref{R_V}) we obtain 
about a few double NS coalescences per one 
year of observations within the assumed advanced GW detectors horizon $D_{hor}=400$ Mpc.

However, one should differentiate between the possible merging rate within some volume
and the \textit{detection rate} of certain type of double compact binaries from this 
volume (see Section \ref{sec:secI:DetectRate} below and \cite{Abadie_al10} for more detail) -- 
the detection rate by one detector can be lower by a factor of $(2.26)^3$. The correction factor takes
into account the averaging over all sky locations and orientations.

The latest results of the search for GWs from coalescing
binary systems within 40~Mpc volume 
using LIGO and Virgo observations between July 7, 2009, and October 20, 2010~\cite{Abadie_al12}
established an observational upper limit to the binary
$1.35\,M_{\odot} + 1.35\,M_{\odot}$ NS
coalescence rate of $< 1.3 \times 10^{-4} \mathrm{\ yr}^{-1}\mathrm{\ Mpc}^{-3}$. 
Adopting the scaling factor from the measured local SFR density~\cite{Botticella_al11}, 
the corresponding Galactic upper limit is ${\cal R}_G<0.05 \mathrm{\ yr}^{-1}$.
This is still too high to put interesting astrophysical bounds, 
but even upper limits from the advanced LIGO detector are expected to be very constraining.


\subsection{Black holes in binary systems}
\label{sec:BH_inbinaries}

Black holes (BH) in binary systems remain on the top of astrophysical 
studies. Most of the experimental knowledge on the BH physics has so far
been obtained through electromagnetic channel (see~\cite{Psaltis08} for a review), 
but the fundamental features of BHs will be studied through GW experiments
\cite{Sathyaprakash_Schutz09}. Stellar mass black holes result from gravitational 
collapse of the cores of the most massive stars~\cite{Woosley_al02} (see  
\cite{Fryer_New11} for the recent progress in the physics of gravitational collapse).   

Stellar mass BH can be observed in close binary systems at the stage of 
mass accretion from the secondary companion as bright 
X-ray sources~\cite{Shakura_Sunyaev73}. 
X-ray studies of BH in binary systems (which are usually referred to as 
`Black Hole Candidates', BHC) are reviewed in~\cite{Remillard_McClintock06}.
There is strong observational evidence of the launch of relativistic jets
from the inner parts of accretion disk in BHC, which can be observed in the range  
from radio to gamma-rays 
(the `galactic microquasar' phenomenon, \cite{Mirabel_Rodriguez99, Fender_Belloni04, Tavani_al09}). 

Optical spectroscopy and X-ray observations of BHC allow measurements of
masses and, under certain assumptions, spins of BH~\cite{McClintock_al11}.
The measured mass distribution of stellar mass BHs is centered on
$\sim 8\,M_\odot$ and appears to be separated from NS
masses~\cite{Ozel_al12} by a gap (the absence of compact star masses
in the  $\sim 2\mbox{\,--\,}5\,M_\odot$ range) \cite{Bailyn_al98,
  Cherepashchuk03, Ozel_al10, Farr_al11} (see, however, discussion of
possible systematic errors leading to overestimation of dynamical BH
masses in X-ray transients in~\cite{Kreidberg_al12}). 
%
(see, for example, \cite{2001ARep...45..899P,2012ApJ...757...91B,
2012ApJ...749...91F}). 
It is also possible that the gap is due to
``failed core-collapse supernovae'' from red supergiants with  masses
$16.5\,M_\odot < M < 25\,M_\odot$, which would produce black holes
with a mass equal to the mass of the helium core
($5\mbox{\,--\,}8\,M_\odot$) before the collapse, as suggested by
Kochanek~\cite{Kochanek13}. Adopting the latter hypothesis would
increase by $\sim$~20\% stellar mass black holes formation
rate. However, it is unclear how binary interaction can change the
latter possibility. We stress here once again that this important
issue remains highly uncertain due to the lack of ``first-principles''
calculations of stellar core collapses.

Most of more than 20 galactic BHC appear as X-ray transients
with rich phenomenology of outbursts and spectral/time variability~\cite{Remillard_McClintock06},
and only several (Cyg X-1, LMC X-3, LMC X-1 and SS433) 
are persistent X-ray sources.  Optical companions in BH transients
are low-mass stars (main-sequence or evolved) filling Roche lobes,
and the transient activity is apparently related to accretion  disk
instability~\cite{Lasota01}. Their evolution is driven by orbital angular
momentum loss due to GW emission and magnetic stellar wind
\cite{Pylyser_Savonije88}. Persistent BHC, in contrast, have 
early-type massive optical companions and belong to the class of High-Mass
X-ray Binaries (HMXB) (see~\cite{Tauris_vandenHeuvel06, vandenHeuvel09} and 
references therein). 

Like in the case of pulsar timing for NS, to estimate the mass of the BH companion in a close binary system 
one should measure the radial velocity curve of the companion (usually optical) star from spectroscopic 
observations. The radial velocity curve 
(i.e. the dependence of the radial velocity of the companion on the binary phase) 
has a form that depends on the orbital 
eccentricity $e$, and the amplitude $K$. In the Newtonian limit for two 
point masses, the binary mass function can be expressed through the observed quantities $P_{b}$ (the binary orbital period), 
orbital eccentricity $e$ and semi-amplitude of the radial velocity curve $K$ as
\begin{equation}
\label{fmv}
f(M_v)=\frac{M_x\sin^3i}{(M_x+M_v)^2}=\frac{P_b}{2\pi G}K_v^3(1-e^2)^{3/2}
\approx 1.038\times 10^{-7}[M_\odot]\left(\frac{K_v}{\hbox{km/s}}\right)^3
\left(\frac{P_b}{1\hbox{d}}\right)(1-e^2)^{3/2}\,.
\end{equation}
From here one readily finds the mass of the unseen (X-ray) companion to be
\begin{equation}
M_x=f(M_v)\left(1+\frac{M_v}{M_x}\right)^2\frac{1}{\sin^3 i}\,.
\end{equation}
 Unless the mass ratio $M_v/M_x$ and orbital inclination angle $i$ are known from independent measurements
(for example, from the analysis of optical light curve and the duration of X-ray eclipse), the mass function
of the optical component gives the lower limit of the BH mass: $M_x\ge f(M_v)$.
 
When using the optical mass function to estimate the BH mass as described above, one should always check
the validity of approximations used in deriving formula (\ref{fmv}) (see a detailed discussion of different
effects related to non-point-like form of the optical companion in, e. g., \cite{1996PhyU...39R...1C}). For example, 
the optical O-B or WR stars in HMXBs have a powerful high-velocity stellar wind, which can affect the 
dynamical BH mass estimate based on spectroscopic measurements (see the discussion \cite{2014arXiv1401.5425M} 
for BH+WR binaries).   

Parameters of known galactic and extragalactic HMXB with black holes
are summarized in Table~\ref{table:HMXB_BH}. 

No PSR\,+\,BH system has been observed so far, despite optimistic
expectations from the early population synthesis
calculations~\cite{Lipunov_al94} and recent examination of possible
dynamical formation of such a binary in the Galactic
center~\cite{FaucherGiguere_Loeb11}. Therefore, it is not possible to
obtain direct experimental constraints on the NS\,+\,BH coalescence rate
based on observations of real systems. Recently, the BH candidate with Be-star MWC 656 in a wide 60-day
orbit was reported \cite{2014Natur.505..378C} (see Table \ref{table:HMXB_BH}). It is a very weak X-ray source,  but 
can be the counterpart of the gamma-ray 
source AGL J2241+4454. The BH mass estimation in this case is based on the measured radial velocity curve of 
the Be-star, whose mass determination relies on its spectral classification. This system can be the progenitor
of the long-sought BH+NS binary. Whether or not this binary system can become a merging BH+NS binary depends on the 
details of the common envelope phase  which is thought to happen after the bright accretion stage in this system, 
when the Be-star will evolve to the giant stage.

\begin{table}
  \renewcommand{\arraystretch}{1.2}
  \caption{Observed parameters of HMXB with black holes.}
  \label{table:HMXB_BH}
  \centering
  {\small
  \begin{tabular}{lcccccc}
    \toprule
    Source & $P_\mathrm{b}$ (hr) & $f(M_v)$ [$M_\odot$] & $M_x$ [$M_\odot$] &  $M_v$ [$M_\odot$] & Sp. type & Ref. \\
    \midrule
\multicolumn{7}{c}{Galactic binaries}\\
\midrule
    \epubtkSIMBAD{Cyg X-1} & 134.4 & 0.244 & $\sim$~15 & $\sim$~19 & O9.7Iab 
& \cite{2011ApJ...742...84O} \\
  \epubtkSIMBAD{SS 433} & 313.9 & 0.268  & $\sim$~5 & $\sim$~18 & A7I:
 & \cite{Cherepashchuk_al09} \\
 \epubtkSIMBAD{Cyg X-3} & 4.8& 0.027 & $\sim$~2.4 & $\sim$~10  & WN &\cite{2013MNRAS.429L.104Z} \\
\epubtkSIMBAD{MWC 656} & 1448.9 & 0.205 & 3.8-6.9 & 10-16  & B1.5--B2 IIIe &\cite{2014Natur.505..378C} \\
\midrule
\multicolumn{7}{c}{Extragalactic binaries}\\
\midrule
\epubtkSIMBAD{LMC X-1} & 93.8 & 0.13 & $\sim$~11 & $\sim$~32 & O7III 
& \cite{2009ApJ...697..573O} \\
\epubtkSIMBAD{M33 X-7} & 82.8 & 0.46 & $\sim$~16 & $\sim$~70 & O7-8III &\cite{2007Natur.449..872O} \\
  \epubtkSIMBAD{IC10 X-1} & 34.4 & 7.64 & $\sim$~23\,--\,33 & $\sim$~17\,--\,35 & WR &
    \cite{2007ApJ...669L..21P, 2008ApJ...678L..17S} \\
\epubtkSIMBAD{NGC 300 X-1} & 32.3 & 2.6 & $\sim$~14.5 & $\sim$~15 & WR &\cite{Crowther_al10} \\
\multicolumn{7}{c}{Extragalactic BH\,+\,WR candidates}\\
\epubtkSIMBAD{CXOU J123030.3+413853} & 6.4 & ? & $\gtrsim$~14.5 & ? & WR &\cite{2013MNRAS.436.3380E} \\
 (in NGC 4490)\\
\epubtkSIMBAD{CXOU J004732.0-251722.1}
  & 14-15 & ? & ? & ? & WR &\cite{2014arXiv1401.5425M}\\
(in NGC 253)\\
\bottomrule
  \end{tabular}}
  \renewcommand{\arraystretch}{1.0}
\end{table}

\subsection{A model-independent upper limit on BH-BH/BH-NS coalescence rate}


Even without using the population synthesis tool, one can search for NS\,+\,BH
or BH\,+\,BH progenitors among known BH in HMXB. This program has been pursued
in~\cite{Bulik_al11, Belczynski_al11}, \cite{Belczynski_al13}. In these papers,
the authors examined the future evolution of two bright HMXB IC10 X-1/NGC300 X-1
found in nearby low-metallicity galaxies. Both binaries consist of massive
WR-stars (about $20\,M_{\odot}$) and BH in close orbits (orbital periods about
30 hours). Masses of WR-stars seem to be high enough to produce the second BH,
so these system may be immediate progenitors of coalescing BH-BH systems. The
analysis of the evolution of the most known Galactic HMXB Cyg~X-1
\cite{Belczynski_al11}, which can be a NS\,+\,BH progenitor, led to the
conclusion that the galactic formation rate of coalescing NS\,+\,BH is likely to
be very low, less than 1 per 100~Myrs. This estimate is rather pessimistic even
for advanced LIGO detectors. Implications of the growing class of
short-period BH+WR binaries (see Table \ref{table:HMXB_BH}) for the
(BH+BH)/(BH+NS) merging rate are discussed in \cite{2014arXiv1401.5425M}.

While for the double NS systems it is possible to obtain the upper limit
for coalescence rate based on the observed binary pulsar statistics (the Bailes
limit, see Section \ref{sec:ns_freq} above), it is not so easy for BH+NS or
BH+BH systems due to the (present-day) lack of their observational candidates.
Still, the crude estimate can be found from the following considerations. A
rough upper limit on the coalescence rate of BH\,+\,NS(BH) binaries is set by
the observed formation rate of high-mass X-ray binaries, their direct
progenitors. The present-day formation rate of galactic HMXBs is about ${\cal
R_{HMXB}}\sim N_{HMXB}/t_{HMXB}\sim 10^{-3}$ per year (here we conservatively
assumed $N_{HMXB}=100$ and $t_{HMXB}=10^5$~yrs). This estimate can be made more
precise considering that only very close double compact system can coalesce in
the Hubble time, which requires a common envelope stage after the HMXB stage to
occur (see Fig. 7 below). The CE stage is likely to happen in sufficiently close
binaries after the bright X-ray accretion stage\footnote{We neglect here a
highly unlikely case of direct collision of a newly born BH with the companion
due to the possible BH natal kick.}. The analysis of observations of different
X-ray source populations in the galaxies suggests (\cite{Gilfanov:2013} and
references therein) that only a few per cent of all BH formed in a galaxy can
pass through a bright accretion stage in HMXBs. For the Galactic NS formation
rate once per several decades, the minimum BH progenitor mass $20 M_\odot$ and
Salpeter initial mass function, this yields the estimate  a few $\times 10^{-5}$
per year. The probability for a given progenitor HMXB to become a merging
NS\,+\,BH or BH\,+\,BH binary is very model-dependent. For example, recent
studies of a unique galactic microquasar SS433 \cite{Blundell_al08,
Cherepashchuk_al09} suggest the BH mass in this system to be at least $\sim
5\,M_\odot$ and the optical star mass to be above $15\,M_\odot$. In SS433, the
optical star fills the Roche lobe and forms a supercritical accretion disk
\cite{Fabrika04}. The mass transfer rate onto the compact star is estimated to
be about $10^{-4}\,M_{\odot}\mathrm{\ yr}^{-1}$. According to the canonical HMXB
evolution calculations, a common envelope must have been formed in such a binary
on a short time scale (thousands of years) (see \S~\ref{kicks} and
\cite{2013AARv..21...59I}), but the observed stability of the binary system
parameters in SS433 over 30~years \cite{Davydov_al08} shows that this is not the
case. This example clearly illustrates the uncertainty in our understanding of
HMXB evolution with BHs. 

Thus we conclude that in the most optimistic case where the
(NS\,+\,BH)/(BH\,+\,BH) merging rate is equal to the formation rate of
their HMXB progenitors, the upper limit for Galactic ${\cal
  R}_{\mathrm{NS+BH}}$ is a few $\times
10^{-5}\mathrm{\ yr}^{-1}$. The detection or non-detection of such
binaries within $\sim$~1000~Mpc distance by the advanced LIGO
detectors can therefore very strongly constrain our knowledge of the evolution of HMXB systems.

\label{sec:2v3}

\newpage
	\section{Basic Principles of Binary-Star Evolution}  
\label{section:physical_principles}

Beautiful early general reviews of the topic can be found, e.g.,
in~\cite{Bhattacharya_vandenHeuvel91, vandenHeuvel83} and more later
ones, e.g., in~\cite{Tauris_vandenHeuvel06, 2006epbm.book.....E,
2011BASI...39....1V}. Here we restrict ourselves to recalling several
facts concerning binary evolution which are most relevant to the
formation and evolution of compact binaries. We also will not discuss
possible dynamical effects which can affect the binary evolution
(like the Kozai--Lidov mechanism of eccentricity change in
hierarchical triple systems, see
e.g., \cite{2013ApJ...766...64S}). The readers who have experience in
the field can skip this section.

\subsection{Keplerian binary system and radiation back reaction}
\label{sec:appA}

We start with some basic facts about Keplerian motion in a binary
system and the simplest case of evolution of two point masses due to
gravitational radiation losses. The stars are highly condensed
objects, so their treatment as point masses is usually adequate for
the description of their interaction in the binary. Furthermore,
Newtonian gravitation theory is sufficient for this purpose as long as
the orbital velocities are small compared to the speed of light
$c$. The systematic change of the orbit caused by the emission of
gravitational waves will be considered in a separate paragraph below.


\subsubsection{Keplerian motion}

Let us consider two point masses $M_1$ and $M_2$ orbiting each other 
under the force of gravity.
It is well known (see~\cite{L_L_v1}) that this problem
is equivalent to the problem of a single body with mass $\mu$ moving 
in an external gravitational potential. The value of the external
potential is determined by the total mass of the system 
\begin{equation}
  M = M_1 + M_2.
  \label{B:M}
\end{equation}
The reduced mass $\mu$ is
\begin{equation}
  \mu = \frac{M_1M_2}{M}.
  \label{B:mu}
\end{equation}
The body $\mu$ moves in an elliptic orbit with eccentricity $e$
and major semi-axis $a$. 
The orbital period $P$ and orbital frequency $\Omega=2\pi/P$
are related to $M$ and $a$ by Kepler's third law
\begin{equation}
  \Omega^2 = \left( \frac{2\pi}{P} \right)^2 = \frac{GM}{a^3}.
  \label{B:3Kepl}
\end{equation}
This relationship is valid for any eccentricity $e$.

Individual bodies $M_1$ and $M_2$ move around the 
barycentre of the system in elliptic orbits with the same
eccentricity $e$. The major semi-axes $a_i$ of the two ellipses
are inversely proportional to the masses
\begin{equation}
  \frac{a_1}{a_2} = \frac{M_2}{M_1},
  \label{B:a1}
\end{equation}
and satisfy the relationship $a=a_1+a_2$. 
The position vectors of the bodies from the system's barycentre are 
$\vec{r}_1 = M_2 \vec{r} /(M_1+M_2)$ and 
$\vec{r}_2 = - M_1 \vec{r} /(M_1+M_2)$, where $\vec{r} = \vec{r}_1 - \vec{r}_2$
is the relative position vector. Therefore, 
the velocities of the bodies with respect to the system's
barycentre are related by 
\begin{equation}
  -\frac{\vec{V}_1}{\vec{V}_2} = \frac{M_2}{M_1},
  \label{B:v1}
\end{equation}
and the relative velocity is $\vec{V}= \vec{V}_1-\vec{V}_2$.

The total conserved energy of the binary system is 
\begin{equation}
  E = \frac{M_1 \vec{V}_1^2}{2} + \frac{M_2 \vec{V}_2^2}{2} - 
  \frac{GM_1M_2}{r} =
  \frac{\mu \vec{V}^2}{2} - \frac{GM_1M_2}{r} =
  -\frac{GM_1M_2}{2a},
  \label{B:E}
\end{equation}
where $r$ is the distance between the bodies.
The orbital angular momentum vector is perpendicular to the orbital plane
and can be written as
\begin{equation}
  \vec{J}_{\mathrm{orb}} =
  M_1\vec{V}_1\times\vec{r}_1 + M_2\vec{V}_2\times\vec{r}_2 =
  \mu\vec{V}\times\vec{r}.
  \label{B:vecJ}
\end{equation}
The absolute value of the orbital angular momentum is
\begin{equation}
  |\vec{J}_{\mathrm{orb}}| = \mu\sqrt{GMa(1-e^2)}.
  \label{B:Je}
\end{equation}

For circular binaries with $e=0$
the distance between orbiting bodies does not depend on time,
\begin{displaymath}
  r(t,e=0) = a,
\end{displaymath}
and is usually referred to as orbital separation.
In this case, the velocities of the bodies, as well as their 
relative velocity, are also time-independent,
\begin{equation}
  V \equiv |\vec{V}| = \Omega a = \sqrt{GM/a},
  \label{B:Vorb}
\end{equation}
and the orbital angular momentum becomes
\begin{equation}
  |\vec{J}_{\mathrm{orb}}| = \mu Va = \mu \Omega a^2.
  \label{B:J}
\end{equation}


\subsubsection{Gravitational radiation from a binary}
\label{sec:gwr_from_binary}

The plane of the orbit is determined by the orbital
angular momentum vector $\vec{J}_{\mathrm{orb}}$. The
line of sight is defined by a unit vector $\vec{n}$. The
binary inclination angle $i$ is defined by the relation
$\cos i = (\vec{n},\vec{J}_{\mathrm{orb}}/J_{\mathrm{orb}})$
such that $i=90^\circ$ corresponds to a system visible edge-on.

Let us start from two point masses $M_1$ and $M_2$ in a circular
orbit. In the quadrupole approximation~\cite{L_L_v2}, the two polarisation
amplitudes of GWs at a distance $r$ from the source are given by
\begin{eqnarray}
  h_+ &=& \frac{G^{5/3}}{c^4}
  \frac{1}{r}\,2(1+\cos^2i)(\pi f M )^{2/3}\mu\cos(2\pi f t),
  \label{A:h_+}
  \\
  h_\times &=& \pm \frac{G^{5/3}}{c^4}
  \frac{1}{r}\,4 \cos i (\pi f M )^{2/3}\mu\sin(2\pi f t).
  \label{A:h_x}
\end{eqnarray}%
Here $f=\Omega/\pi$ is the frequency of the emitted GWs (twice the
orbital frequency). Note that for a fixed distance $r$ and a given
frequency $f$, the GW amplitudes are fully determined by $\mu M^{2/3}
= {\M}^{5/3}$, where the combination
\begin{displaymath}
  \M\equiv\mu^{3/5}M^{2/5}
\end{displaymath}
is called the ``chirp mass'' of the binary. 
After averaging over the orbital period (so that
the squares of periodic functions are replaced by 1/2)
and the orientations of the binary orbital plane, 
one arrives at the averaged (characteristic) GW amplitude
\begin{equation}
  h(f,\M,r) =
  \left( \langle h_+^2 \rangle + \langle h_\times^2 \rangle \right)^{1/2} =
  \left( \frac{32}{5} \right)^{1/2} \frac{G^{5/3}}{c^4}
  \frac{\M^{5/3}}{r} (\pi f)^{2/3}.
  \label{A:meanh}
\end{equation}


\subsubsection{Energy and angular momentum loss}
\label{section:AML}

In the approximation and under the choice of coordinates that we are
working with, it is sufficient to use the Landau--Lifshitz
gravitational pseudo-tensor~\cite{L_L_v2} when calculating the
gravitational waves energy and flux. (This calculation can be
justified with the help of a fully satisfactory gravitational
energy-momentum tensor that can be derived in the field theory
formulation of general relativity~\cite{Babak_Grishchuk00}). The energy
$dE$ carried by a gravitational wave along its direction of
propagation per area $dA$ per time $dt$ is given by 
\begin{equation}
  \frac{dE}{dA\,dt} \equiv F =
  \frac{c^3}{16\pi G}
  \left[ \left( \frac{\partial h_+}{\partial t} \right)^2 +
  \left( \frac{\partial h_\times}{\partial t} \right)^2 \right].
  \label{A:flux} 
\end{equation}
The energy output $dE/dt$ from a localised source in all directions is
given by the integral
\begin{equation}
  \frac{dE}{dt} = \int F(\theta,\phi) r^2 \, d\Omega.
  \label{A:loss}
\end{equation}
Replacing
\begin{displaymath}
  \left( \frac{\partial h_+}{\partial t} \right)^2 +
  \left( \frac{\partial h_\times}{\partial t} \right)^2 = 
  4\pi^2 f^2 h^2(\theta, \phi)
\end{displaymath}
and introducing
\begin{displaymath}
  h^2 = \frac{1}{4 \pi} \int h^2(\theta, \phi) \, d\Omega,
\end{displaymath}
we write Eq.~(\ref{A:loss}) in the form
\begin{equation}
  \frac{dE}{dt} = \frac{c^3}{G} (\pi f)^2 h^2 r^2.
  \label{A:loss2}
\end{equation}

Specifically for a binary system in a circular orbit, one finds
the energy loss from the system (sign minus) with the help of 
Eqs.~(\ref{A:loss2}) and (\ref{A:meanh}):
\begin{equation}
  \frac{dE}{dt} = 
  - \left( \frac{32}{5}\right)\frac{G^{7/3}}{c^5} (\M\pi f)^{10/3}.
  \label{A:dEdt}
\end{equation}
This expression is exactly the same one that can be obtained directly from
the quadrupole formula~\cite{L_L_v2},
\begin{equation}
  \frac{dE}{dt} =
  -\frac{32}{5} \frac{G^4}{c^5} \frac{M_1^2M_2^2M}{a^5},
  \label{A:GW:dEdt}
\end{equation}
rewritten using the definition of the chirp mass and
Kepler's law. Since energy and angular momentum are continuously
carried away by gravitational radiation, the two masses in orbit spiral
towards each other, thus increasing their orbital frequency $\Omega$. 
The GW frequency $f=\Omega/\pi$ and the GW amplitude $h$ 
are also increasing functions of time. The rate of the frequency 
change is\epubtkFootnote{A signal with such an increasing frequency is 
reminiscent of the chirp of a bird. This explains the
origin of the term ``chirp mass'' for the 
parameter $\M$ which fully determines the GW frequency and 
amplitude behaviour.}
\begin{equation}
  \dot{f} = \left( \frac{96}{5} \right)
  \frac{G^{5/3}}{c^5}\pi^{8/3}\M^{5/3}f^{11/3}.
  \label{A:dotf}
\end{equation}

In spectral representation, the flux of energy per unit area
per unit frequency interval is given by the right-hand-side of the
expression 
\begin{equation}
  \frac{dE}{dA\,df} = \frac{c^3}{G} \frac{\pi f^2}{2}
  \left( \left| \tilde h(f)_+ \right|^2 +
  \left| \tilde h(f)_\times \right|^2 \right)
  \equiv \frac{c^3}{G} \frac{\pi f^2}{2} S_h^2(f),
  \label{A:S_h}
\end{equation}
where we have introduced the spectral density $S_h^2(f)$ of the
gravitational wave field $h$. In the case of a binary system, the
quantity $S_h$ is calculable from Eqs.~(\ref{A:h_+}, \ref{A:h_x}):
\begin{equation}
  S_h = \frac{G^{5/3}}{c^3} \frac{\pi}{12} \frac{\M^{5/3}}{r^2}
  \frac{1}{(\pi f)^{7/3}}.
  \label{A:S_h:2}
\end{equation}


\subsubsection{Binary coalescence time}
\label{A:GW_evol}

A binary system in a circular orbit loses energy according to
Eq.~(\ref{A:dEdt}). For orbits with non-zero eccentricity $e$, the
right-hand-side of this formula should be multiplied by the factor 
\begin{displaymath}
  f(e) = \left( 1+\frac{73}{24}e^2+\frac{37}{96}e^4 \right)(1-e^2)^{-7/2}
\end{displaymath}
(see~\cite{Peters64}). The initial binary separation $a_0$ decreases
and, assuming Eq.~(\ref{A:GW:dEdt}) is always valid, it should vanish
in a time
\begin{equation}
  t_0 = \frac{c^5}{G^3} \frac{5 a_0^4}{256 M^2\mu} =
  \frac{5c^5}{256} \frac{(P_0/2\pi)^{8/3}}{(G\M)^{5/3}} \approx
  (9.8 \times 10^6 \mathrm{\ yr})
  \left( \frac{P_0}{1 \mathrm{\ h}} \right)^{8/3}
  \left( \frac{\M}{M_\odot} \right)^{-5/3}\!\!\!\!\!\!\!\!\!\!\!.
  \label{A:GW:t_0}
\end{equation}
As we noted above, gravitational radiation from the binary depends on
the chirp mass $\M$, which can also be written as $\M\equiv
M\eta^{3/5}$, where $\eta$ is the dimensionless ratio
$\eta=\mu/M$. Since $\eta\le 1/4$, one has $\M\lesssim 0.435 M$. For
example, for two NS with equal masses $M_1=M_2=1.4\,M_\odot$, the chirp
mass is $\M\approx 1.22\,M_\odot$. This explains the choice of
normalisation in Eq.~(\ref{A:GW:t_0}).

\epubtkImage{figure03.png}{
  \begin{figure}[htbp]
    \centerline{\includegraphics[width=0.7\textwidth]{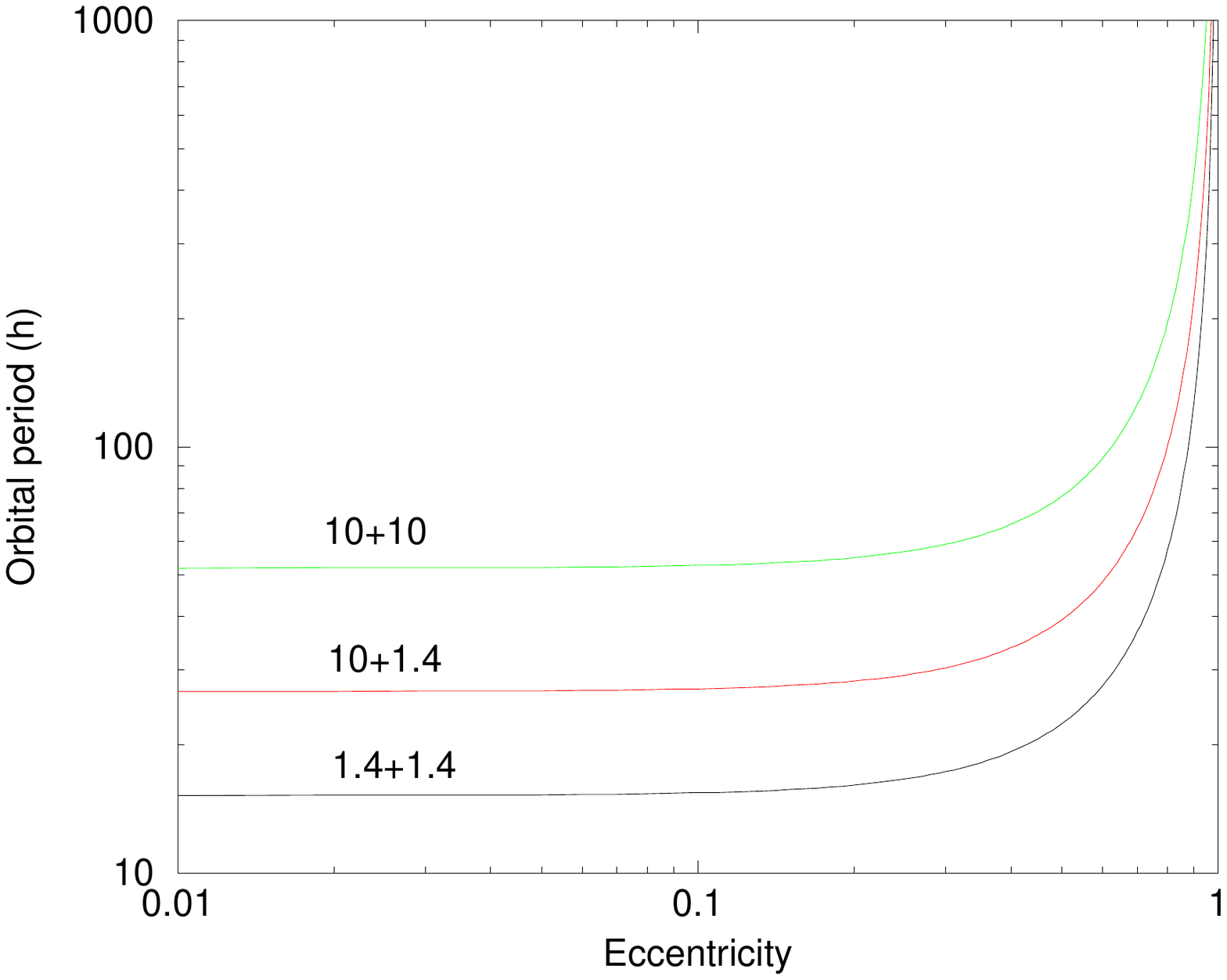}}
    \caption{The maximum initial orbital period (in hours) of two
      point masses which will coalesce due to gravitational wave
      emission in a time interval shorter than 10\super{10}~yr, as a
      function of the initial eccentricity $e_0$. The lines are
      calculated for $10\,M_\odot + 10\,M_\odot$ (BH\,+\,BH),
      $10\,M_\odot + 1.4\,M_\odot$ (BH\,+\,NS), and
      $1.4\,M_\odot + 1.4\,M_\odot$ (NS\,+\,NS).}
    \label{A:GW:p-e}
  \end{figure}
}

The coalescence time for an initially eccentric orbit with $e_0 \ne 0$
and separation $a_0$ is shorter than the coalescence time for a
circular orbit with the same initial separation $a_0$~\cite{Peters64}:
\begin{equation}
  t_\mathrm{c}(e_0) = t_0 \, f(e_0),
  \label{A:GW:t_c}
\end{equation}
where the correction factor $f(e_0)$ is 
\begin{equation}
  f(e_0) = \frac{48}{19} \frac{(1-e_0^2)^4}{e_0^{48/19}
  \left( 1+\frac{121}{304}e_0^2 \right)^{3480/2299}}
  \int_0^{e_0} \frac{\left( 1+\frac{121}{304}e^2 \right)^{1181/2299}}
  {(1-e^2)^{3/2}} \, e^{29/19} \, de.
  \label{A:GW:f(e)}
\end{equation}
To merge in a time interval shorter than 10~Gyr the
binary should have a small enough initial orbital period $P_0\le
P_{cr}(e_0,\M)$ and, accordingly, a small enough initial semi major
axis $a_0\le a_{cr}(e_0,\M)$. The critical orbital period is plotted
as a function of the initial eccentricity $e_0$ in
Figure~\ref{A:GW:p-e}. The lines are plotted for three typical sets of
masses: two neutron stars with equal masses ($1.4\,M_\odot +
1.4\,M_\odot$), a black hole and a neutron star ($10\,M_\odot +
1.4\,M_\odot$), and two black holes with equal masses ($10\,M_\odot +
10\,M_\odot$). Note that in order to get a significantly shorter
coalescence time, the initial binary eccentricity should be $e_0\geq
0.6$.


\subsubsection{Magnetic stellar wind}
\label{sec:msw}

In the case of low-mass binary evolution, there is another important physical
mechanism responsible for the removal of orbital angular momentum, in addition
to GW emission discussed above. This is the magnetic stellar wind (MSW), or
magnetic braking, which is thought to be effective for main-sequence G-M dwarfs
with convective envelopes, i.e., approximately, in the mass interval
$0.3\mbox{\,--\,}1.2\,M_\odot$. The upper mass limit corresponds to the
disappearance of a deep convective zone, while the lower mass limit stands for
fully convective stars. In both cases a dynamo mechanism, responsible for
enhanced magnetic activity, is thought to become ineffective. The idea behind
angular momentum loss (AML) by magnetically coupled stellar wind is that the
stellar wind is compelled by magnetic field to corotate with the star to rather
large distances where it carries away large specific angular
momentum~\cite{schatz_msw62}. Thus, it appears possible to take away substantial
angular momentum without evolutionary significant mass-loss in the wind. In the
quantitative form, the concept of angular momentum loss by MSW as a driver of
the evolution of compact binaries was introduced into analyses of them by
Verbunt and Zwaan~\cite{vz81} when it became evident that momentum loss by GWs
is unable to explain the observed mass-transfer rates in cataclysmic variables
(CVs) and low-mass X-ray binaries, as well as the deficit of cataclysmic
variables with orbital periods between $\approx2$ and $\approx$~3~hr (so-called
``period gap'')\epubtkFootnote{Actually, no CVs were known in the ``gap''
when this fact was first realised \cite{1981ApJ...251..611R}.} Verbunt and Zwaan
based their reasoning on observations of the spin-down of single G-dwarfs in
stellar clusters with age $V=\lambda t^{-1/2}$ (``Skumanich
law''~\cite{Skumanich72}). In the latter formula $\lambda$ is an empirically
derived coefficient $\sim1$. Applying this to a binary component and assuming
tidal locking between the star's axial rotation and orbital motion, one arrives
at the rate of angular momentum loss via MSW
%
%
%
\begin{equation}
  \dot{J}_\mathrm{MSW} = -0.5 10^{-28} \lambda^{-2} k^2 M_2 R_2^4 \omega^3, 
    \label{eq:MSW}
\end{equation}
where 
$M_2$ and $R_2$ are the mass and radius of the optical component of the system 
 and $k^2 \sim 0.1$ is gyration radius of the optical component squared.

Radii of stars filling their Roche lobes should be proportional to binary
separations, $R_\mathrm{o}\propto a$, which means that the time scale of orbital
angular momentum removal by MSW is $\tau_\mathrm{MSW}\equiv
(\dot{J}_\mathrm{MSW}/J_{\mathrm{orb}})^{-1}\propto a$. This should be compared
with AML by GWs with $\tau_\mathrm{GW} \propto a^4$. Clearly, MSW (if it
operates) is more efficient in removing angular momentum from a binary system at
larger separations (orbital periods), and at small orbital periods GWs always
dominate. Magnetic braking is especially important in CVs and in LMXBs with
orbital periods exceeding several hours and is the driving mechanism for mass
accretion onto the compact component. 
\textbf{Indeed, the Doppler tomography reconstruction of the Roche-lobe filling low-mass
K7 secondary star in a well-studied LMXB Cen X-4 revealed the presence of cool
spots on its surface. The latter provide an evidence for the action of magnetic field
at the surface of the star, thus supporting the magnetic braking as the  
driving mechanism of mass exchange in this binary \cite{2014arXiv1402.1639S}}.  

Expression~(\ref{eq:MSW}) with assumed $\lambda=1$ and $k^2=0.07$~\cite{rvj83}
is often considered as a kind of a ``standard''. However, Eqs.~(\ref{A:GW:dEdt})
and (\ref{eq:MSW}) do not allow to reproduce some observed features of CVs, see,
e.g.,
\cite{rvj83,Politano_Weiler06,2008MNRAS.388.1582L,2009ApJ...693.1007T,2011ApJS..194...28K},
but the reasons for discrepant behaviour are not clear as yet. Recently, Knigge
et al.~\cite{2011ApJS..194...28K}, based on the study of properties of donors in
CVs, suggested that a better description of evolution of CVs is provided by
scaling the law (\ref{eq:MSW}) by a factor $(0.66\pm0.05)(R_2/\rs)^{-1}$ above
the period gap and by scaling (\ref{A:GW:dEdt}) by a factor $2.47\pm0.22$ below
the gap (where AML by MSW is not acting). Note that these ``semi-empirical'', as
called by the their authors, numerical factors must be taken with some caution,
since they are based on the fitting of the \porb-$M_2$ relation for observed
stars by a single evolutionary sequence with initially unevolved donor with
$M_{2,0}=1\ms$ and $M_{\mathrm{wd},0}=0.6\ms$.
  
The simplest reason for deviation of AML by MSW from Eq.~(\ref{A:GW:dEdt}) may
be unjustified extrapolation of stellar rotation rates over several orders of
magnitude -- from slowly rotating single field stars to rapidly spinning components of
close binaries. For the shortest orbital periods binaries supposed to evolve due
to AML via GW, the presence of circumbinary discs may enhance their 
orbital angular momentum loss
\cite{2005ApJ...635.1263W}.


\subsection{Mass exchange in close binaries}
\label{sec:mass_exc}

As mentioned in the Introduction, all binaries may be  considered either as 
 ``close'' or as ``wide''.
In the former case, mass exchange between the components can occur. 
This process can be accompanied by the mass and angular momentum loss from the system.

The shape of the stellar surface is determined by the shape of the equipotential
level surface $\Phi=\mathrm{const.}$ 
Conventionally, the total potential which includes gravitational and centrifugal
forces is approximated by the Roche potential (see, e.g. \cite{kopal1959}) which
is defined under the following assumptions:

the gravitational field of two components is approximated by that of two point masses;

the binary orbit is circular; 

the components of the system corotate with the binary orbital period. 

Let us consider a Cartesian reference frame (x,y,z) 
rotating with the binary, with the origin at the primary $M_1$;
the x-axis is directed along the line of centers;
the y-axis is aligned with the orbital motion of the primary component;
the z-axis is perpendicular to the orbital plane.
The total potential at a given point (x,y,z)  is then
\begin{equation}
\label{eq:roche}
\Phi=-\frac{GM_1}{\left[x^2+y^2+z^2\right]^{1/2}}-\frac{GM_2}{\left[(x-a)^2+y^2+z^2\right]^{1/2}}-
\frac{1}{2}\Omega_{\textrm{orb}}^2\left[(x-\mu a)^2+y^2\right],
 \end{equation}
where $\mu=M_2/(M_1+M_2)$, $\Omega=2\pi/\porb$.

A 3-D representation of dimensionless Roche potential in co-rotating frame for
a binary with mass ratio of components $q=2$   
is shown in 
Figure~\ref{fig:roche}.

\epubtkImage{rochepot.png}{
  \begin{figure}[htbp]
     \centerline{\includegraphics[height=0.5\textwidth]{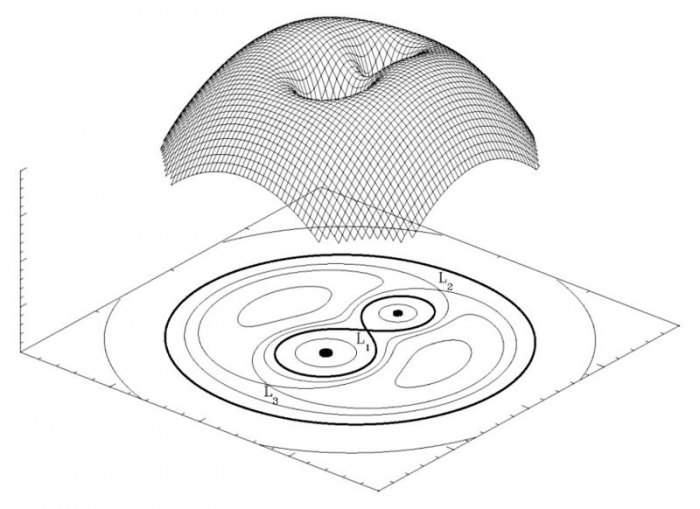}}
         \caption{
 3-D representation of  dimensionless Roche potential in the co-rotating frame for
a binary with mass ratio of components $q=2$. 
}
    \label{fig:roche}
  \end{figure}
}

For close binary evolution the most
important is the innermost  level surface which encloses both
components. It defines the ``critical'', or ``Roche'' lobes of the components.%
\epubtkFootnote{Actually, the Roche potential should also incorporate the
effects of radiation pressure from the binary components. 
Then in some cases the resulting potentials do not exhibit
the  contact surfaces of the classical Roche
potential \cite{1972Ap&SS..19..351S, 2010ASPC..435...85D}.}
Inside these lobes the matter is bound to the respective component. In
the libration point $L_1$ (the inner Lagrangian point) the net force
exerted onto a test particle corotating with the binary vanishes, so the particle
can escape from the surface of the star and can be captured by  the
companion. The matter flows along the surface of the Roche-lobe filling companion in
the direction of $L_1$ and escapes from the surface of the contact
component as a highly inhomogeneous stellar wind.
Next, the wind forms a (supersonic) stream  directed at certain angle respective to
the line connecting the centers of components.  Depending on the size
of the second companion, the stream may hit the latter (the so-called ``direct
impact'') or form a disk orbiting the companion~\cite{ls75}.

Though it is evident that the mass exchange is a complex 3-D gas-dynamical
process which has to take into account also radiation transfer and, in
some cases, even nuclear reactions, as a matter of fact, virtually all
computations of the evolution of non-compact close binaries have been 
performed in 1-D approximation. The Roche lobe
overflow (RLOF) is conventionally considered to begin when the radius of one binary
component (for example, the primary, i.e. initially more massive and hence faster evolving 
star) becomes equal to the radius of a sphere with the volume
equal to that of the Roche lobe. For the latter, an expression
precise to better than 1\% for arbitrary mass ratio of
components $q$ was suggested by Eggleton~\cite{Eggleton83}:
\begin{equation}
\label{eq:roche_eggl}
\frac{R_l}{a}\approx\frac{0.49q^{2/3}}{0.6q^{2/3}+\ln(1+q^{1/3})}.
\end{equation}
For practical purposes like analytical estimates, a more convenient is
the expression suggested by Kopal \cite{kopal1959} (usually called
the ``Paczy\'nski formula'' who introduced it into close binary modeling):
\begin{equation}
\label{eq:roche_kopal}
\frac{R_l}{a} \approx 0.4623 \left(\frac{q}{1+q}\right)^{1/3}\,,
\end{equation}
which is accurate to $\lesssim 2\%$ for $0<q\aplt0.8$.

In close binaries, the zero-age main sequence (ZAMS) mass ceases to be the sole
parameter determining stellar evolution. The nature of compact remnants 
of close binary components depends also on their evolutionary stage at
RLOF, i.e., on the component separation and their mass-ratio $q$. Evolution
of a star may be considered as consumption of nuclear fuel accompanied by
increase of its radius. Following pioneering work of R.~Kippenhahn and his
collaborators on the evolution of close binaries in the late 1960s, the
following basic cases of mass exchange are usually considered: 
\textbf{A} -- RLOF at the core hydrogen-burning stage;
\textbf{B} -- RLOF at the  hydrogen-shell burning stage;
\textbf{C} -- RLOF after exhaustion of He in the stellar core.
Also, more ``fine'' gradations exist: case \textbf{AB} -- RLOF at the late
stages of core H-burning, which continues as Case~\textbf{B} after a short break
upon exhaustion of H in the core; case \textbf{BB} -- RLOF by the star which
first filled its Roche lobe in the case \textbf{B}, contracted under the Roche
lobe after the loss of the hydrogen envelope, and resumed the mass loss due to the
envelope expansion at the helium shell burning stage. Further, one may consider
the modes of mass-exchange, depending, e.g., on the nature of the envelope of
the donor (radiative vs.\ convective) its relative mass, reaction of accretor
etc., see, e.g., \cite{2006epbm.book.....E,2008ASSL..352..233W} and
Section~\ref{A:mass_transf}.

In the cases \textbf{A} and \textbf{B} of mass exchange the remnants
of stars with initial masses lower than  $(2.3\mbox{\,--\,}2.8)\ms$
are degenerate He WD -- a type of objects not produced by single
stars.%
\epubtkFootnote{Hydrogen burning time for solar composition single
stars with $M_0\lesssim 0.95\,M_\odot$ exceeds the age of the
Universe.}

In the cases \textbf{A} and \textbf{B} of mass exchange the remnants of stars with initial mass  $\apgt 2.5\ms$ 
are helium stars.

If the mass of helium remnant star does not exceed $\simeq0.8\ms$, 
after exhaustion of He in the core it does not expand and transforms
directly into 
a WD of the same mass \cite{it84a}. Helium stars with masses between 
$\simeq0.8\ms$
and $\simeq(2.3\mbox{\,--\,}2.8)\ms$
expand in the helium shell burning stage, re-fill their Roche lobes,
lose the remnants of the helium envelopes  and also transform into CO WD.

Stellar radius may be taken as a proxy for the evolutionary state of 
a star. 
We plot in Figure~\ref{f:remn} 
the types of the stellar
remnant as a function of both initial mass and the radius of a star at
the instant of RLOF. In Figure~\ref{fig:popsynthremn} initial-final mass relations
for components of close binaries are shown. Clearly, relations presented
in the former two Figures are only approximate, reflecting current uncertainty
in the theory of stellar evolution (in this particular case, relations used in
the population synthesis code \textsf{IBiS} \cite{ty02} are shown.

\epubtkImage{radii.png}{%
  \begin{figure}[htbp]
    \centerline{\includegraphics[width=0.9\textwidth]{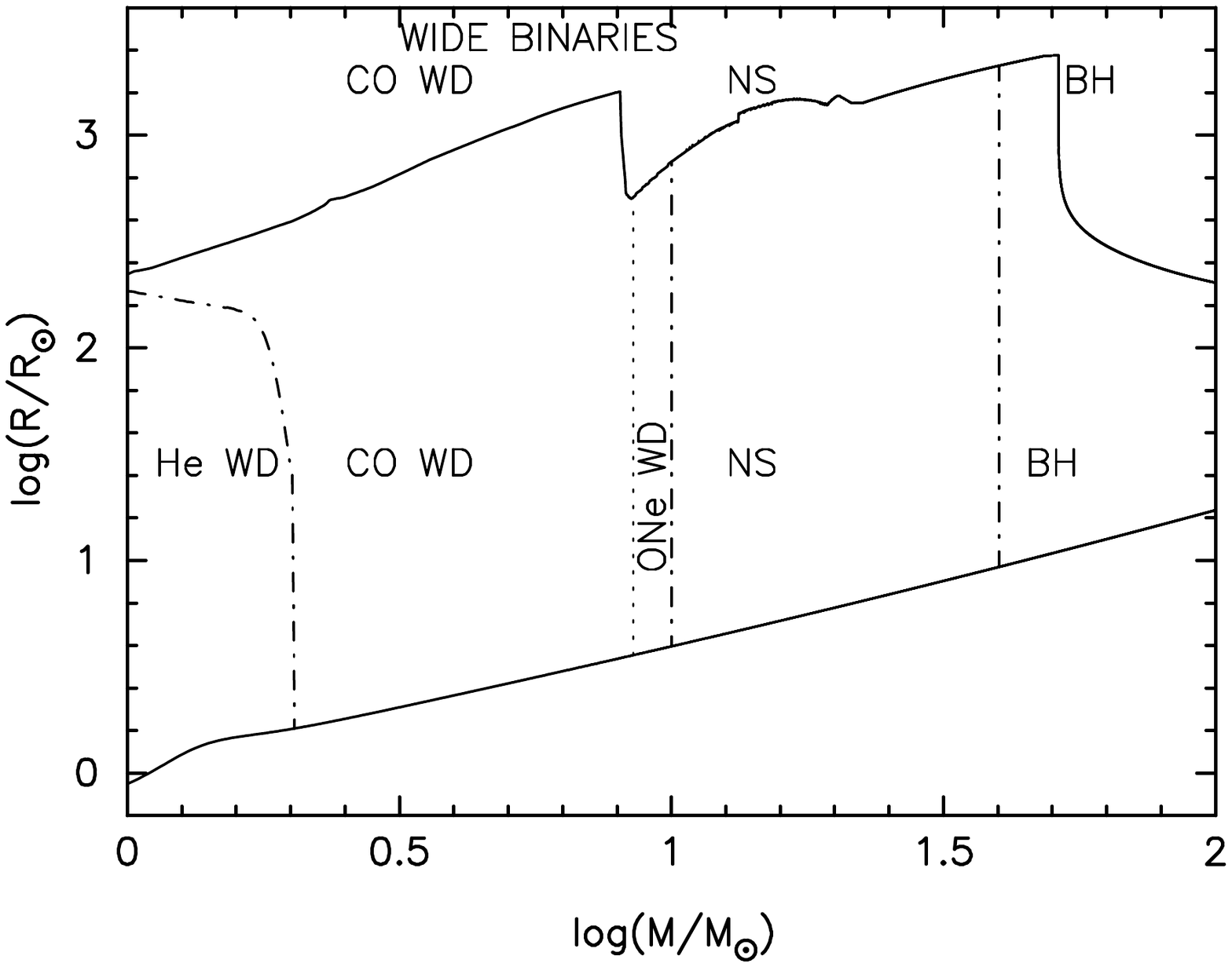}}
    \caption{Descendants of components of close binaries depending on
    the radius of the star at RLOF. The upper solid line separates
    close and wide binaries (after \cite{hpt00}). The boundary between
    progenitors of He- and CO-WDs is uncertain by several
    $0.1\,M_\odot$, the boundary between CO and ONe varieties of WDs
    and WD and NSs -- by $\sim 2\,M_\odot$. The boundary between
    progenitors of NS and BH is shown at $40\ms$
    after \cite{2010AA...520A..48R}, while it may be possible that it
    really is between $20\ms$ and $50\ms$ (see the text for discussion
    and references.)}
    \label{f:remn}
  \end{figure}
}

\epubtkImage{remnants.png}{%
  \begin{figure}[htbp]
  \centerline{\includegraphics[width=0.9\textwidth]{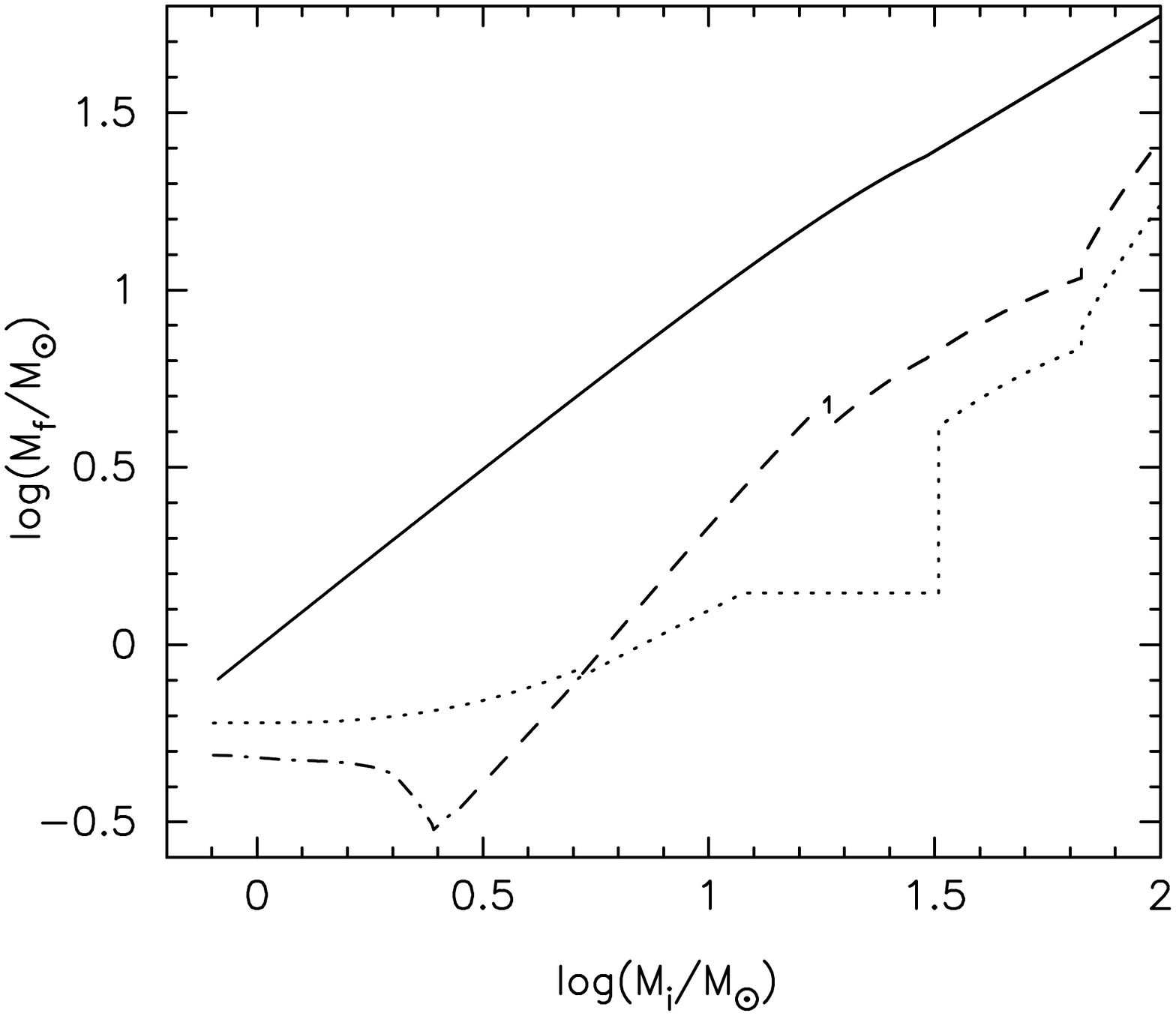}}
  \caption{Relation  between ZAMS masses of stars $M_i$ and their
  masses at TAMS  (solid line), masses of helium stars (dashed line),
  masses of He WD (dash-dotted line), masses of CO and ONe WD (dotted
  line). For stars with $M_i \lesssim 5\ms$ we plot the upper limit
  of WD masses for case \textbf{B} of mass
  exchange. After \cite{ty02}.}
  \label{fig:popsynthremn}
\end{figure}
}

\subsection{Mass transfer modes and mass and angular 
momentum loss in binary systems}
\label{A:mass_transf}

GW emission is the sole factor responsible for the change of orbital
parameters of a detached pair of compact (degenerate) stars. However,
it was quite early recognised that in the stages of evolution
preceding formation of compact objects,  the mode of mass
transfer between the components and the loss of matter and orbital
angular momentum by the system as a whole play a dominant dynamical
role and define the observed features of the binaries,
e.g., \cite{1967AcA....17....7P,1971NInfo..20...86T,rrw74, 1975MmSAI..46..217M,
1978NInfo..42...45P, 1980Ap&SS..68..433D}.
 
Strictly speaking, as we mentioned already above, 
these processes should be treated hydrodynamically
and they require complicated numerical calculations. However,
binary evolution can also be described semi-qualitatively,
using a simplified description in terms of point-like bodies. 
The change of their integrated physical quantities, such as masses,
orbital angular momentum, etc., governs the evolution of the orbit.
This description turns out to be successful in reproducing the
results of more rigorous numerical calculations
(see, e.g., 
\cite{1994inbi.conf....1S} for more detail and references). In this approach, 
the key role is allocated to the total orbital angular momentum $J_{\mathrm{orb}}$
of the binary. 

Let star 2 lose matter at a rate $\dot{M}_2<0$ and let $\beta$
$(0\le\beta\le1)$ be the fraction of the ejected matter which leaves
the system (the rest falls on the first star),
i.e., $\dot{M}_1=-(1-\beta)\dot{M}_2\ge 0$.
Consider circular orbits with orbital angular momentum given by
Eq.~(\ref{B:J}). Differentiate both parts of Eq.~(\ref{B:J}) by time $t$ 
and exclude $d\Omega/dt$ with the help of Kepler's third
law (\ref{B:3Kepl}). This gives  the rate of change of the
orbital separation:
\begin{equation}
  \frac{\dot{a}}{a} = -2 \left( 1+(\beta-1)\frac{M_2}{M_1} -
  \frac{\beta}{2} \frac{M_2}{M} \right)
  \frac{\dot{M}_2}{M_2} + 2\frac{\dot{J}_{\mathrm{orb}}}{J_{\mathrm{orb}}}.
  \label{A:dotaa}
\end{equation}
In Eq.~\ref{A:dotaa} $\dot{a}$ and $\dot{M}$ are not independent
variables if the donor fills its Roche lobe. One defines the mass transfer
as conservative if both $\beta=0$ and $\dot{J}_{\mathrm{orb}}=0$. The mass
transfer is called non-conservative if at least one of these
conditions is violated.

It is important to distinguish some specific cases (modes) of mass
transfer:
\begin{enumerate}
\item conservative mass transfer,
\item non-conservative Jeans mode of mass loss (or fast wind mode),
\item non-conservative isotropic re-emission,
\item sudden mass loss from one of the components during supernova
  explosion, and
\item common-envelope stage.
\end{enumerate}

As specific cases of angular momentum loss we consider GW emission (see
Section~\ref{section:AML} and~\ref{A:GW_evol}) and the magnetically coupled
stellar wind (see Section~\ref{sec:msw}) which drive the orbital evolution for
short-period binaries. For non-conservative modes, one can also consider some
less important cases, such as, for instance, formation of a circumbinary ring by
the matter leaving the system (see, e.g.,
\cite{Soberman_al97,2001ApJ...548..900S}). Here, we will not go into details
of such sub-cases.


\subsubsection{Conservative accretion}

In the case of conservative accretion, matter from $M_2$ is fully
deposited onto $M_1$. The transfer process preserves the total mass
($\beta=0$) and the orbital angular momentum of the system. It follows
from Eq.~(\ref{A:dotaa}) that 
\begin{displaymath}
  M_1 M_2 \sqrt{a} = \mathrm{const},
\end{displaymath}
so that the initial and final binary separations are related as
\begin{equation}
  \frac{a_\mathrm{f}}{a_\mathrm{i}} =
  \left( \frac{M_{1\mathrm{i}}\,M_{2\mathrm{i}}}
  {M_{1\mathrm{f}}\,M_{2\mathrm{f}}} \right)^2\!\!\!.
  \label{A:conserv}
\end{equation}
%
The orbit
shrinks when the more massive component loses matter, and the orbit
widens in the opposite situation. During such a mass exchange, the
orbital separation passes through a minimum, if the masses become
equal in the course of mass transfer.


\subsubsection{The Jeans (fast wind) mode}

In this mode 
ejected matter completely escapes from the system,
that is, $\beta =1$. 
Escape of the matter can 
occur either via
spherically-symmetric wind 
or in the form of bipolar jets moving from
the system at high velocity. In both cases, 
the matter escapes with specific angular momentum of mass-losing star
$J_2 = (M_1/M) J_{\mathrm{orb}}$ (we
neglect a possible proper rotation of the star,
see~\cite{vandenHeuvel83}). For the loss of orbital momentum
$\dot{J}_{\mathrm{orb}}$ it is reasonable to take
\begin{equation}
  \dot{J}_{\mathrm{orb}} = \frac{\dot{M}_2}{M_2} J_2.
  \label{A:jspecif}
\end{equation}
In the case $\beta =1$, Eq.~(\ref{A:dotaa}) can be written as
\begin{equation}
  \frac{\dot{\Omega} a^2}{\Omega a^2} =
  \frac{\dot{J}_{\mathrm{orb}}}{J_{\mathrm{orb}}} -
  \frac{M_1\dot{M}_2}{M M_2}.
  \label{A:dotaa2} 
\end{equation}
Then Eq.~(\ref{A:dotaa2}) in conjunction with Eq.~(\ref{A:jspecif}) 
gives $\Omega a^2 = \mathrm{const}$, that is, 
$\sqrt{GaM}=\mathrm{const}$. Thus, as a result of Jeans mode of mass loss,
the change in orbital separation is 
\begin{equation}
  \frac{a_\mathrm{f}}{a_\mathrm{i}} =
  \frac{M_\mathrm{i}}{M_\mathrm{f}}.
  \label{A:Jeans}
\end{equation}
Since the total mass decreases, the orbit always widens.

Mass loss may be considered as occurring in Jeans mode
in situations in which
hot white dwarf components of cataclysmic
variables lose mass by optically thick winds~\cite{kathac94} or when   
time-averaged mass loss from novae is considered~\cite{y73a}.


\subsubsection{Isotropic re-emission}
\label{sec:reemission}

The matter lost by star 2 can first accrete onto star 1, and then, a fraction
$\beta$ of the accreted matter, can be expelled from the system. This happens,
for instance, when a massive star transfers matter onto a compact star on the
thermal timescale ($<$~10\super{6}~years). Accretion luminosity may exceed the
Eddington luminosity limit, and the radiation pressure pushes the infalling
matter away from the system, in a manner similar to the spectacular example of
the \epubtkSIMBAD{SS~433} binary system. Other examples may be systems with
helium stars transferring mass onto relativistic objects~\cite{lyhnp05,
ghosh_cyg3analog06}. In this mode of mass-transfer, the binary orbital momentum
carried away by the expelled matter is determined by the orbital momentum of the
accreting star rather than by the orbital momentum of the mass-losing star,
since mass loss happens from the vicinity of accretor. The assumption that all
matter in excess of accretion rate can be expelled from the system, thus
avoiding the formation of a common envelope, will only hold if the liberated
accretion energy of the matter falling from the Roche lobe radius of the
accretor star to its surface is sufficient to expel the matter from the
Roche-lobe surface around accretor, i.e. $\md \aplt \md_{\max} =
\medd(r_{\mathrm{L,a}}/r_{\mathrm{a}})$, where $r_{\mathrm{a}}$ is the radius of
accretor \cite{kb99}.
 
The orbital momentum loss can be written as
\begin{equation}
  \dot{J}_{\mathrm{orb}} = \beta \frac{\dot{M}_2}{M_1} J_1,
  \label{A:jspecif_1}
\end{equation}
where $J_1=(M_2/M)J_{\mathrm{orb}}$ is the orbital momentum of the star $M_1$.
In the limiting case when all the mass initially accreted 
by $M_1$ is later expelled from the system, $\beta=1$, Eq.~(\ref{A:jspecif_1})
simplifies to 
\begin{equation}
  \frac{\dot{J}_{\mathrm{orb}}}{J_{\mathrm{orb}}} =
  \frac{\dot{{}M_2} M_2}{M_1 M}.
  \label{A:jreemiss}
\end{equation}
After substitution of this formula into Eq.~(\ref{A:dotaa})
and integration over time, one arrives at
\begin{equation}
  \frac{a_\mathrm{f}}{a_\mathrm{i}} =
  \frac{M_\mathrm{i}}{M_\mathrm{f}}
  \left( \frac{M_{2\mathrm{i}}}{M_{2\mathrm{f}}} \right)^2
  \exp{\left( -2 \, \frac{M_{2\mathrm{i}}-M_{2\mathrm{f}}}{M_1} \right)}.
  \label{A:re-em}
\end{equation}
The exponential term makes this mode of the mass transfer
very sensitive to the components mass ratio.
If $M_1/M_2\ll 1$, the separation $a$ between the stars
may decrease so much that the approximation of point masses becomes
invalid. The tidal orbital instability
(Darwin instability) may set in, and the compact star
may start spiralling toward the companion star centre
(the common envelope stage; see Section~\ref{A:CE} below). 
On the other hand, ``isotropic reemission'' may stabilise mass-exchange if
$M_1/M_2 > 1$ \cite{ynh02}.


\subsection{Supernova explosion}
\label{A:SN}

Supernovae explosions in binary systems occur in a timescale
much shorter than the orbital period, so the loss of mass is
practically instantaneous. This case can be treated analytically
(see, e.g., \cite{Blaauw61,1961BAN....15..291B,tautak98}).

Clearly, even a spherically symmetric sudden mass loss due to SN explosion
will be asymmetric in the center of mass of the binary system, leading to the 
system recoil (the 'Blaauw-Boersma' recoil).
In general, the loss of matter and radiation is non-spherical, so that
the remnant of the supernova explosion (neutron star or black hole)
acquires some recoil velocity called kick velocity $\vec{w}$. In a
binary, the kick velocity should be added to the orbital velocity of the
pre-supernova star.

The usual treatment proceeds as follows. Let us consider a pre-SN
binary with initial masses $M_1$ and $M_2$. The stars move in a
circular orbit with orbital separation $a_\mathrm{i}$ and relative velocity
$\vec{V}_\mathrm{i}$. The star $M_1$ explodes leaving a compact remnant of mass
$M_\mathrm{c}$. The total mass of the binary decreases by the amount $\Delta M
= M_1 - M_\mathrm{c}$. 
Unless the binary is disrupted, it will end up in
a new orbit with eccentricity $e$, major semi axis $a_\mathrm{f}$, and
angle $\theta$ between the orbital planes before and after the
explosion. In general, the new barycentre will also receive some
velocity, but we neglect this motion. The goal is to evaluate the
parameters $a_\mathrm{f}$, $e$, and $\theta$.
 
It is convenient to work in an instantaneous 
reference frame centred on $M_2$ right at the time of explosion. 
The $x$-axis is the line from $M_2$ to $M_1$, the $y$-axis points in
the direction of $\vec{V}_\mathrm{i}$, and the $z$-axis is perpendicular to the 
orbital plane. In this frame, the pre-SN relative velocity is
$\vec{V}_\mathrm{i} = (0, V_\mathrm{i}, 0)$, where 
$V_\mathrm{i}=\sqrt{G(M_1+M_2)/a_\mathrm{i}}$ (see Eq.~(\ref{B:Vorb})).
The initial total orbital momentum is $\vec{J}_\mathrm{i} = \mu_\mathrm{i} a_\mathrm{i} (0, 0, -V_\mathrm{i})$.
The explosion is considered to be instantaneous. 
Right after the explosion, the position vector of the exploded 
star $M_1$ has not changed: $\vec{r}= (a_\mathrm{i}, 0, 0)$. However,
other quantities have changed: $\vec{V}_\mathrm{f}= (w_x, V_\mathrm{i}+w_y, w_z)$
and $\vec{J}_\mathrm{f} = \mu_\mathrm{f} a_\mathrm{i} (0, w_z, -(V_\mathrm{i}+w_y))$, where 
$\vec{w} = (w_x, w_y, w_z)$ is the kick velocity and
$\mu_\mathrm{f}= M_\mathrm{c}M_2/(M_\mathrm{c}+M_2)$ is the reduced mass of the system after explosion.
The parameters $a_\mathrm{f}$ and $e$ are being found 
by equating the total
energy and the absolute value of orbital momentum of the initial
circular orbit to those of the resulting elliptical orbit
(see Eqs.~(\ref{B:E}, \ref{B:J}, \ref{B:Je})): 
\begin{eqnarray}
  \mu_\mathrm{f} \frac{V_\mathrm{f}^2}{2} -
  \frac{GM_\mathrm{c}M_2}{a_\mathrm{i}} &=&
  -\frac{GM_\mathrm{c}M_2}{2a_\mathrm{f}},
  \label{A:SN:E}
  \\
  \mu_\mathrm{f} a_\mathrm{i}
  \sqrt{w_z^2+(V_\mathrm{i}+w_y)^2} &=&
  \mu_\mathrm{f} \sqrt{G(M_\mathrm{c}+M_2)a_\mathrm{f}(1-e^2)}.
  \label{A:SN:J}
\end{eqnarray}%
For the resulting $a_\mathrm{f}$ and $e$ one finds 
\begin{equation}
  \frac{a_\mathrm{f}}{a_\mathrm{i}} =
  \left[ 2-\chi \left( \frac{w_x^2+w_z^2+(V_\mathrm{i}+w_y)^2}
  {V_\mathrm{i}^2} \right) \right]^{-1}
  \label{A:SN:afai}
\end{equation}
and
\begin{equation}
  1-e^2 = \chi \frac{a_\mathrm{i}}{a_\mathrm{f}}
  \left( \frac{w_z^2+(V_\mathrm{i}+w_y)^2}{V_\mathrm{i}^2} \right),
  \label{A:SN:ecc}
\end{equation}
where $\chi \equiv (M_1+M_2)/(M_\mathrm{c}+M_2) \ge 1$. The angle $\theta$ is 
defined by
\begin{displaymath}
  \cos \theta =
  \frac{\vec{J}_\mathrm{f} \cdot \vec{J}_\mathrm{i}}
  {|\vec{J}_\mathrm{f}|~|\vec{J}_\mathrm{i}|},
\end{displaymath}
which results in
\begin{equation}
  \cos\theta = \frac{V_\mathrm{i}+w_y}{\sqrt{w_z^2+(V_\mathrm{i}+w_y)^2}}.
  \label{A:SN:theta} 
\end{equation}

The condition for disruption of the binary system depends on the absolute
value $V_\mathrm{f}$ of the final velocity, and on the parameter $\chi$.
The binary disrupts if its total energy defined by the left-hand-side
of Eq.~(\ref{A:SN:E}) becomes non-negative or, equivalently, if its eccentricity
defined by Eq.~(\ref{A:SN:ecc}) becomes $e \ge 1$. From either of these
requirements one derives the condition for disruption:
\begin{equation}
  \frac{V_\mathrm{f}}{V_\mathrm{i}} \ge \sqrt{\frac{2}{\chi}}.
  \label{A:disrupt}
\end{equation}
The system remains bound if the opposite inequality is
satisfied. Eq.~(\ref{A:disrupt}) can also be written in terms
of the escape (parabolic) velocity $V_\mathrm{e}$ defined by the
requirement
\begin{displaymath}
  \mu_\mathrm{f} \frac{V_\mathrm{e}^2}{2} -
  \frac{GM_\mathrm{c}M_2}{a_\mathrm{i}} = 0. 
\end{displaymath}
Since $\chi = M/(M - \Delta M)$ and $V_\mathrm{e}^2 = 2 G(M - \Delta M)/ a_\mathrm{i} =
2V_\mathrm{i}^2 / \chi$, one
can write Eq.~(\ref{A:disrupt}) in the form
\begin{equation}
  V_\mathrm{f} \ge V_\mathrm{e}.
  \label{A:disrupt2}
\end{equation}
The condition of disruption simplifies 
in the case of a spherically symmetric SN explosion, that is, when there is
no kick velocity, $\vec{w}= 0$, and, therefore, $V_\mathrm{f} = V_\mathrm{i}$. In this case,
Eq.~(\ref{A:disrupt}) reads $\chi \ge 2$, which is equivalent to
$\Delta M \ge M/2$. Thus, the system unbinds if more
than a half of the mass of the binary is lost. In other words, 
the resulting eccentricity 
\begin{equation}
  e = \frac{M_1-M_\mathrm{c}}{M_\mathrm{c}+M_2}
  \label{A:SN:symm-ecc}
\end{equation}
following from Eqs.~(\ref{A:SN:afai}, \ref{A:SN:ecc}), and $\vec{w} = 0$
becomes larger than 1, if $\Delta M > M/2$. 

So far, we have considered
an originally circular orbit. If the pre-SN star moves in an 
originally eccentric orbit, the condition of disruption of the system 
under symmetric explosion reads
\begin{displaymath}
  \Delta M = M_1-M_\mathrm{c} > \frac{1}{2} \frac{r}{a_\mathrm{i}},
\end{displaymath}
where $r$ is the distance between the components at the moment of explosion.


\subsection{Kick velocity of neutron stars}
\label{kicks}

The kick  imparted to a NS at birth is one of the 
major problems in the theory of stellar evolution. By itself it is an
additional parameter, the introduction of which has been motivated first of
all by high space velocities of radio pulsars inferred from the
measurements of their proper motions and distances. Pulsars were
recognised as a high-velocity Galactic population soon after their
discovery in
1968~\cite{Gunn_Ostriker70}. Shklovskii~\cite{Shklovskii70} put
forward the idea that high pulsar velocities may result from
asymmetric supernova explosions. Since then this hypothesis has been
tested by pulsar observations, but no definite conclusions on its
magnitude and direction have been obtained as yet.

Indeed, the distance to a pulsar is usually derived from the dispersion
measure evaluation and crucially depends on the assumed model of the electron
density distribution in the Galaxy. In the middle of the 1990s, Lyne and
Lorimer~\cite{Lyne_Lorimer94} derived a very high mean space velocity of
pulsars with known proper motion of about 450~\kms. This value was
difficult to adopt without invoking an additional natal kick velocity of
NSs. 

The high mean space velocity of pulsars, consistent with earlier
results by Lyne and Lorimer, was confirmed by the analysis of a
larger sample of pulsars ~\cite{Hobbs_al05}. The recovered
distribution of 3D velocities is well fit by a Maxwellian distribution
with the mean value $w_0=400\pm 40 \mathrm{\ km\ s}^{-1}$ and a 1D rms
$\sigma=265 \mathrm{\ km\ s}^{-1}$.

 Possible physical reasons for natal NS kicks due to hydrodynamic
effects in core-collapse supernovae are summarised in~\cite{Lai_al01,
  Lai01}. Large kick velocities ($\sim$~500~\kms and even more) imparted to nascent NSs are generally confirmed by detailed numerical simulations (see, e.g., \cite{Nordhaus_ea11, 2013AA...552A.126W}). Neutrino effects in the strong magnetic field of a young NS
may be also essential in explaining kicks up to
$\sim$~100~\kms~\cite{Chugai84, Dorofeev_al85, Kusenko04}.
Astrophysical arguments favouring a kick velocity are also summarised
in~\cite{Tauris_vdHeuvel00}. To get around the theoretical difficulty
of insufficient rotation of pre-supernova cores in single stars to
produce rapidly spinning young pulsars, Spruit and
Phinney~\cite{Spruit_Phinney98} proposed that random off-centre kicks
can lead to a net spin-up of proto-NSs. In this model, correlations
between pulsar space velocity and rotation are possible and can be
tested in further observations.

Here we should note that the existence of some kick follows not only
from the measurements of radio pulsar space velocities, but also from
the analysis of binary systems with NSs. The impact of a kick
velocity $\sim$~100~\kms explains the precessing binary pulsar orbit 
in \epubtkSIMBAD{PSR~J0045--7319}~\cite{Kaspi_al96}. The evidence of the kick velocity
is seen in the inclined, with respect to the orbital plane,
circumstellar disk around the Be star \epubtkSIMBAD[LS~2883]{SS~2883} -- an optical component
of a binary \epubtkSIMBAD{PSR~B1259--63}~\cite{Prokhorov_Postnov97}.
Evidence for $\sim 150$~km/s natal kicks has also been inferred from the
statistics of the observed 
short GRB distributions relative to their host galaxies \cite{2014arXiv1401.7986B}.

Long-term pulse profile changes interpreted as geodetic precession are
observed in the relativistic binary pulsars
\epubtkSIMBAD{PSR~1913+16}~\cite{Weisberg_Taylor02}, \epubtkSIMBAD{PSR~B1534+12}~\cite{Stairs_al04},
\epubtkSIMBAD{PSR~J1141--6545}~\cite{Hotan_al05}, and
\epubtkSIMBAD{PSR~J0737--3039B}~\cite{Burgay_al05}. These observations indicate that
in order to produce the misalignment between the orbital angular momentum
and the neutron star spin, a component of the kick velocity
perpendicular to the orbital plane is required~\cite{Wex_al00,
  Willems_Kalogera04, Willems_al04}. This idea seems to gain
observational support from recent thorough polarisation
measurements~\cite{Johnston_al05} suggesting alignment of the
rotational axes with pulsar's space velocity. Detailed discussion of the spin-velocity alignment in young pulsars and implications for the SN kick mechanisms can be found in 
Noutsos et~al.~\cite{Noutsos_ea13}. Such an alignment
acquired at birth may indicate the kick velocity directed preferably
along the rotation of the proto-NS. For the first SN explosion in a close
binary system this would imply the kick to be mostly perpendicular to the orbital
plane. Implications of this effect for the formation and coalescence rates of binary NSs were discussed by Kuranov et al. \cite{2009MNRAS.395.2087K}.

It is worth noticing that the analysis of the formation of the double
relativistic pulsar \epubtkSIMBAD{PSR~J0737--3039}~\cite{Podsiadlowski_al05} may
suggest, from the observed low eccentricity of the system $e\simeq
0.09$, that a small (if any) kick velocity may be acquired if the
formation of the second NS in the system is associated with the collapse
of an ONeMg WD due to electron-captures. The symmetric nature of
electron-capture supernovae was discussed in~\cite{Podsiadlowski_al04}
and seems to be an interesting issue requiring further studies (see,
e.g., \cite{Pfahl_al02a, 2006AstL...32..393K} for the analysis of the
formation of NSs in globular clusters in the frame of this
hypothesis). Note that electron-capture SNe are expected to be weak
events, irrespective of whether they are associated with the core-collapse
of a star which retained some original envelope or with the AIC of a
WD~\cite{ritossa_berro_one99, kitaura_one06, dessart_onecoll06}. 
In the case of AIC, rapid rotation of collapsing object along with flux freezing and dynamo action 
can grow the WD's magnetic
field to magnetar strengths during collapse. Further, magnetar generates outflow of
the  matter and formation of a pulsar wind nebula which may be observed as a 
radio-source for a few month  \cite{2013ApJ...762L..17P}.

We also note the hypothesis of Pfahl et al.~\cite{prps02}, based
on observations of high-mass X-ray binaries with long orbital
periods ($\gtrsim$~30~d) and low eccentricities ($e<0.2$), that
rapidly rotating precollapse cores may produce neutron stars with
relatively small kicks, and vice versa for slowly rotating
cores. Then, large kicks would be a feature of stars that retained
deep convective envelopes long enough to allow a strong magnetic
torque, generated by differential rotation between the core and the
envelope, to spin the core down to the very slow rotation rate of
the envelope. A low kick velocity imparted to the second (younger)
neutron star ($<$~50~\kms) was inferred from the analysis of
large-eccentricity binary pulsar
\epubtkSIMBAD{PSR~J1811--1736}~\cite{Corongiu_al06}. The large orbital period of
this binary pulsar (18.8~d) then may suggest an evolutionary
scenario with inefficient (if any) common envelope
stage~\cite{Dewi_Pols03}, i.e., the absence of deep convective shell
in the supernova progenitor (a He-star). This conclusion can be
regarded as supportive to ideas put forward by Pfahl et al.~\cite{prps02}.
A careful investigation of radio profiles of double \epubtkSIMBAD{PSR~J0737-3039A/B} \cite{Ferdman_al13} and \textit{Fermi} detection of gamma-ray emission from the recycled 22-ms pulsar \cite{Guillemot_al13} imply its spin axis to be almost aligned with the orbital angular momentum, which lends further credence to the hypothesis that the second supernova explosion in this system was very symmetric.

Small kicks in the case of $e$-capture in the O-Ne core leading to NS
formation are justified by hydrodynamical considerations. Indeed,
already in 1996, Burrows and Hayes~\cite{Burrows_Hayes96} noted that
large scale convective motions in O and Si burning stages preceding
formation of Fe-core may produce inhomogeneities in the envelope of
protostellar core. They in turn may result in asymmetric
neutrino transport which impart kicks up to 500~\kms. Such a violent
burning does not precede the formation of O-Ne cores and then small or even zero kicks
can be expected.

In the case of asymmetric core-collapse supernova explosion, it is natural
to expect some kick  during BH
formation as well~\cite{Lipunov_al97, fbb98, py98,
  Postnov_Prokhorov00, nyp01, bel_kal02, yungelson_bh06}. The similarity
of NS and BH distribution in the Galaxy suggesting BH kicks was noted
in~\cite{jonker_nelemans04}.
An evidence for a moderate (100\,--\,200~\kms) BH kick
 has been derived from kinematics of several BH X-ray transients (microquasars): \epubtkSIMBAD{XTE~J1118+180} \cite{Fragos_al07}, \epubtkSIMBAD{GRO~1655--40} \cite{Willems_al05}, \epubtkSIMBAD{MAXI~J1659-152} \cite{Kuulkers_al13}.
However, no kicks or small ones  seem to
be required to explain the formation of other BH candidates \epubtkSIMBAD{Cyg~X-1} \cite{Mirabel_Rodriguez03}, \cite{Wong_al12},
\epubtkSIMBAD[Nova Sco 1994]{X-Nova Sco} \cite{Nelemans_al99}, \epubtkSIMBAD{V404 Cyg} \cite{Miller-Jones_al09}. Population synthesis modeling of Galactic distribution of BH binaries supports the need for
(possibly, bimodal) natal BH kicks \cite{Repetto_al12}. Janka~\cite{2013MNRAS.434.1355J} argued
that  the similarity of BH kick distribution with NS kick distribution as inferred from the
analysis by Repetto et~al.~\cite{Repetto_al12} favours BH kicks being due to  gravitational
interaction with asymmetric mass ejection (the ``gravitational tug-boat mechanism''), and disfavours neutrino-induced kicks (in the last case, by momentum conservation, BH kicks are expected to be reduced by the NS to BH mass ratio relative to the NS kicks). 
Facing current uncertainties in SN explosions  and BH formation mechanisms, it is not excluded that low-kick BHs can be 
formed without associated SN explosion due to 
neutrino asymmetry, 
while high-velocity galactic BHs in LMXBs 
analysed by Repetto et~al.~\cite{Repetto_al12} can be formed by the gravitational tug-boat mechanism suggested by Janka~\cite{2013MNRAS.434.1355J}.

To summarise, the kick velocity remains one of the 
important unknown parameters of binary evolution with NSs and BHs, and 
further phenomenological input here 
of great importance. Large kick velocities will significantly affect the 
spatial distribution of coalescing compact binaries (e.g., \cite{Kelley_ea10}) 
and BH kicks are extremely important for BH spin misalignment 
in coalescing binary BH-BH systems (e.g., Gerosa et~al.~\cite{2013PhRvD..87j4028G} 
and reference therein). 
Further constraining this parameter from various observations and
theoretical understanding of possible asymmetry of 
core-collapse supernovae seem to be of paramount importance for 
the formation and evolution of close compact binaries.


\subsubsection{Effect of the kick velocity on the evolution of a binary system}

The collapse of a star to a BH, or its explosion leading to the
formation of a NS, are normally considered as instantaneous. 
This assumption is well justified in binary systems, 
since typical orbital velocities before the explosion 
do not exceed a few hundred km/s, while most of the mass 
is expelled with velocities about several thousand km/s. 
The exploding star $M_1$ leaves
the remnant $M_\mathrm{c}$, and the binary loses a portion of its mass: 
$\Delta M = M_1 - M_\mathrm{c}$. The
relative velocity of stars before the event is 
\begin{equation}
  V_\mathrm{i} = \sqrt{G(M_1+M_2)/a_\mathrm{i}}.
\end{equation}
Right after the event, the relative velocity is
\begin{equation}
  \vec{V}_\mathrm{f} = \vec{V}_\mathrm{i}+\vec{w}.
  \label{V+W}
\end{equation}
Depending on the direction of the kick velocity vector $\vec{w}$, the
absolute value of $\vec{V}_\mathrm{f}$ varies in the interval from
the smallest $V_\mathrm{f} = |V_\mathrm{i} - w|$ to the largest
$V_\mathrm{f} = V_\mathrm{i} + w$. 
The system gets disrupted
if $V_\mathrm{f}$ satisfies the condition (see Section~\ref{A:SN})
\begin{equation}
  V_\mathrm{f} \ge V_\mathrm{i} \sqrt{\frac{2}{\chi}},
  \label{I:disrupt}
\end{equation}
where $\chi \equiv (M_1 + M_2)/(M_\mathrm{c}+M_2)$.

Let us start from the limiting case when the mass loss is 
practically zero ($\Delta M = 0$, $\chi =1$), while a non-zero kick 
velocity can still be present. This situation can be relevant to 
BH formation.
It follows from Eq.~(\ref{I:disrupt}) that, for relatively small
kicks, $w< (\sqrt{2}-1)V_\mathrm{i}$, the system always (independently of the
direction of $\vec{w}$) remains bound, while for $w> (\sqrt{2}+1)V_\mathrm{i}$
the system always decays. By averaging over equally probable
orientations of $\vec{w}$ with a fixed amplitude $w$, one can show that
in the particular case $w= V_\mathrm{i}$ the system disrupts or survives with
equal probabilities. If $V_\mathrm{f} < V_\mathrm{i}$, the semi major axis of the system
becomes smaller than the original binary separation, $a_\mathrm{f} < a_\mathrm{i}$ (see
Eq.~(\ref{A:SN:afai})). This means that the system becomes more hard
than before, i.e.\ it has a greater negative total energy than the
original binary. If $V_\mathrm{i} <V_\mathrm{f}
<\sqrt{2}V_\mathrm{i}$, the system remains bound,
but $a_\mathrm{f} > a_\mathrm{i}$. For small and moderate kicks $w
\gtrsim V_\mathrm{i}$, the
probabilities for the system to become more or less bound are
approximately equal.

In general, the binary system loses some fraction of its mass
$\Delta M$. 
In the absence of the kick, the system remains bound if $\Delta M <
M/2$ and gets disrupted if $\Delta M \ge M/2$ (see
Section~\ref{A:SN}). Clearly, a ``properly'' oriented kick velocity
(directed against the vector $\vec{V}_\mathrm{i}$) can keep the system bound,
even if it would have been disrupted without the kick. And, on the
other hand, an ``unfortunate'' direction of $\vec{w}$ can disrupt the
system, which otherwise would stay bound.

Consider, first, the case $\Delta M < M/2$. The parameter $\chi$ varies
in the interval from 1 to 2, and the escape velocity $V_\mathrm{e}$ varies in
the interval from $\sqrt{2} V_\mathrm{i}$ to $V_\mathrm{i}$. 
It follows from Eq.~(\ref{A:disrupt2}) that the binary always remains
bound if $w< V_\mathrm{e} - V_\mathrm{i}$, and always unbinds if $w>
V_\mathrm{e} + V_\mathrm{i}$.
This is a generalisation of the formulae 
derived above for the limiting case $\Delta M = 0$. Obviously, 
for a given $w$, the probability for the system to disrupt or become 
softer increases when $\Delta M$ becomes larger. Now turn 
to the case $\Delta M > M/2$. The escape velocity of the compact 
star becomes $V_\mathrm{e}<V_\mathrm{i}$. The binary is always disrupted if the
kick velocity is too large or too small: $w > V_\mathrm{i} + V_\mathrm{e}$ or
$w < V_\mathrm{i} - V_\mathrm{e}$. However, for all intermediate values of $w$,
the system can remain bound, and sometimes even more hard than before,
if the direction of $\vec{w}$ happened to be approximately
opposite to $\vec{V}_\mathrm{i}$. A detailed calculation of the probabilities
for the binary survival or disruption requires integration
over the kick velocity distribution function $f (\vec{w})$ 
(see, e.g., \cite{Brandt_Podsiadlowski95}).

\subsection{Common envelope stage}
\label{A:CE}




\subsubsection{Formation of common envelope}

Common envelopes (CE) are, definitely, the most important (and
not solved as yet) problem in the evolution of close binaries. In
the theory of stellar evolution, it was recognised quite early
that there are several situations when formation of an envelope
that engulfs entire system seems to be inevitable. It can happen 
when the mass transfer rate from the mass-losing star is so high
that the companion cannot accommodate all the accreting
matter~\cite{1970BAAS....2S.295B,1973NInfo..27...93Y,rrw74,1977PASJ...29..249N,1977ApJ...211..881W,1985MNRAS.216...37P,2013IAUS..281..209P}.
Another instance is encountered when it is impossible to keep
synchronous rotation of a red giant and orbital revolution of a
compact companion~\cite{sparks-74,alexander-76}. Because of tidal
drag forces, the revolution period decreases, while rotation
period increases in order to reach synchronism (Darwin
instability). If the total orbital momentum of the binary $J_{orb} <
3J_{RG}$, the synchronism cannot be reached and the companion
(low-mass star, white dwarf or a neutron star) spirals into the
envelope of the red giant. Yet another situation is the formation of
an extended envelope which enshrouds the system due to unstable
nuclear burning at the surface of an accreting WD in a compact
binary~\cite{1974ApJS...28..247S,1978ApJ...222..604P,1979PASJ...31..287N}.
It is also possible that a compact remnant of a supernova
explosion with ``appropriately'' directed kick velocity finds itself
in an elliptic orbit whose minimum periastron distance
$a_\mathrm{f}(1-e)$ is smaller than the stellar radius of the
companion.
Common envelope stage
appears unavoidable on observational grounds. The evidence for a
dramatic orbital angular momentum decrease  
in the course of evolution follows immediately
from observations of 
cataclysmic variables, in which a
white dwarf accretes matter from a small red dwarf main-sequence
companion, close binary nuclei of planetary nebulae, 
low-mass X-ray
binaries, and X-ray transients (neutron stars and black holes accreting
matter from low-mass main-sequence dwarfs). At present, typical separation of
components in these systems is $\sim \rs$, while the formation  
of compact stars 
requires progenitors with radii
$\sim(100\mbox{\,--\,}1000)\rs$.
\epubtkFootnote{In principle, it is possible to form a close binary
system with compact star (NS or BH) from a hierarchical triple system
without an initial CE phase due to dynamical interactions
(Kozai--Lidov mechanism), see~\cite{2013ApJ...766...64S}.}
An indirect evidence
for reality of the common envelope stage comes, for instance, from  
X-ray and FUV observations of a prototypical 
pre-cataclysmic binary \epubtkSIMBAD{V471~Tau}
showing anomalous C/N
contamination of the K-dwarf companion to white dwarf~\cite{Drake_Sarna03,2012ApJ...751...66S}. 

Let us consider the following mass-radius exponents 
which describe the response of a star to the mass loss in a binary system:
\begin{equation}
\zeta_L=\left(\frac{\partial \ln R_1}{\partial \ln M_1}\right),~~
\zeta_{th}=\left(\frac{\partial \ln R_L}{\partial \ln M_1}\right)_{th},~~
\zeta_{ad}=\left(\frac{\partial \ln R_L}{\partial \ln M_1}\right)_{ad},
\end{equation}
where $\zeta_L$ is response of the Roche lobe to the mass loss,
$\zeta_{th}$ -- thermal-equilibrium response, and
 $\zeta_{ad}$ -- adiabatic hydrostatic response.  
If  $\zeta_{ad} > \zeta_L > \zeta_{th}$, the star retains hydrostatic equilibrium, but does not remain in thermal equilibrium and mass loss occurs in the thermal time scale of the star.
If  $\zeta_{L} > \zeta_{ad}$ the star cannot retain hydrostatic equilibrium and 
mass loss proceeds in the dynamical time scale \cite{1969ASSL...13..237P}.
If both  $\zeta_{ad}$ and $\zeta_{th}$ exceed $\zeta_L$, mass loss occurs due to expansion of the star caused by nuclear burning or due to the shrinkage of 
Roche lobe owed to angular momentum loss. 

A high rate of mass overflow onto a compact star from
a normal star is always expected when the normal star goes off the
main sequence and develops a deep convective envelope. The physical
reason for this is that convection tends to make entropy constant
along the radius, so the radial structure of convective stellar
envelopes is well described by a polytrope (i.e., the equation of
state can be written as $P=K\rho^{1+1/n}$) with an index $n=3/2$. The
polytropic approximation with $n=3/2$ is also valid for degenerate
white dwarfs with masses not too close to the Chandrasekhar limit. For
a star in hydrostatic equilibrium, this results in 
inverse mass-radius relation, $R\propto M^{-1/3}$, first 
found for
white dwarfs. Removing mass from a star with a negative power of the
mass-radius relation increases its radius. On the other hand, the
Roche lobe of the more massive star should shrink in response to the
conservative mass exchange between the components. This further
increases the mass loss rate from the Roche-lobe filling star leading to
dynamical mass loss and eventual formation of a
common envelope. If the star is completely convective or completely 
degenerate, dynamically unstable mass loss 
occurs if the mass ratio of components (donor to accretor)  is $\apgt 2/3$. 
  
In a  more realistic case  when  the star is not completely convective
and has a radiative core,  certain insight may be gained by the analyses of composite 
polytropic models.      
Conditions for the onset of dynamical mass loss become less rigorous 
\cite{1972AcA....22...73P,hw87,Soberman_al97}: contraction of   
a star replaces expansion if relative mass of the core $m_c > 0.214$ and in a binary with mass ratio of components close to 1, 
a Roche-lobe filling star with deep convective envelope may remain 
dynamically stable if  $m_c > 0.458$. As well, stabilising effect upon mass loss may have mass and momentum loss from the system,
if it happens in a mode which results in increase of specific angular momentum of the binary (e.g., in the case of
isotropic reemission by a more massive component of the system in CVs or UCXBs).  
 
Criteria for thermal or dynamical mass loss upon RLOF and  formation of a common envelope 
need systematic exploration of response of stars to mass removal in different evolutionary stages and at different  rates.  The response also depends on the 
mass of the star. While computational methods for such an analyses are elaborated
\cite{2010ApJ...717..724G,2010Ap&SS.329..243G}, 
calculations of respective grids of models with full-fledged evolutionary codes
at the time of writing of this review were not completed.

\subsubsection{``Alpha''-formalism}
\label{ss:alphaCE}

Formation of the common envelope and evolution of a binary inside the former is a 3D hydrodynamic process which
may also include nuclear reactions and current understanding of the process as a whole, as well as computer 
power, are not sufficient for solution of the problem 
(see for detailed discussion \cite{2013AARv..21...59I}).
Therefore, the outcome of this stage is, most commonly,  evaluated in a simplified 
way, based on the balance of the binding energy of the stellar envelope 
and the binary orbital energy, following independent suggestions of van den 
Heuvel~\cite{1976IAUS...73...35V}, Tutukov and Yungelson~\cite{ty79},
and Webbink~\cite{Webbink84} and commonly named after the later author as ``Webbink's''
or ``$\alpha$''-formalism. 
The orbital evolution of the compact star
$m$ inside the envelope of the donor-star $M_1$ is driven by the
dynamical friction drag \cite{pac76}. This leads to a gradual spiral-in process of
the compact star. 
%
%
The above-mentioned energy condition 
may be written as 
\begin{equation}
  \frac{GM_1(M_1-M_{c})}{\lambda R_\mathrm{L}} =
  \alpha_\mathrm{ce}
  \left( \frac{GmM_{c}}{2a_\mathrm{f}}-\frac{GM_1m}{2a_\mathrm{i}} \right),
  \label{A:CE:eq}
\end{equation}
where $a_\mathrm{i}$ and $a_\mathrm{f}$ are the initial and the final orbital
separations, 
$\alpha_\mathrm{ce}$ is common envelope parameter~\cite{ty79,1988ApJ...329..764L} which describes efficiency of
expenditure of orbital energy on expulsion of the envelope
and $\lambda$ is a numerical coefficient that depends on the
structure of the donor's envelope, introduced by de~Kool et~al.~\cite{dek90},
 $R_\mathrm{L}$ is the Roche lobe radius of
the normal star that can be approximated, e.g., by 
Eqs.~(\ref{eq:roche_eggl}) or (\ref{eq:roche_kopal}).
%
From Eq.~(\ref{A:CE:eq}) one derives 


%
\begin{equation}
  \frac{a_\mathrm{f}}{a_\mathrm{i}} =
  \frac{M_c}{M_1} \left( 1+\frac{2a_\mathrm{i}}
  {\lambda\ace R_\mathrm{L}}
  \frac{M_1-M_c}{m} \right)^{-1} 
\lesssim
  \frac{M_c}{M_1} \frac{m}{\Delta M},
  \label{A:CE:afai}
\end{equation}
where $\Delta M=M_1-M_c$ is the mass of the ejected envelope.
For instance, the mass $M_{c}$ of the helium core of a massive star can be
approximated as~\cite{ty73}
\begin{equation}
  M_\mathrm{He} \approx 0.1 (M_1/M_\odot)^{1.4}.
  \label{A:MHe}
\end{equation}
Then, $\Delta M \sim M$ and, e.g., in the case of $m \sim \ms$ the
orbital separation during the common envelope stage can decrease as
much as by factor 40 and even more.

This treatment does not 
take into account possible transformations of the binary components during the CE-stage.
In addition, the pre-CE evolution of the binary components may  
be important for the onset and the outcome of the CE, as calculations of synchronisation of 
red giant stars in close binaries carried out by Bear and Soker \cite{2010NewA...15..483B} indicate.
The outcome of common envelope stage is considered as a merger of components
if $a_f$ is such
that the donor core comes into contact with the companion. 
Otherwise, it is assumed that the core and companion 
form a detached system with the orbital separation $a_f$.  
Note an important issue raised recently by Kashi and Soker~\cite{2011MNRAS.417.1466K}: while Eq.~(\ref{A:CE:afai}) can formally imply that the system 
remains detached at the end 
of the CE stage,
in fact, some matter 
of the envelope can not reach the escape velocity and remain bound to the system and form a 
circumbinary disk. Angular momentum loss due to the interaction with the disk may result in further reduction of binary separation and
merger of components. This may influence the features  of such  stellar  populations as close binary WD, hot subdwarfs, cataclysmic variables) 
 and the rate of \sna (see Section~\ref{section:wd_formation} below). 


In the above formalism for the common envelope, the outcome of the  CE stage 
depends, in fact, on the
product of two parameters: $\lambda$, which is the measure of the binding
energy of the envelope to the core prior to the mass transfer in a binary
system, and $\ace$, which is the common envelope efficiency itself.
Evaluation of both parameters suffers from large physical uncertainties.
For example, for $\lambda$-parameter the most debatable issues are accounting of
the internal energy in the binding energy of the envelope and definition of core/envelope 
boundary itself. Some authors argue (see e.g., \cite{2011ApJ...731L..36I}) that 
enthalpy rather than internal energy should be included in the calculation
of $\lambda$, which seems physically justifiable for convective
envelopes\epubtkFootnote{Indeed, to put a gas element of unit mass with specific volume $v$
at a distance $R$ around mass $M$ from infinity, the work equal to the
gravitational potential $\phi$ should be done; but then one should
heat up the gas element, which is equivalent to give it the specific
internal energy $\epsilon$, and also do $Pv$-work to empty space in
the already present gas, hence $\phi+H$ is relevant in calculating the
binding energy of an isentropic envelope, where $dH=Pdv$.}.
We refer the interested reader to the detailed discussion 
in~\cite{2011MNRAS.411.2277D,2013AARv..21...59I}. 

There exist several sets of fitting formulae for
$\lambda$ \cite{2010ApJ...716..114X,2010ApJ...722.1985X}  or binding 
energy of the envelopes \cite{2011ApJ...743...49L} 
based on detailed evolutionary computations for a range of stellar models 
at different evolutionary stages. But we note, that in both studies the 
same specific evolutionary code {\sl ev} \cite{pol_98} was used and 
the core/envelope interface at location in the star with hydrogen abundance
in the hydrogen-burning shell $X=0.1$ was assumed;
  in our opinion, the latter assumption 
is not justified neither by any physical assumptions nor by evolutionary computations 
for Roche-lobe filling stars. 
Even if the inaccuracy of the latter assumption is 
neglected, it is possible to use these formulae only in population 
synthesis codes
based on the evolutionary tracks obtained by   {\sl ev} code.     
   
Our test calculations do not confirm the recent claim 
\cite{2011ApJ...730...76I} that core/envelope interface, which also defines the masses of the remnants
in all cases of mass-exchange, is close to the radius of the sonic velocity maximum 
in the hydrogen-burning shell  of the mass-losing star at RLOF.
As concerns parameter $\ace$, the most critical issue is whether
the sources other than orbital energy can contribute to the energy balance, which include, 
in particular, recombination
energy \cite{2003MNRAS.343..456S}, nuclear energy \cite{il93}, or 
energy released by accretion onto the compact star. 
It should then be noted that sometimes imposed  to Eq.~(\ref{A:CE:eq}) 
restriction $\ace\leq1$ can not
reflect the whole complexity of the processes occurring in common envelopes.  
Although full-scale
hydrodynamic calculations of a common-envelope evolution exist,
(see, e.g., \cite{2002ASPC..263...81T,2003RMxAC..18...24D,2008ApJ...672L..41R,2012ApJ...746...74R,2012ApJ...744...52P}
and references therein), we stress again that 
the process is still very far from comprehension, especially, at the final  
stages. 

\textit{Actually,  the state of the common envelope problem is  currently such that 
 it is possible only to estimate the product \al\ by modeling specific systems
or well defined samples of objects corrected for observational selection effects,} like it is done in \cite{2013AA...557A..87T}.



\subsubsection{``Gamma''-formalism}
\label{ss:gammaCE}

Nelemans et.~al.~\cite{Nelemans_al00} noted that the 
$\alpha$-formalism failed to reproduce parameters of several close
binary  white dwarfs known circa 2000. In particular, for close helium
binary white dwarfs  
with known masses of both components, one can 
reconstruct the evolution of the 
system ``back'' to the pair of main-sequence progenitors of components, since
there is a unique relation between the mass of a white dwarf and the
radius of its red giant progenitor~\cite{rrw74} which is almost independent of the total mass of the star.
Formation  of close binary white dwarfs should
definitely involve spiral-in phase in the common envelope during the second episode
of mass loss (i.e., from the red
giant to the white dwarf remnant of the original primary in the
system).  In the observed systems, mass ratios of components tend to concentrate to
$M_{\mathrm{bright}}/M_{\mathrm{dim}} \simeq 1$.
This means, that, if the first stage of mass transfer occurred through a common envelope, the separation of components did not reduce much, contrary to what is expected from Eq.~(\ref{A:CE:afai}). The values of $\al$ which appeared necessary for 
reproducing the observed systems turned out to be negative, 
which means that simple energy conservation law 
(\ref{A:CE:afai}) in this case is violated.
Instead, Nelemans et al. suggested  that the  first stage of mass exchange, between a giant and a main-sequence star with comparable masses ($q\ge0.5$), can be described by what they called  ``$\gamma$-formalism'',
in which not the energy but the angular momentum $J$
is balanced and conservation of energy is implicitly implied, though formally 
this requires $\al > 1$:
\begin{equation}
  \frac{\delta J}{J} = \gamma \frac{\Delta M}{M_\mathrm{tot}}\,.
  \label{ce_gamma}
\end{equation}
Here $\Delta M$ is the mass 
lost by the donor,
$M_\mathrm{tot}$ is the total
mass of the binary system before the common envelope, and $\gamma$ is a
numerical coefficient.   
Similar conclusions were later reached by Nelemans and Tout~\cite{nelemans_tout05}
and van der Sluys et al.~\cite{sluys_wd06} who analysed larger samples of binaries and used
detailed fits to 
evolutionary models instead of simple ``core-mass -- stellar radius'' relation applied by 
Nelemans et al.~\cite{Nelemans_al00}.
It turned out 
that combination of $\gamma=1.75$ for the first stage of mass exchange with $\al=2$ for 
the second episode of mass exchange enables, after taking into account selection effects, a satisfactory model 
for population of close binary WD with known masses. In fact, findings of Nelemans et al. 
and van der 
Sluys et al. confirmed  the need for a loss of mass and momentum from the system
to explain observations of binaries mentioned in Section~\ref{A:mass_transf}.
(Note that $\gamma=1$ corresponds to the loss of the angular momentum
by a fast stellar wind, which always increases the 
orbital separation of the binary.) We stress that no physical process behind 
$\gamma$-formalism has been suggested as yet, 
and the model remains purely phenomenological and should 
be further investigated.
\epubtkFootnote{Attempts to apply the $\gamma$-formalism
to \emph{both} stages of mass exchange ($\gamma,\gamma$-formalism),
which failed to reproduce evolution of close binary WD and 
criticized the $\gamma$-formalism, are unjustified
extrapolation of the limited application of the formalism suggested by~\cite{Nelemans_al00}.}

We should note that ``$\gamma$-formalism'' was introduced under 
the assumption that stars with deep 
convective envelopes (giants) always lose mass unstably, i.e., \mdot\ is high.
As mentioned above, if the mass exchange is nonconservative,  
the associated angular momentum loss 
can increase the specific orbital angular momentum of the system $\propto \sqrt{a}$
and hence the orbital separation $a$.
Then the increasing Roche lobe 
can accommodate the expanding donor,  
and the formation of a common envelope can be 
avoided. In fact, this happens if part of the energy released by accretion 
is used to expel the matter from the vicinity
of accretor (see, e.g., computations presented by King and Ritter~\cite{kr99}, Beer et~al.~\cite{2007MNRAS.375.1000B}, Woods et~al.~\cite{2012ApJ...744...12W} 
who assumed that  reemission occurs).  It is not excluded that such 
nonconservative mass-exchange with mass and momentum loss from the system may be the process 
underlying the ``$\gamma$-formalism''.   
     
Nelemans and Tout~\cite{nelemans_tout05}, Zorotovic et~al.~\cite{2010AA...520A..86Z} and
De Marco et~al.~\cite{2011MNRAS.411.2277D}
attempted to estimate \ace\ using samples of close WD+M-star binaries.
It turned out that their formation can be explained both if  $\al >0$ in Eq.~(\ref{A:CE:eq}) or
$\gamma\approx1.5$ in Eq.~(\ref{ce_gamma}). 
This is sometimes considered as an argument against 
the $\gamma$-formalism. But we note 
that in progenitors of WD+M-star binaries the companions to giants are low-mass  small-radius 
objects, which are quite different from $M\geq\ms$ companions to WD in progenitors of 
WD+WD binaries and, therefore, the energetics
of the evolution in CE in precursors of two types of systems can differ. In a recent 
population synthesis study of the 
post-common-envelope binaries aimed at comparing to SDSS stars
and taking into account selection effects, Toonen and Nelemans 
\cite{2013AA...557A..87T} concluded that the best fit to the observations can be obtained
for small universal $\al=0.25$ without the need to invoke 
the $\gamma$-formalism. However, they also note
that for almost equal mass precursors of binary WD the widening of the orbit in the course of mass 
transfer is needed (like in  $\gamma$-formalism), while the formation of low mass-ratio WD+M-star
systems requires diminishing of the orbital separation (as suggested by $\alpha$-formalism).

\subsection{Other notes on the CE problem}

A particular case of common envelopes (``double spiral-in'') occurs when both components are evolved.
It was suggested~\cite{nyp_01} that common envelope is formed by the envelopes of both components of the system and binding energy of both of them should
be taken into account, i.e., the outcome of CE must be found from solution of an equation
\begin{equation}
 E_{\mathrm{bind}, 1}+E_{\mathrm{bind},2}=\ace(E_{\mathrm{orb},i}-E_{\mathrm{orb},f}).
  \label{eq:double_ce}
\end{equation}

Formulations of the common envelope equation
different from Eq.~(\ref{A:CE:eq}) are met in the literature
(see, e.g., \cite{2011MNRAS.411.2277D} for a review);
$a_\mathrm{f}/a_\mathrm{i}$ similar to the values produced by
Eq.~(\ref{A:CE:eq}) are then obtained for different
$\alpha_{\mathrm{ce}}\lambda$ values.

Common envelope events are expected to be rare ($\lesssim
0.1\pyr$~\cite{ty02}) and short-lasting ($\sim$~years). Thus,
the binaries at the CE-stage itself are difficult to be observed, 
despite the energy released at this stage should be comparable to the 
binding
energy of a star, and the evidence for them comes from the very existence of
binary compact objects, as described above. Recently, it was
suggested that luminous red transients --- the objects with 
peak luminosity intermediate between the brightest Novae and
\sna (see, e.g., \cite{2012PASA...29..482K}) --- can be associated
with CE \cite{2013Sci...339..433I,2013A&A...555A..16T}.

We recall also that the stability and timescale of mass-exchange in a
binary depends on the mass ratio of components $q$, the structure of
the envelope of Roche-lobe filling star, and possible stabilising
effects of mass and momentum loss from the system~\cite{tfy82, hw87,
  yl98, hknu99, Han_al02, ivanova_taam04, fty04,
  bunning_ritter_me06,2012ApJ...744...12W}. We note especially recent
claims \cite{2011ApJ...739L..48W, 2012ApJ...760...90P} that the local
thermal timescale of the super-adiabatic layer existing over
convective envelope of giants may be shorter than the donor's
dynamical timescale. As a result, giants may adjust to a high mass
loss and retain their radii or even contract (instead of dramatically
expanding, as expected in the adiabatic approximation
\cite{2010ApJ...717..724G, 2010Ap&SS.329..243G}). As well, response of
stars to mass loss evolves in the course of the latter depending on
whether super-adiabatic layer may be removed.
Recent discovery of young helium WD companions to blue stragglers
in wide binaries (\porb\ from 120 to 3010 day) in open cluster NGC~188
 \cite{2014arXiv1401.7670G} supports the idea that in certain cases 
RLOF in wide systems may be stable).
These considerations,
which need further systematic exploration, may change the ``standard''
paradigm of stability of mass exchange formulated in this Section.


\newpage
\section{Evolutionary Scenario for Compact Binaries with Neutron Star
  or Black Hole Components} 
\label{section:evolution_high_mass}

\subsection{Compact binaries with neutron stars}
\label{sec:compactNS}

Compact binaries with NS and BH components are descendants of
initially massive binaries with  
$M_1 \gtrsim 8\,M_{\odot}$.
The  scenario of evolution of massive binaries from a pair of
main-sequence stars to a relativistic binary consisting of NS or BH
produced by core-collapse SN 
was independently elaborated by Tutukov and
Yungelson~\cite{ty73,mty76}%
\epubtkFootnote{Tutukov and Yungelson paper~\cite{ty73} was published
  in 1973 prior to discovery of the first binary pulsar
  \epubtkSIMBAD{PSR~1913+16} in 1974~\cite{1974IAUC.2704....1T,
    Hulse_Taylor75} and, though the evolutionary path was traced to a
  binary NS (or BH), since no such objects were known at that time, it
  was suggested that all pairs of NS are disrupted at the second NS
  formation.}
and van den Heuvel et~al.~\cite{vdHeuvel_deLoore73,
  Flannery_vdHeuvel75}. This scenario is fully confirmed by more than
30 years of astronomical observations and is now considered as
``standard''.

\epubtkImage{fignew-lrr3.png}{
  \begin{figure}[htbp]
      \centerline{\includegraphics[scale=0.7]{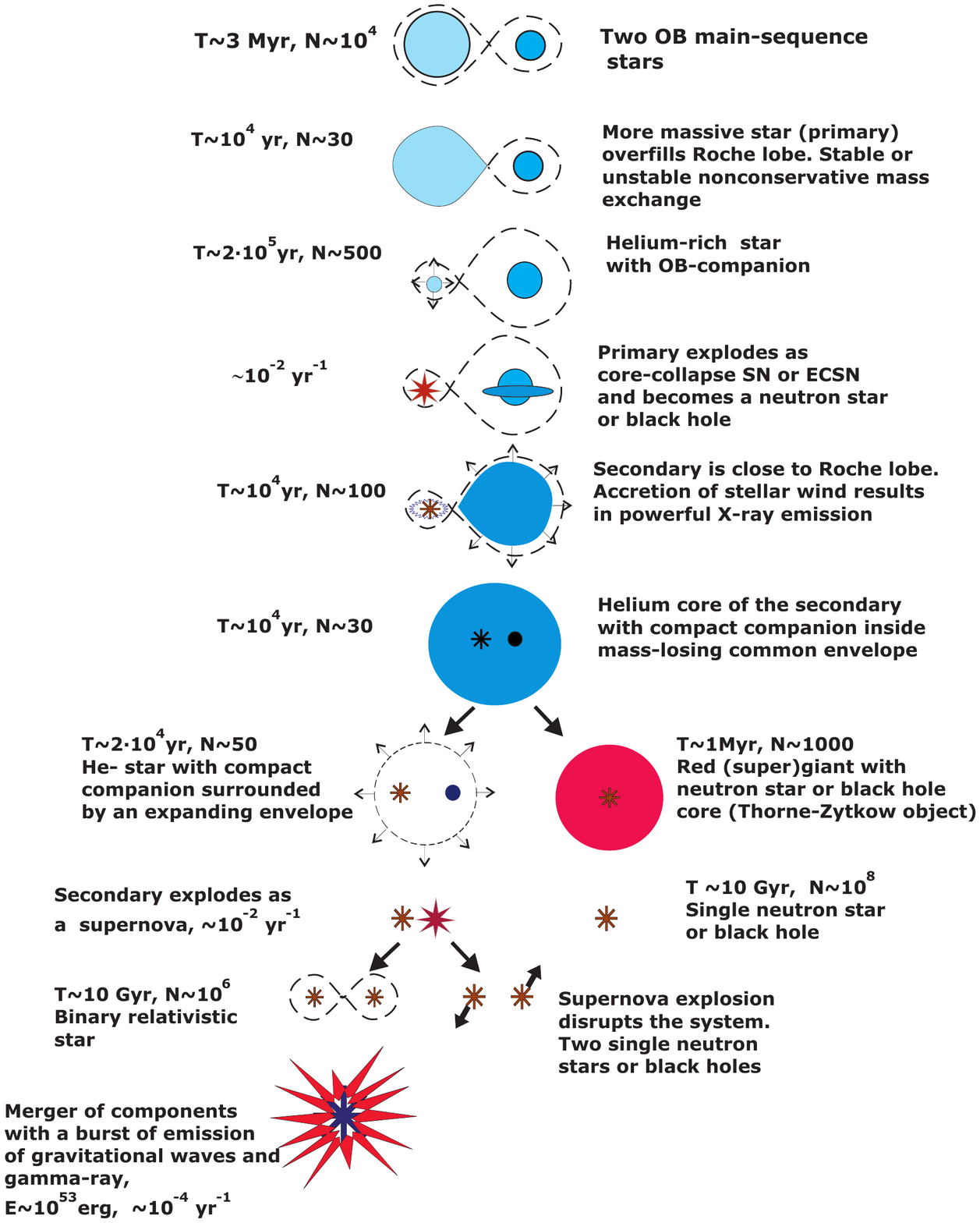}}
    \caption{Evolutionary scenario for formation of neutron
      stars or black holes in close binaries. T is typical time scale of an evolutionary stage, 
N is the estimated number of objects in the given evolutionary stage.}
    \label{figure:massive_flow}
  \end{figure}
}

Certain modifications to the original scenario were introduced after Pfahl etal.~\cite{prps02} paid attention to the fact that Be/X-ray binaries harbouring
 NS fall into two classes: objects with highly eccentric orbits
($e=0.3\mbox{\,--\,}0.9$) and $\porb \geq 30$\,day and objects with $e < 0.2$
and $\porb \leq 30$\,day. Almost simultaneously, van den
Heuvel~\cite{2004ESASP.552..185V} noted that 5 out of 7 known double neutron
stars (DNS) and one massive WD\,+\,NS system in the Galactic disk have 
$e<0.27$
and the measured or estimated masses of second-born neutron stars in most of
these systems are close to $1.25\ms$ (for recent confirmation of these
observations see, e.g.,
\cite{2010ApJ...719..722S,2011Natur.479..372K,2011BASI...39....1V,Ferdman_al13}).
Neutron star would have $1.25\ms$ mass if it is a product of an electron-capture
supernova (ECSN), which was accompanied by the loss of binding energy equivalent
to $\approx0.2\ms$. These discoveries lead van den
Heuvel~\cite{2004ESASP.552..185V} to inclusion of ECSN in the evolutionary
scenario for formation of DNS. Specifics of NS formed via ECNS is also their low natal kicks (typically, a Maxwellian velocity distribution with $\sigma
=(20\mbox{\,--\,}30)$\,\kms\ is inferred instead of $\sigma \sim 200$\,\kms). 

In Figure~\ref{figure:massive_flow} we present the ``standard'' evolutionary scenario for formation of DNS. Other versions of 
this scenario may differ by the types of supernovae occurring in them or in the order of their formation, which is defined by the 
initial masses of components and orbital period of the binary.   

It is convenient to consider subsequent stages of the evolution of a binary system 
according to the physical state of its
components, including phases of mass exchange between them.
\begin{enumerate}
\item Initially, the pair of high-mass OB main-sequence stars is detached and stars 
are inside their Roche lobes. Tidal interaction is very effective
  and the possible initial eccentricity vanishes before the primary
  star $M_1$ fills its Roche lobe. The duration of this stage is
  determined by the hydrogen burning time of the primary, more massive component, and  typically
  is $ < 10$\,Myr (for massive main-sequence stars, the time
  of core hydrogen burning is $t_{\mathrm{nucl}}\propto M^{-2}$). The star burns
  out hydrogen in its central parts, so that a dense central helium
  core with mass $M_{\mathrm{He}}\simeq 0.1 (M/M_{\odot})^{1.4}$ forms by the
  time when the star leaves the main sequence. The expected number of
  such binaries in the Galaxy is about 10\super{4}.
\item After the core hydrogen exhaustion, the primary leaves the 
main-sequence and starts to expand rapidly. When its radius approaches
  the Roche lobe, 
mass transfer onto the secondary, less massive star which still resides on the 
main-sequence, begins. 
  Depending on the masses of components and evolutionary state of the donor, the mass-transfer may   
proceed via non-conservative but stable Roche-lobe overflow (RLOF) or via a common envelope. 
Even if the common envelope is avoided and the first mass exchange event proceeds on the thermal time scale of the donor 
  $\tau_\mathrm{KH}\approx GM_1^2/R_1L_1$, its
  duration for typical stars   is rather  short, of the order of 10\super{4}~yr, so only several dozens of
  such binaries are expected to be present in the Galaxy.  

\item  
       Mass transfer ends when most of the primary's hydrogen envelope is lost,
       so a naked helium core is left. This core can be observed as a
       Wolf--Rayet (WR) star with an intense stellar wind if its mass exceeds
       $(7\mbox{\,--\,}8)\,M_{\odot}$~\cite{nl00, fadeyev03a, fadeyev04}. The
       duration of the WR stage is about several 10\super{5}~yr, so the Galactic
       number of such binaries should be several hundreds.
  
  During the mass-exchange episode  
  the secondary star acquires large angular
  momentum carried by the infalling matter, so that its outer envelope can
  be spun up to an angular velocity close to the limiting (Keplerian)
  value. Such massive rapidly rotating stars are observed as
  Be-stars. 

\item 
Stars more massive than $\simeq8\ms$ 
end their evolution by formation of a NS. ZAMS mass range $8\mbox{\,--\,}12(\pm1)\ms$,
is a ``transitional'' one in which NS are formed via ECSN  at the lower masses and via core collapses at the higher masses, as discussed in more detail in 
Section~\ref{section:introduction}.   

Supernovae associated with massive naked He stars (almost devoid of H-envelopes) are usually
associated with SN~Ib/c. 
The inferred Galactic type SN~Ib/c rate is 
$(0.76\pm0.16)\times10^{-2}\pyr$~\cite{2011MNRAS.412.1473L}. At least half of them
should be in binaries. 
As mentioned in Section~\ref{section:introduction} ECSN may be progenitors of 
the faintest type~II-P supernovae,
because they produce only little amount of radioactive Ni. 
Expected properties of core-collapse SN and ECSN are compared in 
Table~\ref{tab:comp_sn} \cite{2010ApJ...719..722S}. Peculiarly, the historical Crab SN
in our Galaxy is suggested  to be a  ECSN \cite{2013MNRAS.434..102S}.

\begin{table}[htb]
\renewcommand{\arraystretch}{1.2}
\caption[Comparison of Fe-Core Collapse and e-Capture
  Supernovae.]{Comparison of Fe-Core Collapse and e-Capture
  Supernovae~\cite{2010ApJ...719..722S}.}
\label{tab:comp_sn}
  \centering
  \small{
  \begin{tabular}{lll}
    \toprule
Properties & Iron Core Collapse & ECSN\\  
\midrule
\emph{Supernova Properties} & & \\
Explosion energy & $\sim 10^{51}\,$ergs & $\aplt 10^{50}\,$ergs  \\
Ejecta & rich in heavy elements (Fe, Si, O) & few heavy elements \\
&& \\
\emph{Neutron Star Properties} & & \\
Masses & range of masses & characteristic mass $\simeq 1.25\,M_{\odot}$ \\
Neutron star kick & large standard kick ($\sigma\simeq 265\,$km\,s$^{-1}$) & low kick \\
&& \\
\emph{Binary Properties} & & \\
Occurrence & single or binaries & preferentially in binaries \\
Eccentricity & high & low \\
Recycled pulsar spin & misaligned-aligned with orbit & aligned with orbit  \\
& (e.g., geodetic precession) & \\
 \bottomrule
  \end{tabular}}
  \renewcommand{\arraystretch}{1.0}
\end{table}

Disruption of the binary due to the second SN in the system is very likely (e.g., if the mass lost during the symmetric SN explosion
  exceeds 50\% of the total mass of the pre-SN binary, or it is even smaller
  in the presence of the kick; see Section~\ref{A:SN}
  above). Population synthesis estimates show that (4\,--\,10)\% of initial binaries
  survive the first core-collapse SN explosion in the system, depending on
  the assumed kick distribution 
\cite{1997AA...318..812D,py98,1998astro.ph..1127Y,2009MNRAS.395.2087K}.
Some runaway Galactic OB-stars must have been formed in this way. 
Currently,
only one candidate O-star with non-interacting NS companion is known, thanks to multi-wavelength observations
 -- HD~164816, a late O-type  spectroscopic
binary \cite{2012MNRAS.427.1014T}. 
Null-results of earlier searches for similar objects
\cite{1996AJ....111.1220P,1996ApJ...461..357S}, though being dependent on assumptions on the beaming factor of pulsars and their magnetic field evolution,
are consistent with very low fraction of surviving systems, but may be
also due to obscuration of radio emission by the winds of massive stars.

\item If the system survives the first SN explosion, a rapidly
  rotating Be star in orbit with a young NS appears. Orbital evolution
  following the SN explosion is described above by
  Eqs.~(\ref{A:SN:afai}\,--\,\ref{A:SN:symm-ecc}). The orbital
  eccentricity after the SN explosion is high, so enhanced accretion onto
  the NS occurs at the periastron passages. Most of about 100 Galactic
  Be/X-ray binaries~\cite{Raguzova_Popov05} 
  may be  formed in this way.   
Post-ECSN binaries have a larger chance for survival thanks to low kicks. It is possible that a significant fraction 
of Be/X-ray binaries belong to this group of objects. 
  The duration of Be/X-ray stage depends on the binary parameters, but in
  all cases it is limited by the time left for the (now more massive)
  secondary to burn hydrogen in its core.
    
 An important parameter of NS evolution is the surface magnetic field
  strength. In binary systems, magnetic field, in combination with NS
  spin period and accretion rate onto the NS surface, determines the
  observational manifestation of the neutron star (see~\cite{Lipunov92}
  for more detail). Accretion of matter onto the NS can reduce the surface
  magnetic field and spin-up the NS rotation (pulsar
  recycling)~\cite{BK_Komberg74, Romani90, Romani95, B-K06}.
\item Evolving  secondary expands to engulf the NS in its own turn. Formation of a common
envelope is, apparently, inevitable due to the large mass ratio of components.
  The common envelope
  stage after $\sim$~10\super{3}~yr ends up with the
  formation of a WR star with a compact companion surrounded by an
  expanding envelope (\epubtkSIMBAD{Cyg~X-3} may serve as an example), or the NS
  merges with the helium core during the common envelope to form a
  still hypothetical Thorne--\.{Z}ytkow (TZ) object \cite{tz75}. 

Cygnus~X-3 -- a WR star with black hole or neutron star companion is unique in the Galaxy,
because of high probability of merger of components in CE, short lifetime of surviving
massive WR-stars and high velocity ($\sim$~1000~\kms) of their stellar winds which prevents formation of accretion 
disks~\cite{ey98,lyhnp05,2012ApJ...748..114L}. On the other hand, it is suggested that in the Galaxy
may exist a population of $\sim$~100 He-stars of mass between  1 and $7\ms$ with relativistic companions which do not 
reveal  themselves, because these He-stars do not have strong enough winds \cite{lyhnp05}.  
 
 The 
possibility of existence of TZ-stars
  remains unclear (see~\cite{Barkov_al01}).
It was suggested that the merger products first become supergiants, but rapidly lose their envelopes due to heavy winds and become WR stars.
Peculiar Wolf--Rayet stars of WN8 subtype were suggested as observed 
counterparts of them \cite{2002ASPC..263..123F}. These stars tend to have large spatial velocities, overwhelming majority of them are single and they are the most
variable among all single WR stars. Estimated observed number of them in the Galaxy is $\sim$~10.  
Single (possibly, massive) NS or BH should descend from  them.

  A note should be made concerning the phase when a common envelope
  engulfs the first-formed NS and the core of the
  secondary. Colgate~\cite{colgate71} and Zel'dovich
  et~al.~\cite{zeld72} have shown that hyper-Eddington accretion onto
  a neutron star is possible if the gravitational energy released in
  accretion is lost by neutrinos. Chevalier~\cite{chevalier93}
  suggested that this may be the case for the accretion in common
  envelopes. Since the accretion rates in this case may be as high as
  $\sim 0.1 \, M_{\odot} \mathrm{\ yr}^{-1}$, the NS may collapse
  into a BH inside
  the common envelope. An essential caveat is that the accretion in
  the hyper-Eddington regime may be prevented by the angular momentum of
  the captured matter. Magnetic field of the NS may also be
  a complication. The possibility of hyper-critical accretion still
  has to be studied. Nevertheless, implications of this hypothesis for
  different types of relativistic binaries were explored in great
  detail by Bethe and Brown and their coauthors (see,
  e.g., \cite{brown_bh00} and references therein). Also, the possibility
  of hyper-Eddington accretion was included in several population
  synthesis studies with evident result of diminishing the population
  of NS\,+\,NS binaries in favour of neutron stars in pairs with low-mass
  black holes (see, e.g., \cite{py98, bel_kal02}). 
  
Recently, Chevalier \cite{2012ApJ...752L...2C} pointed to the
possible connection of CE with neutron star with some luminous and
peculiar type~IIn supernova. The neutron star engulfed by the massive
companion may serve as a trigger of the SN explosion in dense
environments due to a violent mass-loss in the preceding CE
phase. However, presently our understanding of the evolution of
neutron stars inside CE is insufficient to test this hypothesis.

\item The secondary He-star ultimately explodes as a 
supernova
  leaving behind a double NS binary, or the system  disrupts to
  form two single high-velocity NSs or BHs. Even for a symmetric SN
  explosion the disruption of binaries after the second SN explosion could
  result in the observed high average velocities of radio pulsars (see
  Section~\ref{kicks} above). In the surviving close binary NS system,
  the older NS is expected to have faster rotation velocity (and
  possibly higher mass) than the younger one because of the recycling
  at the preceding accretion stage. The subsequent orbital evolution
  of such double NS systems is entirely due to GW emission (see
  Section~\ref{A:GW_evol}) and ultimately leads to the coalescence of
  the components.
\end{enumerate}

Detailed studies of possible evolutionary channels which produce
merging binary NS can be found in the literature, e.g.,
\cite{Tutukov_Yungelson93a, ty93b,Lipunov_al97,py98, Bagot97,
  1997AA...318..812D, Wettig_Brown96, bel_kal02, Hurley_al02,
  Ivanova_al03, voss_taurisns03, Dewi_vdHeuvel04, Willems_Kalogera04,
  2007ARep...51..308B,2007PhR...442...75K, 2008MNRAS.388..393K,
  2010MNRAS.406..656K, 2011MNRAS.413..461O,
  2011AA...527A..70K,2011MNRAS.413..461O, Dominik_al12,
  Ferdman_al13}). We emphasize that above-described scenario applies
only to \emph{close}
binaries, that have components massive enough to produce ECSN or
core-collapse SN, but not so massive that the loss of hydrogen
envelope by stellar wind and associated widening of the orbit via the
Jeans mode of mass ejection may prevent RLOF by the primary. This
limits the relevant mass range by
$M_1\aplt(40\mbox{\,--\,}50)\,\ms$ \cite{mpty79,vanbev93}. 

There exists also a population of NSs accompanied by low-mass
$[\sim (1\mbox{\,--\,}2)\,M_{\odot}]$ companions. A scenario similar to the one
presented in Figure~\ref{figure:massive_flow} may be sketched for them
too, with the difference that the secondary component stably transfers
mass onto the companion (see, e.g., \cite{ity95b, kw96, kw98,
  tf_bh02}). This scenario is similar to the one for low- and
intermediate-mass binaries considered in
Section~\ref{section:wd_formation}, with the WD replaced by a NS or a
BH. Compact low-mass binaries with NSs may be dynamically formed in 
dense stellar environments, for example in globular clusters. The
dynamical evolution of binaries in globular clusters is beyond the
scope of this review; see, e.g., \cite{Benacquista_LRR02} and
\cite{B-K06} for more detail and further references.

At the end of numerical modeling of the evolution of massive binaries outlined above,
one arrives to the population of binary NS, the main parameter of which is 
distribution over orbital periods \porb\ and eccentricities $e$. 
In the upper panel of Figure~\ref{fig:nspe} we show, as an example, a model of  the probability distribution 
of Galactic NS\,+\,NS binaries with $\porb\leq10^{4}$\,day in \porb-$e$ plane at birth; 
 the lower panel of the same figure shows model probability distribution
of the present day numbers of NS\,+\,NS systems younger than 10~Gyr \cite{py98}. 
It is clear, that a significant fraction of systems born with 
$\porb\leq1$\,day merge. A detailed discussion of general-relativity 
simulations of NS\,+\,NS mergers including effects of
magnetic fields and  micro-physics is 
presented in \cite{2012LRR....15....8F}.  

\epubtkImage{figure_nspe.png}{
  \begin{figure}[htbp]
    \centerline{\includegraphics[width=0.8\textwidth]{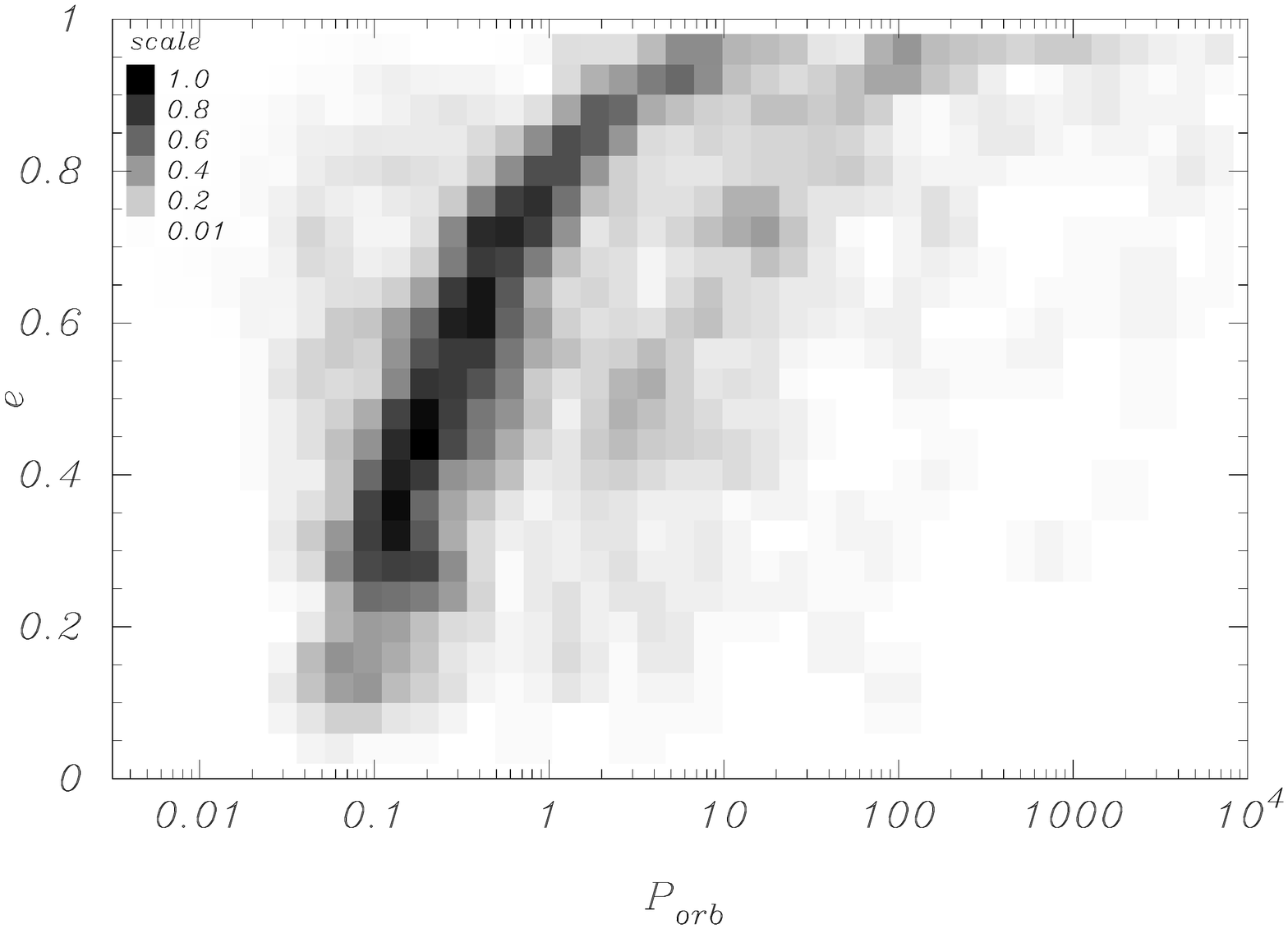}}
    \centerline{\includegraphics[width=0.8\textwidth]{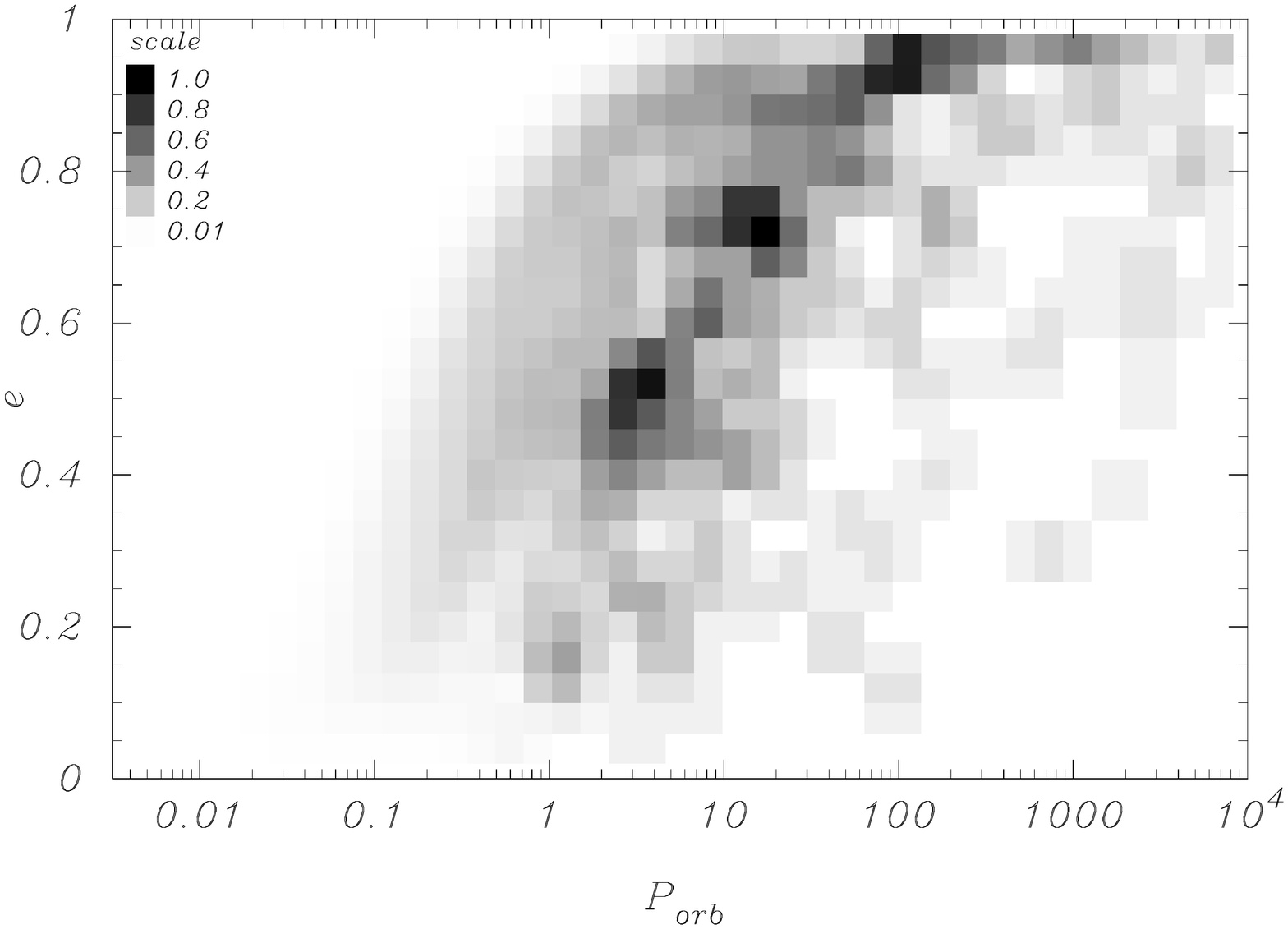}}
    \caption{\emph{Upper panel:} the probability distribution for the
      orbital parameters of the NS\,+\,NS binaries with $\porb\leq10^4$
      day at the moment of birth. The darkest shade corresponds to a
      birthrate of $1.2 \times 10^{-5}\pyr$. \emph{Lower panel:} the
      probability distribution for the present day orbital parameters
      of the Galactic disc NS\,+\,NS binaries younger than 10\,Gyr. The
      grey scaling represents numbers in the Galaxy. The darkest shade corresponds to 1100 binaries with given combination of \porb~ and $e$.
    }
    \label{fig:nspe}
\end{figure}}

As we noted before, important phases in the above described scenario
are the stages in which one of the components is a WR-star. Evolution
in close binaries is a channel for formation of a significant fraction
of them, since RLOF is able to remove hydrogen envelopes from much
lower mass stars than stellar wind. As WR-stars are important
contributors to the UV-light of the galaxies (as well as their lower
mass counterparts -- hot subdwarfs), 
this points to necessity to include binary evolution effects in the
spectrophotometric population synthesis models, see, e.g.,
\cite{2013arXiv1302.0927L}.


\clearpage
\subsection{Black hole formation parameters}

So far, we have considered the formation of NSs and binaries with
NSs. It is believed that very massive stars end up their evolution by
formation of stellar mass black holes. We will discuss now their
formation.

In the analysis of BH formation, new important parameters appear. The first
one is the threshold mass $M_{\mathrm{cr}}$ beginning from which a main-sequence star,
after the completion of its nuclear evolution, can collapse into a BH. This
mass is not well known; different authors suggest different values: van den
Heuvel and Habets~\cite{Heuvel_Habets84} -- $40\,M_{\odot}$; Woosley
et~al.~\cite{Woosley_al95} -- $60\,M_{\odot}$; Portegies Zwart, Verbunt,
and Ergma~\cite{PortegiesZwart_al97}, 
Ergma \& van den Heuvel \cite{1998AA...331L..29E}, 
 Brown et al.~\cite{2001NewA....6..457B} --  $20\,M_{\odot}$. 
A simple physical
argument usually put forward in the literature is that the mantle of the
main-sequence star with $M>M_{\mathrm{cr}}\approx 30\,M_{\odot}$ before the collapse has
a binding energy well above 10\super{51}~erg (the typical supernova
energy observed), so that the supernova shock is not strong enough to
expel the mantle~\cite{Fryer99, Fryer04}.

The upper mass limit for BH formation (with the caveat that the
role of magnetic-field effects is not considered) is, predominantly,
a function of stellar-wind mass loss in the core-hydrogen,
hydrogen-shell, and core-helium burning stages. For a specific
combination of winds in different evolutionary stages and
assumptions on metallicity it is possible to find the types of
stellar remnants as a function of initial mass (see, for
instance~\cite{heger_death03}). Since stellar winds are mass (or
luminosity) and metallicity-dependent, a peculiar consequence of
mass-loss implementation in the latter study is that for $Z \simeq
Z_{\odot}$ the mass-range of precursors of black holes is constrained
to $M\approx (25\mbox{\,--\,}60)\,M_{\odot}$, while more massive stars
form NSs because of heavy mass loss. The recent discovery of the possible
magnetar in the young stellar cluster
\epubtkSIMBAD{Westerlund~1}~\cite{muno_wester_magn06} hints to the reality of such
a scenario. Note, however, that the estimates of $\dot{M}$ are rather
uncertain, especially for the most massive stars, mainly because of
clumping in the winds (see, e.g., \cite{kudr_urban_winds06,
  crowther_araa06, hamann_wn06}). Current reassessment of the role of
clumping generally results in the reduction of previous mass-loss
estimates. Other factors that have to be taken into account in the
estimates of the masses of progenitors of BHs are rotation and
magnetic fields.

The second parameter is the mass $M_{\mathrm{BH}}$ of the nascent BH. There are
various studies as for what the mass of the BH should be
(see, e.g., \cite{Timmes_al96, Bethe_Brown98, Fryer99,
  Fryer_Kalogera01}). In some papers a typical BH mass was found to be
not much higher than the upper limit for the NS mass
(Oppenheimer--Volkoff limit $\sim(1.6\mbox{\,--\,}2.5)\,M_{\odot}$, depending on
the unknown equation of state for NS matter) even if the
fall-back accretion onto the supernova remnant is
allowed~\cite{Timmes_al96}. Modern measurements of black hole
masses in binaries suggest a broad range of them ---  
 $(4\mbox{\,--\,}17)\,M_{\odot}$~\cite{orosz_bh03, McClintock_Remillard03,
  Remillard_McClintock06}. 
A continuous range of BH masses up to $10\mbox{\,--\,}15\,M_{\odot}$ was derived in
calculations~\cite{Fryer_Kalogera01}.  However, as we stressed
in Section \ref{sec:BH_inbinaries}, in view of the absence of robust
``first-principle''
calculations of stellar core collapses, both the BH progenitor's mass and
that of the formed BH itself remain major parameters. Here many possibilities
remain, including
the interesting suggestion by Kochanek \cite{Kochanek13} that the absence of
high mass giants 
($16.5\,M_{\odot} < M < 25\,M_{\odot}$) as the progenitors of Type~IIP
supernovae may indicate
the BH formation starting from the progenitor masses as low as 16
$M_{\odot}$.

There is observational evidence that the dynamically determined
BH masses in extragalactic HMXBs
(M~33 X-7, NGC~300 X-1, and IC10 X-1) residing in
low-metallicity galaxies are higher (16-30 $M_{\odot}$)  than
in the Milky Way surroundings \cite{Crowther_al10}. Unless being a selection
effect
(the brightest X-ray sources are observed first), this may indicate the
dependence of
BH formation on the progenitor's metallicity. This dependence is actually
expected in the current models of evolution of single
stars~\cite{Georgy_al09} and HMXB evolution~\cite{Linden_al10}.

It is still a challenge to reproduce successful supernova explosions in numerical calculations, especially in 3D 
(see, for example,  B. M\"uller's talk at  the conference  in Kyoto in October 2013,  
\url{https://dl.dropboxusercontent.com/u/239024933/Programs_for_Conferences.html} and the discussion in \cite{2014arXiv1402.4362P}). 
In the current numerical calculations, spectra of BH masses and kicks received by nascent BH are still
model dependent, see, e. g., \cite{2012ApJ...749...91F, 2013MNRAS.434.1355J}. 
Therefore, in the further discussion we will parameterize
the BH mass $M_{\mathrm{BH}}$ by the fraction of the pre-supernova mass $M_{*}$
that collapses into the BH: $k_{\mathrm{BH}}=M_{\mathrm{BH}}/M_{*}$. In fact, the
pre-supernova mass $M_{*}$ is directly related to $M_{\mathrm{cr}}$, but the form
of this relationship is somewhat different in different scenarios for
massive star evolution, mainly because of different mass-loss
prescriptions. According to our parameterization, the minimal BH mass
can be $M_{\mathrm{BH}}^{\min}=k_{\mathrm{BH}}M_{*}$, where $M_{*}$ itself depends on
$M_{\mathrm{cr}}$. The parameter $k_{\mathrm{BH}}$ can vary in a wide range.

The third parameter, similar to the case of NS formation, is the
possible kick velocity $\mathbf{w}_{\mathrm{BH}}$ imparted to the newly formed BH
(see the end of Section~\ref{kicks}). In general, one expects
that the BH should acquire a smaller kick velocity than a NS, as black
holes are more massive than neutron stars. A possible relation (as
adopted, e.g., in calculations~\cite{Lipunov_al97}) reads
\begin{equation}
  \frac{w_{\mathrm{BH}}}{w_{\mathrm{NS}}} =
  \frac{M_{*}-M_{\mathrm{BH}}}{M_{*}-M_{\mathrm{OV}}} =
  \frac{1-k_{\mathrm{BH}}}{1-M_{\mathrm{OV}}/M_{*}}\,,
  \label{BH_kick}
\end{equation}
where $M_{\mathrm{OV}}=2.5 \,M_{\odot}$ is the maximum NS mass.
When $M_{\mathrm{BH}}$ is close to $M_{\mathrm{OV}}$, the ratio $w_{\mathrm{BH}}/w_{\mathrm{NS}}$ 
approaches 1, and
the low-mass black holes acquire kick velocities similar to those of 
neutron stars. When $M_{\mathrm{BH}}$ is significantly larger than $M_{\mathrm{OV}}$,
the parameter $k_{\mathrm{BH}}=1$, and the BH kick velocity becomes vanishingly
small.
The allowance for a quite moderate $w_{\mathrm{BH}}$ can 
increases the coalescence rate of binary BH~\cite{Lipunov_al97}.

The possible kick velocity imparted to newly born black holes
makes the orbits of survived systems highly eccentric. It is
important to stress that some fraction of such binary BH can retain
their large eccentricities up to the late stages of their
coalescence. This signature should be reflected in their emitted
waveforms and should be modeled in templates.

Asymmetric explosions accompanied by a kick change the space
orientation of the orbital angular momentum. On the other hand, the
star's spin axis remains fixed (unless the kick was off-centre). As
a result, some distribution of the angles between the BH spins and
the orbital angular momentum (denoted by $J$) will be
established~\cite{Postnov_Prokhorov00, Kalogera00}. It is interesting
that even for small kicks of a few tens of km/s an appreciable
fraction (30\,--\,50\%) of the merging binary BH can have $\cos
J<0$. This means that in these binaries the orbital angular momentum
vector is oriented almost oppositely to the black hole spin. This is
one more signature of imparted kicks that can be tested
observationally. The BH spin misalignment can have important
consequences for the BH-NS mergings \cite{Foucart_al13}. The link
between forthcoming observations by second- and third-generation GW
detectors with astrophysical scenarios of BH spin formation and
evolution in compact binaries is discussed in more detail in
\cite{2013PhRvD..87j4028G}.

\newpage
\section{Formation of Double Compact Binaries} 
\label{section:binary_NS}





\subsection{Analytical estimates}

A rough estimate of the formation rate of double compact binaries can
be obtained ignoring many details of binary evolution. To do this, we
shall use the observed initial distribution of binary orbital
parameters and assume the simplest conservative mass transfer
($M_1 + M_2= \mathrm{const}$) without kick velocity imparted to the
nascent compact stellar remnants during SN explosions.

\paragraph*{Initial binary distributions.}
From observations of binaries it is possible to derive
the formation rate of binary stars with initial masses $M_1$, $M_2$
(with mass ratio $q=M_2/M_1\le 1$), orbital semi-major axis $A$, and
eccentricity $e$. According to~\cite{Popova_al82,vtykp88}, the present
birth rate of binaries in our Galaxy can be written in factorised
form as 
\begin{equation}
  \frac{dN}{dA\,dM_1\,dq\,dt} \approx
  0.087 \left( \frac{A}{R_\odot} \right)^{-1}
  \left( \frac{M_1}{M_\odot} \right)^{-2.5} f(q),
  \label{E:bin_brate}
\end{equation}
where $f(q)$ is a poorly constrained distribution over the initial mass
ratio of binary components.

One usually assumes a mass ratio distribution law in the form 
$f(q)\sim q^{-\alpha_q}$ where $\alpha_q$ is a
a parameter, also derived observationally, see 
\S~\ref{section:introduction}; another often used form of the $q$-distribution was
suggested by Kuiper~\cite{Kuiper35}:
\begin{displaymath}
  f(q) = 2/(1+q)^2.
\end{displaymath} 
The range of $A$ is $10 \leq A/R_\odot \leq 10^6$. In deriving the above
Eq.~(\ref{E:bin_brate}), Popova et al.~\cite{Popova_al82} took
into account selection effects to convert the ``observed'' distribution of
stars into the true one. An almost flat logarithmic distribution of
semimajor axes was also found in~\cite{Abt83}. 


Taking Eq.~(\ref{E:bin_brate}) at face value, assuming 100\% binarity, the mass range of the primary 
components $M_1=0.08 M_\odot$  to $100 M_\odot$, a flat distribution over semimajor 
axes  (contact at ZAMS)$\leq \log (A/R_\odot)\leq$~(border between close and wide binaries), $f(q)=1$ 
for close binaries, and $f(q)\propto q^{-2.5}$ for $0.3\leq q \leq 1$ and  $f(q)=2.14$ 
for $q<0.3$ for wide binaries with $\log(A/R_\odot)\leq 6$,
as accepted in Tutukov and Yungelson's BPS code \textsf{IBiS} \cite{ty02}, 
we get SFR~$\approx 8 M_\odot$ per yr, which is several times as high as the modern
estimates of the current Galactic SFR. However, if 8 $M_\odot$ per year 
is used as a constant average SFR for 13.5~Gyr, we get the right mass of the Galactic 
disc. 
Clearly, Eq.~(\ref{E:bin_brate}) is rather approximate, since most of the stellar  
mass resides in low-mass stars for which IMF, $f(q)$, $f(A)$, binary fraction, etc., 
are poorly known. However, if we consider only solar chemical composition 
stars with $M_1>0.95 M_\odot$ (which can evolve off the main-sequence in the Hubble 
time), we get, under the ``standard'' assumptions in the \textsf{IBiS}-code, 
e. g., a WD formation rate of 0.65 per yr, which is reasonably 
consistent with observational estimates (see  Liebert et al. \cite{2005ApJS..156...47L}), 
the SNII+SNIb/c rate about 1.5/100 per yr, which is consistent with the inferred 
Galactic rate \cite{2001MmSAI..72..863C}) or with the pulsar formation rate 
\cite{2004ApJ...617L.139V}. We also get the ``proper'' rate of WD+WD mergers for SN Ia 
(a few per thousand years).        

\paragraph*{Constraints from conservative evolution.}
For this estimate we shall
assume that the primary mass should be at least $10\,M_\odot$. 
Equation~(\ref{E:bin_brate}) tells
that the formation rate of such binaries is about 1 per 50 years. We
shall restrict ourselves by considering only close binaries, in which
mass transfer onto the secondary is possible. This narrows the binary
separation interval to $10\mbox{\,--\,}1000\,R_\odot$ (see
Figure~\ref{f:remn}); the birth rate of close massive
($M_1>10\,M_\odot$) binaries is thus $1/50 \times 2/5 \mathrm{\ yr}^{-1} =
1/125 \mathrm{\ yr}^{-1}$. The mass ratio $q$ should not be very small
to make the formation of the second NS possible. The lower limit for $q$ is
derived from the condition that after the first mass transfer stage,
during which the mass of the secondary increases, $M_2+\Delta M\ge
10\,M_\odot$. Here $\Delta M=M_1-M_{\mathrm{He}}$ and the mass of the helium
core left after the first mass transfer is $M_{\mathrm{He}}\approx 0.1
(M_1/M_\odot)^{1.4}$. This yields
\begin{displaymath}
  m_2+(m_1-0.1m_1^{1.4}) > 10,
\end{displaymath}
where we used the notation $m=M/M_\odot$, or in terms of $q$:
\begin{equation}
  q \ge 10/m_1+0.1 m_1^{0.4}-1.
  \label{E:ll_q}
\end{equation}

An upper limit for the mass ratio is obtained from the requirement
that the binary system remains bound after the sudden mass loss in the
second supernova explosion\epubtkFootnote{For the first supernova
  explosion without kick this is always satisfied.}. From
Eq.~(\ref{A:SN:symm-ecc}) we obtain
\begin{displaymath}
  \frac{0.1[m_2+(m_1-0.1m_1^{1.4})]^{1.4}-1.4}{2.8} < 1,
\end{displaymath}
or in terms of $q$:
\begin{equation}
  q \le 14.4/m_1+0.1m_1^{0.4}-1.
  \label{E:ul_q}
\end{equation}

Inserting $m_1=10$ in the above two equations yields the appropriate
mass ratio range $0.25< q< 0.69$, i.e.\ 20\% of the binaries for Kuiper's
mass ratio distribution. So we conclude that the birth rate of binaries
which potentially can produce double NS system is $\lesssim 0.2 \times
1/125 \mathrm{\ yr}^{-1} \simeq 1/600 \mathrm{\ yr}^{-1}$.

Of course, this is a very crude upper limit -- we have not taken into
account the evolution of the binary separation, ignored initial binary
eccentricities, non-conservative mass loss, etc. However, it is not
easy to treat all these factors without additional knowledge of
numerous details and parameters of binary evolution (such as the
physical state of the star at the moment of the Roche lobe overflow,
the common envelope efficiency, etc.). All these factors should
decrease the formation rate of double NS. The coalescence rate of
compact binaries (which is ultimately of interest for us) will be even
smaller -- for the compact binary to merge within the Hubble time, the
binary separation after the second supernova explosion should be less
than $\sim 100\,R_\odot$ (orbital periods shorter than $\sim$~40~d) 
for arbitrary high orbital eccentricity $e$ (see Figure~\ref{A:GW:p-e}).
The model-dependent distribution of NS kick velocities provides another
strong complication. We also stress that this upper limit was obtained
assuming a constant Galactic star-formation rate and normalisation 
of the binary formation by Eq.~(\ref{E:bin_brate}).

Further (semi-)analytical investigations of the parameter space of binaries
leading to the formation of coalescing binary NSs are still possible
but technically very difficult, and we shall not reproduce them here.
The detailed semi-analytical approach to the problem of formation of NSs in
binaries and evolution of compact binaries has been developed by Tutukov and
Yungelson~\cite{Tutukov_Yungelson93a, ty93b}.


\subsection{Population synthesis results}

A distinct approach to the analysis of binary star evolution is based
on the population synthesis method -- a Monte Carlo simulation of the
evolution of a sample of binaries with different initial
parameters. 
This approach was first applied to model various
observational manifestations of magnetized NSs in massive binary
systems~\cite{Kornilov_Lipunov83a, Kornilov_Lipunov83b,
  Dewey_Cordes87} and generalised to binary systems of arbitrary mass
in~\cite{Lipunov_al96} (The Scenario Machine code). To achieve a
sufficient statistical significance, such simulations usually involve
a large number of binaries, typically of the order of a million. The
total number of stars in the Galaxy is still four orders of magnitude
larger, so this approach cannot guarantee that rare stages of the
binary evolution will be adequately reproduced\epubtkFootnote{Instead
  of Monte Carlo simulations one may use a sufficiently dense grid in
  the 3D space of binary parameters and integrate over this grid (see,
  e.g., \cite{ty02} and references therein).}.

Presently, there are several population synthesis codes used for
massive binary system studies, which take into account with different
degree of completeness various aspects of binary stellar evolution
(e.g., the codes by Portegies Zwart et
al.~\cite{py98, yungelson_bh06}, Bethe and
Brown~\cite{Bethe_Brown98}, Hurley, Tout, and Pols~\cite{Hurley_al02},
Belczynski et al.~\cite{belcz_startrack08}, Yungelson and
Tutukov~\cite{ty02}, De Donder and Vanbeveren~\cite{2004NewAR..48..861D}). A review of applications of the population
synthesis method to various types of astrophysical sources and further
references can be found in~\cite{Popov_Prokhorov04,
  yungelson05b}. Some results of population synthesis
calculations of compact binary mergers carried out by different groups
are presented in Table~\ref{table:mergings}.

\begin{table}
  \renewcommand{\arraystretch}{1.2}
  \caption{Examples of the estimates for Galactic merger rates of
    relativistic binaries calculated under different assumptions on
    the parameters entering population synthesis.}
  \label{table:mergings}
  \centering
  \begin{tabular}{lrrrr}
    \hline \hline
    Authors & Ref. & NS\,+\,NS & NS\,+\,BH & BH\,+\,BH \\
    & & \multicolumn{1}{c}{[yr\super{-1}]} &
    \multicolumn{1}{c}{[yr\super{-1}]} &
    \multicolumn{1}{c}{[yr\super{-1}]} \\
    \hline
    Tutukov and Yungelson (1993) & \cite{ty93b} &
    3~\texttimes~10\super{-4} & 2~\texttimes~10\super{-5} & 1~\texttimes~10\super{-6} \\
    Lipunov et~al.\ (1997) & \cite{LPP97} &
    3~\texttimes~10\super{-5} & 2~\texttimes~10\super{-6} & 3~\texttimes~10\super{-7} \\
    Portegies Zwart and Yungelson (1998) & \cite{py98} &
    2~\texttimes~10\super{-5} & 10\super{-6} \\
    Nelemans et~al.\ (2001) & \cite{nyp01} &
    2~\texttimes~10\super{-5} & 4~\texttimes~10\super{-6} & \\
    Voss and Tauris (2003) & \cite{voss_taurisns03} &
    2~\texttimes~10\super{-6} & 6~\texttimes~10\super{-7} & 10\super{-5} \\
    De Donder and Vanbeveren (2004) & \cite{DeDonder_Vanbeveren04} &
    3~\texttimes~10\super{-3}\,--\,10\super{-5} & 3~\texttimes~10\super{-5} & 0 \\
    O'Shaughnessy et al.\ (2005) & \cite{Oshaughnessy_al05} &
    7~\texttimes~10\super{-6} & 1~\texttimes~10\super{-6} & 1~\texttimes~10\super{-6} \\
    de Freitas Pacheco et al.\ (2006) & \cite{freitas_ns06} &
    2~\texttimes~10\super{-5} \\
    Dominik et al. (2012) & \cite{Dominik_al12} &
(0.4-77.4)~\texttimes~10\super{-6} & (0.002-10.6)~\texttimes~10\super{-6} & (0.05-29.7)~\texttimes~10\super{-6}		\\
    Mennekens and Vanbeveren (2013) & \cite{Mennekens_Vanbeveren13} &
10\super{-7}-10\super{-5} & 10\super{-6}-10\super{-5}& 0 \\
    \hline \hline
  \end{tabular}
  \renewcommand{\arraystretch}{1.0}
\end{table}

Actually, the authors of the studies mentioned in
Table~\ref{table:mergings} make their simulations for a range of
parameters. We list in the table the rates for the models which the
authors themselves consider as ``standard'' or ``preferred'' or
``most probable'', calculated for solar metallicity (or give the ranges, as for 
Brussels code results \cite{Mennekens_Vanbeveren13}). Generally, for the NS\,+\,NS merger rate
Table~\ref{table:mergings} shows the scatter within a factor $\sim$~4,
which may be considered quite reasonable, having in mind the
uncertainties in input parameters. There are several outliers,
\cite{ty93b}, \cite{voss_taurisns03}, and \cite{Mennekens_Vanbeveren13}. The high rate
in~\cite{ty93b} is due to the assumption that kicks to
nascent neutron stars are absent. The low rate in~\cite{voss_taurisns03}
is due to the fact that these authors apply in the common envelope
equation an evolutionary-stage-dependent structural constant
$\lambda$. Their range for $\lambda$ is 0.006\,--\,0.4, to
be compared with the ``standard'' $\lambda=0.5$ applied in most of the
other studies. Low $\lambda$ favours mergers in the first critical
lobe overflow episode and later mergers of the first-born neutron
stars with their non-relativistic companions\epubtkFootnote{Note that
  similar low values of $\lambda$ for $20$ to $50\,M_\odot$ stars were
  obtained also in~\cite{Podsiadlowski_al03}. If confirmed, these
  results may have major impact on the estimates of merger rates for
  relativistic binaries.}. A considerable scatter in the rates of
mergers of systems with BH companions is due, mainly, to
uncertainties in stellar wind mass loss for the most massive
stars. For instance, the implementation of winds in the code used
in~\cite{py98, nyp01} resulted in the absence
of merging BH\,+\,BH systems, while a rather low $\dot{M}$ assumed
in~\cite{voss_taurisns03} produced a high merger rate of BH\,+\,BH
systems.
We note an extreme scatter of the estimates of the merger rate of NS+NS binaries
in \cite{Dominik_al12}: the lowest estimate is obtained assuming
very tightly bound envelopes of stars (with parameter $\lambda=0.01$), while the upper estimate ---
assuming completely mass-conservative evolution.
The results of the Brussels group \cite{DeDonder_Vanbeveren04, Mennekens_Vanbeveren13} differ from \textsc{StarTrack}-code results \cite{Dominik_al12}
and other codes in predicting an insignificant BH-BH merging rate. This is basically due to assumed
enhanced mass loss  in the red supergiant stage (RSG) of massive star evolution. 
In this scenario, 
unlike, e.g., Voss and Tauris' assumptions~\cite{voss_taurisns03},
the allowance for the enhanced
mass loss at the Luminous Blue Variable (LBV) phase of evolution 
for stars with an initial mass $\gtrsim 30\mbox{\,--\,}40\,M_{\odot}$ leads to a 
significant orbital increase and hence the avoidance of the second
Roche-lobe overflow and spiral-in process in the common envelope,
which completely precludes
the formation of close double BH systems merging within the Hubble time.
The Brussels code also takes into account the time evolution of galactic 
metallicity enrichment by massive single and binary stars. A more detailed 
comparison of different population synthesis results of NS\,+\,NS, NS\,+\,BH and BH\,+\,BH
formation and merging rates can be found in review \cite{Abadie_al10}.

A word of caution should be said here. It is hardly possible to trace
the  detailed evolution of each binary, so approximate 
descriptions of evolutionary tracks of stars, 
their interaction, effects of supernovae, etc. are invoked.
Thus, fundamental uncertainties
of stellar evolution mentioned above are complemented by 
(i) uncertainties of the scenario and (ii) uncertainties in the
normalisation of the calculations to the real galaxy (such as the
fraction of binaries among all stars, the star formation history,
etc.). The intrinsic uncertainties in the population synthesis results
(for example, in the computed event rates of binary mergers etc.) are
in the best case not less than $\simeq$(2\,--\,3\,).
This should always be borne in mind when using the population
synthesis calculations. However, we emphasize again the fact that the
double NS merger rate, as inferred from binary pulsar statistics with
account for the double pulsar observations~\cite{Burgay_al03,
 Kalogera_al04, Kim:2013tca}, is very close to the population syntheses 
estimates assuming NS kicks about (250\,--\,300)~km/s.

\newpage
\section{Detection Rates}      
\label{sec:secI:DetectRate}




From the point of view of detection, a gravitational wave signal from merging close
binaries is characterized by the signal-to-noise ratio $S/N$, which
depends on the binary masses, the distance to the binary, the
frequency, and the noise characteristics. A pedagogical derivation of
the signal-to-noise ratio and its discussion for different detectors
is given, for example, in Section~8 of the review~\cite{Grishchuk_al01}.
 
Coalescing binaries emit gravitational wave signals with a well known
time-dependence (waveform) $h(t)$ (see Section~\ref{sec:appA} above). This
allows one to use the technique of matched
filtering~\cite{Thorne87}. The signal-to-noise ratio $S/N$ for a
particular detector, which is characterized by the noise spectral density 
$S_n(f)$ [$\mathrm{Hz}^{-1/2}]$ or the dimensionless noise
rms amplitude $h_\mathrm{rms}$ at a given frequency $f$, depends
mostly on the ``chirp'' mass of the binary system
$\M=(M_1+M_2)^{-1/5}(M_1M_2)^{3/5}=\mu^{3/5} M^{2/5}$ (here $\mu=M_1M_2/M$ is 
the reduced mass and $M=M_1+M_2$ is the total mass of the system)
and its (luminosity) distance $r$: $S/N\propto \M^{5/6}/r$ \cite{Thorne87, Flanagan_Hughes98}. 
%
%
For a given type
of coalescing binaries (NS\,+\,NS, NS\,+\,BH or BH\,+\,BH), the signal-to-noise ratio
will also depend on the frequency of the innermost stable circular orbit
$f_{\mathrm{ISCO}}\sim 1/M$, as well as on the orientation of the binary with respect to the 
given detector and its angular sensitivity (see, e.g.,  
Section~8 in \cite{Grishchuk_al01} and \cite{Sathyaprakash_Schutz09} for more detail). 
Therefore, from the point of view of detection of the specific type of 
coalescing binaries at a pre-requisit signal-to-noise 
ratio level, it is useful to determine the detector's maximum
 (or ``horizon'') distance $D_{\mathrm{hor}}$, 
which is calculated for an optimally oriented (``ideal'') coalescing binary with a
given chirp mass $\M$.   
For a secure
detection, the $S/N$ ratio is usually raised up to 7\,--\,8 to avoid false
alarms over a period of a year (assuming Gaussian
noise)\epubtkFootnote{If a network of three detectors, such as two
  LIGOs and VIRGO, runs simultaneously, the $S/N$ ratio in an
  individual detector should be $>7/\sqrt{3}\approx 4$)}. This
requirement determines the maximum distance from which an event can be
detected by a given interferometer~\cite{Cutler_Thorne02, Abadie_al10}. The horizon distance 
$D_{\mathrm{hor}}$ of
LIGO~I/VIRGO (advanced LIGO, expected) interferometers for relativistic binary inspirals with
account of the actual noise curve attained in the S5 LIGO scientific run
are given in~\cite{Abadie_al10}: 33(445)~Mpc for
NS\,+\,NS ($1.4\,M_\odot + 1.4\,M_\odot$, $\M_{\mathrm{NS}}=1.22\,M_\odot$), 70(927)~Mpc for NS\,+\,BH
($1.4\,M_\odot + 10\,M_\odot$), and 161(2187)~Mpc for BH\,+\,BH
($10\,M_\odot + 10\,M_\odot$, $\M_{\mathrm{BH}}=8.7\,M_\odot$). The distances increase for a
network of detectors.

It is worth noting that the dependence of the $S/N$ for different types 
of coalescing binaries on $f_{\mathrm{ISCO}}$ rather slightly (to within 10\%) changes
as $S/N\propto \M^{5/6}$, so, for example, the ratio of BH and NS
detection horizons scales as $S/N$, i.e.  
$D_{hor, BH}/D_{hor, NS}=(\M_{\mathrm{BH}}/\M_{\mathrm{NS}})^{5/6}\approx 5.14$.
This allows us to estimate the relative detection ratio for different 
types of coalescing binaries by a given detector (or a network of detectors).
Indeeed, at a fixed level of $S/N$, the detection volume is proportional to
$D_{\mathrm{hor}}^3$ and therefore it is proportional to $\M^{5/2}$. The detection
rate $\D$ for binaries of a given class (NS\,+\,NS, NS\,+\,BH or BH\,+\,BH) is the
product of their coalescence rate $\R_\mathrm{V}$ and the detector's
horizon volume $\propto \M^{5/2}$ for these binaries.

It is seen from Table~\ref{table:mergings} that the model Galactic
rate $\R_\mathrm{G}$ of NS\,+\,NS coalescences is typically higher than the rate of
NS\,+\,BH and BH\,+\,BH coalescences. However, the BH mass is 
significantly larger than the NS mass. So a binary involving one or
two black holes, placed at the same distance as a NS\,+\,NS binary,
produces a significantly larger amplitude of gravitational waves. With
the given sensitivity of the detector (fixed $S/N$ ratio), a BH\,+\,BH
binary can be seen at a greater distance than a NS\,+\,NS binary. Hence,
the registration volume for such bright binaries is significantly 
larger than the registration volume for relatively weak binaries. The
detection rate of a given detector depends on the interplay between
the coalescence rate and the detector's response to the sources of one
or another kind.

If we assign some characteristic (mean) chirp mass to different types
of double NS and BH systems, the expected ratio of their detection
rates by a given detector is
\begin{equation}
  \frac{\D_{\mathrm{BH}}}{\D_{\mathrm{NS}}} =
  \frac{\R_{\mathrm{BH}}}{\R_{\mathrm{NS}}}
  \left( \frac{\M_{\mathrm{BH}}}{\M_{\mathrm{NS}}} \right)^{5/2}\!\!\!\!\!\!\!\!,
\label{DBH/DNS}
\end{equation}
where $\D_{\mathrm{BH}}$ and $\D_{\mathrm{NS}}$ refer to BH\,+\,BH and NS\,+\,NS pairs,
respectively. 
Taking $\M_{\mathrm{BH}}=8.7\,M_\odot$ (for $10\,M_\odot + 10\,M_\odot$) 
and $\M_{\mathrm{NS}}=1.22\,M_\odot$ (for $1.4\,M_\odot + 1.4\,M_\odot$), 
Eq.~(\ref{DBH/DNS}) yields
\begin{equation}
   \frac{\D_{\mathrm{BH}}}{\D_{\mathrm{NS}}} \approx
   140 \frac{\R_{\mathrm{BH}}}{\R_{\mathrm{NS}}}.
   \label{DBH/DNS.2}
\end{equation}
As $\frac{\R_{\mathrm{BH}}}{\R_{\mathrm{NS}}}$ is typically 0.1\,--\,0.01
(see Table~\ref{table:mergings}), this relation suggests that the
registration rate of BH mergers can be \textit{higher} than that of NS
mergers. This estimate is, of course, very rough, but it can serve as
an indication of what one can expect from detailed calculations. We
stress that the effect of an enhanced detection rate of BH binaries is
independent of the desired $S/N$ and other characteristics of the
detector; it was discussed, for example,
in~\cite{ty93b, LPP97, Grishchuk_al01, Dominik_al12}.



Unlike the ratio of the detection rates, 
the expected value of the detection rate of specific type of compact coalescing binaries 
by a given detector (network of detectors)
requires detailed evolutionary calculations and the knowledge of the 
actual detector's noise curve, as we discussed
above.
To calculate a realistic detection rate of binary mergers the
distribution of galaxies should be taken into account within the
volume bounded by the detector's horizon (see, for
example, the earlier attempt to take into account only bright galaxies
from Tully's catalog of nearby galaxies in~\cite{Lipunov_al95_gwsky},
and the use of LEDA database of galaxies to estimate the detection
rate of supernovae explosions~\cite{Baryshev_Paturel01}). In this context, a
complete study of galaxies within 100~Mpc was done by Kopparapu et al. 
\cite{Kopparapu_al08}. Based on their results, 
Abadie et al. \cite{Abadie_al10} 
derived the approximate formula for the number of the equivalent Milky-Way type galaxies 
within large volumes, which is applicable for distances $\gtrsim 30$~Mpc:
\begin{equation}
\label{sec:det_NG}
N_G(D_{\mathrm{hor}})=\frac{4\pi}{3}\left(\frac{D_{\mathrm{hor}}}{\hbox{Mpc}}\right)^3(0.0116)(2.26)^{-3}\,.
\end{equation}
Here the factor 0.0116 is the local density of the equivalent Milky-Way type
galaxies derived in \cite{Kopparapu_al08}, and the factor 2.26 takes into 
account the reduction in the detector's horizon value when averaging over 
all sky locations and orientations of the binaries. Then the expected detection rate
becomes $\D(D_{\mathrm{hor}})=\R_G\times N_G(D_{\mathrm{hor}})$.

However, not
only the mass and type of a given galaxy, but also the star formation
rate and, better, the history of the star formation rate in that
galaxy are needed to estimate the
expected detection rate $\cal D$ (since the coalescence rate 
of compact binaries in the galaxies strongly evolves
with time~\cite{Lipunov_al95_evol, Mennekens_Vanbeveren13, Dominik_al13}).

So to assess the merger rate from a
large volume based on the Galactic values, the best one can do at
present appears to be using formulas like Eq.~(\ref{sec:det_NG}) or (\ref{R_V}) given
earlier in Section~\ref{sec:ns_freq}. This, however, adds another
factor two of uncertainty in the estimates. Clearly, a more accurate
treatment of the transition from Galactic rates to larger volumes with
an account of the galaxy distribution is very desirable.

To conclude this Section, we will briefly comment on the possible
electromagnetic counterparts of compact binary coalescences.
It is an important issue, since the localization error boxes of NS+NS
coalescences by GW detectors network only are expected to be, 
in the best case, about several square degrees
(see the detailed analysis in \cite{2013arXiv1309.3273R}, as well as 
\cite{Aasi_al13} for discussion of the likely evolution of sensitivity and sky localization of the sources for the advanced detectors), which 
is still large for precise astronomical identification. Any associated electromagnetic signal can greatly help to pinpoint the source.
Double NS
mergings are the most likely progenitors of short gamma-ray bursts
(\cite{Nakar07,FongBerger13} and references therein). Indeed, 
recently, a short-hard GRB~130603B was found to be followed by a 
rapidly fading  IR afterglow
\cite{Berger_al13}, which is most likely due to `macronova' or
`kilonova' produced by decaying radioactive heavy elements expelled during a double
NS merging \cite{Li_Paczynski98, Rosswog05, Metzger_al10, Hotokezaka_al13, 2014MNRAS.tmp..182G}. Detection of
electromagnetic counterparts to GW signals from coalescing binaries is an
essential part of the strategy of the forthcoming advanced LIGO/VIRGO
observations \cite{2014MNRAS.437..649S}. Different aspects of this multi-messenger GW astronomy are further 
discussed in papers \cite{2013PhRvD..88d3011P,
2013ApJ...767..124N,2013arXiv1309.1554K, Aasi_al13,2013arXiv1312.5133M}, etc.

\newpage
\section{Short-Period Binaries with White-Dwarf Components} 
\label{section:wd_formation}

Binary systems with white dwarf components that are interesting for general 
relativity and cosmology come in several flavours:
\begin{itemize}
\item Detached binary white dwarfs or ``double
  degenerates'' (DDs, we shall use both terms as synonyms below).
\item Cataclysmic variables (CVs) -- a class of variable semidetached
  binary stars containing a white dwarf and a companion star that is
  usually a red dwarf or a slightly evolved star, a subgiant.
\item A subclass of the former systems in which the Roche lobe
  is filled by another white dwarf or low-mass partially degenerate
  helium star (AM~CVn-type stars or ``interacting
  double-degenerates'', IDDs). They appear to be important 
  ``verification sources'' for planned space-based
 low-frequency GW detectors.
\item Detached systems with a white dwarf accompanied by a
  low-mass non-degenerate helium star (sd\,+\,WD systems).
\item Ultracompact X-ray binaries (UCXBs) containing a 
  NS and a Roche lobe overflowing WD or low-mass partially degenerate helium star.
\end{itemize}


As Figure~\ref{f:GW_sources} shows, compact binary stars emit
gravitational waves within the sensitivity limits for space-based
detectors if their orbital periods range from $\sim$~20~s to
$\sim$~20\,000~s. This means that in principle 
gravitational wave radiation may play a pivotal role in the evolution 
of all AM~CVn-stars, UCXBs, a considerable fraction of CVs,
and some DD and SD\,+\,WD systems and they would be observable in GWs if
detectors would be sensitive enough and confusion noise absent.

Though General Relativity predicted that binary stars should be the
sources of gravitational waves as early as in the 1920s, this prediction became a
matter of actual interest only with the discovery of the 
cataclysmic variable \epubtkSIMBAD{WZ~Sge} with orbital period 
$P_{\mathrm{orb}} \approx 81.5 \mathrm{\ min}$  by Kraft, Mathews, and
Greenstein in 1962~\cite{kmg62}, who immediately recognised the
significance of short-period binary stars as testbeds for
gravitational waves physics. Another impetus to the study of binaries
as sources of gravitational wave radiation (GWR) was imparted by the
discovery of ultra-short period variability of a faint blue star
HZ~29\,=\,\epubtkSIMBAD{AM~CVn} ($P_{\mathrm{orb}} \approx 18 \mathrm{\ min}$) by Smak in
1967~\cite{sma67}. Smak~\cite{sma67} and Paczy\'{n}ski~\cite{pac67a}
speculated that the latter system is a close pair of white dwarfs,
without specifying whether it is detached or semidetached. Faulkner et
al.~\cite{ffw72} inferred the status of \epubtkSIMBAD{AM~CVn} as a
``double-white-dwarf semidetached'' nova. \epubtkSIMBAD{AM~CVn} was later classified
as a cataclysmic variable after flickering typical for CVs was found for
\epubtkSIMBAD{AM~CVn} by Warner and Robinson~\cite{wr72}%
\epubtkFootnote{Flickering is a fast intrinsic brightness
  scintillation occurring on time scales from seconds to minutes with
  amplitudes of 0.01\,--\,1~mag and suggesting an ongoing mass
  transfer.}
and it became the prototype of a subclass of binaries.%
\epubtkFootnote{Remarkably, a clear-cut confirmation of the binary
  nature of \epubtkSIMBAD{AM~CVn} and the determination of its true
  period awaited for almost 25 years~\cite{nsg_am00}.}

An evident milestone in GW studies was the discovery of a binary
pulsar by Hulse and Taylor~\cite{1974IAUC.2704....1T,Hulse_Taylor75}
and determination of the derivative of its \porb, consistent with
GR predictions \cite{2005ASPC..328...25W}.
Equally significant is the recent discovery of a \porb=12.75\,min detached 
eclipsing binary white dwarf \epubtkSIMBAD{SDSS~J065133.338+284423.37} (J0651, for simplicity) 
by Kilic, Brown et~al.~\cite{2011MNRAS.413L.101K,2011ApJ...737L..23B}.
The measured change of the orbital period of J0651 over 13 months of observations 
is $(-9.8\pm2.8)\times10^{-12}\mathrm{\ s\ s}^{-1}$, which is 
consistent, within 3$\sigma$\ errors, with expectations from 
GR -- $(-8.2\pm1.7)\times10^{-12}\mathrm{\ s\ s}^{-1}$ \cite{2012ApJ...757L..21H}.
The strain amplitude of gravitational waves 
from this object at a frequency of $\sim$~2.6~mHz 
should be $1.2\times10^{-22}\mathrm{\ Hz}^{-1/2}$, 
which is about 10\,000 times as high as that from the Hulse-Taylor binary pulsar. 
It is currently the second most powerful GW source known and it is expected 
to be discovered in the first weeks of operation of eLISA. 
\epubtkFootnote{It is expected that recently launched astrometric 
space mission GAIA will detect about 200
eclipsing double degenerates \cite{2013ASPC..467...27N}.}

The origin of all above-mentioned classes of short-period binaries was
understood after the notion of common envelopes and the formalism for
their treatment were suggested in the 1970s (see Section~\ref{A:CE}). A
spiral-in of components in common envelopes allowed us to explain how
white dwarfs -- former cores of highly evolved stars with radii of
$\sim\,100\,R_{\odot}$ -- may acquire companions separated by $\sim\,R_{\odot}$
only.
We recall, however, that most studies of the formation
of compact objects through common envelopes are based on a simple
formalism of comparison of binding energy of the envelope with the orbital
energy of the binary, supposed to be the sole source of energy for the loss
of the envelope, as was discussed in Section~\ref{A:CE}. 


Although we mentioned above that some currently accepted features of binary
evolution may be subject to certain revisions in the future, one may
expect that, for example, changes in the stability criteria for mass
exchange will influence mainly the parameter space ($M_1$, $M_2$,
$a_0$) of progenitors of particular populations of binary stars, but
not the evolutionary scenarios for their formation. If the objects of
a specific class may form via different channels, the relative
``weight'' of the latter may change. While awaiting for detailed
evolutionary calculations of different cases of mass exchange with
``new'' physics,%
\epubtkFootnote{Advent of new fast and stable stellar evolutionary
  codes like MESA~\cite{2011ApJS..192....3P,2013arXiv1301.0319P},
  \url{http://mesa.sourceforge.net} will, hopefully, facilitate
  achieving this aim in the nearest future.}
we present below currently accepted scheme of the evolution of
binaries leading to the formation of compact binary systems with WD
components.

For stars with radiative envelopes, to the first approximation, the mass exchange
from $M_1$ to $M_2$ is stable for binary mass ratios $q=M_2/M_1 \lesssim 1.2$;
for $1.2 \lesssim q \lesssim 2$ it proceeds in the thermal time scale of the
donor, $t_{KH}=GM^2/RL$; for $q \gtrsim 2$ it proceeds in the dynamical time
scale $t_d= \sqrt{R^3/GM}$. The mass loss occurs in the dynamical time scale,
$\dot M\sim M/t_d$, if the donor has a deep convective envelope or if it is
degenerate and conditions for stable mass exchange are not satisfied. Note, however, 
that in the case of AGB stars the stage when the photometric (i.e. measured at
the optical depth $\tau=2/3$) stellar radius becomes equal to the Roche lobe
radius, is preceded by RLOF by {atmospheric} layers of the star, and the
dynamical stage of mass loss may be preceded by a quite long stage of stable
mass loss from the radiative atmosphere of the
donor~\cite{1972AcA....22...73P,1973PASP...85..769P,2012BaltA..21...88M}. It is
currently commonly accepted, despite the lack of a firm observational proof,
that the distribution of binaries over $q$ is even or rises to small $q$ (see
Section~\ref{section:introduction}). Since the accretion rate is limited either
by the rate that corresponds to the thermal time scale of the accretor or its
Eddington accretion rate, both of which are typically lower than the mass-loss
rate by the donor, the overwhelming majority ($\sim$~90\%) of close binaries
pass in their evolution through one to four common envelope stages.

An ``initial donor mass -- donor radius at RLOF'' diagram showing
descendants of stars after mass-loss in close binaries is presented above in
Figure~\ref{f:remn}. We remind here that solar metallicity
stars with $M\lesssim 0.95\,M_{\odot}$ do not evolve past the core-hydrogen
burning stage in the Hubble time.

\epubtkImage{fig_scen_low6.png}{%
  \begin{figure}[htbp]
     \centerline{\includegraphics[width=0.8\textwidth]{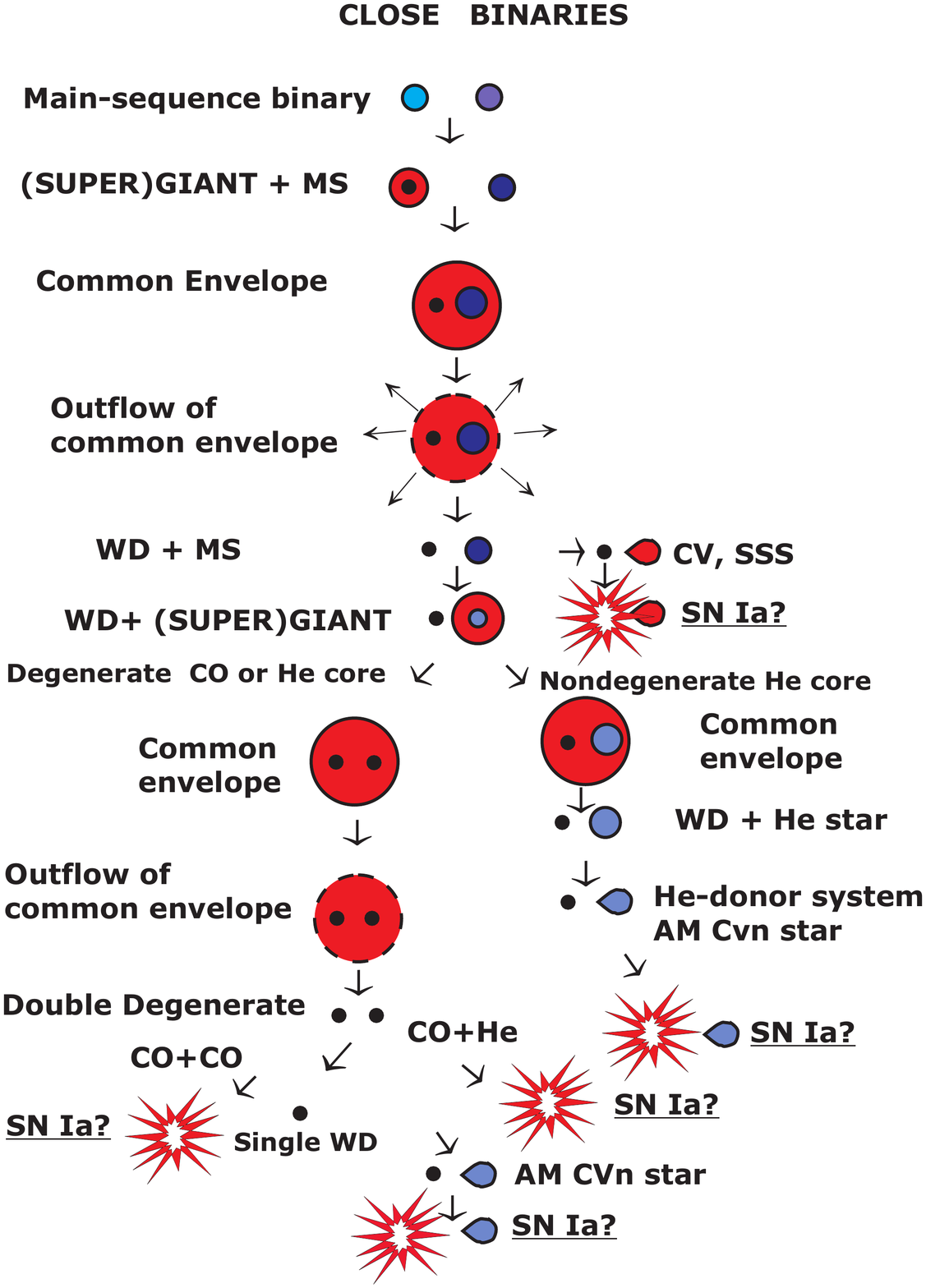}}
    \caption{Formation of close binary dwarfs and their
      descendants (scale and colour-coding are arbitrary).}
    \label{figure:wd_flow}
  \end{figure}
}

\subsection{Formation of compact binaries with white dwarfs}
\label{sec:formation_dd}

A flowchart schematically presenting the typical scenarios for formation of
low-mass compact binaries with WD components and some
endpoints of evolution is shown in Figure~\ref{figure:wd_flow}. 
Evolution of low- and intermediate mass close binary stars ($M_1\leq(8-10)\ms$)
is generally 
much more complex and less straightforward than the evolution of more massive binaries 
(Fig.~\ref{figure:massive_flow}).
For this reason, 
not all possible scenarios are plotted, but only the most probable
routes to SNe~Ia and to systems that may emit potentially detectable
gravitational waves. For simplicity, we consider only the most general
case when the first RLOF results in the formation of a common envelope.

The overwhelming majority of stars in binaries fill their Roche lobes when they
have He- or CO-cores, i.e., in cases \textbf{B} or \textbf{C} of mass 
exchange (Section~\ref{sec:mass_exc}). As noted in \S~\ref{A:CE}, the 
models of common envelopes are, in fact absent. 
It is usually \textit{assumed} that it proceeds in dynamical or thermal time scale and
is definitely so short that its duration can be neglected compared to other evolutionary 
stages.   

\subsubsection{Post-common envelope binaries}

We recall, for convenience, that in stars with solar metallicity with a ZAMS
mass below $(2.3\mbox{\,--\,}2.8) M_{\odot}$, helium cores are degenerate, and if
these stars overflow the Roche lobe prior to He core ignition, they produce 
$M\aplt 0.47\ms$ helium white dwarfs. Binaries with non-degenerate He-core
donors ($M\gtrsim (2.3\mbox{\,--\,}2.8)\,M_{\odot}$) first form a close
He-star\,+\,MS-star pair that can be observed as a subdwarf (sdB or sdO) star
with a MS companion \cite{ty90}. The minimum mass of He-burning stars is
close to $0.33\ms$ \cite{it85}.
The binary hypothesis for the origin of hot subdwarf stars was
first suggested by Mengel et~al.~\cite{1976ApJ...204..488M}, but it envisioned a
stable Roche lobe overflow.  Apparently, there exist  populations of
close sdB/sdO stars formed via common envelopes and of wide systems with
A-F type companions to subdwarfs, which may be post-stable-mass-exchange stars;
similarly, the merger products and genuinely single stars can be found among sdB/sdO stars. 
For the overview of 
properties of binary sdB/sdO stars, models of the population and current
state of the problem of the origin of these stars see, e.g.,
\cite{sj00,green+01,saio_jef02,Han_al02,
2003MNRAS.341..669H,ahmad_jeffery04,ty05,2006BaltA..15..183F,2009ARAA..47..211H,
2010Ap&SS.329...91G,2010Ap&SS.329...25N,
2012MNRAS.425.1013G,2013EPJWC..4304001G,2014MNRAS.438...14D,2014arXiv1401.0446G}.
When a He-star 
with the mass $\aplt 2.3\ms$
completes its evolution, a pair
harbouring a CO white dwarf and a MS-star appears. A large number of post-common
envelope binaries or ``WD+MS stars'' is known (see, e.g., SDSS-sample and its
analysis in~\cite{2011AA...536A..43N,2011AA...536L...3Z}). The most recent
population synthesis model of this class of binaries was published by Toonen and
Nelemans~\cite{2013AA...557A..87T}. Of course, the WD\,+\,MS population is
dominated by systems with low-mass MS-stars. Population synthesis studies
suggest that about 1/3 of them never evolve further in the Hubble time (e.g.,
\cite{yungelson05b}), in a reasonable agreement with above-mentioned
observations. 

\subsubsection{Cataclysmic variables}

If after the first common-envelope stage the orbital separation of the binary
$a\simeq \mathrm{several\ } R_{\odot}$ and the WD has a low-mass ($\lesssim
1.4\,M_{\odot}$) main-sequence companion, the latter can overflow its Roche lobe
during the hydrogen-burning stage or very shortly after it because of loss
of angular momentum by a magnetically coupled stellar wind and/or GW radiation, see \S~\ref{A:GW_evol} and \ref{sec:msw}. 
If  the mass ratio of the
components allows to avoid their merging in the CE $(q\aplt1.2)$, a
cataclysmic variable (CV) can form. Variability in cataclysmics can be due to
thermal-viscous 
instability of accretion disks \cite{1974PASJ...26..429O}
or/and unstable burning of accreted hydrogen on the WD surface
\cite{1952MNRAS.112..583M,1952MNRAS.112..598M,1968ApJ...152..245R,1971MNRAS.152..307S,
1971MNRAS.155..129S,1972ApJ...176..169S}; see B.~Warner's monograph
\cite{2003cvs..book.....W} for a comprehensive review of CVs in general,
\cite{Lasota01} for a review of the disk instability model and, e.g.,
\cite{1978ApJ...222..604P,1982ApJ...259..244I,2013ApJ...762....8D,2007MNRAS.374.1449E,
1993ApJ...406..220S,prkov95,2004ApJ...600..390T,
yaron05,2010ApJ...725..831S,2007ApJ...663.1269N,2009ApJ...705..693S,2007ApJ...660.1444S,
piers_99,piers+00,it96symb,gilpsetal03,2013ApJ...777..136W,idan_shaviv_he13} for
dependence of hydrogen burning regimes on the  WD surface and characteristics of
outbursts, depending on WD masses, their chemical composition (He, CO,
ONe), temperature and accretion rate. Outbursts produced by thermonuclear burning of
accreted hydrogen are identified with Novae
\cite{1971MNRAS.152..307S,1971MNRAS.155..129S} and are able to explain different
classes of them, see, e.g., \cite{yaron05}. As an example, we present in
Figure~\ref{f:nomotorates} the dependence of limits of different burning regimes
on mass and accretion rate for CO WD from Nomoto's paper
\cite{2007ApJ...663.1269N}. The stable burning limits found in this paper, by
stability analysis of a steady-burning WD, within factor $\simeq 2$
agree with those obtained in a similar way, e.g., by Shen and Bildsten
\cite{2007ApJ...660.1444S} and with the results of 
time-dependent
calculations by Wolf et~al.~\cite{2013ApJ...777..136W} (see Fig.~9 in the latter paper.)
\epubtkImage{f4_nomoto.png}{%
  \begin{figure}[htb]
     \centerline{\includegraphics[width=0.7\textwidth]{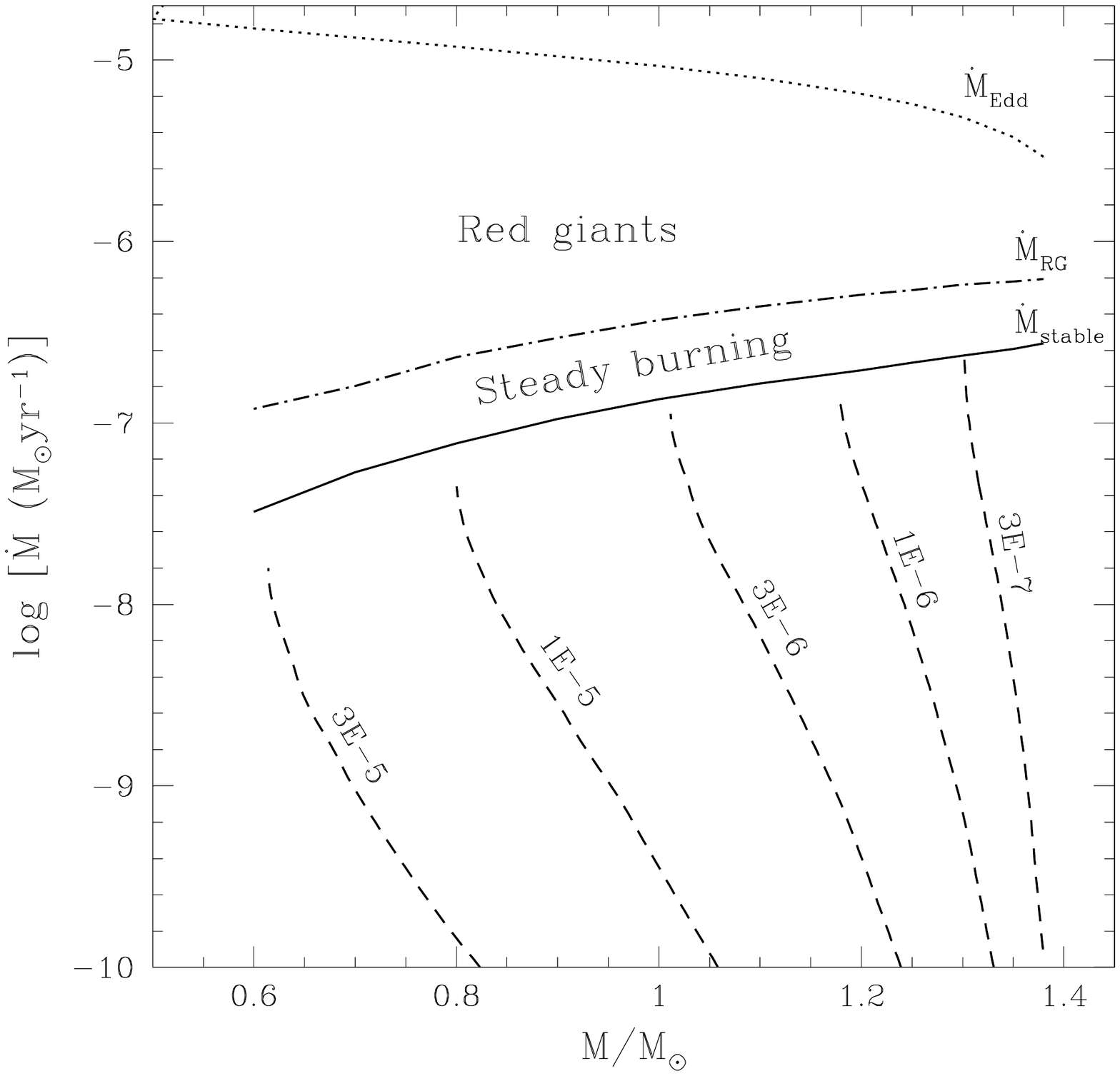}}
     \caption{Limits of different burning regimes of accreted hydrogen 
onto a CO WD as a function of mass of the WD and accretion rate \mdot\
\cite{2007ApJ...663.1269N}. If $\dot{M}_{\rm stable} \aplt \mdot
\aplt \dot{M}_{\mathrm{RG}}$, hydrogen burns steadily. If $\mdot \aplt
\dot{M}_{\mathrm{stable}}$, H-burning shells are thermally unstable;
with decrease of \mdot\ the strength of flashes increases. For
$\mdot \apgt \dot{M}_{\mathrm{RG}}$, hydrogen burns steadily but
the excess of unburnt matter forms an extended red-giant size envelope. In the 
latter  case 
white dwarfs may lose matter by wind or due to Roche lobe overflow. 
Dotted line shows the Eddington accretion rate
\medd\ as a function of \mwd. The dashed lines are the loci of 
the hydrogen envelope mass at hydrogen ignition.}
     \label{f:nomotorates}
  \end{figure}
}
\clearpage

Binaries with WDs and similar systems with subgiants ($M \lesssim
2\,M_{\odot}$) that steadily burn accreted hydrogen are usually identified
with supersoft X-ray sources -- SSS \cite{hbnr92}. 
Note, however, that the actual fraction of the stable burning stage when these stars
can be observed as SSS is still a matter of debate, see 
\cite{2010ApJ...724L.212H,2013arXiv1310.2170N}.
Post-Novae  in the stage of residual hydrogen burning also can be observed as
SSS \cite{1995ASSL..205..453T,2013IAUS..281..105H,2013A&A...559A..50N}.
In these systems, the duration of the SSS stage is debated as well, see, e.g.,
\cite{2013ApJ...777..136W}.
 A review of SSS may be found, e.g., in \cite{kh97}; 
the population synthesis models for SSS where computed, for instance, in
\cite{rdss94,kahabka_sss95,ylttf96,2010AstL...36..780Y}. 

An accreting CO WD in a binary system can accumulate enough mass to explode 
as a \sna\ if hydrogen burns stably or in mild flashes.
This is the so-called ``single-degenerate'' (SD) scenario
for SNe~Ia originally suggested by Schatzman as early as in 1963 
\cite{1963stev.conf..389S} and ``rediscovered'' and elaborated numerically 
10 years later by Whelan and Iben~\cite{wi73}. We shall consider these possible
progenitors of \sne\ in more detail below (\S~\ref{sec:snia}). 

If the WD belongs to the ONe-variety, it may experience an AIC into a
neutron star due to electron captures on
Ne and Mg, and a low-mass X-ray binary may be formed \cite{1980SSRv...27..595C}.

The final stages of the evolution of  CVs are not clear (and, in fact, poorly investigated). 
It was hypothesised \cite{rs83} that,  when the donor  mass decreases
below several hundredths of $M_{\odot}$, the disk-orbit tidal 
interaction becomes unefficient
and the disk turns into a sink of orbital momentum. More orbital momentum is drained from the orbit than
returned back, the mass loss rate increases and the donor can be tidally disrupted.

As noted by G. Dubus (private communication) at 
$q\lesssim 0.02$, which may be attained in the Hubble time, the conventional
picture of mass exchange may become invalid: the 
circularisation radius (the minimum radius of accretion disk defined by 
the specific angular momentum of the matter at the 
$L_1$ point and its interaction with the donor) exceeds the outer radius of the
disk (determined by tidal effects \cite{1977ApJ...216..822P,1977MNRAS.181..441P,vr88}). 
The resonance effects come into play and gaps may appear in the outer disk, since
orbital resonances confine the motion of disk particles 
to certain radius ranges, similarly to the formation of gaps in the asteroid belt and 
Saturn rings. It is not clear how the disk behaves then and how it interacts 
with still inflowing matter.    

\subsection{Binary white dwarfs}

The second common envelope stage may happen when the giant companion
to the WD overfills its Roche lobe. Before proceeding to discussion of
later stages, we note the following. Usually, this second
common envelope is thought to result, based merely on the final separation
of the components as given by Eqs.~(\ref{A:CE:eq}) or (\ref{ce_gamma}) --
in a merger, if the sum of the radii of Roche lobes of the WD and the core
of the giant (considered as another WD or a He-star) is $\leq a_f$, or
in the formation of a detached system otherwise. However, as it was
speculated by Livio and Riess \cite{2003ApJ...594L..93L} after the
first detection of \sna with hydrogen in the spectrum (SN~2002ic
\cite{hamuy03}), there is a very small probability ($\aplt1\%$) that WD
merges with the core of an AGB star when part of the AGB envelope
(common envelope) is still not ejected. These considerations are
corroborated by results of 1D 
\cite{2011ApJ...730...76I,2011MNRAS.417.1466K} and
and 3-D simulations 
\cite{2012ApJ...744...52P} which show that only a fraction of the common
envelope is ejected in the dynamical time scale, and the rest of the
matter remains in the vicinity of the close pair formed by the core of
the AGB star and the spiralling-in WD. As we mentioned in Section~\ref{A:CE},
interaction with this matter may result in further shrinking of the
system and ultimate merger of the components. 
If both cores are carbon-oxygen and the total mass
of the merger product exceeds \mch, it may explode as a \sna\ ,
see brief discussion of this ``core-degenerate'' \cite{2011MNRAS.417.1466K}
 \sne\ scenario in \S~\ref{sec:snia}.
    
If the system avoids merging inside CE and the donor had a degenerate
core, a close binary WD (or double-degenerate, DD) is formed 
(the left branch of the scenario shown in Figure~\ref{figure:wd_flow}). In
Figure~\ref{fig:marsh_dd} we show an indicative plot of possible
combinations of components in double-degenerate systems
\cite{2011CQGra..28i4019M} (it is assumed that the bulk chemical
composition of WD is uniquely related to its mass).

\epubtkImage{marsh_dd.png}{%
  \begin{figure}[htbp]
    \centerline{\includegraphics[scale=0.8]{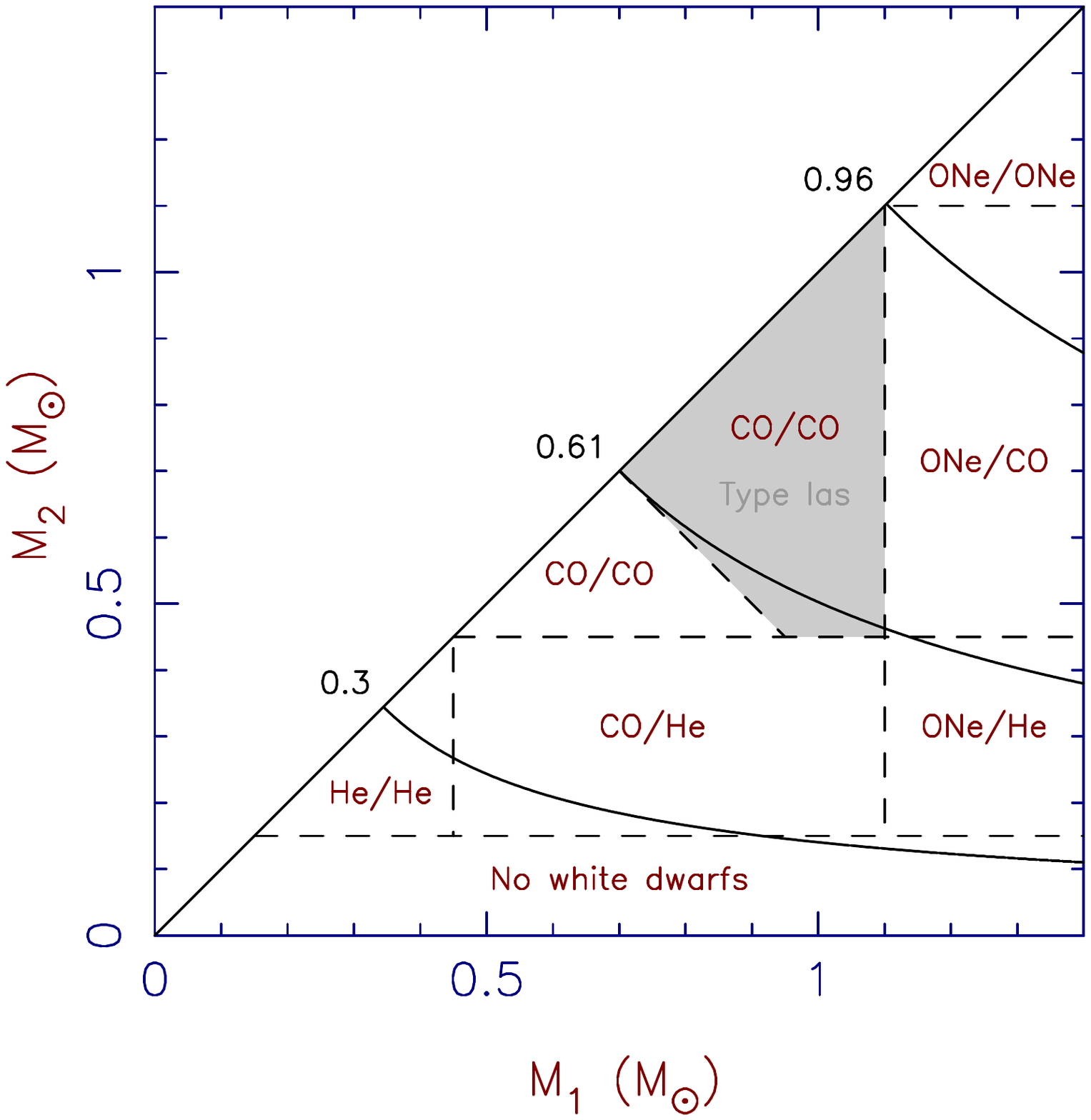}}
    \caption{Possible combinations of masses and chemical compositions
      of components in close binary WD
      \cite{2011CQGra..28i4019M}. Solid curves are lines of constant
      chirp mass (see Sections~\ref{sec:gwr_from_binary} and
      \ref{section:waves}).}
    \label{fig:marsh_dd}
  \end{figure}
}
\clearpage

In Figure~\ref{fig:WD_SEBA} we compare the most recent simulated
distribution of the total masses of binary WD vs.\ their \porb\ with
the distribution of $\mathrm{M_{tot}}$ for 46 observed binary WDs known 
at the time of writing
of the quoted study~\cite{2012AA...546A..70T}. The limiting stellar magnitude
was assumed to be $V_{\mathrm{lim}}=21$. 
It is difficult to judge about agreement with observations because
the sample of observed WDs is quite random, but the ``naked eye'' estimate suggests
that the model that uses $\gamma$-formalism fits observations better than 
the model that uses $\alpha$-formalism. 

\epubtkImage{WD_SEBA.png}{%
  \begin{figure}[htbp]
    \centerline{
      \includegraphics[width=0.45\textwidth, angle=-90]{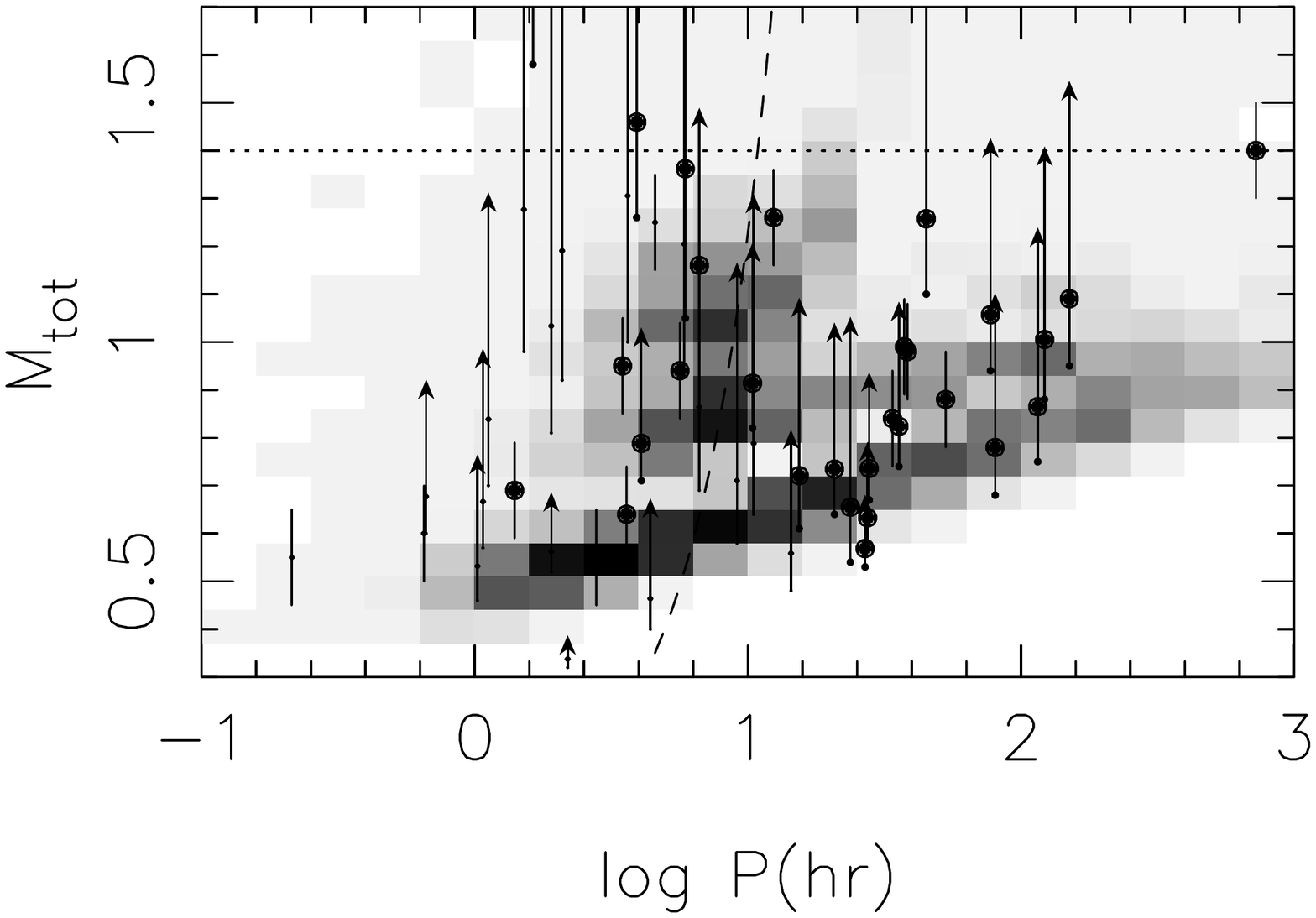}
      \includegraphics[width=0.45\textwidth, angle=-90]{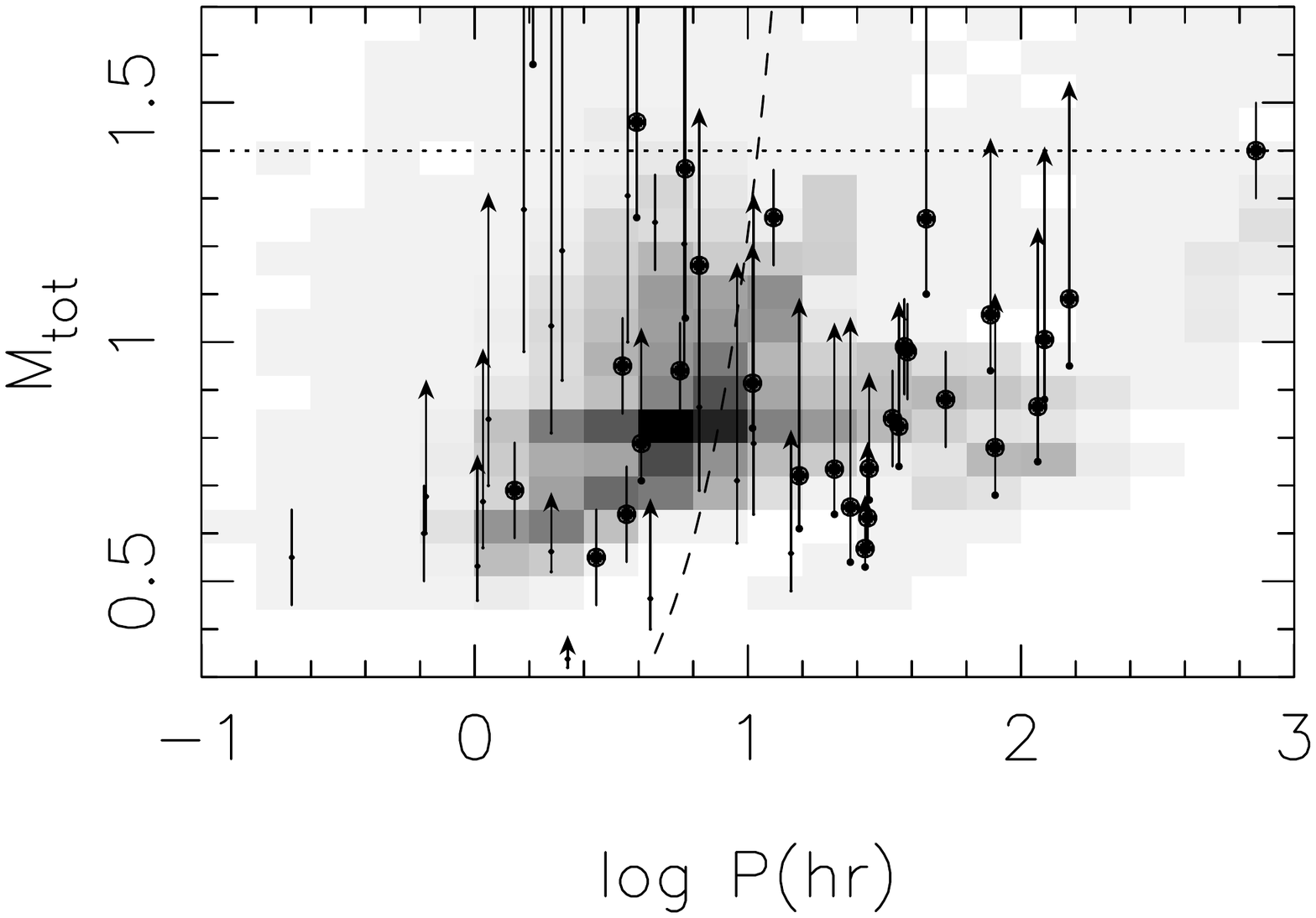}
    }
    \caption{The total masses of binaries in the simulated population of
      double WD with $V_{\mathrm{lim}}=21$ as a function of system's orbital
      period. In the left plot the outcome of the first common
      envelope stage is described by $\gamma$-formalism,
      Eq.~(\ref{ce_gamma}), and in the right plot -- by
      $\alpha$-formalism, Eq.~(\ref{A:CE:afai}). In both plots the
      latter equation describes both common envelope stages. The grey
      scale (the same for both plots) corresponds to the density of
      objects in the linear scale. Observed binary white dwarfs are
      shown as filled circles (see the original paper for
      references). The Chandrasekhar mass limit is shown by the dotted
      line. To the left of the dashed line the systems merge within
      13.5~Gyr. From \cite{2012AA...546A..70T}.}
    \label{fig:WD_SEBA}
  \end{figure}
}
\clearpage

The fate of DDs is solely determined by GWR. The closest of them may be brought
into contact by AML via GWR. For instance, a $(0.6+0.6)\ms$ WD pair can merge in
10~Gyr if the post-CE separation of the components is about $2\rs$. We shall
discuss the population of observed detached DDs below in
Section~\ref{section:observations}. If one of the DD components fills its Roche lobe, 
the outcome of the
contact depends on the chemical composition of the stars and their masses. There
are several possible endpoints: merger leading to a supernova (the left and
the right-most branches at the bottom of Fig.~\ref{figure:wd_flow}, see
\S~\ref{sec:snia}), stable mass exchange with the formation of an AM~CVn system (as
indicated in the Figure, see \S~\ref{section:am-evol}), direct formation of a
single massive WD (the central branch, see \S~\ref{sec:dd_snia} and discussion of
Fig.~\ref{fig:dd_outcomes}) or the formation 
of a R~CrB type star with an extended helium envelope 
\cite{Webbink84,1996ApJ...456..750I,2012MNRAS.426L..81Z,2013ApJ...772...59M},
 which will also end its evolution as a single WD (not shown in
Figure~\ref{figure:wd_flow}). White dwarfs produced by the merger process may be
hidden among single WDs without traces of H or He, like those with oxygen-rich
atmospheres \cite{2010Sci...327..188G}.

If the donor has a non-degenerate He-core (ZAMS 
mass $M_{\mathrm{ZAMS}} \gtrsim
 (2.3\mbox{\,--\,}2.8)\,M_{\odot}$) and the system does not merge, after
the second CE-stage a helium star\,+\,WD system can arise (the ``Non-degenerate
He-core'' branch of evolution shown to the right in Figure~\ref{figure:wd_flow}).
From the point of GW detection, 
important can be low-mass He-stars ($M_{\mathrm He} \aplt 0.8\ms$ for Z=0.02). 
The lifetime of He-remnants of the close binary components can be as long as that of their main-sequence progenitors \cite{it85}:
\begin{equation}
\label{eq:t_he}
t_{\mathrm He}\approx10\super{7.15}m_{\mathrm He}\super{-3.7}~{\mathrm yr}
\end{equation}
($m$ is the mass in solar units). 
These stars expand only slightly at the core-helium burning stage.  
If the separation of components is sufficiently small 
($\porb \aplt (120-140)$ min), AML via GWR may bring
He-star to RLOF while He is still burning in its core.
If $M_{\mathrm{He}}/M_{\mathrm{wd}} \lesssim 2$, a stable mass exchange
is possible \cite{skh86,2009MNRAS.395..847W}. 
A detailed study of the evolution of semidetached low-mass 
He-stars with WD companions may be found in \cite{2008AstL...34..620Y}, see also 
Section~\ref{section:am-evol}. 
These stars overfill their Roche lobes at orbital periods ranging from 16 to 50 min 
and first evolve to shorter periods  
with a typical $\dot{M}$ of several  $10\super{-8}\pyr$. The mass loss quenches the central 
nuclear burning, and the helium star becomes ``semi-degenerate''.
An AM~CVn-type system may be formed in this way,
see Section~\ref{section:am-evol}) for more 
details. One
cannot exclude that a Chandrasekhar mass may be accumulated by the WD in
this channel of evolution, but the probability of such a scenario seems to
be very low, $\sim$~1\% of the inferred Galactic rate of
SNe~Ia~\cite{sol_yung_am05}.
If the He-star completes the core He-burning
before RLOF, it becomes a CO-WD. In Figure~\ref{figure:wd_flow} it
``jumps'' into the ``double degenerate'' branch of evolution.

 If the mass 
of the helium remnant-star is $\approx(0.8-2.3)\ms$, it hardly can overfill the Roche lobe at the core He-burning stage, but  
after exhaustion of He in the core it expands and may refill the Roche lobe \cite{it85}. 
Mass-loss rates of these ``helium-giants'' occur in the thermal time scale of their envelopes: 
$\dotm \approx 10^{-6 \pm 1}$ Myr. They may lose 
several $0.1\ms$ and, if companion WD is initially sufficiently massive, the latter  may accumulate
\mch \cite{yoon_lang_sn03,2009MNRAS.395..847W}.  If the ``massive'' 
He-star completes core He-burning before RLOF or its companion does not explode, the
He-star turns into a CO-WD. In Figure~\ref{figure:wd_flow} it
``jumps'' into the ``double degenerate'' branch of evolution.
  



\subsection{Type Ia supernovae}
\label{sec:snia}

The most intriguing possible outcomes of the evolution of compact binaries with
WDs are explosions of type Ia~SNe. Presently, the most stringent
constraints on the nature of the exploding object in \sna are placed
by the recent SN~2011fe.
We reproduce from the paper by Bloom et~al.~\cite{bloom_2011fe}
Figure~\ref{fig:2011fe} 
showing the 
constraints on the parameters of
the exploded object in SN~2011fe, which were obtained by excluding certain regions
of progenitors space on observational grounds. The blank space in
Figure~\ref{fig:2011fe} leaves only a compact object (WD or NS) as the progenitor to this SN, but the possible NS phase transition to a quark star is highly unlikely. 

\epubtkImage{cons2011fe.png}{%
  \begin{figure}[htbp]
    \centerline{\includegraphics[width=0.9\textwidth]{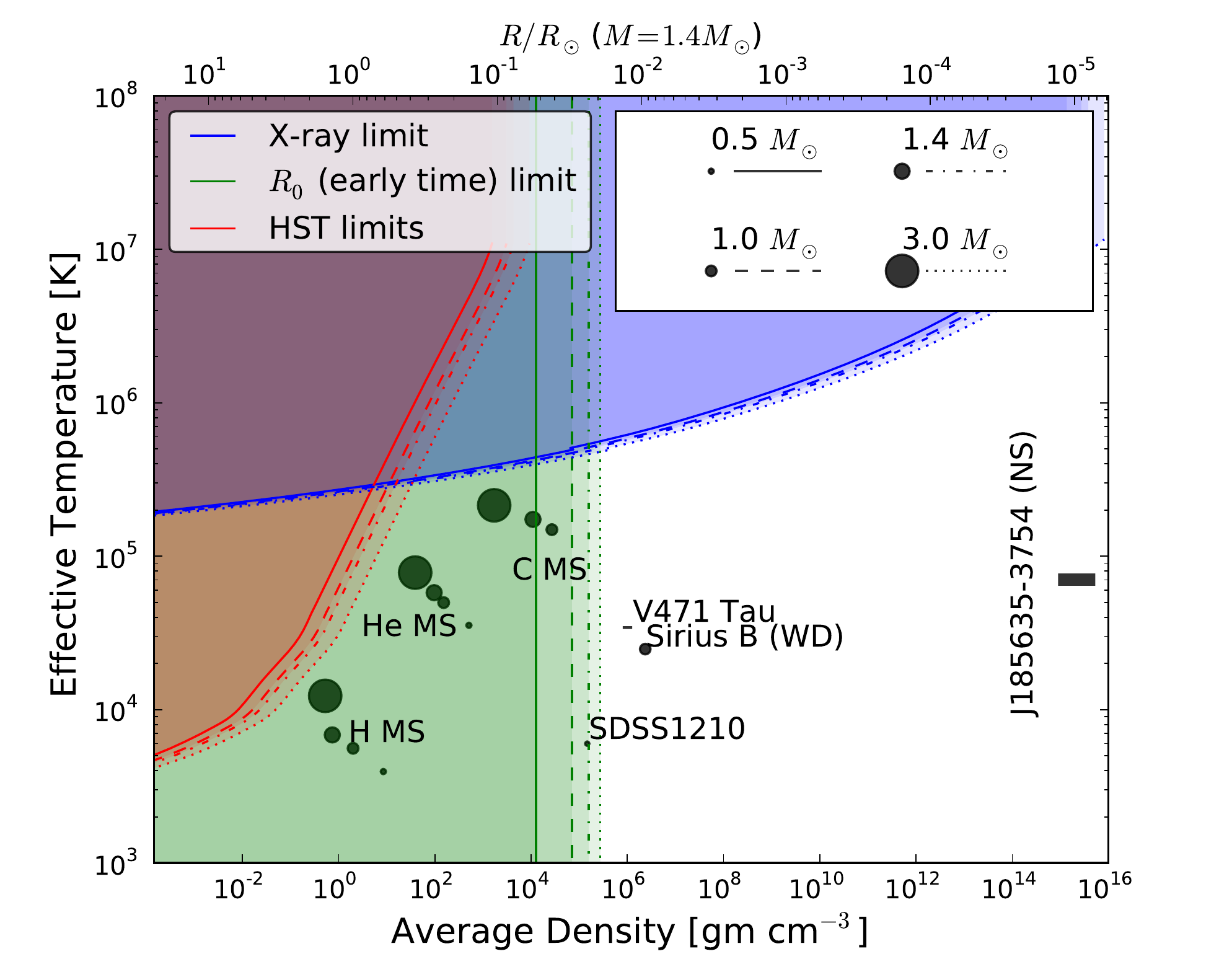}}
    \caption{Constraints on mass, effective temperature, radius and
      average density of the primary star of SN~2011fe. The shaded red
      region is excluded by non-detection of an optical quiescent
      counterpart in the Hubble Space Telescope imaging. The shaded
      green region is excluded from considerations of the
      non-detection of a shock breakout at early times. The blue
      region is excluded by the non-detection of a quiescent
      counterpart in the Chandra X-ray imaging. The location of the H,
      He, and C main-sequence is shown, with the symbol size scaled
      for different primary masses. Several observed WDs and NSs are
      shown. The primary radius in units of $R_{\odot}$ is shown for
      mass $M_p = 1.4\,M_{\odot}$. From \cite{bloom_2011fe}.}
  \label{fig:2011fe}
  \end{figure}
}
\clearpage 

 The exploding star radius estimate in Figure~\ref{fig:2011fe} is
derived by assuming that SN~2011fe was discovered about 11 hr after the explosion, as inferred by Nugent et al. \cite{nsct+11} from a simple shock-breakout model
which gives a measure of the stellar radius. However, as noted by Piro and
Nakar~\cite{2013ApJ...769...67P}, since the early luminosity
of \sna\ is powered by radioactive decay of $^{\rm 56}$Ni, if there is no Ni
mixed in the outermost layers of the ejecta, there can be a "dark phase"
lasting from a few hours to several days, where the only source of emission is
radiative cooling of shock-heated gas that is too dim to be detected. Duration
of the ``dark stage'' depends on radial distribution of $^{\rm 56}$Ni. Piro and Nakar suggested that the explosion of SN~2011fe
occurred earlier than estimated by Nugent et al.,
 and bounds on the radius of the exploding object are less stringent 
than those suggested by Bloom et al.
However, even several times larger limit hardly leaves place for any 
exploding object different from WD.
A comprehensive analysis of data on SN~2011fe obtained untill mid-2013
and their implications for
the problem of progenitors of \sna\ can be found in the review paper by
Chomiuk~\cite{2013PASA...30...46C}. Early time observations of SN2011fe can be explained in the frame of ``pulsational-delayed detonations'' paradigm for
\sne\ explosions \cite{2013arXiv1310.7747D}.

The theory of stellar evolution should be able to explain the formation of \sne progenitors
which fit the basic observational constraints: the explosion energy, the
frequency of explosions in the galaxies with different morphology, the observed
light curves and spectra, chemical yields, etc. 

Until very recently, the most popular paradigms for \sne used the accumulation of the Chandrasekhar mass $\mch\approx 1.38\ms$
by a CO WD via accretion in semidetached binaries or in 
a merging of binary WD with the total mass exceeding the Chandrasekhar limit. We discuss them below and then proceed to some other ideas
actively debated at present.

\subsubsection{Single-degenerate scenario}
\label{sec:ss_snia}

The formation of an $M=\mch$\ WD prone to a thermonuclear explosion seems to be a natural
outcome of accretion of hydrogen in a semidetached binary. The
masses of CO WDs are limited by $\approx 1.2\ms$ (see \S~\ref{section:introduction}).  It is
easily envisioned that, if hydrogen burning on the surface of an accreting CO WD is stable or occurs in mild
flashes, as well as the subsequent helium burning, and the WD is not eroded, 
the matter can be 
accumulated on the WD surface to increase the WD mass of WD to \mch. The WD mass growth via the 
hydrogen burning (``CV, SSS'' branch of evolution in
Figure~\ref{figure:wd_flow}) is considered to be equivalent to the direct
accretion of helium at the rate determined by hydrogen burning~\cite{nom82a}. 
Strictly speaking, the latter assumption is not completely accurate, since H-burning
modifies the thermal state of He-layer that accumulates beneath the H-burning shell, and in the case of direct accumulation of helium on the WD surface, 
the WD evolution can be different.
The efficiency of matter retention on the WD surface can also differ \cite{cassisi_etal98}. However, 
this issue has not been systematically studied as yet.

\epubtkImage{he_regimes_pty.png}{%
  \begin{figure}[htb]
    \centerline{\includegraphics[width=0.85\textwidth]{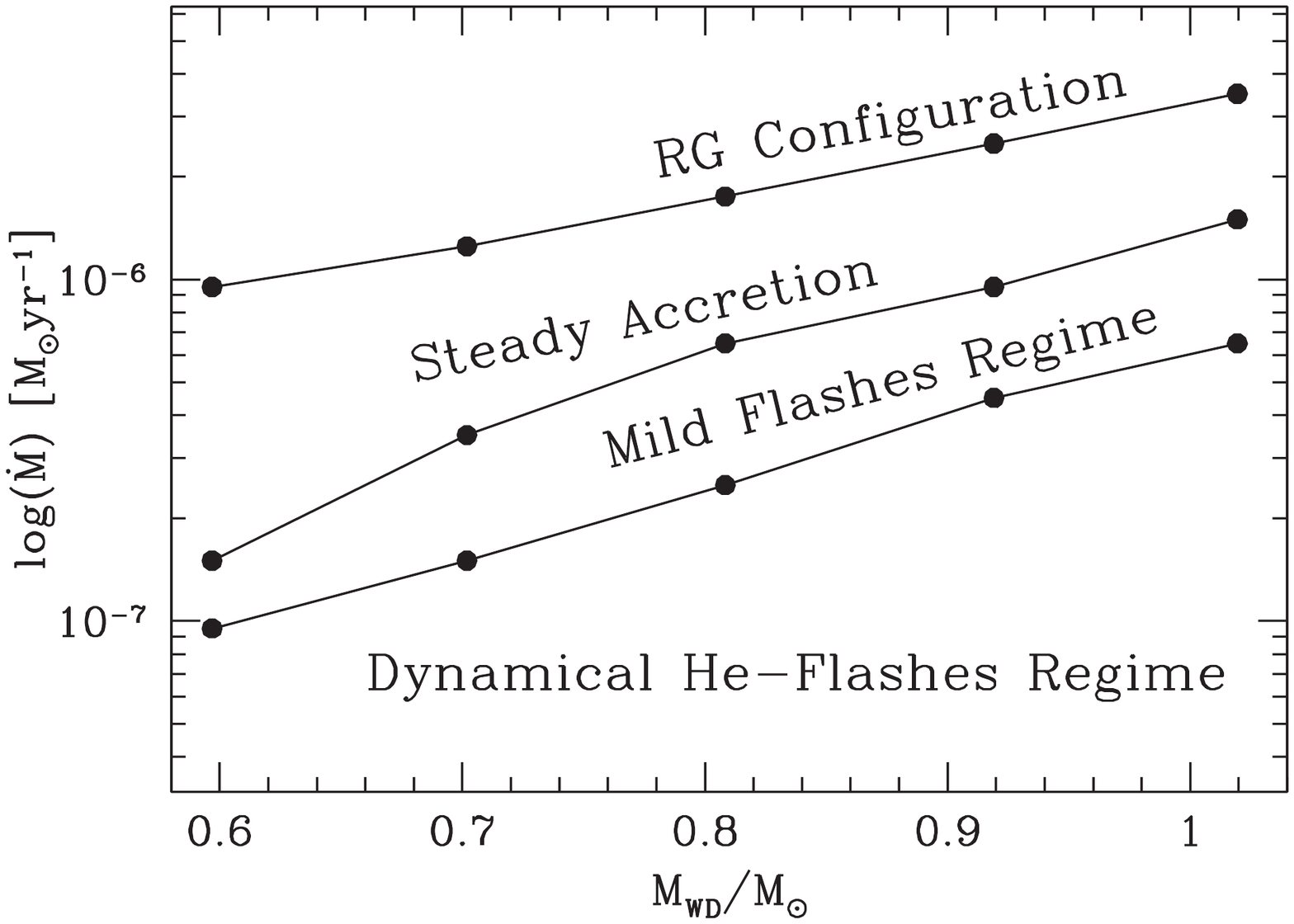}}
    \caption{Limits on different burning regimes of helium accreted
      onto a CO WD as a function of the WD mass and accretion rate
      \cite{2013IAUS..281..209P}, Piersanti, Tornamb\'e \& Yungelson
      (\textit{in prep.}). Above the ``RG Configuration'' line, 
      accreted He forms an extended red-giant like envelope. Below
      $(3\mbox{\,--\,}5)\times10^{-8}\myr$, accreted He detonates
      and the mass is lost dynamically. In strong flashes regime, He-layer
      expands and mass is lost due to RLOF and interaction with the
      companion. Helium accumulation efficiency for this regime is
      shown in Figure~\ref{f:heefficiency}, and the critical He masses 
      for WD detonation are shown in Figure~\ref{fig:detmashe}.}
    \label{f:heregimes}
  \end{figure}
}
\clearpage

\epubtkImage{accum1.png}{%
  \begin{figure}[htb]
    \centerline{\includegraphics[width=0.85\textwidth]{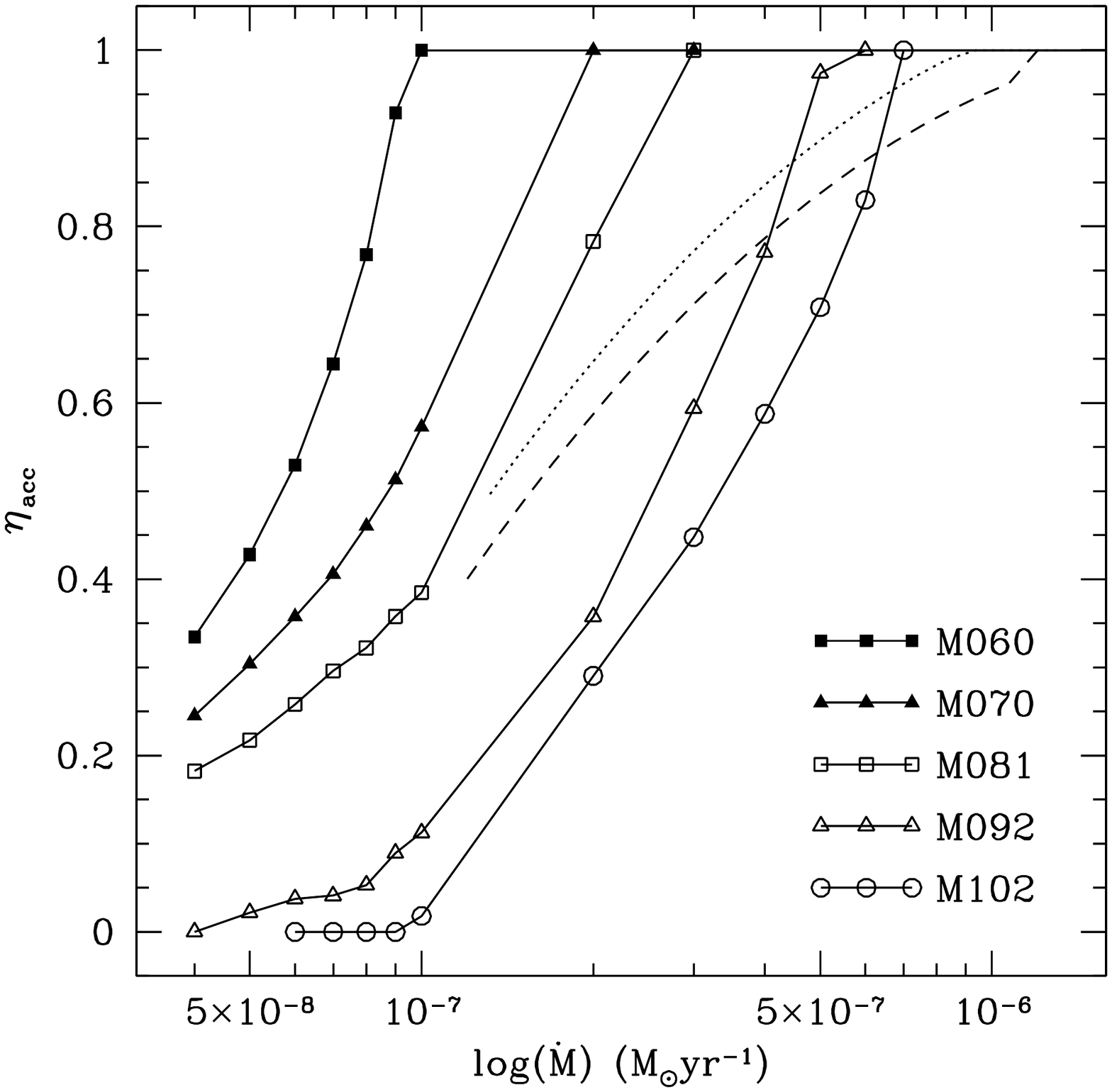}}
    \caption{The accumulation efficiency of helium  as a function of the
      accretion rate for CO WD models of 0.6, 0.7, 0.81, 0.92,
      $1.02\ms$ (top to bottom), as calculated by Piersanti
      et~al. (\textit{in prep.}). Dotted and dashed lines represent
      the accumulation efficiency for CO WDs with initial masses
      $0.9$ and $1.0\ms$, respectively, after Kato \& Hachisu
      \cite{1999ApJ...513L..41K}.}
    \label{f:heefficiency}
  \end{figure}
}
\clearpage

Another version of this scenario, which is sometimes discussed, involves a WD
accreting matter from the stellar wind of the companion in wide binary systems (in symbiotic stars). However, the following considerations immediately suggest low
efficiency of this channel. In symbiotic binaries, the components evolve independently,
hence the typical mass of the WD accretor should be small: $\sim 0.6\ms$. The
efficiency of accretion from the stellar wind of the optical companion is of the order of 10\%
\cite{1952MNRAS.112..195B,2009ApJ...700.1148D}. Thus, there is little chance to
accumulate \mch, which is also confirmed by numerical simulations~\cite{yltk95,
lu_yung_han_symb06, 2010AstL...36..780Y}.

Note that the ``single-degenerate''(SD)  channel to \sna encounters 
severe problems. Stable burning of hydrogen or burning in mild flashes
on the surface of a massive WD may occur 
within a narrow range of accretion rates $\mdot\sim(10^{-6}\mbox{\,--\,}10^{-7})\myr$
(Figure~\ref{f:nomotorates}). 
Mass-loss rates by MS-stars in cataclysmic variables driven by the angular momentum loss 
via magnetic stellar wind are below several $10^{-8}\myr$ and result in 
Nova explosions, which limit the growth of $M_{\mathrm{wd}}$ or even cause their erosion 
\cite{prkov95,yaron05,2007MNRAS.374.1449E}.   
 Donors in the Hertzsprung gap or on the early red-giant branch may provide higher \mdot\ 
if they lose mass in  thermal time scale $\dot{M}_{d,th}$. Masses of companions
should be commensurable to the masses of WD by virtue of the  mass-exchange stability conditions. 
However, the mass loss in these stars in the initial phase of mass exchange
proceeds on a time scale that is even shorter than the thermal one
\cite{1972AcA....22...73P} and \mdot\ can exceed $\dot{M}_{\mathrm{RG}}$ 
(Figure~\ref{f:nomotorates}) by 1\,--\,2 orders of magnitude and even be
higher than \medd.

 It is possible that the precursor of a SD-\sna is observed, e.g., in the supersoft
X-ray source WX~Cen \cite{2013ApJ...772L..18Q}. If for this 
$\porb=0.4169615(\pm22)$\,d  system \cite{2004MNRAS.351..685O},  
$dP/dt = -5.15\times10^{-7} \mathrm{d\ yr^{-1}}$ reflects the real secular decrease   
of the orbital period, then the donor is more massive than the accretor, and the time scale of mass exchange is  
thermal. The estimated mass of WD in this system 
 is $0.9\ms$, and there is a chance that the mass of WD will grow up to \mch. 
{ It is sometimes claimed, see, e. g., \cite{2012BASI...40..393K} and
references therein, that recurrent Novae may be also precursors of \sne.
These stars are presumably cataclysmic variables with 
high-mass ($\sim 1.3\ms$) WD-components, accreting at the rates slightly lower than the $\dot{M}_{\mathrm{stable}}$ (see Fig.~\ref{f:nomotorates}).
In this burning regime some mass accumulation is possible, e.g.,
 \cite{prkov95,yaron05}. However, since it is expected that accumulation of very little mass is necessary for a massive dwarf 
to initiate a thermonuclear runaway ($\sim 10\super{-6}\ms$), the  
appropriate ``theoretical'' Novae rate will far exceed the observed rate
\cite{2014arXiv1401.6148S}.  
Apparently, the mass function of CO~WD in cataclysmic variables is such, that the number of massive WD is vanishingly small.


As a remedy for the deficit of the candidate SD \sne and lack, at that time, 
of obvious candidates for \sna from merging dwarfs, Hachisu et~al. 
\cite{1996ApJ...470L..97H} suggested
 that the excess of matter over $\dot{M}_{\mathrm{RG}}$ can be
blown out from the system by optically thick winds, similar to
 the winds blowing from Novae after eruptions, which were introduced into
 consideration by the same authors
 \cite{kathac94}\epubtkFootnote{Note that the existence of these winds
 has never been proved rigorously by radiative transfer theory.}. The
 efficiency of accretion then becomes
 $\dot{M}_{d,th}/\dot{M}_{\mathrm{RG}}$. The angular momentum is lost
 in the ``isotropic reemission'' mode, and this also rises the upper
 limit of the components ratio thus allowing a stable mass exchange
 (see the discussion in
 \cite{1996ApJ...470L..97H,yl98}). The strong wind
 stops when the mass-loss rate drops below $\dot{M}_{\mathrm{RG}}$.
 However, even in this regime of steady accretion WD must lose mass via 
 strong wind, since its surface is hot ($(1-3)\times10^5$\,K) and, hence, 
 the efficiency of mass capture by the accretor is $<1$. It drops further in the flaring
 regime, when the expanding envelope of WD may exceed the 
 corresponding Roche lobe radius \cite{it96symb}. Finally, at
 $\dot{M}_{\mathrm{accr}} \lesssim 10^{-8}\myr$ all accumulated 
mass is lost in novae explosions~\cite{prkov95, yaron05}. 

The same authors introduced into the model of SD \sne
progenitors the ``stripping effect''. It assumes that a strong hot
wind from WD irradiates the companion and induces an additional mass loss. The combined effect of the additional mass and angular
momentum losses rises further the masses of WD companions which
still allow a stable mass exchange: to about $3\ms$ and $6\ms$ for
MS- and red-giant companions, respectively  
(see \cite{2008ApJ...683L.127H,2012ApJ...744...69H} for
the latest versions of the model). However, the ``stripping effect''
needs a very strong fine tuning and is hardly feasible. 

The next major complication of the SD \sne model is associated with mismatch 
of regimes of stable and unstable burning of H and He, noted 
by Iben and Tutukov \cite{it96symb}; see the paper by Bours et al. 
\cite{2013AA...552A..24B} for the latest
study of the influence of H and He retention efficiency upon the  rate of \sne with 
SD-progenitors.
In 
Figure~\ref{f:heregimes} we show the limits of different burning 
regimes of helium accreted onto a CO WD as a function of the WD mass 
and accretion rate \cite{2013IAUS..281..209P}, Piersanti, Yungelson and 
Tornamb\'{e}  (\textit{in prep.}).
Comparison of Figures~\ref{f:nomotorates} and 
\ref{f:heregimes} clearly suggests that, if even some He can be accumulated via H-burning 
in the regimes of steady burning or mild flashes, it can be almost entirely
lost in strong flashes of He-burning (Figure~\ref{f:heefficiency}).
In fact, the issue of He-accumulation is still not settled.
In the latter Figure we also show for comparison He-accumulation efficiencies computed 
by  Kato \& 
Hachisu~\cite{2004ApJ...613L.129K} under different assumptions. The most important
source of difference is, in our opinion, the ignorance of the interaction of the 
expanding envelope of the flaring 
WD with the companion in \cite{2004ApJ...613L.129K}. 

Results of the most recent calculations of accretion of H up to accumulation of He-layer prone to explosion 
through thousands of small-scale outbursts which do not erode the dwarf,
are controversial. Idan et~al.~\cite{idan_shaviv_he13} found that
accretion onto massive  $(\geq1)\ms$ WD at $\dotm \sim 10^{-6}\myr$ leads
to a strong flash removing all accumulated
helium. On the other hand, Newsham et~al.~\cite{2013arXiv1303.3642N} claim that WD with 0.70, 1.00, and $1.35\ms$ accreting at similar \mdot\ can continuously grow in mass despite recurrent helium flashes 
and, hence, can be progenitors of \sne. The main difference 
between two model is in the assumptions on the mass-loss in outbursts: while Idan et~al. treat 
it as dynamic ejection, Newsham et~al. consider it as a less efficient (and less sophisticated in treatment)
super-Eddington wind, occurring if the luminosity exceeds the Eddington limit. 
As a result, in the latter
simulations the models experience a series of weak He-outbursts and retain some He. Note that in the computations
of Newsham et~al.\ the accreted matter does not interact with the core  material; 
this seems to be at odds with general wisdom suggesting that diffusion or other mixing processes (e.g., shear mixing) should occur at the interface.

As we already mentioned, Bours et~al.~\cite{2013AA...552A..24B} attempted to compare the rates of \sne produced via the SD channel using in the population synthesis code several sets of retention efficiencies and values of the common
envelope parameter $\al$ or $\gamma$-parameter of Eq.~(\ref{ce_gamma}) employed
by different groups. For a Milky-Way like galaxy, none of the computations
produced the \sne rate larger than $\sim$~10\% of that inferred for the Galaxy
-- (4~\textpm~2)~\texttimes~10\super{-3}~yr\super{-1}~\cite{captur01}. 
Claeys et al. \cite{claeys_sn_14} have shown that SD \sne start to dominate SN
Ia rate if common envelopes are inefficient (\ace=0.2); the rate of \sne also
becomes then comparable to the rate in the field-galaxy dominated sample, but
remains by factor 3 lower than the rate in galaxy clusters. Only unrealistic
unlimited accretion onto WDs allows one to obtain an \sne rate compatible with the
observed galaxy-cluster rate.
 
Yoon and Langer \cite{2004A&A...419..623Y,2005A&A...435..967Y} and Hachisu et al.
\cite{2012ApJ...756L...4H} have shown that if 
accreting WDs rotate differentially, their mass can grow 
without explosion up to $M\approx 2.3\ms$, unless they lose the angular momentum through secular instabilities. 
Supernova explosion happens during spin-down of such a WD (``spin-up/spin-down'' scenario). 
If true, this is a possible way to accumulate enough mass for \sna\ in the SD-scenario and 
even provide an explanation for some of ``super-Chandrasekhar'' \sna.
Conclusions of Yoon and Langer were questioned by Piro \cite{2008ApJ...679..616P} who claims that 
baroclinic instabilities and/or the shear growth of small magnetic fields provide sufficient torque to 
bring the WD very close to rigid-body rotation in a time shorter than the typical accretion time scale; in the latter case the mass of non-exploding WD is limited by $\approx1.5\ms$ \cite{1973ApJ...183..215D}, leaving no room for efficient mass
accumulation. Thus, ``spin-up/spin-down'' scenario is still highly questionable.  

There is ample observational evidence that at least a fraction of \sna is not produced by 
semi-detached systems or systems with a detached giant (symbiotic stars). 
The major objection to the SD scenario comes from the fact that no hydrogen is
observed in the SNe~Ia spectra, while it is expected that up to several tenth
of $\ms$ of H-rich matter may be stripped from the companion by the
SN shell, see \cite{2012ApJ...750..151P,2012AA...548A...2L,2013ApJ...765..150S}
and references to earlier work therein.
Hydrogen, if present, may be discovered both in very early
and late optical spectra of SNe and in radio- and X-ray~\cite{boffi_branch95,lentz_02}. 
Radio emission is produced by synchrotron mechanism from the region of interaction between 
the high-velocity SN ejecta and much slowly moving circumstellar medium
formed by the wind of the SN precursor or the wind of its companion 
\cite{1982ApJ...258..790C,1982ApJ...259..302C}.  The peak monochromatic emission depends
on the ratio of the rate of mass loss in the wind and its velocity $\dot{M}_w/u_w$ and allows one to estimate the pre-explosion $\dot{M}_w$. 
Panagia et~al.~\cite{panagia_radio06} find a 2$\sigma$ upper limit 
of $\sim3 \times 10^{-8}\myr$ on a steady pre-supernova 
mass-loss rate for 27 SNe. This excludes relatively massive companions to WD. On the other hand, 
such \mdot\ do not exclude systems, in which a fraction of mass lost by low-mass companions 
is accreted onto WD and retained (see above). 

Non-accreted material blown away from the system before the explosion
should become the circumstellar matter (CSM), hence the detection of CSM in \sna spectra would lend credence to the SD-model. 
Ionisation and latter recombination of CSM should manifest itself as time-variable
absorption features detectable (in principle) via multi-epoch
high-spectral-resolution observations. 
It was noted, however,  
that  for confidential differentiation
between CSM and ISM features for an individual SN 
multi-epoch observations are needed, while this condition was not always 
satisfied in early CSM detection reports. Currently, it is estimated that 
$18\pm11$\% of events exhibit time-variability that can be associated with CSM
\cite{2013arXiv1311.3645S}. However, 
immediately after the very first report of CSM-discovery in SN~2006X 
\cite{2007Sci...317..924P}
Chugai \cite{2008AstL...34..389C} noted that for the  wind densities typical for red giants, the expected optical depth of the wind in Na~I lines is too small for their detection, while under the same conditions the optical depth of the 
Ca~II~3934\,\AA\ absorption line is sufficient for detection, and concluded that the
Na~I and Ca~II absorption lines detected in SN~2006X could not be formed in the
red giant wind and are most likely related to clouds at distances exceeding the
dust evaporation radius ($> 10^{17}\mathrm{\ cm}$). The problem of
interpretation of sodium lines variability observations still remains open.   

\epubtkImage{kasen.png}{%
  \begin{figure}[htbp]
   \centerline{\includegraphics[width=0.5\textwidth]{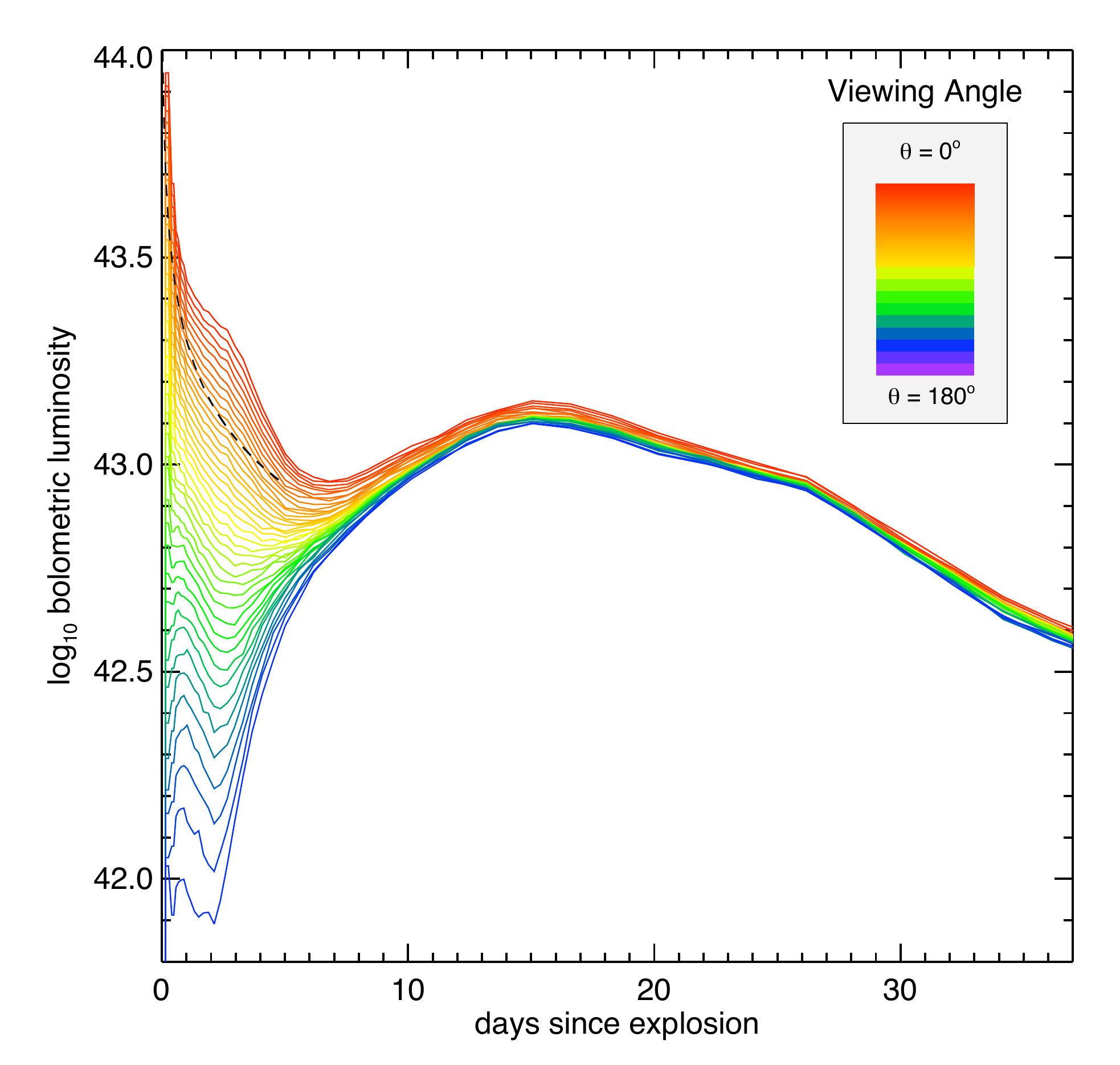}}
   \vfill
   \centerline{\includegraphics[width=0.75\textwidth]{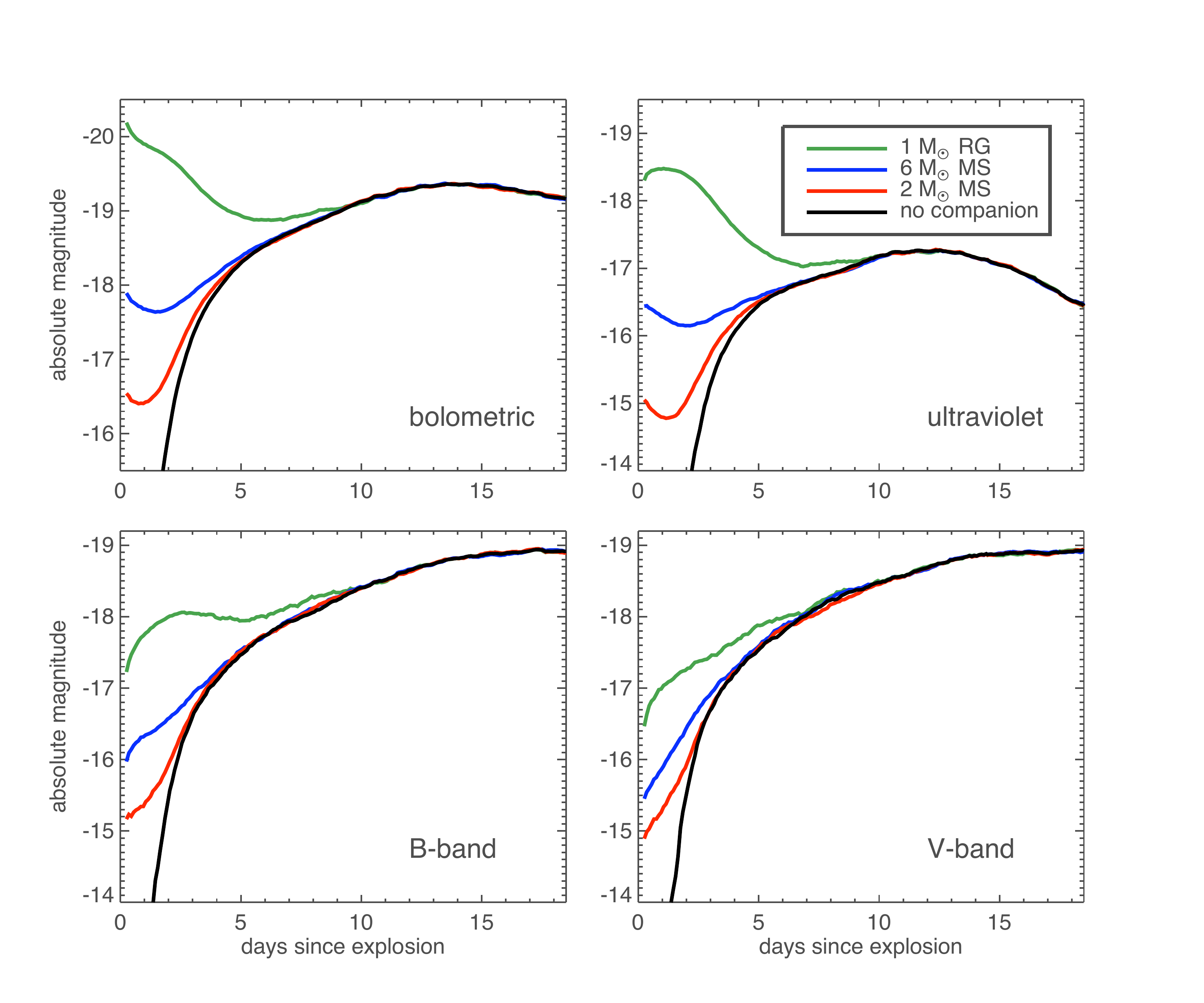}}
    \caption{\emph{Upper panel:} model light curve of a \sna having collided with a red
giant companion separated by $2\times10^{13}\mathrm{\ cm}$. The luminosity due
to the collision is prominent at times $t < 8$ days. The black dashed line shows the analytic prediction for the early phase luminosity. 
\emph{Lower panel:} Signatures of  interaction in the early broadband light curves of \sna 
for a red-giant companion at $2 \times 10^{13}\mathrm{\ cm}$ (green lines), a $6\ms$ main-sequence
 companion at $2 \times 10^{12}\mathrm{\ cm}$ (blue lines), and a $2\ms$ main-sequence companion at
$5\times10^{11}\mathrm{\ cm}$ (red lines). The
ultraviolet light curves are constructed by integrating the flux in the region 1000\,--\,3000~\AA\ and converting to the AB magnitude system. For all light curves shown, the
viewing angle is 0. From \cite{Kasen:2009si}.}}
  \label{f:kasen_effect}
  \end{figure}

Recently, Kasen \cite{Kasen:2009si} suggested an important test of the possible collisions of SN ejecta with the 
companion stars by observing  emission from the impact of shocks and SN debris which is due to dissipation of the kinetic
energy and reheating of the gas and occurs in the immediate vicinity to the star, at $\sim(10^{11}\mbox{\,--\,}10^{13})\mathrm{\ cm}$.  Immediately after the explosion, an X-ray outburst  with 
$L\sim 10^{44}\mathrm{\ ergs/s}$ at (0.1\,--\,2)~keV lasting from minutes to hours is produced.  Later,  radiative 
diffusion from deeper layers of the shock-heated ejecta produces optical/UV emission, which
exceeds the radioactively powered luminosity of the SN for a few days after the explosion. 
The properties of the emission provide a straightforward measure of the distance between the stars
and  the companion's radius (in the case of RLOF). The light-curves 
modeled by Kasen for different companions are shown in Figure~\ref{f:kasen_effect}.
Sure, the effect is prominent only for viewing angles looking down upon the
shocked region ($\theta\sim 0$) and, thus, may be found in statistical studies.
Bianco et~al.~\cite{2011ApJ...741...20B} applied Kasen's models to a sample of 87 spectroscopically
confirmed 
\sne from the Supernova  Legacy Survey \cite{2006AA...447...31A} and  
ruled out the contribution from white dwarf -- red giant binary systems
(separation $a\sim10^{13}\mathrm{\ cm}$) to \sne
explosions greater than 10\% at the 2$\sigma$ level, and greater than 20\% at the 3$\sigma$ level.
As noted by Kasen, optical detection of $\sim1\ms$ main-sequence companions 
($a\sim10^{11}\mathrm{\ cm}$) will be challenging, requiring measurement of subtle differences in
the light curves at times $\leq$~2~days.
Hayden et~al.~\cite{2010ApJ...722.1691H} applied the \sna light-curve template to 108 objects with well-confirmed
early-light curves from Sloan Digital Sky Survey (SDSS-II \cite{2008AJ....135..338F}) and found no 
shock signatures. This study limited mass of putative companions to exploding objects by about $6\ms$ and ``strongly disfavours red giant companions''.

In the SD scenario, the former companion of the exploding WD is expected to survive. 
It must have a high spatial velocity -- (100\,--\,300)~\kms \cite{2013AA...554A.109L} and a high luminosity, see  Table~\ref{table:schaefer}. 

%

In Table~\ref{table:schaefer} we present a summary of the expected
properties of companions compiled by Schaefer and
Pagnotta~\cite{2012Natur.481..164S} (see this paper for a brief review
of attempts to find ex-companions). As for now, all attempts to find
ex-companions failed (in the particular case of SNR~0509-67.5 in SMC
studied by Schaefer and Pagnotta, the non-detection was at the
5-$\sigma$ level).

A very ``popular'' object for the search of the former companion is
Tycho's SN, but even for this close remnant of a relatively recent
event (AD~1572) all claimed detections were disproved, see, e.g.,
\cite{kerzendorf_tycho13}. For another Galactic \sna\ candidate --- Kepler 
SN (AD~1604), the possible candidates with $L>\ls$ are excluded \cite{2014ApJ...782...27K}. 
\begin{table}
\renewcommand{\arraystretch}{1.2}
\caption[Properties of companions to exploding WD.]{Properties of
  companions to exploding WD. After \cite{2012Natur.481..164S}.}
\label{table:schaefer}
  \centering
{\small
\begin{tabular}{lcclc}
\toprule
Candidate Class                       & $P_{\mathrm{orb}}$ & $V_{\mathrm{ex-comp}}$ & Surviving companion  & $M_V$ \\
  & (days)    &   (km/s)      &   & (mag) \\
\midrule
Double-degenerate      &  ---      &  ---          & ---                   & ---   \\
Recurrent nova         & 0.6\,--\,520    &   50\,--\,350      & Red giant or subgiant & --2.5 to +3.5 \\
Symbiotic star         & 245\,--\,5700   &   50\,--\,250      & Red
giant             & --2.5 to +0.5 \\
Supersoft source       & 0.14\,--\,4.0   &   170\,--\,390     & Subgiant or $>1.16\ms$ MS & +0.5 to +4.2 \\
Helium star donor      & 0.04\,--\,160   &   50\,--\,350      & Red
giant or subgiant core & --0.5 to +2.0\\ 
\bottomrule
  \end{tabular}}
  \renewcommand{\arraystretch}{1.0}
\end{table}

Badenes et~al.~\cite{2007ApJ...662..472B} noted that most models for
\sna in the SD-scenario predict optically thick outflows from the
white dwarf surface  with velocities above 200~\kms. Such outflows
should excavate large low-density cavities around the
progenitors. However, Badenes et~al.\ found that such cavities are
incompatible with the dynamics of the forward shock and the X-ray
emission from the shocked ejecta in the known \sne remnants in the Galaxy
(Kepler, Tycho, SN~1006), LMC (0509-67.5, 0519-69.0, N103B), and in M31
(SN~1885). No sources corresponding to the pre-explosion objects have been
found in the archival Chandra data as yet
\cite{2013MNRAS.435..187N}. On the other hand, e.g.,
spectropolarimetric observations of type Ia SN~2012fr suggest a simple
axial symmetry of the explosion, which, together with the observed
features of Ca and Si in the spectrum, is inconsistent with the DD
merger scenario \cite{2013MNRAS.433L..20M}.

 It was speculated that in the case of the above-mentioned 
``spin-up/spin-down'' scenario for semidetached progenitors of \sne,
  if the remnant of the donor star becomes a WD, during the spin-down phase it can 
become too dim for detection and lose traces of hydrogen 
\cite{2011ApJ...738L...1D,2011ApJ...730L..34J,2012ApJ...759...56D}.  
However, the viscous time scale of a WD is 
still too uncertain for definite conclusions \cite{2013ApJ...778L..35M}.

As noted first by Canal et~al.~\cite{1996ApJ...456L.101C},
combinations of masses of accretors and donors enabling 
stable mass transfer and H-burning on the WD surface 
exist only for $\sim$~10\super{9}~years after the star-formation burst.
This means that the current rate of \sne with semidetached progenitors in early-type galaxies should be small, much lower than observed.
This conclusion is
apparently supported by the deficit of observed supersoft X-ray sources in the latter galaxies 
\cite{2010Natur.463..924G,2010ApJ...712..728D}: while the number of supersoft sources necessary to be consistent with the rate of \sne in MW or nearby galaxies is $\sim$~1000, less than about 100  are observed.
We note with caution that, although WDs which accrete H at  the
rates allowing its stable burning are conventionally, following  \cite{hbnr92}, 
identified with SSS, this 
inference still needs to be confirmed by the rigorous modeling of the spectra of accreting WDs.
Another test of contribution of accreting WDs to the rate of \sne in the early-type galaxies was
suggested by Woods and Gilfanov~\cite{2013MNRAS.432.1640W}: since accreting WDs, even if bloated, 
have photospheric temperatures $\sim$~10\super{5}\,--\,10\super{6}~K, they are powerful sources of ionising UV emission. 
 Then, if a significant population of steadily burning hydrogen WD exists, 
strong emission in He~II~$\lambda4686$ recombination lines should be expected 
from the interstellar medium. Application of this test to the
sample of about 11\,500 passively evolving galaxies allowed one 
to limit the contribution  of SD-scenario to the total \sne rate by (5-10)\%  
\cite{2014arXiv1401.1344J}.

\subsubsection{Merger of white dwarfs and double-degenerate scenario for \sna}
\label{sec:dd_snia}

A compelling scenario for accumulation of \mch\ involves two CO WDs in a close binary system that can  merge in the Hubble time due to the orbital angular momentum loss via GWR \cite{ty79a,ty81,it84a,Webbink84}. 
Theoretical models suggested that a proper
Candidate WD pair with the total mass $\apgt \mch$\ may be found among  
about 1000 field WDs with stellar magnitudes $V\lesssim 16\mbox{\,--\,}17$ \cite{nyp_01}. This number of (super)Chandrasekhar merging pairs also appeared sufficient to 
obtain the observed rate of \sne in a MW-like galaxy.  
It was speculated in~\cite{ty79a}, that, due to the fact that degenerate dwarfs expand with  mass loss 
($R \propto M^{-1/3}$), if the mass ratio of components 
is $<2/3$, the merger of pairs of WDs occurs on the 
dynamical time scale (in a time comparable to 
few orbital periods) and it was suggested that the 
lighter of the two dwarfs transforms into a ``heavy disc'' or ``envelope'' from which the matter accretes onto the
central object.
This inference was confirmed by SPH calculations of Benz et~al.~\cite{bbc_90} and many other studies.
However, it was shown for 1D non-rotating models
that the central C-ignition and SN~Ia explosion are possible only for accretion rate onto the central object
$\dot{M}_{\mathrm{a}}\lesssim(0.1\mbox{\,--\,}0.2)
\dot{M}_{\mathrm{Edd}}$~\cite{nomoto_iben85}, while
it was
expected that in the merger products of binary dwarfs $\dot{M}_{\mathrm{a}}$ is close to $\dot{M}_{\mathrm{Edd}}\sim 10^{-5}\,M_{\odot}
\mathrm{\ yr}^{-1}$ because of high viscosity in the
transition layer between the core and the disk \cite{mochko90}.
For such a high $\dot{M}_{\mathrm{a}}$, the nuclear
burning will start at the core edge and  propagate inward.
 If this is not the case, an off-centre ignition may be due to   
the Kelvin--Helmholtz contraction of the inner region of the envelope on a thermal timescale of 10\super{3}\,--\,10\super{4}~yr which compresses the base of the envelope. Nuclear burning 
produces an ONeMg WD \cite{2012ApJ...748...35S}. 
The latter will collapse without producing a 
SN~Ia~\cite{isern_83}. Note, however, that the analysis of the role of the angular momentum deposition into the central object and its thermal response to accretion in the ``canonical'' model led Piersanti, Tornamb\'{e} and
coauthors
\cite{piers_03b,2013MNRAS.431.1812T} to the conclusion that, as
a result of the WD spin-up, instabilities associated with
rotation, deformation of the WD, and AML by a distorted
configuration via GWR, the accretion  occurs not as
a continuous process but  episodically and, hence, the resulting
``effective'' accretion rate onto the WD decreases
to a few 
$10^{-7}\myr$. At this
$\dot{M}_{\mathrm{a}}$, a close-to-centre ignition of carbon becomes
possible. However, as noted by Isern et~al.~\cite{2012ASPC..453...99I},
the difficulty of this scenario is the avoidance of mass loss from the system.
The off-centre C-ignition can be also avoided if, after the disruption of the secondary,
several conditions are fulfilled: the local maximum temperature at the interface between the core (the former
primary) and the envelope (the former secondary) must
be lower than the critical limit for carbon ignition,  
the time scale of the neutrino cooling at the core/envelope interface is shorter than the 
angular momentum loss time scale, the mass-accretion rate from the disc must be lower than 
$5\times10^{-6}\mbox{\,--\,}10^{-5}\myr$ \cite{2007MNRAS.380..933Y}.

The problems of the SD scenario mentioned above, the discovery of several pairs of WDs with total mass close to \mch\ and the merger time shorter than the Hubble one
(see Section~\ref{section:observations}), and the discovery of several \sne of
super-Chandrasekhar mass
\cite{howell_2003fg_superch,2007ApJ...669L..17H,2009ApJ...707L.118Y,2010ApJ...715.1338Y,
2010ApJ...713.1073S,2011MNRAS.410..585S,2011MNRAS.412.2735T,2012ApJ...757...12S}
lent more credence to the DD-scenario. However, current formulation of this
scenario is quite different from the early versions due to several
circumstances. First, the accurate posing of initial conditions for the mass
transfer and the angular momentum conservation during it allowed one to follow
the merger process for $\sim$~100 orbits prior to the coalescence, to trace
variation of the accretion rate during this phase, and to show that explosive
events are possible before the coalescence or shortly after the merging
\cite{Motl_07,Dsouza_06,Guillochon_10}. Second, it was realised that the stability of the mass transfer also depends on the efficiency of spin-orbit coupling \cite{mns04}. In addition, if the circularisation radius of the accretion stream is less than the accretor's radius, a direct impact occurs. Otherwise, the accretion proceeds via disk. Analytical estimates of different regimes of mass exchange in binary WD systems are presented in Figure~\ref{fig:dd_stab_dan}.   

\epubtkImage{dd_stab_dan.png}{%
 \begin{figure}[htb]
    \centerline{\includegraphics[width=0.8\textwidth]{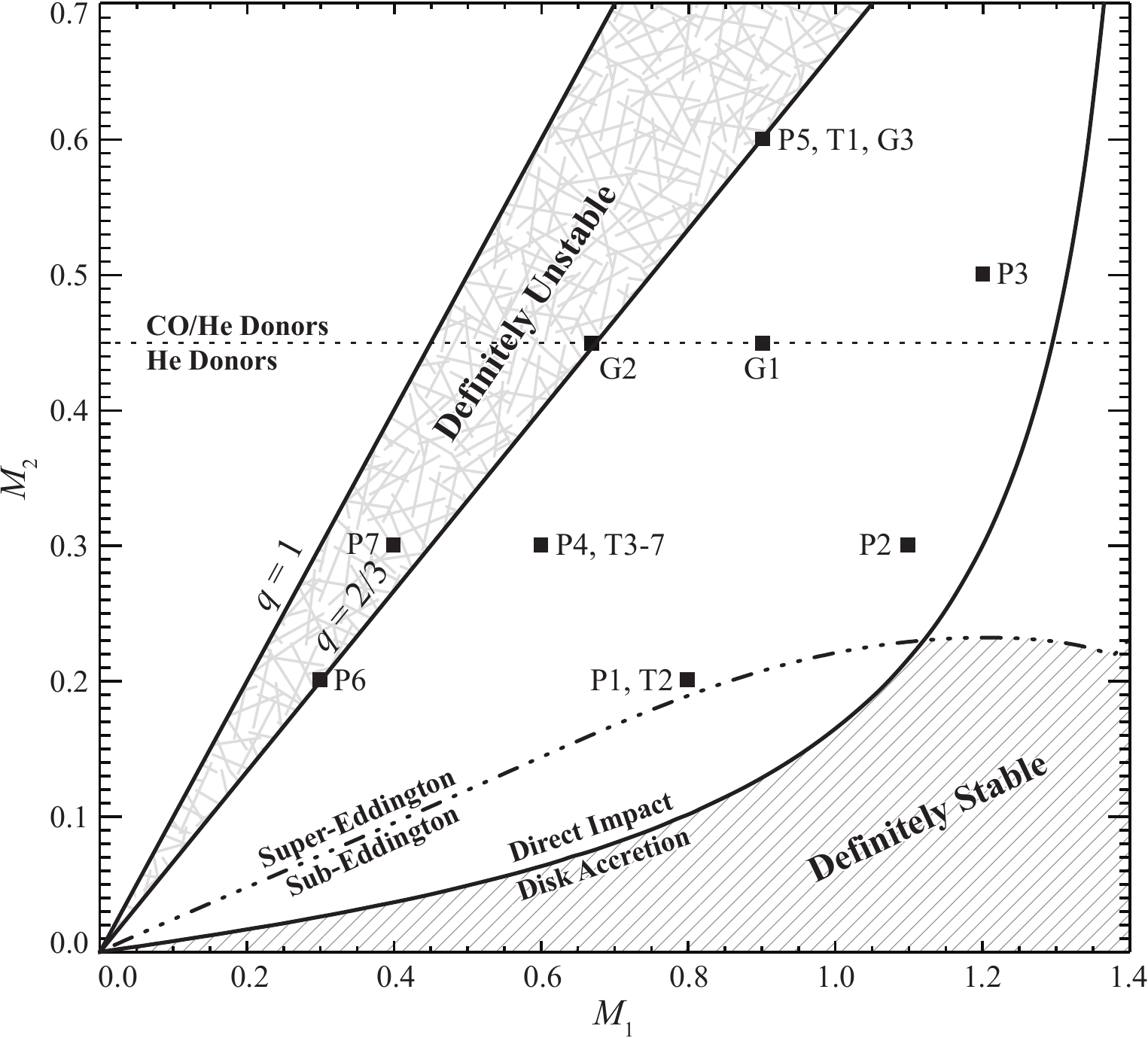}}
    \caption{Estimates of regimes of mass transfer in binary WD. The
      instantaneous tidal coupling is assumed. For longer time scales
      of tidal coupling, the stability limit in the plot shifts down
      \cite{mns04}. From \cite{dan_prelude11}. The filled squares mark
      initial positions of the models studied in the quoted paper.}
    \label{fig:dd_stab_dan}
  \end{figure}
}


Yet another very important, even crucial, fact concerns with the role of possible detonation of He. 
It was found by Livne and Glasner~\cite{livne90,lg91} that accretion of He onto a  
$(0.6\mbox{\,--\,}0.9)\,M_{\odot}$ CO WD at a rate close to 
$10^{-8}\myr$ results in the accumulation of a degenerate He layer which detonates when 
its mass becomes $\sim 0.1\,M_{\odot}$. The helium burning sweeps around WD and converging shock waves  initiate a compression wave that can
result in the central carbon detonation.
For a certain time, this model involving sub-Chandrasekhar mass accretors
(nicknamed ``edge-lit detonations'', ELD or ``double-detonations'') that may
occur in a MW-like galaxy 
at a rate of $\sim$~10\super{-3}~yr\super{-1}, was considered as one of
the alternative mechanisms for SD or DD scenarios for SNe~Ia explosions.
But it was shown by H\"offlich and
coauthors~\cite{hk96,1996ApJ...457..500H} that the behaviour of light curves
produced in this model does not resemble any of the known SNe~Ia, and this mechanism
was rejected. However, there exists a rather numerous class of events,  called SN~Iax, observational properties of which are, to a certain extent, consistent with
the model in which a CO WD accretes from a He-star, the burning starts in the centre,
the flame propagates to the surface, ejects some matter, but then damps out
(the so-called ``failed deflagration'' scenario). 
Such SNe presumably do not consume the whole WD and leave
bound remnants \cite{2013ApJ...767...57F}.

The interest to the ``double-detonation'' model re-surged in relation
to the outcome of He accretion on a WD in interacting double
degenerates~\cite{bild_accr06}. The latest studies by Fink et al. 
\cite{fink_dd10} show that He
detonation triggers core detonations in models ranging in
core mass from $0.810$ up to $1.385\ms$ with the corresponding shell
masses from $0.126$ down to $0.0035\ms$. Similar
estimates were obtained by Moll et all. \cite{moll_woosley_eld13}. It is possible
that even the least massive CO WD ($0.45\ms$) may experience double
detonations if accreted layers of He are sufficiently massive, close
to $0.2\ms$, as it is shown by Sim et al.~\cite{2012MNRAS.420.3003S}. 
 We note, however,
 that the robustness of these 
results was questioned by Waldman et al.~\cite{2011ApJ...738...21W}, since triggering a 
detonation 
must be investigated on scales much
smaller than those used, e.g.,  by Fink et al.~\cite{fink_dd10}.
Nevertheless, it is quite possible that 
the primary objects
for the DD-scenario may be pairs of CO-accretors and helium or hybrid donors.

By the present time, systems with direct impact are the best studied ones  (Figure~\ref{fig:dd_stab_dan}). As found by Guillochon
et~al.~\cite{Guillochon_10} and later explored in more detail by Dan
et~al.~\cite{dan_how_12,2014MNRAS.438...14D}, in the direct impact systems helium detonations in the torus of the accreted He with mass $\lesssim 0.1\ms$ can occur several orbits before the merging, if the nuclear
timescale $\tau_{3\alpha}$ becomes shorter than the local dynamical
timescale  $\tau_{\mathrm{dyn}}$.  If the subsequent shock compression
is strong enough, the second detonation may follow in the CO core and
lead to \sna. Another possibility is the ``surface detonation'' in the
layer of accreted matter immediately after the merging with the
subsequent central detonation. 

In Figure~\ref{fig:dd_outcomes} we show a ``map'' of the outcomes of
mergers of white dwarfs after Dan et~al.~\cite{2014MNRAS.438...14D}. These authors explored systematically 225 WD models with different mass and chemical composition.  Referring the interested reader for more  
details about the assumptions, computations and the results to the original  paper, 
we briefly consider the merger
products shown in the Figure (moving from bottom left to upper
right).  Note that some of the results were well known earlier.

\epubtkImage{dd_outcomes.png}{
  \begin{figure}[htb]
    \centerline{\includegraphics[width=\textwidth]{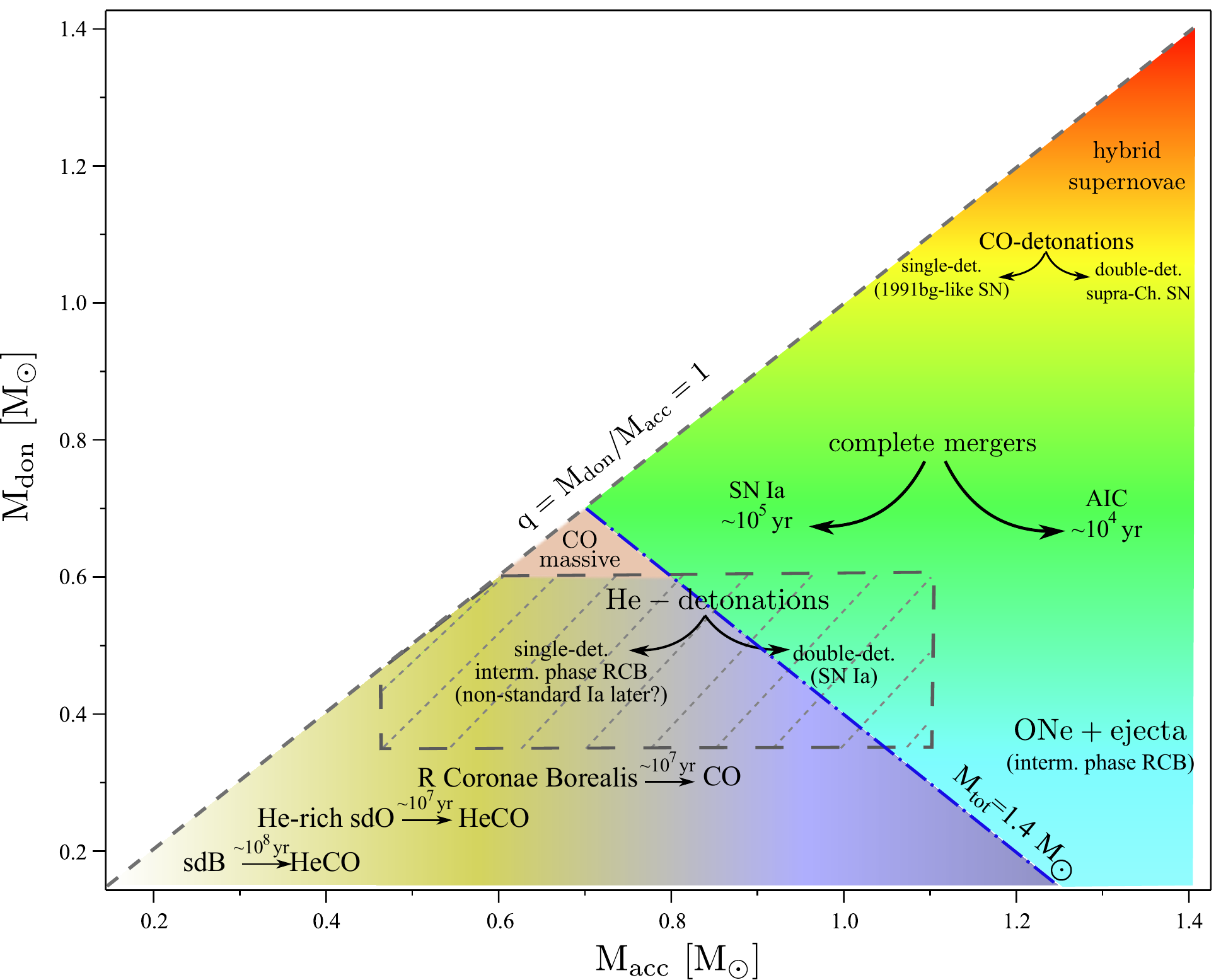}}
    \caption{Outcomes of merger of a binary WD depending on the mass
      and chemical composition of the components \cite{2014MNRAS.438...14D}.
 Systems in the hatched region are expected to experience He-detonations 
during the mass transfer or at the time of merger. 
      The numbers near the arrows indicate relevant timescales.}
    \label{fig:dd_outcomes}
  \end{figure}
}

If the mass of the merger product of two He WD exceeds $(0.38\,--\,0.45)\ms$, He burns in flashes
propagating inward and, finally, He ignites in the
core~\cite{sj00}. The object turns into a low-mass compact He star. If
the mass is larger than $\approx 0.8\ms$, an extended envelope forms
after the core He exhaustion ~\cite{pac_he71}. The
final product is a CO WD with relatively thick He mantle. Low-mass He
stars formed via merger are identified with single sdB stars. Mergers
of two hybrid or a CO WD and a He WD with the total mass lower than
$\approx 0.8\ms$ are suggested to produce sdO 
subdwarfs~\cite{2007AA...462..269S}. As a confirmation of the merger
scenario, one can consider the discovery of a rapidly rotating sdB star
SB~290 \cite{geier_fast_rot_sd13}. On the other hand, recently it was
shown that the merger of two He WDs with mass as small as $0.24\ms$
may result in the dynamical He-burning and, possibly, in 
detonation \cite{2013ApJ...771...14H}.
Incomplete He-burning is expected in the majority of WD
explosions in this case, and  \super{40}Ca, \super{44}Ti, or \super{48}Cr,
rather than \super{56}Ni will be the predominant burning products. 

The merger products of CO and He WDs with total mass exceeding 
$\approx0.8\ms$ form He giants. As mentioned above, it is suggested that this is one of the ways to form R~CrB stars \cite{Webbink84,1996ApJ...456..750I}. 
Another path to R~CrB stars may involve the merger of a CO WD and a hybrid WD
\cite{saio_jef02,pandey_extremehe06}. The fate of R~CrB stars is
determined by the ``competition'' between the core growth due to
He-burning in the shell and the mass loss from an extended
envelope. In rare cases where CO core accumulates \mch, a peculiar
supernova explosion can be expected. Otherwise, a massive  CO WD
forms.

There is a small patch in Figure~\ref{fig:dd_outcomes} occupied by
binary CO WDs with the total mass lower than \mch. It is expected that
the merger of their components will result in the formation of a
massive single WD (see \cite{ji_mgn_wd13} and references
therein). Several unusually massive WDs -- GD~362
\cite{2007ApJ...661L.179G}, RE~J0317--853 \cite{2010AA...524A..36K},
which have mass estimates close to the Chandrasekhar mass, can be
potential descendants of these pairs.

Helium detonations can result from the merging of WD pairs with
parameters lying within the hatched region in
Figure~\ref{fig:dd_outcomes}. These include CO accretors and most
massive helium and hybrid WD. Double-detonations are possible for most
massive objects. The known problem of this mechanism of production of
sub-Chandrasekhar \sna is that to make the modelled values consistent
with observations of \sna, a fine tuning of He and C abundances in
the accreting matter is required \cite{2010ApJ...719.1067K}. It is
important to note that the newly-formed He WDs have thin hydrogen
envelopes which may be stably transferred onto the companion prior to
the merger. As found by Shen et~al.~\cite{Shen_abs_13}, this material
is ejected from the binary system and sweeps up the surrounding
interstellar medium  $10^3\mbox{\,--\,}10^4\mathrm{\ yr}$ before the
\sna explosion. The resulting circumstellar medium profiles closely
match the ones found in some \sna.

As mentioned above, the merger of CO WD pairs with
$M_{\mathrm{tot}} \geq \mch$ may result in \sna if certain conditions
on the accretion rate in the core/envelope are fulfilled
\cite{piers_03b, 2013MNRAS.431.1812T, 2007MNRAS.380..933Y}. According to
Dan et al.~\cite{2014MNRAS.438...14D}, the merging CO WDs detonate only if $M_{\mathrm{tot}}
\apgt 2.1\ms$. Then they may be identified  with super-Chandrasekhar
\sna. However, we note that the consideration of abundances of
elements synthesised in \sne places certain constraints on the
exploding object mass. For instance, the solar abundance of Mn is
consistent with the assumption that about 50\% of  all \sne are
explosions of near-Chandrasekhar mass WDs~\cite{2013arXiv1309.2397S}.

Finally, it is possible that the merger involves a CO and an ONe
WDs. Such (very rare) mergers can give rise to ``hybrid SNe'' which,
in the presence of He layer, will manifest themselves as a kind of
SN~Ib or, if He-layer is absent, as Ia.

 We should mention that there exist merger calculations 
leading to \sna for configurations which did not produce explosions in the 
simulations by Dan et al. \cite{2014MNRAS.438...14D}.   

Pakmor et~al.~\cite{2013ApJ...770L...8P} simulated a merger of
$1.1\ms$ and $0.9\ms$ CO WD with $0.01\ms$ surface layers of He and
found  that the merger is violent, and He detonation forms close to
the surface of the core of the primary in the region compressed by a
shock in the He layer after only 7 orbits, i.e., much earlier than in
the study of Dan and coauthors, and also triggers detonation of C, not
encountered for this combination of dwarfs in the study of Dan
et~al. Based on their result, Pakmor et~al. claim that the merger of
CO WDs with low-mass He envelopes or a CO and a He WD thus provide a
unified model for normal and fast-declining \sna. 
Kromer et al. \cite{kromer_violent13} were able to reproduce 
the early-time observables of a subluminous \sna with slowly-evolving
light curve by violent merger of ``naked'' CO WD of $0.9\ms$ and $0.76\ms$
which leads initially to the local C-ignition at the hottest (compressed and heated) region at the  
accretor surface; then detonation develops around the hot bubble and consumes the entire merged structure. 

 We note that Scalzo et al. \cite{2014arXiv1402.6842S} estimated that 
$(0.9\,–\,1.4)\ms$
were ejected by  13 \sna\ studied by them. The   
correlation of ejected mass 
and $\mathrm{\super{56}Ni}$ in this  set of  \sna\ in  most of the cases did not completely comply with double detonation model (e. g.
\cite{fink_dd10}). Conclusion of Scalco et al. is that at least two formation channels, one for 
sub-Chandrasekhar mass \sna\  and another for Chandrasekhar mass and more massive 
\sna, are required. An interesting new channel to \sna\ discussed in this respect
are direct collisions of WD, which are possible in triple systems
consisting of two white dwarfs accompanied by a third star, thanks to Kozai resonances which can reduce
the time to the merger or collision 
\cite{2012arXiv1211.4584K,2013ApJ...778L..37K,2010ApJ...724..111R}. Both sub-Chandrasekhar-mass and
super-Chandrasekhar-mass SNe~Ia could arise through this 
channel. This inference finds support in the discovery of doubly-peaked 
(``two-horn'') line profiles
in several \sna\ \cite{2014arXiv1401.3347D}; such profiles are expected if two WD detonate \cite{2013ApJ...778L..37K}.

Having in mind still uncertain conditions for initiation of C-burning, Moll et al. \cite{moll_prompt13} and Raskin et al. \cite{Raskin:2013wqa}
considered \sna explosions initiated in the merger process of CO WD, 
either before disruption
of the secondary (dubbed by them ``peri-merger'' detonations) or  after complete 
disruption of the latter when the exploding object is surrounded by an envelope (``disk'').
An important conclusion of \cite{moll_prompt13} is that in the case of ``peri-merger'' detonations, a model for configuration with masses $0.96\ms$+$0.81\ms$ generally 
reproduced normal \sna.  Explosions of more massive configurations ($1.20\ms$+$1.06\ms$ 
and $1.06\ms$+$1.06\ms$) yielded over-luminous \sna.  Post-merger explosions produced peculiar \sna with features defined, to the great extent, by the disk structure; 
these   \sna do not conform with Phillips \cite{phillips_rel93} relation between the peak B-band magnitude 
and the decline rate of the light curve found for the ``standard'' \sna. In both cases, the observed brightness of events
strongly depends on the viewing angle. The mergers of massive systems may be invoked to explain super-Chandrasekhar \sne.       
      
The  difference in the results of simulations by different authors may be due to different initial conditions
of models~\cite{dan_how_12,raskin_remnants12} and resolution of 
the codes used~\cite{dan_how_12,2013ApJ...770L...8P}. Especially important is whether 
the binary components corotate at the merger onset.
These problems clearly warrant further 
examination.

Note in addition, that the intermediate mass components of close
binaries, before becoming white dwarfs, pass through the stage of a
helium star. If the mass of the helium star exceeds $\simeq
0.8\,M_{\odot}$, it expands to giant dimensions after exhaustion of He
in the core and can overfill the Roche lobe and, under proper
conditions, can stably transfer mass to the secondary
companion~\cite{it85}. For the range of mass-accretion rates expected
for these stars, both the conditions for stable and unstable helium
burning may be fulfilled. In the former case, the accumulation of
$M_\mathrm{Ch}$ and a SN~Ia become possible, as it was shown
explicitly by Yoon and Langer~\cite{yoon_lang_sn03}. However, the
probability of such a SN~Ia is only $\sim10^{-5}\pyr$.




\paragraph*{Core-degenerate scenario.}
\label{sec:core_degenerate}

As a subclass of the merger scenarios for \sna, one may consider the
so-called ``core-degenerate'' (CD) scenario. In the original form this
scenario involves the merger of a WD with the core of  an AGB star  \cite{sparks-74}
accompanied by the explosion in the presence of remainders of the
 AGB star, or a common envelope, or nearby shells of 
matter presumably ejected
shortly before the explosion. It was aimed at explaining SN~2002ic
with signatures of circumstellar matter in the
spectrum~\cite{2003ApJ...594L..93L, cy04}. Only 16 objects of SN~2002ic class were known at the time of writing of this review
\cite{2013ApJS..207....3S}. Further
development of 
the model resulted in a picture in which the merger
product is a rapidly rotating WD with mass close to or above
\mch. The fast rotation prevents the immediate explosion, which
happens only after a long spin-down stage, 
like in the spin-up/spin-down scenario 
\cite{2012MNRAS.419.1695I}.
The core-degenerate scenario is claimed to explain successfully,
 the basic features of SN~2011fe 
\cite{2014MNRAS.437L..66S}.

\vskip 0.3cm
\noindent
To summarise, the problem of progenitors of SNe~Ia is still
unsettled. Large uncertainties in the model parameters involved in the
computation of the evolution leading to the SN~Ia explosion and in
calculations of the explosions themselves, do not allow us to exclude
any type of progenitors. 
The existence of at least two families of
progenitors is 
not excluded by observations (see, e.g.,
\cite{mmn_snia_rev13} for the latest discussion).
A high proportion of
``peculiar'' SN~Ia $(36 \pm 9)\%$~\cite{li_01} suggests a large
spread in the ignition conditions in the exploding objects that also
may be attributed to the diversity of progenitors. Note that a high
fraction of ``peculiar'' SNe~Ia observed and their environmental dependence
\cite{2013arXiv1309.1182R} cast certain doubts on their accuracy  as
standard candles for cosmology.

\subsection{Ultra-compact X-ray binaries}
\label{sec:ucxb}

\enlargethispage{\baselineskip}
The suggested channels for the formation of UCXBs in the field are, in fact,
``hybrids'' of scenarios presented in Figures~\ref{figure:massive_flow}
and \ref{figure:wd_flow}. In UCXB progenitors, the primary forms a
neutron star directly via core collapse 
or via accretion-induced collapse (AIC) of a white dwarf. 
while the secondary is not massive enough to form a NS.
Then, several scenarios similar to the scenarios for the systems with the
first-formed white dwarf and driven by systemic AML via GWR are open. Usually,
two main scenarios are considered in which either white dwarf or a low-mass
helium star companion to NS overflows the Roche lobe. Also, a low-mass companion
to a neutron star may overfill the Roche lobe at the end of the main sequence
and become a He-rich donor. The dominant scenario differs from one population
synthesis study to another, depending, mainly, on assumptions on the initial
binary parameters, common envelopes, the kick velocity of nascent neutron stars,
conditions for the onset of mass transfer, retention efficiency of the matter by
NS, etc. In fact, in all cases evolution of the donors leads to their
transformation into similar low-mass degenerate objects.
  
For  the most recent studies of the origin, evolution and stability of
UCXB and review of earlier work we refer the reader to
papers~\cite{2012AA...537A.104V, 2012AA...543A.121V,
2013AA...552A..69V, 0004-637X-768-2-183, 0004-637X-768-2-184}. We
specially note that in~\cite{bel_taam_ucb04} it was found  that the
dominant accretors in UCXB can be black holes formed via AIC of
a neutron star. This result, however, can be a consequence of the choice
of the upper limit for the neutron star mass of $2\ms$, close to the
lower bound of current theoretical estimates of maximum mass of NS,
see, e.g., \cite{chamel_max_mass_ns13}.  

For UCXBs, like for \am stars, an analysis of details of the chemical
composition of donors, which show up in their optical or X-ray
spectra, seems to be a promising way for distinguishing between
the possible progenitors~\cite{2003whdw.conf..359N, 2010MNRAS.401.1347N,
  2013MNRAS.432.1264K}. For UCXBs, different chemical composition of
donors may manifest itself in properties of thermonuclear explosions
(type~I X-ray bursts) that occur on the surface of the accreting
neutron stars with low magnetic fields~\cite{1995ApJ...438..852B}.
We discuss this point in more detail in Section~\ref{section:am-evol} 
both for UCXB and AM~CVn stars.

In globular clusters, UCXBs are most likely formed by dynamical
interactions, as first suggested by Fabian et~al.~\cite{fpr_xray_gc75}; 
see, e.g., \cite{ivanova_ultra_gc05, lombardi_ucb_gc06} and references 
therein for the latest studies on the topic.


\newpage
\section{Observations of Double-Degenerate Systems} 
\label{section:observations}
                                                                                                                                                        
Interrelations between observations and theoretical
interpretations are different for different groups of compact
binaries. Cataclysmic variables like Novae have been observed for
centuries, their lower-amplitude cousins (including AM~CVn-type stars)
for decades, see 
monograph \cite{2003cvs..book.....W}
for a comprehensive historical review.  
Their origin and evolution found theoretical
explanation after effects of common envelopes and the orbital angular momentum loss,
in particular  via gravitational waves
radiation and magnetic braking, were recognized in the late 1960\,--\,1980s \cite{pac67a,pac76,ty79a,1980MNRAS.190..801W,vz81}. 
As of February~1, 2006 when the atlas and 
catalogue of CVs by Downes et al.~\cite{downes06,url03} were frozen, they contained about 1600 objects, but the number of known CV is permanently growing. 
Orbital periods were measured for about 1100~CVs; see the online catalogue by Kolb and Ritter~\cite{2011yCat....102018R}.
In particular, very recent discoveries \cite{carter13a,carter13b,kupfer13,levitan13}
brought the number of confirmed and candidate AM~CVn stars and related helium-rich CVs to over 40.
  
Ultracompact X-ray binaries were discovered with the advent of the X-ray
astronomy era in the late 1960s and can be found already in the first published
catalogues of X-ray sources (see for instance~\cite{giacconi74}). Their
detailed optical study became possible only with 8~m-class
telescopes. Currently, known population of UCXB consists of 13 systems with
reliably measured orbital periods, three candidates with uncertain orbital periods,
five without period measurement but very low optical-to-X-ray ratios, and eight faint
but persistent X-ray sources  \cite{2007AA...465..953I}. 
The place of UCXBs in the scenarios of evolution
of close binaries, their origin and evolution were studied already
before optical identification~\cite{ty81, rj84, tfey85}. Below, we briefly consider information on the known population of detached close binary white dwarfs and subdwarfs with WD companions. 
Interacting double-degenerate stars will be discussed in 
Sections~\ref{section:am-evol} and \ref{section:opt+x}.

\subsection{Detached binary white dwarfs and subdwarfs}
\label{sec:binwd}

Unlike CVs and UCXBs, the existence of close detached white dwarfs (double degenerates, DDs) was
first deduced from the analysis of scenarios for the evolution of close
binaries~\cite{webbink-79, ty79a, ty81, it84a, Webbink84}. It was
also suggested that DDs may be precursors of SNe~Ia. This theoretical
prediction stimulated optical surveys for close DDs, and the first such a binary was
detected in 1988 by Saffer, Liebert, and
Olszewski~\cite{slo88}. However, a series of surveys for
DDs performed over a decade~\cite{rob_shaft87, bgr_90,
fwg91, mdd95, sly98} resulted in only about a dozen of definite
detections~\cite{marsh00}.



\epubtkImage{spy080214.png}{
  \begin{figure}[htbp]
    \centerline{\includegraphics[angle=-90,width=\textwidth]{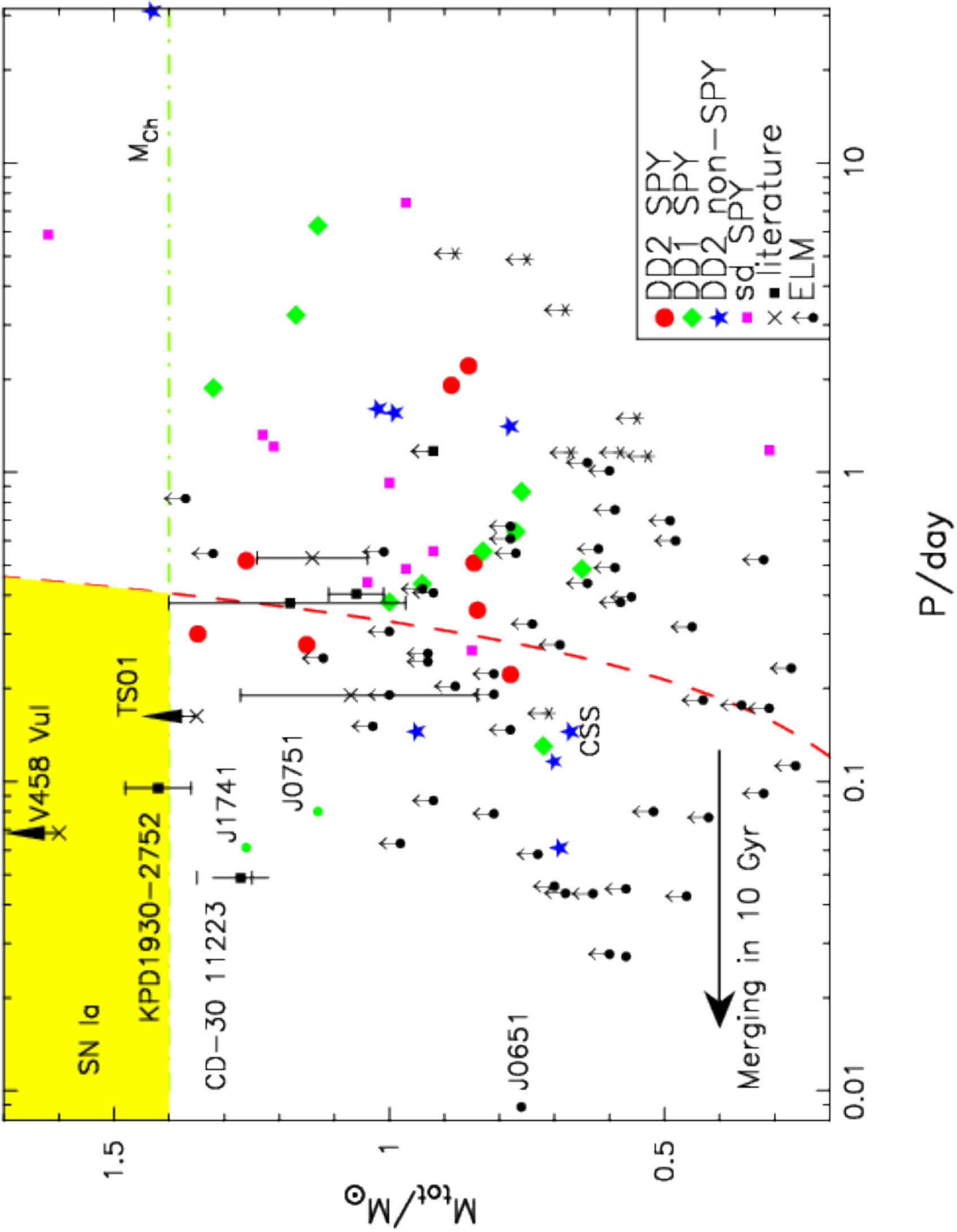}}
    \caption{Known close binaries with two WD components, or a WD and
    a sd component. Red  circles mark double-line WD found by
    SPY. Green diamonds are single-line WD found by SPY. Blue
    asterisks  mark double-line WD discovered in surveys other than
    SPY. Magenta squares are sd+WD systems from SPY. Black crosses and
    small squares are single-line WD and sd found by different
    authors. Filled black circles are extremely low-mass WD (ELM) for
    which, typically, only one spectrum is observed. For single-line
    systems from SPY we assume inclination of the orbit
    $i=60^{\circ}$, for other single-line systems we present lower
    limits of the total mass and indicate this by arrows. 
    Green circles are double-lined ELM WD suggested to be definite
    precursors of AM CVn stars
    Several
    remarkable systems are labeled (see the text for details). 
   The ``merger'' line is plotted assuming equal masses of the components.
An update of the plot provided by R.~Napiwotzki for the previous version of this review.}
    \label{figure:pm_spy}
  \end{figure}}

The major effort to discover close DDs was undertaken by the ``ESO Supernovae Ia
Progenitors surveY (SPY)'' project: a systematic radial velocity
survey for DDs with the UVES spectrograph at the ESO VLT (with PI R.~Napiwotzky;
see~\cite{nap_01,koe+01,nap_spy_03}
for the design of the project and~\cite{2007ASPC..372..387N} 
for the latest summary of its results). 
The project was aimed at discovering DDs as potential
progenitors of \sna,
but brought, as a by-product, an
immense wealth of data on physical parameters, kinematics etc. 
of white dwarfs and subdwarfs, see e. g., 
\cite{koe+01,
2009A&A...505..441K,2012ApJ...757..116F,2011A&A...528L..16G,2006A&A...447..173P,
2007A&A...470.1079V,2005ASPC..334..369K,2008MmSAI..79..723G}.
Theoretical models predicted that it can 
be necessary to search for binarity of up to 1000 field WDs with stellar magnitude $V
\lesssim 16\mbox{\,--\,}17$ to find a WD with mass close to \mch\ \cite{nyp_01}. 
More than 1000 white dwarfs
and pre-white dwarfs were observed (practically all white dwarfs
brighter than $V\approx16.5$\, available for observations from the ESO
site in Chile). SPY tremendously increased the number of detected DDs
to more than 150. Their system parameters are continuously determined from
follow-up observations. As well, among the objects detected by SPY there appeared a 
large number of subdwarf stars 
with massive WD companions which also present a substantial interest to the \sna problem (see papers by S.~Geier and his coauthors).

Figure~\ref{figure:pm_spy} shows orbital periods and  total masses
of the currently known (February 2014)
close DDs vs.\ orbital periods and compares them with
the Chandrasekhar mass and the critical periods necessary for the binary 
components to merge in 10\,Gyr for given $M_\mathrm{tot}$ 
and by assuming equal masses of components.
In fact, only one DD with $M_\mathrm{tot}$ close to \mch\ was was detected by SPY (see the Figure). On the other hand, more than hundred new close double WDs were discovered by SPY (most of them 
are still awaiting follow-up observations for determination of parameters). 


Meantime, several other interesting  objects were discovered, including, first of all, \epubtkSIMBAD{PN~G135.9+55.9} = TS01, 
first detected
as the nucleus
of a planetary nebula \cite{tovmassian-04}. Later, its binarity was discovered. 
A study of its parameters led to the total mass estimate
possibly exceeding \mch\ 
\cite{2010AA...511A..44S,2010ApJ...714..178T}. 
Another remarkable system is V458~Vul (Nova Vul 2007 No. 1) -- 
the nucleus of a planetary nebula with a massive WD companion which exploded as a Nova in 2007. 
Masses of the components are estimated to be $0.58\ms$ and $\apgt \ms$ \cite{2010MNRAS.407L..21R,2012MmSAI..83..811C}.
A high mass of PN in this system $(\sim0.2)\ms$ points to its formation via ejection
of a common envelope about 14\,000~yr ago \cite{2008ApJ...688L..21W}. 
Components of the system will merge due to the orbital angular momentum loss via GWR
in $\sim 10^{7}$\,yr. By this time, the current nucleus of PN will turn into a WD. 
Mass exchange after the contact is expected to be unstable (see Figure~\ref{fig:dd_stab_dan})
and this system is currently considered to be the most likely precursor of \sna.

Another possible \sna precursor is \epubtkSIMBAD{KPD1930-2752} -- an sdB star with 
unseen massive WD companion. The estimated total mass of the system ranges from $1.36\ms$ to $1.48\ms$ 
\cite{2007ASPC..372..393G}. In $\simeq 200$\,Myr prior to the merger, the present 
sdB star can turn
into a WD, and then a mass-transfer phase will ensue. Its stability depends on the efficiency of 
of tidal coupling (Figure~\ref{fig:dd_stab_dan}).  If the sdB-star will still remain in the  core He-burning stage 
at RLOF, an AM~CVn system will be formed.   

The binary system \epubtkSIMBAD{CD-30~11223} 
(\epubtkSIMBAD{GALEX J1411-3053}) is the shortest \porb\ 
sdB+WD  pair known to date \cite{2012ApJ...759L..25V}. Estimated masses of components are
$M_{\mathrm{sdB}} = 0.44\mbox{\,--\,}0.48\ms$ and $M_{\mathrm{WD}} = 0.74\mbox{\,--\,}0.77\ms$. The time before
RLOF by sdB star ($\approx 30$)\,Myr is apparently too short for 
its transformation into a WD, and 
stable mass transfer onto WD upon RLOF may be expected. Thus, this system is a 
potential precursor of an \am star.    

The most numerous population (58 objects) in Figure~\ref{figure:pm_spy} are
extremely low mass white dwarfs (ELM) -- helium white dwarfs with $M \aplt
0.25\ms$ found in a targeted spectroscopic survey (see
\cite{2013ApJ...769...66B} for the summary of results; data for the plot are
taken from this paper and \cite{2013MNRAS.tmpL.207K,2013arXiv1312.1665G}. Most of them have unseen heavier companions, most
probably another WDs, while in some cases even NS are suspected. Some may have
brown-dwarf companions. About half of ELM WD will merge with the companions in
less than the Hubble time. As Figure~\ref{fig:elm_q} shows, some of them, upon
the merger (which may take longer than the Hubble time) can start a stable mass
transfer, i.e., can form \am stars. 
The most remarkable of ELM WDs is
\epubtkSIMBAD{SDSS~J065133.338+284423.37} (J0651)
\cite{2011MNRAS.413L.101K,2011ApJ...737L..23B}, the shortest known detached WD,
for which relativistic decay of the orbital period is detected. It is expected
that it will be possible to detect some of ELM stars by space-born GW detectors
\cite{2013ASPC..467...47K}.

 Yet another important recently discovered objects are an eclipsing ELM WD  
\epubtkSIMBAD{SDSS~J075141.18−014120.9}  (J0751) and ELM WD 
\epubtkSIMBAD{SDSS~J174140.49+652638.7} (J1741) \cite{2013MNRAS.tmpL.207K}.
The masses of components of J0751 are $0.168\ms$ and $0.97^{+0.06}_{-0.01}\ms$, the orbital period is 
 \porb=0.08 day. The binary components will merge in about 270 Myr. 
Parameters of J1741 are $M_1=0.19\pm0.02\ms$, $M_2\geq1.11\ms$, \porb=0.06 day, the merging time is 160 Myr. With such a low mass ratios,
one can almost definitely expect a stable RLOF (see Figs.~\ref{f:solheim_stab}) for
J1741. For J0751, a stable mass exchange is also highly probable. This makes these two systems be  
the first found almost certain progenitors of AM~CVn stars and, possibly,  He-Novae
(see Section~\ref{section:am-evol}).

The observed precursors of ELM WD can be hidden about 
EL~CVn type binaries, which was 
recently identified 
as a separate class of objects -- eclipsing binaries with short orbital periods 
(0.7 -- 2.2) day,  
spectral type A primaries and very hot low-mass pre-He-WD \cite{2014MNRAS.437.1681M}.
Due to large mass ratios of components, 
RLOF by the A-stars in these binaries  
will result in common envelopes. If components will not merge in CE, theses systems will turn into ELM WD binaries.

Evolutionary sequences for ELM WD were computed by Althaus et al. 
\cite{2013A&A...557A..19A}, who also provided interpolation formulae 
for fitting masses and cooling ages of ELM WD as functions of $T_{eff}$ and $\log g$.

Binary WD \epubtkSIMBAD{CSS~41177} is an eclipsing \porb=2.78~hr  system with
helium WD components:$M_1=0.38\pm0.02\ms$, $M_2=0.32\pm0.01\ms$ 
\cite{2014MNRAS.tmp..132B}. The merging time of components is $1.14\pm0.05$ Gyr
and this system is the  perfect candidate to form a single sdB star \cite{ty90,sj00}, see 
previous Section.

The most massive pair of binary WD with parameters estimated from astrometric data and
optical and near-IR photometry is \epubtkSIMBAD{LHS 3236} \cite{2013ApJ...779...21H}.
It may be a pair of DA WD or a DA WD in pair with a DC WD.  Masses of its 
components are either  $0.93\ms$ and $0.91\ms$ or 
$0.98\ms$ and $0.69\ms$ (depending on types assigned to WD). However, a binary orbital
period of 4.03 yr corresponds to the tremendously long 
merging time  $\simeq 2\times10\super{13}$ Myr.

\epubtkImage{marsh_talk_warwick.png}{%
  \begin{figure}[htbp]
    \centerline{\includegraphics[width=\textwidth]{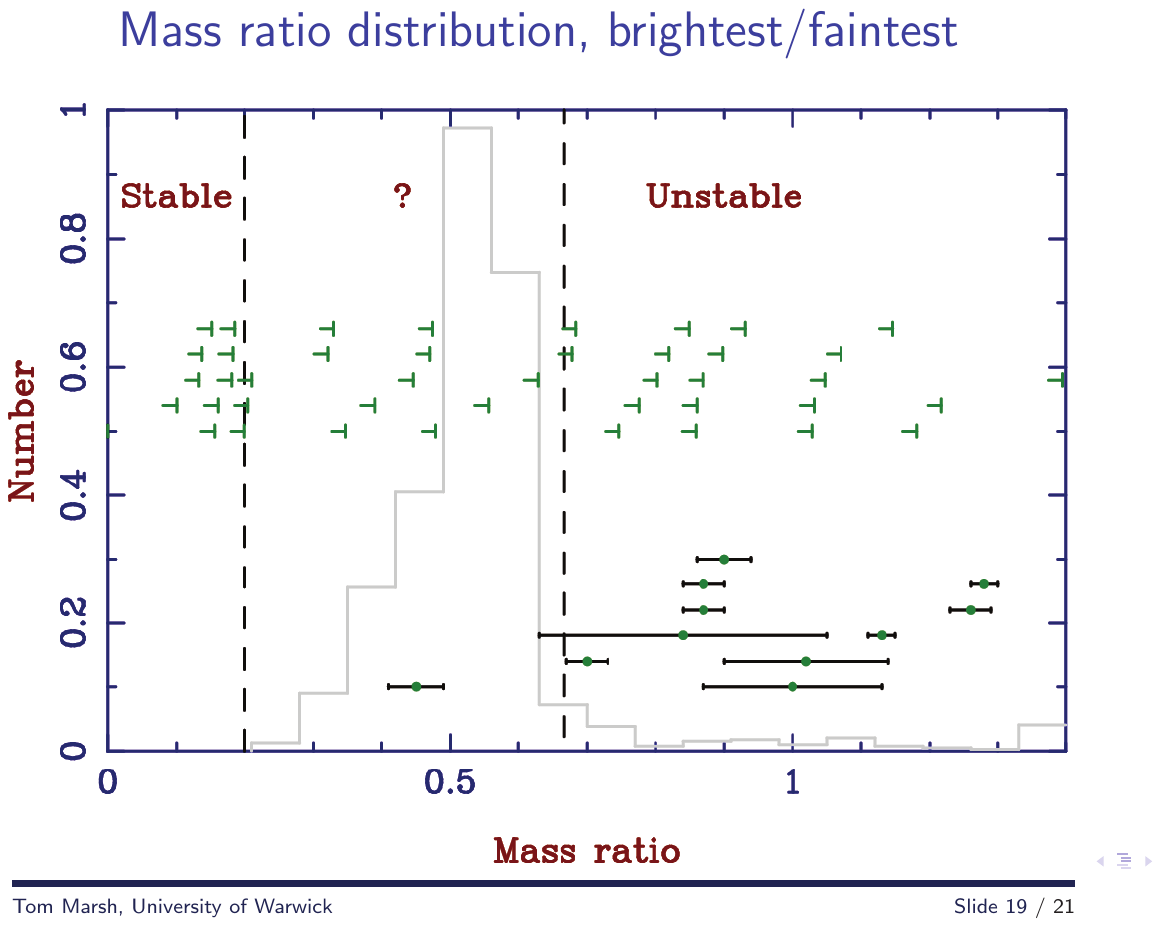}}
    \caption{Mass ratios of ELM WD. Vertical lines separate binaries
    which will, upon RLOF, exchange mass definitely stably (possible
    progenitors of \am stars), WD for which stability of
    mass-exchange will depend on the efficiency of tidal interaction, and definitely unstable stars. The latter systems
may be progenitors of \sne, as discussed in Section~\ref{sec:dd_snia}. 
Courtesy T.~Marsh~\cite{marsh12}.}
    \label{fig:elm_q}
   \end{figure}}


\newpage
\section{Evolution of Interacting Double-Degenerate Systems}     
\label{section:am-evol}


As shown in the flowchart in Figure~\ref{figure:wd_flow} and mentioned
above, there are several possibilities to form a semidetached system in which a WD stably accretes matter from another WD or a helium star. This happens due to the orbital angular momentum losses via GWR, since for the binary WD the orbital angular momentum loss is the only driver of the orbital evolution, and low-mass helium stars
($M\aplt 0.8\ms$) almost do not expand in the course of core He-burning. These
stars, also referred to as ``interacting double-degenerates'' (IDD), are
observationally identified with the ultra-short-period cataclysmic variables 
(\porb = (5\,--\,65)\,min) of the \am class. We shall use below both the terms ``\am
stars'' and ``IDD''. A distinctive peculiarity of \am stars is the absence of 
signs of hydrogen in their spectra (with one possible exception) which suggests
that donor stars in these binaries are devoid of hydrogen.
The absence of hydrogen seems to be reliable, since the threshold for
its detection is quite low: $(10^{-5} - 10^{-2})$ by the number of
atoms~\cite{1982ApJ...257..672W,solheim_nasser01,mhr91}.  

Note that apart from ``double-degenerate'' and ``helium-star'' channels for the
formation of AM~CVn-stars, there exists the third,
``CV''-channel~\cite{tfey85,tfey87,nelson86,phr_am03}. 
In the latter, the donor star fills its
Roche lobe at the very end of the main-sequence stage ($X_c\sim0.01$) or just
after its completion. For such donors the chemical inhomogeneity inhibits
complete mixing at $M\simeq0.3\,M_\odot$ typical for initially non-evolved
donors. The mixing is delayed to lower masses, and as a result the donors become
helium dwarfs with some traces of hydrogen. 
After reaching the minimum orbital periods $\simeq$~(5\,--\,7)~min, these binaries start to
evolve to longer orbital periods. 
 However, the birth rate of systems that can penetrate the
range of periods occupied by the observed AM~CVn-stars,
especially \porb$\aplt25$\,min, is much lower than the birth
rate in the ``double-degenerate'' and ``helium-star'' channels, and below we shall neglect this channel.

Schematically, the tracks of three families of \am stars are shown in Figure~\ref{fig:P_mdot_am} 
 and compared to the track for an ``ordinary'' CV. 

\epubtkImage{P_Mdot_min.png}{
  \begin{figure}[htbp]
   \centerline{
\includegraphics[width=0.8\textwidth]{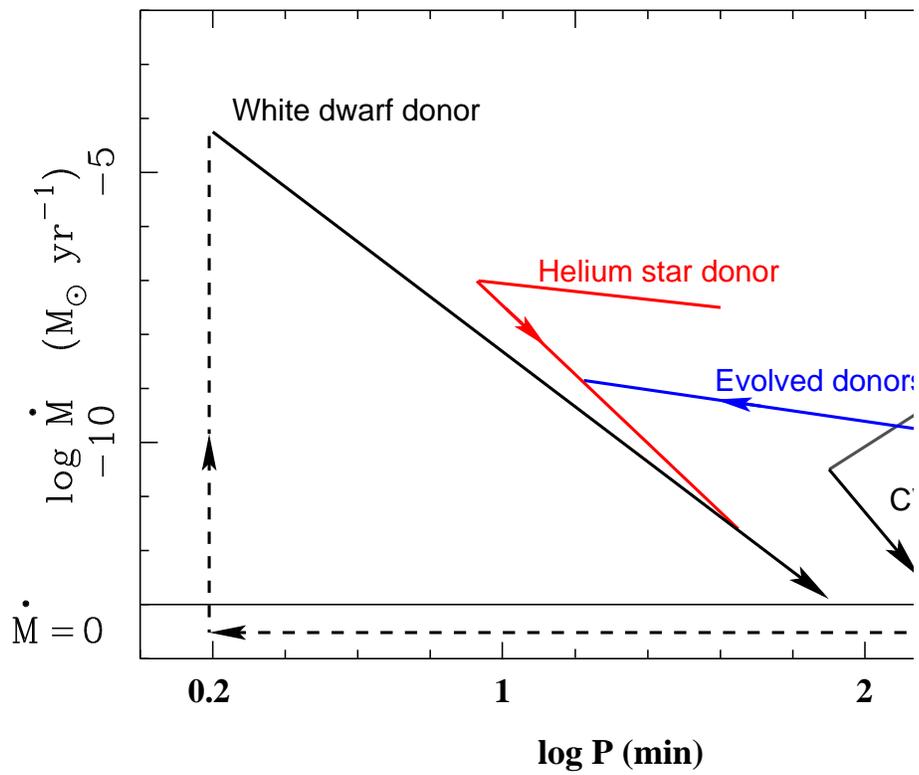}}
    \caption{Sketch of the period -- mass transfer rate evolution of the
    binaries in the three proposed formation channels of \am stars
   (the dashed line shows the detached phase of the white dwarf
   channel). For comparison, the evolutionary path of an ordinary
   hydrogen-rich CV or low-mass X-ray binary is
   shown. From~\cite{2010MNRAS.401.1347N}.}
    \label{fig:P_mdot_am}
  \end{figure}
}
\clearpage

\subsection{``Double-degenerate family'' of \am stars}

The importance of \am stars both for astrophysics and for physics in general
stems from the fact that, because of very short orbital periods and 
relative brightness due to proximity to the
Sun, they are expected to be strong GWR sources and the first galactic objects discovered
by space GW interferometers like LISA (or eLISA), i.e., they will serve as ``verification binaries''
\cite{stroeer06}.
Besides, \am stars can be precursors of
\sna \cite{ty79a,sol_yung_am05,2009ApJ...699.2026R} and still hypothetical
order-of-magnitude weaker explosive events called SN~.Ia
\cite{2007ApJ...662L..95B,2009ApJ...699.1365S,2010ApJ...715..767S,2010Natur.465..322P}.

A comprehensive review paper on \am stars covering their observational
properties and some theoretical aspects was published  recently by
Solheim~\cite{2010PASP..122.1133S}. Therefore, we shall provide below
only basic information about these binaries.


In a binary with stable mass transfer, change of the radius of the
donor exactly matches the change of its Roche lobe. This condition
combined with approximate size of the Roche lobe 
given by Eq.~(\ref{eq:roche_kopal}) (which is valid for low $q$)
%
%
provides
%
%
the following relation between the orbital period of the binary $P_{\mathrm{orb}}$
and the mass $M_2$ and radius $R_{2}$ of the donor:
  \begin{equation}
    P_{\mathrm{orb}} \simeq 101 \mathrm{\ s}
    \left( \frac{R_2}{0.01\,R_\odot} \right)^{3/2}
    \left( \frac{0.1\,M_\odot}{M_2} \right)^{1/2}\!\!\!\!\!\!\!\!.
    \label{eq:pdrel}
  \end{equation}
%
and the expression for the rate of mass transfer for a semidetached
binary in which the mass transfer is driven by the orbital angular
momentum losses:
\begin{equation}
    \frac{\dot{m}}{m} =
    \left ( \frac{\dot{J}}{J} \right )_\mathrm{GWR}
    \!\! \times \left( \frac{\zeta (m)}{2} + \frac{5}{6} -
    \frac{m}{M} \right)^{-1}\!\!\!\!\!\!\!,
    \label{eq:mdot}
\end{equation}
where $\zeta (m) = d \ln r/ d \ln m$.
%
%
For the mass transfer to be stable, the term in the brackets must be
positive, i.e.,
\begin{equation}
 \label{eq:eqdmdt} 
 \frac{m}{M} < \frac{5}{6} + \frac{\zeta(m)}{2}. 
\end{equation}
Violation of this criterion results in mass loss by the WD donor on a
time scale comparable to 10\,--\,100 orbital periods, as discussed in
Section~\ref{section:wd_formation} and, most probably, in the merging
of components, if a supernova explosion is avoided. Of course,
Eq.~(\ref{eq:eqdmdt}) oversimplifies the conditions for a stable mass
exchange. A rigorous treatment must include tidal effects, the
angular momentum exchange, and the possible super-Eddington $\dot{M}$
immediately after RLOF by the donor and the associated common envelope
formation, as well as the possible ignition of accreted
helium on the WD surface~\cite{web_iben87a, hw99, mns04, 2007ApJ...655.1010G}.

\epubtkImage{fig3_solheim.png}{
  \begin{figure}[htb]
     \centerline{\includegraphics[width=0.7\textwidth]{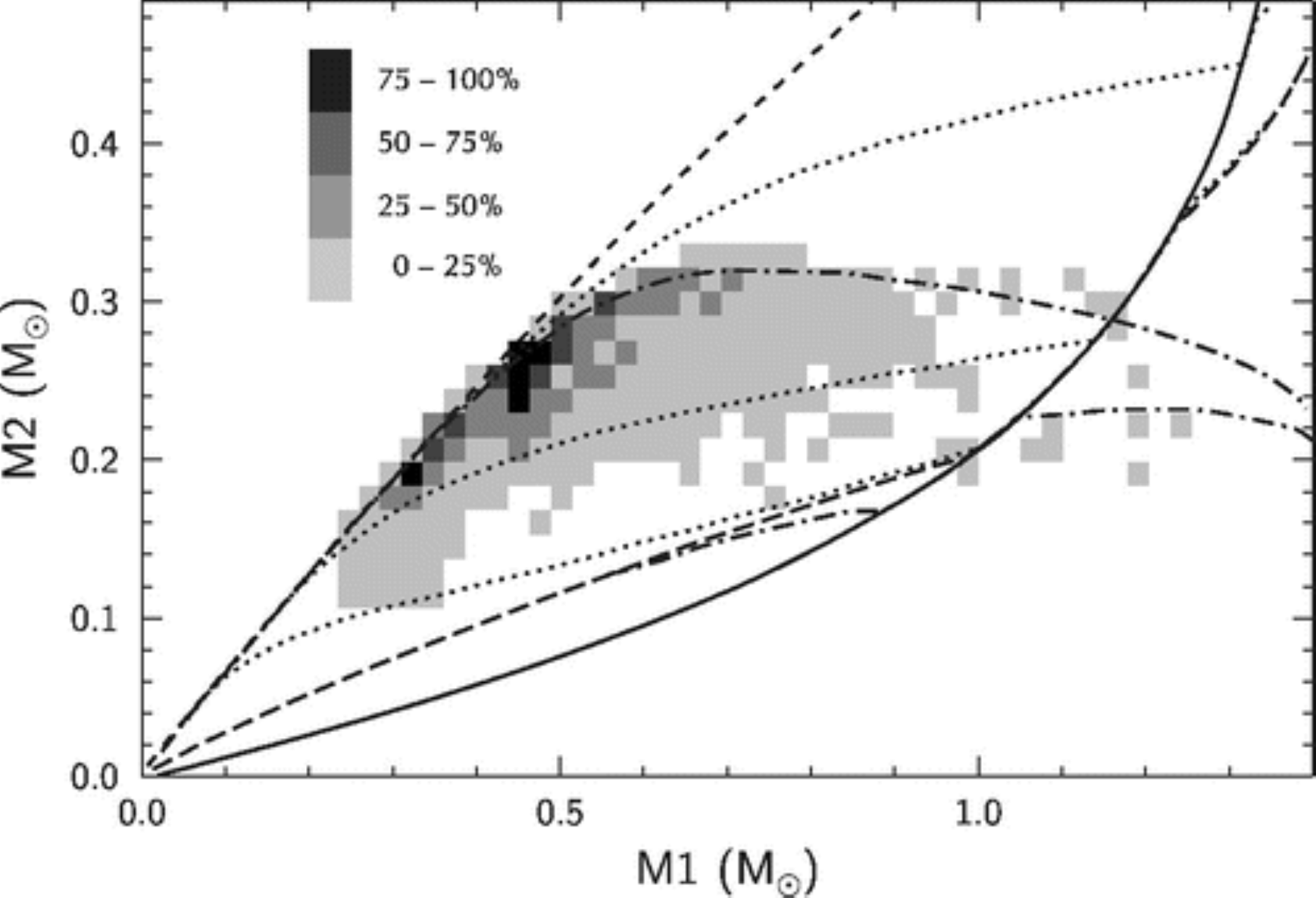}}
     \caption{Birthrates and stability limits for mass transfer between close double white dwarfs.
The shaded areas show the birth probability of progenitors of \am stars in double-degenerate
channel scaled to the maximum birth rate per bin
of $9\times10^{-5}\pyr$ \cite{nyp01}. The upper dashed line corresponds to the upper limit for stable mass transfer. The
lower solid line is the lower limit for direct accretion. The upper dash-dot line is the limit
set by the Eddington luminosity for stable mass transfer at synchronisation time limit $\tau \rightarrow 0$.
The lower dash-dot line is the limit set by the Eddington luminosity for $\tau \rightarrow \infty$, and the lower
broken line is the strict stability limit for the same. The
three dotted lines show how the strict stability limit is raised for shorter synchronisation
time-scales ranging from 1000~yr (bottom), 10~yr (centre), and 0.1~yr (top). 
From~\cite{2010PASP..122.1133S}.}
\label{f:solheim_stab}
  \end{figure}}

\clearpage

Currently, only some of these effects have been studied to certain extent. In Figure~\ref{f:solheim_stab} the distribution of birthrates of the 
possible progenitors of \am stars in the DD channel from~\cite{nyp01} 
is over-plotted with constraints imposed by the mass exchange stability conditions, 
the Eddington accretion rate, and the efficiency of tidal interaction (in 
fact, this is the bottom  
part of Figure~\ref{fig:dd_stab_dan} applied to the population of \am stars). 
It is clear that the substantial number of 
\am stars formed via ``double-degenerate'' scenario may exist only in the case of efficient tidal coupling. As shown in~\cite{nyp01}, the difference in the expected numbers of stars between two limiting cases of synchronisation time 
$\tau \rightarrow 0$ and $\tau \rightarrow \infty$ amounts to two orders of magnitude.  

 AM CVn systems are so close that in some of them the accretion stream may hit the 
accretor directly, and the accretion disk does not form \cite{nyp01a}. The candidate ``direct impact'' 
systems currently include the shortest orbital period known systems HM~Cnc (\porb=5.36 min), 
V407~Vul (\porb=9.48 min), ES~Cet (\porb=10.36 min), see \cite{2010PASP..122.1133S} 
for discussion and references.

Until quite recently, evolution of systems with 
WD donors was calculated using Eq.~\ref{eq:mdot}) 
and applying different $M$--$R$ relations for zero-temperature WD
close to $R\propto M^{-1/3}$, such as given in~\cite{1969ApJ...158..809Z,nau72,vr88},
ignoring detailed models of dwarfs. For He-star donors, radii were approximated 
by fits obtained in the evolutionary models of  semidetached systems (e.g.,
$R \approx 0.043 \, m^{-0.062}$ from~\cite{tf89}).
Early calculations of this kind may be found in~\cite{pac67a, vil71,
faulkner-71, ty79a, rj84, taam_wade_aml85, vr88, ty96, nyp01,
nyp04,2010ApJ...717.1006R}. 
Figure~\ref{figure:dmdtam} shows examples of the evolution of 
systems with a helium degenerate donor or a low-mass
``semidegenerate'' helium star donor and a carbon-oxygen accretor
with initial masses that are currently thought to be typical for
progenitors of AM~CVn systems -- $0.2$ and $0.6\ms$ for double degenerates and
$0.4$ and $0.6\ms$ for He-star systems, respectively. In 
Figure~\ref{figure:dmdtam} the mass--radius relation~\cite{vr88} is
used; for the low-mass He-star -- $M\,-\,R$ relation from 
\cite{tf89}.
The same
equations are applied to obtain the model of the population of
AM~CVn-stars discussed in Sections~\ref{section:waves} and
\ref{section:opt+x}.

\epubtkImage{H2552f04.png}{%
  \begin{figure}[htbp]
    \centerline{\includegraphics[width=\textwidth]{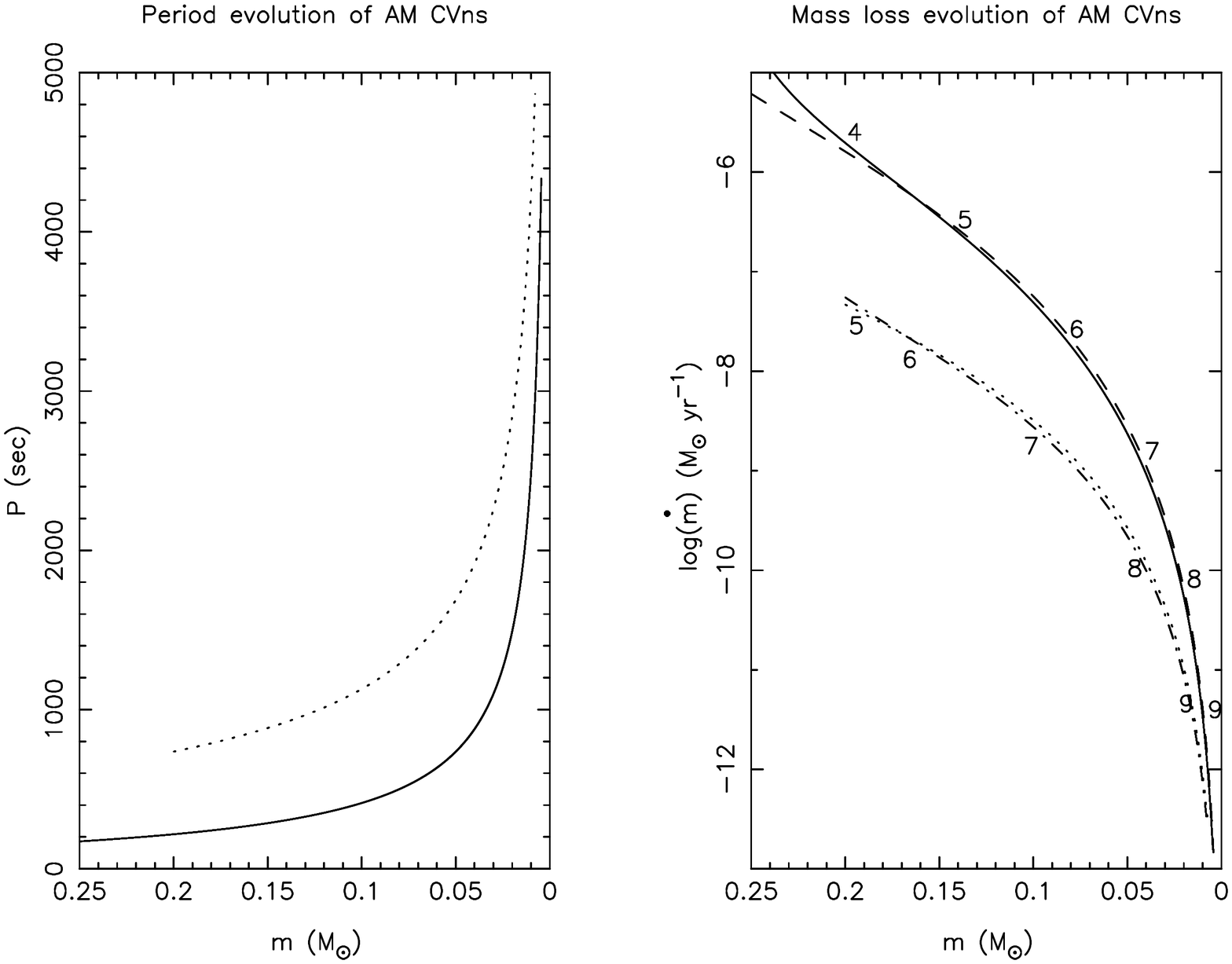}}
    \caption{Examples of the evolution of AM~CVn systems. \emph{Left
    panel:} The evolution of the orbital period as a function of the
    mass of the donor star. \emph{Right panel:} The change of the mass
    transfer rate during the evolution.  The solid and dashed lines
    are for zero-temperature white dwarf donor stars with initial mass
    $0.25\ms$ transferring matter to a primary with initial mass of 0.4
    and $0.6\ms$, respectively, assuming efficient coupling between
    the accretor spin and the orbital motion. The dash-dotted and
    dotted lines are for a helium star donor, starting when the helium
    star becomes semi-degenerate (with a mass of $0.2\ms$). Primaries
    are again $0.4$ and $0.6\ms$. The numbers along the lines indicate
    the logarithm of time in years since the beginning of mass
    transfer. From~\cite{nyp01a}.}
    \label{figure:dmdtam}
  \end{figure}}

\clearpage

Equations~(\ref{A:GW:dEdt}, \ref{eq:roche_kopal}, \ref{eq:mdot})
imply that for $m \ll M$ the mass loss rate scales as $M^{1/3}$. As
a result, for all combinations of donor and accretor masses, the $P$--$\dot{m}$
lines form two rather narrow strips within which they converge with
decreasing $m$. 

The ``theoretical'' model of evolution from shorter periods to longer ones is
supported by observations which found that the UV luminosity of AM~CVn-stars is
increasing as the orbital period gets shorter, since shorter periods are
associated with higher $\dot{M}$~\cite{ramsay_am06}. The orbital period change
$\dot{P}_{\mathrm{orb}}=(3.07\pm0.56)\times10^{-13} \mathrm{day\,day}^{-1}$ at
the $5.4\sigma$ level was quite recently found for the first discovered
eclipsing \am type binary SDSS~J0926+3624 (\porb=28.3~min.
\cite{2013arXiv1309.4087S}). This value of $\dot{P}$ is consistent with
mass-exchange rate close to $1.8\times10^{-10}\myr$, expected both in the
``double-degenerate'' and ``helium-star'' channels of evolution of \am stars for
the period of SDSS~J0926+3624.  Sion et al. \cite{2011ApJ...741...63S}
estimated accretion rates in several \am binaries by means of analyses of their
far- and near-UV spectra and found that in their sample, really, the system
ES~Ceti with the shortest \porb=5.4 min has the highest
$\mdot=2.5\times10\super{-9}\myr$, while the longest orbital period system GP~Com
(46.5 min) has the lowest $\mdot=(3 - 4)\times10\super{-11}\myr$. 

About 1/3 of \am stars demonstrate outbursts~\cite{ramsay_am_outbursts12}. The properties 
of an outburst are well described by thermal-viscous disk instability model,
see~\cite{2012AA...544A..13K} and references therein. Crudely, disks are unstable for
mass accretion rates 
$10^{-12} \aplt \dot{M}_{\mathrm{accr}} \aplt 10^{-9}\myr$.

We should note, however, that the time span between the formation of a
pair of WDs and their contact may range from several Myr to several
Gyr~\cite{ty96,nyp01}. This means that the approximation of the
zero-temperature white dwarfs is not always valid. $T_c$ of WD at the
onset of mass transfer may be as high as almost $10^8\mathrm{\ K}$, as
estimated by  Deloye et~al.~\cite{2007MNRAS.381..525D} for WD
progenitors of donors of \am stars from~\cite{nyp01}. This produces
differences in period distributions of stars at the onset of mass
transfer: while pairs with zero-temperature WDs have narrow
distribution $1.5 \leq P_{\mathrm{cont}} \leq 6.5\mathrm{\ min}$,
systems with finite entropy WDs have $1.5 \leq P_{\mathrm{cont}} \leq
17.5\mathrm{\ min}$ and for about 80\% of the model stars
$P_{\mathrm{cont}} > 6.5\mathrm{\ min}$.

\epubtkImage{deloye.png}{%
  \begin{figure}[htbp]
    \centerline{
     \includegraphics[scale=0.38]{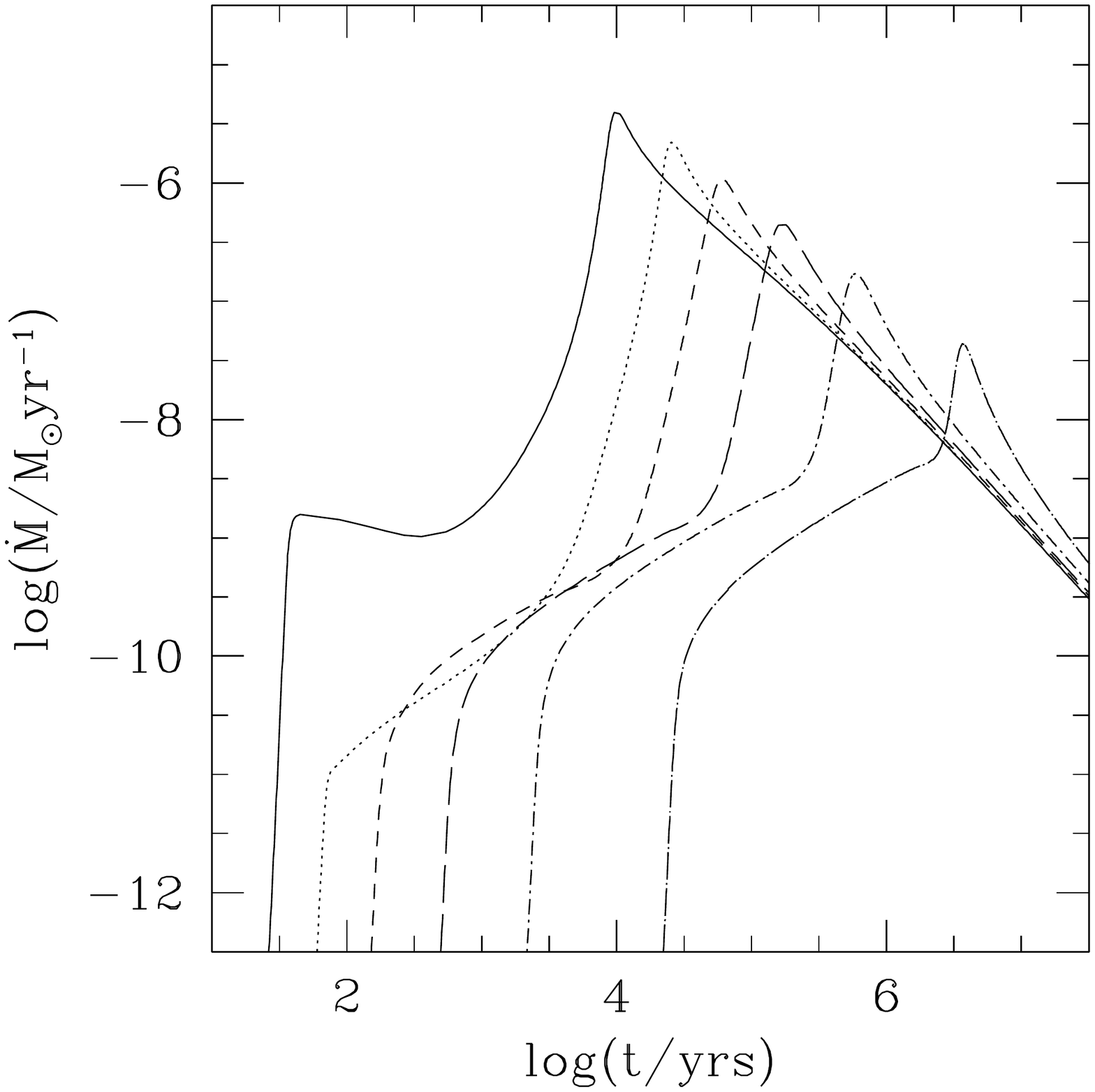}
     \includegraphics[scale=0.38]{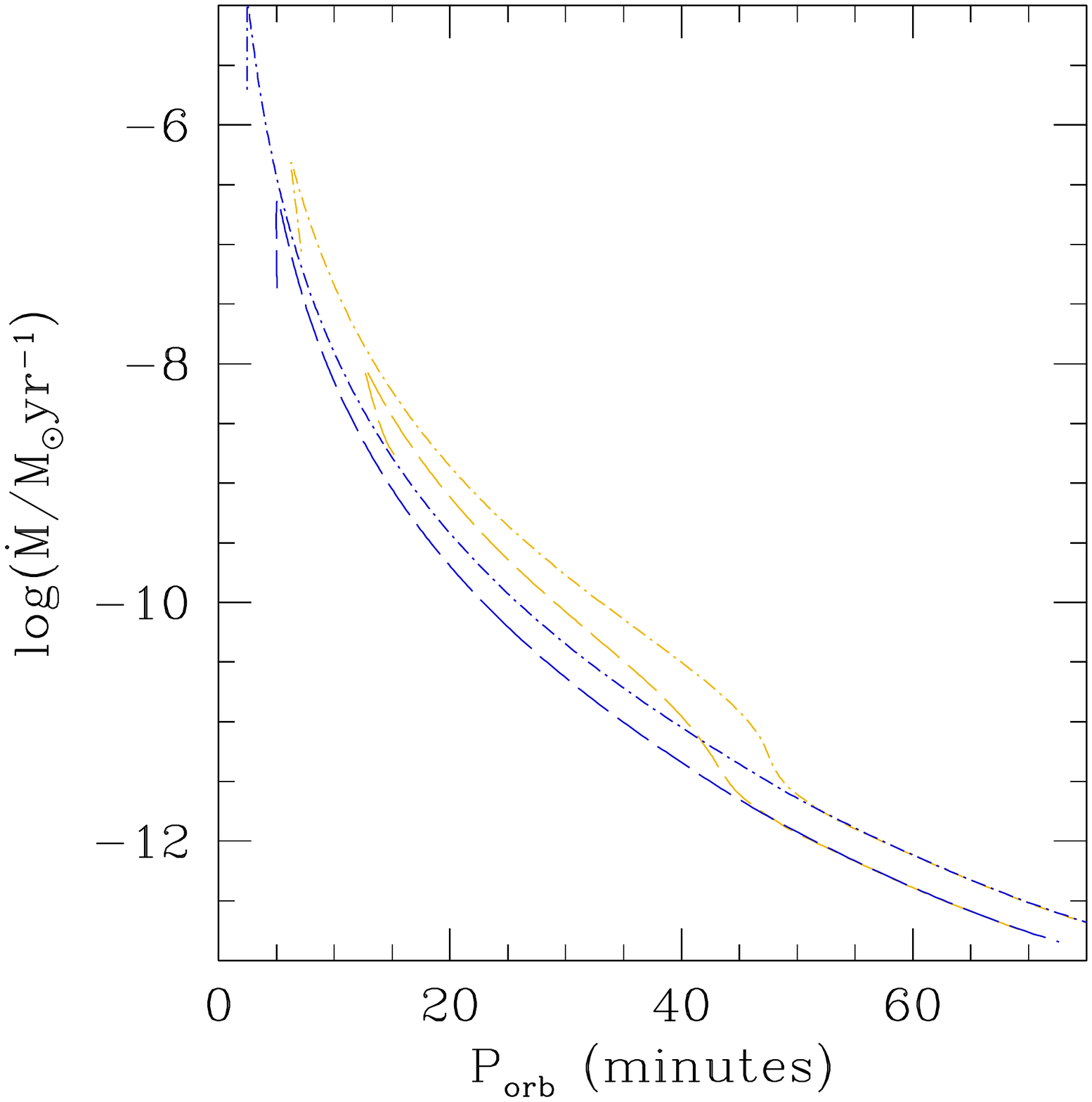}
     }
    \caption{\emph{Left panel:} the $\dot{M}(t)$  evolution of AM CVn systems with 
$M_{2,i}=0.2\ms$ and
$M_{1,i}=0.3\ms$. Systems with donors having the degeneracy parameter 
$\log(\psi_{c,i})$ = 3.5, 3.0, 2.5, 2.0, 1.5 and 1.1 are shown by the solid, dotted, 
short-dashed, dashed, 
shortdash-dotted and dash-dotted lines, respectively.
\emph{Right panel:} the dependence of the $\dot{M} - \porb$ relation on the 
initial binary parameters for pairs 
$M_{1,i}+M_{2,i}=(0.35+0.15)\ms$, $\log(\psi_{c,i})=1.1$ (yellow dashed line),
$M_{1,i}+M_{2,i}=(1.025+0.30)\ms$, $\log(\psi_{c,i})=3.0$ (yellow dot-dashed line),
$M_{1,i}+M_{2,i}=(0.35+0.15)\ms$, $\log(\psi_{c,i})=1.1$ (blue dashed line),
$M_{1,i}+M_{2,i}=(1.025+0.30)\ms$, $\log(\psi_{c,i})=3.0$ (blue dot-dashed line).
From~\cite{2007MNRAS.381..525D}.}
     \label{fig:deloye}
  \end{figure}}

\clearpage

Deloye et~al.\ incorporated the pure helium equation of state  and the
low temperature opacities in their computations of (still unique)
series of full evolutionary models of mass-losing donors with
different initial mass and electron degeneracy.%
\epubtkFootnote{The latter is characterised by the parameter
$\psi=E_{\mathrm{F},c}/kT_c$, where $E_{\mathrm{F},c}$ is the electron
Fermi energy, $T_c$ -- the central temperature of a WD.}
Their main finding is that the evolution of the donor includes three
distinct phases: (i) the onset of mass transfer in which the WD
contracts, mainly due to the fact that very outer layers are never
degenerate, (ii) the phase of adiabatic expansion, and (iii) the final
phase when thermal time scale of the donor becomes shorter than the
mass-loss time scale and white dwarf cools, contracts, and evolves to
a nearly degenerate configuration. 
Early computations were not able to account for these transitions, 
since mass-radius relations for zero-temperature WD were used. 
 
In Figure~\ref{fig:deloye} we show, after Deloye et~al., $\dot{M}(t)$ evolution for 
AM~CVn systems with 
initial masses $M_{2,i} = 0.2\ms$ and
$M_{1,i} = 0.3\ms$ and the dependence of the $\dot{M} - \porb$ relation on the initial binary parameters and the degeneracy parameter for pairs with different initial component masses and electron degeneracy.
Note that the initial stage of decreasing \porb\ is very short and hardly observable
(however, we do observe HM Cnc). At the final stage of evolution, the tracks with
different initial degeneracy merge. The comparison with Figure~\ref{figure:dmdtam}
shows that the tracks for ``intermediate'' adiabatic mass loss stage based on the 
zero-temperature $M\mbox{\,--\,}R$ relation and on the finite entropy models do not differ
significantly, justifying early calculations, while in the stage of cooling at
$\porb > (40\mbox{\,--\,}45)\mathrm{\ min}$, the mass loss rate becomes so small that stars will again
be hardly observable. The main difference between the approximate and full
computations arises from taking into account the initial non-zero temperature and should
be reflected in $P_{\mathrm{orb},\max}$ after the contact. This may affect the 
expected GW signal from AM CVn stars, which must be dominated by the shortest period
systems (and the closest ones), see Section~\ref{section:waves}.          
 
Yet another point, which may slightly change the 
evolution of double-degenerate family of
\am stars based on the ``traditional''
$M\,-\,R$-relation,  is the possible presence of a low-mass ($\sim10^{-3}\ms$) but
geometrically thick (several $0.01\rs$) nondegenerate layer of hydrogen
supported by $p\,-\,p$ nuclear burning at the surface of the donor at RLOF. Models
of mass-transfer in such systems~\cite{2006ApJ...653.1429D,2012ApJ...758...64K}
show that (i) they may make contact at $\porb\approx(6 - 10)\mathrm{\ min}$ and (ii) for
the first $\simeq10^6\mathrm{\ yr}$ of evolution radii of the donors decrease and
systems evolve to shorter periods, until H-rich layer is almost shed. This may
be the case of \am-type star HM Cancri (=~RX~J0806.3+1527), which is also the
star with the shortest known \porb~=~321.5\,s \cite{roelofs_hmcnc_period10} and
the only member of the \am class of binaries with suspected presence of
hydrogen. 

\subsection{``Helium-star family'' of \am stars}

Precursors of He-star components of AM~CV stars with initial mass
$\approx(0.35\mbox{\,--\,}0.80)\ms$ are
$M_{\mathrm{ZAMS}} \approx(2\mbox{\,--\,}5)\ms$ red giants with non-degenerate helium cores
(Figure~\ref{figure:wd_flow}). Helium stars with $M\aplt
0.80\ms$ almost do not expand in the course of evolution (by less than
30\% \cite{pac_he71}). Their lifetimes are comparable to the
main-sequence lifetimes of their precursors (Eq.~\ref{eq:t_he})
and the orbital angular
momentum loss via gravitational waves radiation may bring them to
RLOF before exhaustion of He in their cores
\cite{skh86,it91,ty96}. Mass transfer proceeds steadily if the initial
mass ratio of the He star and the WD is less than approximately 
{\bf 3 \cite{ivanova_taam04}.}
Evolution of low-mass semidetached systems with
He-donors was explored in details by
Yungelson~\cite{2008AstL...34..620Y}. Referring the reader for details
to this paper, we show in Figure~\ref{fig:hetracks} as an example a
set of tracks for the system with initial masses of components
$(M_{\mathrm{He}}+M_{\mathrm{wd}})=(0.40+0.60)\ms$ which have at the formation
(immediately after the completion of common envelope evolution)
orbital periods from 20 to 130~min. The initial abundances of
He and heavy elements at ``He-ZAMS'' 
are assumed to be 0.98 and 0.02, respectively. RLOF in
these systems happens at orbital periods of about 20 to 30~min when He
abundance in their cores ranges from 0.98 to 0.066, from the least
evolved models at RLOF to the most evolved ones. The initial stage of
increasing \mdot\ lasts for $\sim$~10~Myr only. The stars reach
minimum periods of $\sim$~10~min. Their mass at this time is
close to $0.2\ms$. As shown in~\cite{2008AstL...34..620Y}, $\porb
- \dot{M}$ relation very weakly depends on the total mass of the
system. Mass loss leads to the cooling of stellar interiors and
quenching of nuclear burning by the time when the minimum of
\porb\ is reached, see Figure~\ref{fig:hest_hist}. After the minimum of
\porb\ chemical composition of the matter transferred to the
companion is ``frozen'' and does not change during the 
subsequent evolution. In the course of further evolution the
matter of donors becomes increasingly degenerate, and the their
mass-ratio relation approaches that of WD. 

\epubtkImage{mdot_he_lrr.png}{
  \begin{figure}[htbp]
    \centerline{\includegraphics[width=\textwidth]{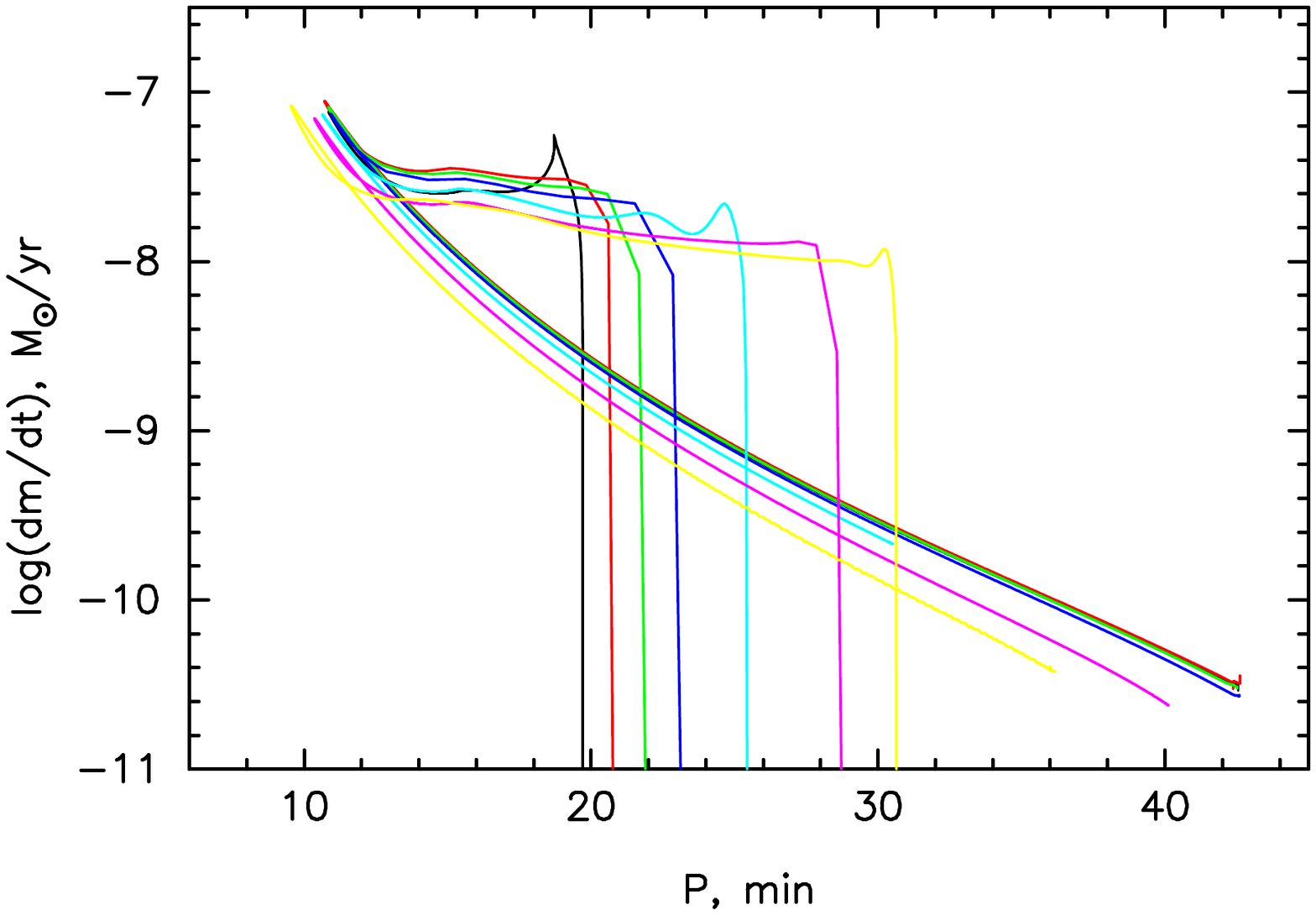}}
    \caption{Mass-loss rate vs.\ orbital period dependence for
    semidetached systems with He-star donors and WD accretors, having
    post-common envelope \porb=20 to 130 min. Initial masses of
    components are $(M_{\mathrm{He}}+M_{\mathrm{wd}})=(0.40+0.60)\ms$. He abundance
    in the cores of the donors at RLOF  ranges from 0.98 to 0.066 (left
    to right).}
    \label{fig:hetracks}
  \end{figure}
}
\clearpage

\epubtkImage{mdot_he_lrr.png}{%
  \begin{figure}[htbp]
\centerline{
\includegraphics[width=0.35\textwidth,clip]{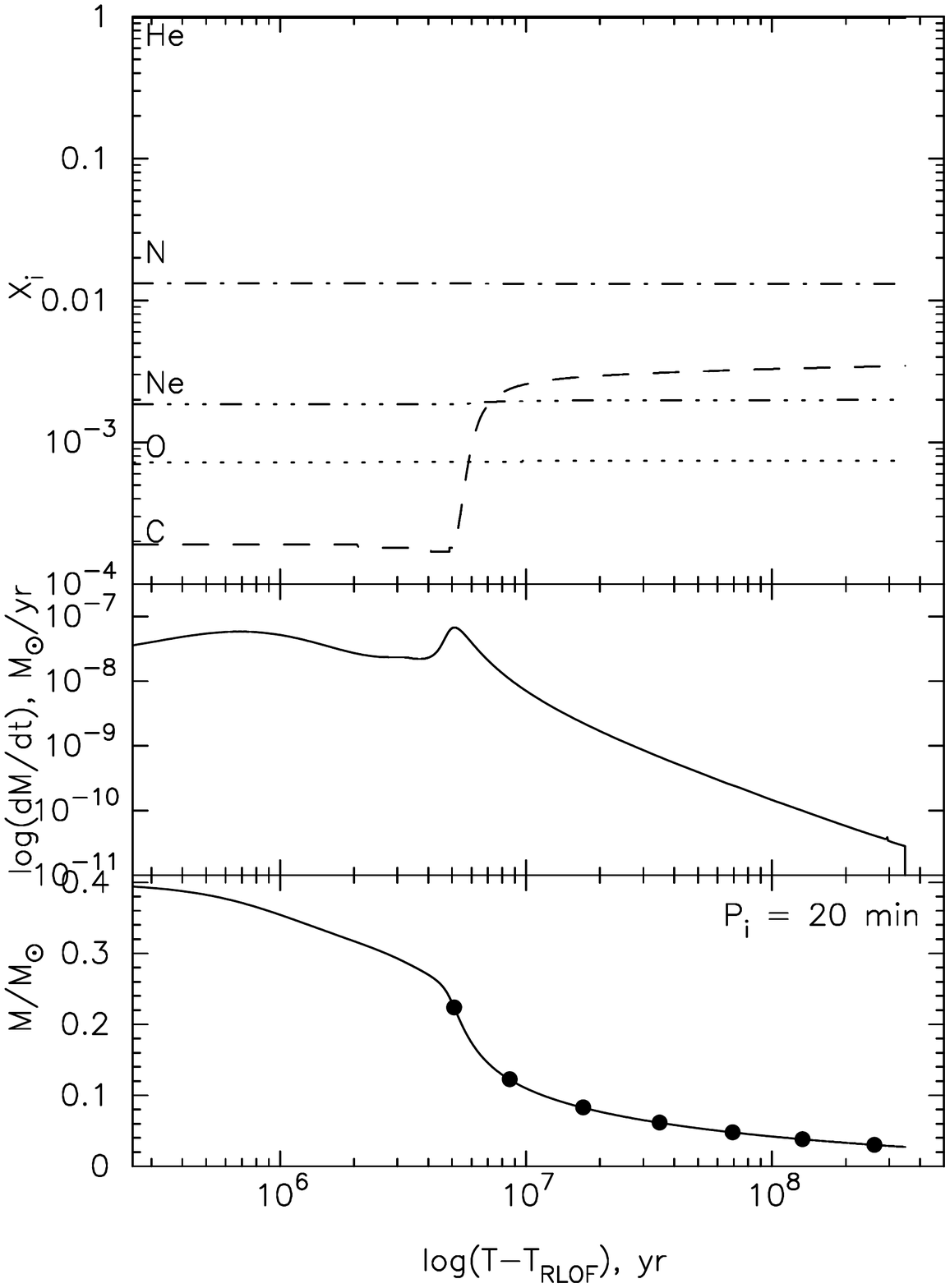}\quad
\includegraphics[width=0.35\textwidth,clip]{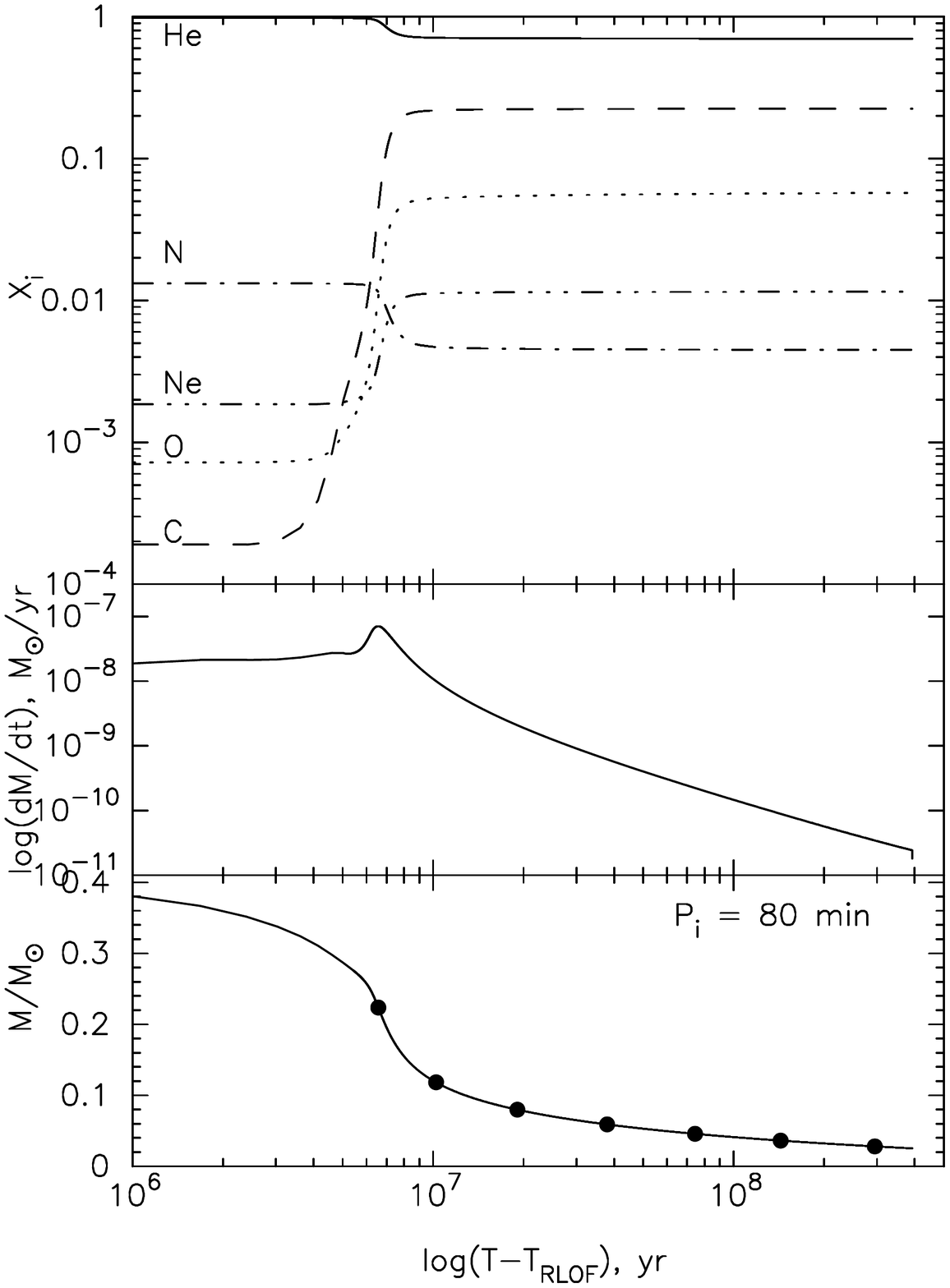}\quad
\includegraphics[width=0.35\textwidth,clip]{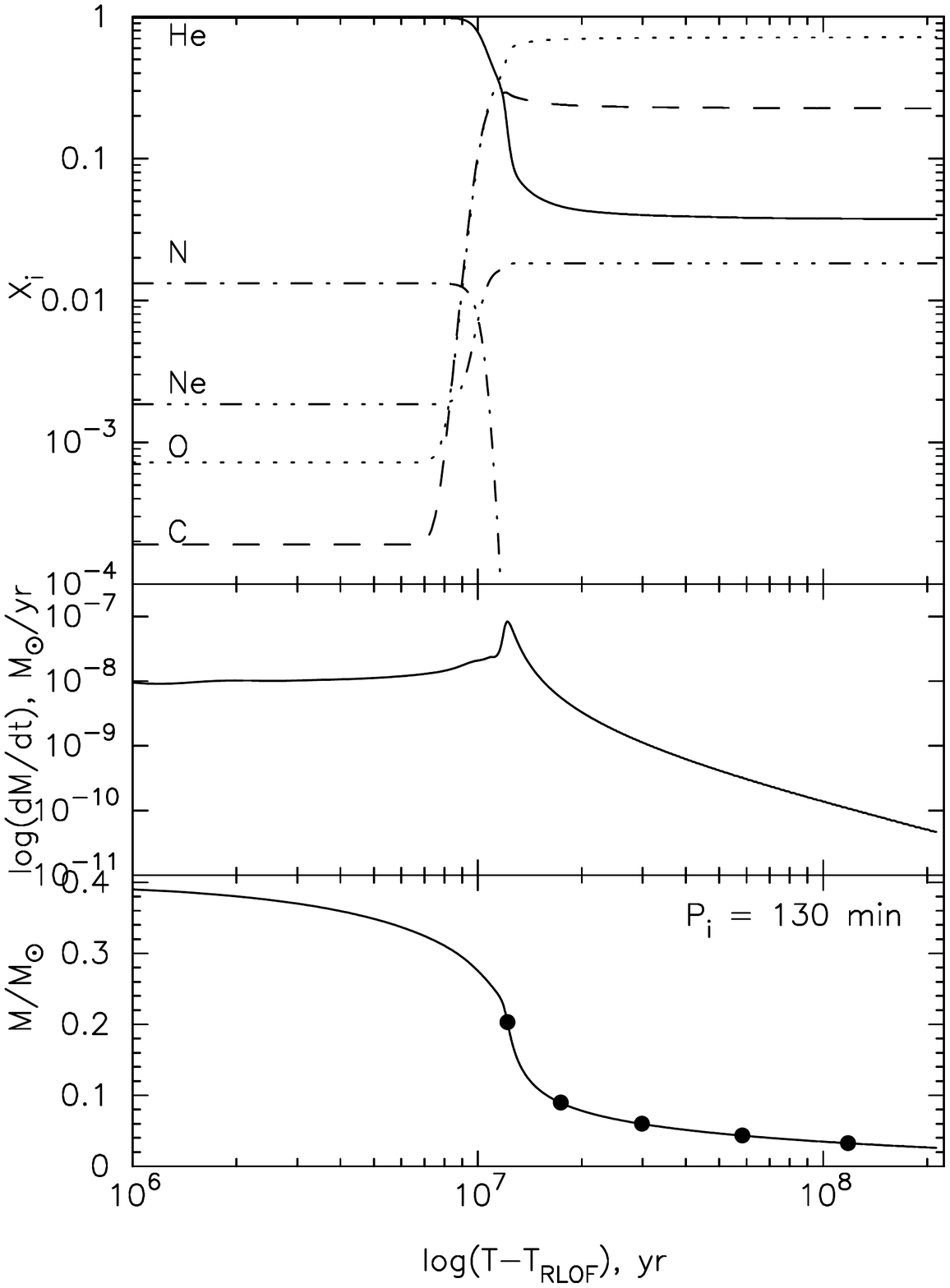}
}
\caption{An overview of the evolution and chemical abundances in the transferred matter
   for helium star
   donors in ultra-compact binaries. We show abundances (top),
   mass-transfer rate (middle) and donor mass (bottom) as a function
   of time since the start of the Roche lobe overflow. The binary period
   is indicated by the solid circles in the bottom panels for 
    \porb\ = 15, 20, 25, 30, 35, and 40 min. The
   initial (post-common-envelope) orbital
   periods are indicated in the bottom panels. The 
   sequences differ by the amount of nuclear processing before
   RLOF. }
    \label{fig:hest_hist}
  \end{figure}}

\clearpage

As mentioned in Section~\ref{section:wd_formation}, until recently it was
deemed that, at accretion rates of He onto a WD below several units
of $10^{-8}\myr$, $(0.1\mbox{\,--\,}0.3)\ms$ of He accumulates at the WD
surface and then detonates; converging detonation waves propagate
to the center of accretor and ignite carbon detonation close to the
center. Prior to the minimum \porb\ He-donors provide $\dotm \sim
10^{-8}\myr$. Correspondingly, Nelemans et~al.~\cite{nyp01,nyp04}
considered ``optimistic'' and ``pessimistic'' scenarios for
the formation of \am stars in the He channel, depending on the amount of He
that may accrete, detonate and destroy the \am star precursor:
$0.15\ms$ or $0.30\ms$. The difference in the predicted number of
observed systems reached factor $\sim$~2 in favour of the
first scenario. However, a comparison of the accretion and mass loss
rates in Figs.~\ref{fig:hetracks} and \ref{f:heregimes} suggests
that the donors rather experience strong flashes of He-burning
than detonations, and the model space density of \am stars in 
\cite{nyp01,nyp04} is, most probably, an underestimate. 
This exacerbates the known
problem of possibly almost factor $\sim$~50  deficit of the observed
\am star spatial density compared to theoretical predictions
\cite{2007MNRAS.379..176R,carter13a}. However, as noted in \cite{carter13a},
the ``observational'' estimate of the number of \am stars depends on the assumed
intrinsic distribution of their magnitudes, and uncertainty in this distribution also
may contribute to the mismatch of theoretical expectations and observations. 
   
An accurate knowledge of
the formation channels and, respectively, the space density of
interacting double-degenerates is necessary for better
understanding of the stability of mass exchange, processes in common
envelopes and proper modeling of gravitational waves foreground
(Section~\ref{section:waves}). The same is true for ultra-compact X-ray
sources.

Observations of chemical composition of IDD may serve as a tool for
diagnostic of their origin~\cite{2003whdw.conf..359N,
2010MNRAS.401.1347N}. In the double degenerate channel, the matter
transferred is a product of the low-temperature CNO-burning, and the
abundances of He, C, N, O yields depend on the WD progenitor's
mass. In the He-star channel, the abundances depend on the initial
mass of the donor and the extent of He-depletion by the time of RLOF
(Figure~\ref{fig:hest_hist}). In the ``evolved CV-channel'', the
matter may contain also the traces of hydrogen. As concerns UCXB, it
was shown in~\cite{ynh02} that the initial masses of their donors are
limited from above by $\approx 0.45\ms$. This means that some of the
donors may descend from ``hybrid'' WDs. However, the latter lose their
He-rich envelopes at very early phases of mass exchange, and,
since they are descendants of helium stars, the matter transferred by
them should not contain He. Nelemans et~al.~\cite{2010MNRAS.401.1347N}
constructed a set of diagnostic diagrams with the abundance ratios (in
mass) $X_{\mathrm{N}}/X_{\mathrm{He}}$,
$X_{\mathrm{N}}/X_{\mathrm{C}}$, $X_{\mathrm{N}}/X_{\mathrm{O}}$,
$X_{\mathrm{O}}/X_{\mathrm{He}}$, $X_{\mathrm{O}}/X_{\mathrm{C}}$ as
functions of \porb\ for different groups of donors. As an example, we
show in Figure~\ref{fig:n_to_c} such a diagram for the ratio
$X_{\mathrm{N}}/X_{\mathrm{O}}$.
 
\epubtkImage{n_to_c.png}{
  \begin{figure}[htb]
   \centerline{\includegraphics[width=0.8\textwidth]{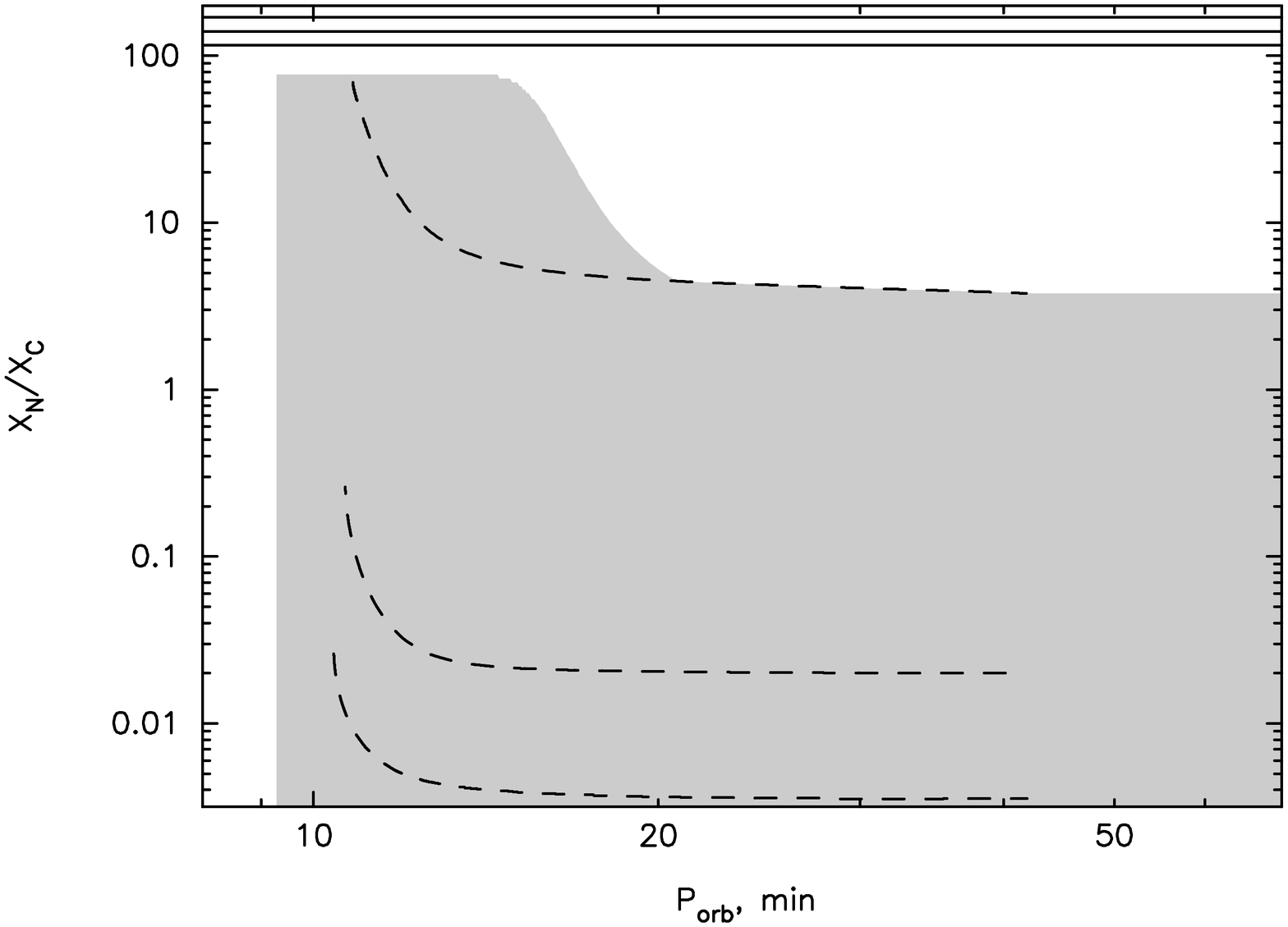}}
   \caption{Abundance ratios (by mass) N/C for He-WD (the solid line) and
   He-star donors (the shaded region with dashed lines)  as a function
   of the orbital period.  For the helium star donors we indicate the
   upper part of the full range of abundances which extends to 0. The
   dashed lines are examples of tracks shown in detail in
   Figure~\ref{fig:hest_hist}. The helium white dwarfs are descendants
   of 1, 1.5 and $2\,M_\odot$ stars (top to bottom).}
    \label{fig:n_to_c}
  \end{figure}
}

The above-described diagnostic of abundances in the spectra of known IDD 
and other means of analysis 
suggest that HM~Cnc, CP~Eri, GP~Com, V396~Hya, SDSS~J124058.03-015919.2,
SDSS~J080449.49+161624.8 almost definitely belong to the WD-family, while 
AM~CVn, HP~Lib, CR~Boo, V~803~Cen, SDSS J173047.59+554518.5 may have 
helium-star donors \cite{2010MNRAS.401.1347N,2010PASP..122.1133S,2014MNRAS.437.2894C}. 
Significantly evolved He-star donors are ruled out for 
SDSS J113732.32+405458.3 and SDSS J150551.58+065948.7 \cite{2013arXiv1312.3335C}.

Among the UCXBs there are two
systems (4U~1626-67 and 4U~0614+09) in which detection of C and O
lines but no He lines suggests hybrid white dwarf or very evolved
helium star donors. For 4U~1916-05, the detected He and N
lines suggest a He white dwarf or an unevolved helium star donor 
\cite{2010MNRAS.401.1347N}.
For the rest of sources the data are not sufficient for classification. 

\subsection{Final stages of evolution of interacting double-degenerate systems}
\label{sec:amfinal}

\epubtkImage{det_mas.png}{
  \begin{figure}[htb]
   \centerline{\includegraphics[width=0.8\textwidth]{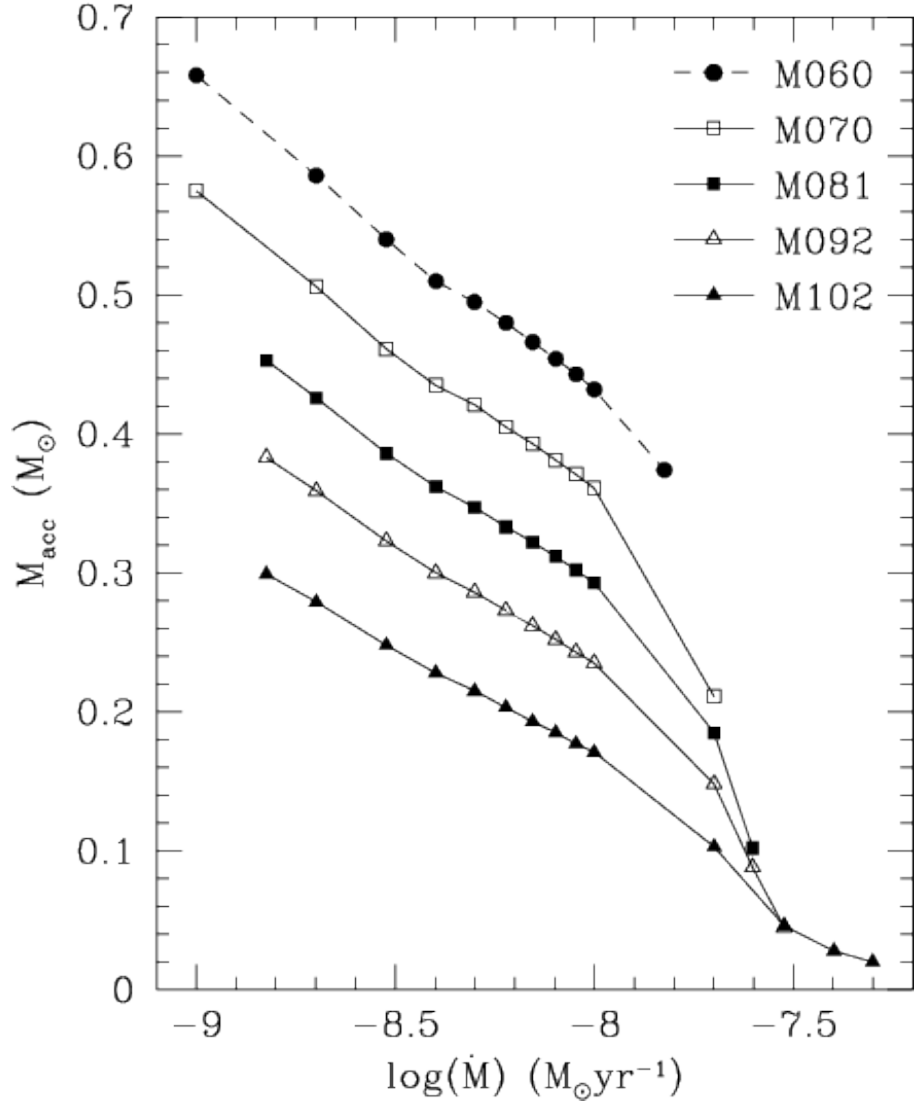}}
    \caption{Accreted mass as a function of the accretion rate for models 
experiencing a 
  dynamical He-flash. The lines refer to WD with different initial masses --- 0.6, 0.7, 
  0.81, 0.92, 1.02 $M_\odot$.   
 Piersanti, Tornamb\'{e} and Yungelson \textit{ in prep.}}
         \label{fig:detmashe}        
  \end{figure}
}
\clearpage
By means of population synthesis modeling, Solheim and
Yungelson~\cite{sol_yung_am05} found that accumulation of \mch\ by WD
in the double-degenerate channel of formation and evolution of IDD is
possible at a rate $\sim 10^{-5}\pyr$, while Ruiter
et~al.~\cite{2009ApJ...699.2026R} provided an estimate which is an
order of magnitude higher. Note that in both studies \sna require massive
WD accretors, which are not typical for observed IDD.   

Yet another kind of explosive events may characterise final stages of evolution
of IDD~\cite{2007ApJ...662L..95B,2009ApJ...699.1365S,2010ApJ...715..767S}
 -- faint thermonuclear supernovae dubbed SN~.Ia. 

 As described above, in the WD-family of AM~CVn stars, the mass transfer rate
initially is as high as $\sim(1\mbox{\,--\,}3)\times10^{-6}\myr$ and  decreases at later times.
While during the initial phase of accretion the nuclear burning in the accreted He
envelope of WD is thermally stable, it becomes thermally unstable as
$\dot{M}_{\mathrm{accr}}$ drops (Fig.~\ref{f:heregimes}). In the He-star family, the mass transfer rate before reaching the orbital period minimum is several
$10^{-8}\myr$, i. e., accretors are in the strong He-flashes regime \textit{ab
ovo}. The ``ignition'' mass for He-outbursts increases with decreasing
$\dot{M}_{\mathrm{accr}}$. This yields $\sim$~10 He-flashes in the $\sim
(10\super{6}-10\super{7})\mathrm{\ yr}$ before the amount of accreted mass
required to achieve He-ignition $(\sim 0.1)\ms$ becomes larger than the mass of
the donor (Fig.~\ref{fig:detmashe}), while \textit{de facto} the binaries are already in the 
``detonation'' regime of He-burning (Fig.~\ref{f:heregimes}). Thus, there must
be the ``last'' most powerful outburst with an ignition mass of $\sim 0.1\,M_\odot$
which can occur in hydrodynamic regime.  It can lead to He-detonation which can be  
potentially observable as a faint thermonuclear
supernova~\cite{2007ApJ...662L..95B}. Since the mass transfer rate drops after
RLOF very rapidly, it is plausible that the observed AM~CVn stars already
experienced their last strong He-flash (which could be of the SN .Ia scale) and are
now slowly transferring He without any expected nuclear-powered phenomena.
Supernovae .Ia probably do not initiate detonations of low-mass accretors
of AM~CVn stars since the masses of the exploding shells are insufficient for this
\cite{2012MNRAS.420.3003S,shen_bildsten_dd13}.

As estimated in \cite{2007ApJ...662L..95B}, SN~.Ia  may happen
every 5000 to 15\,000~yr in an E/S0 galaxy. 
Their brightness and duration are about 1/10 of the ordinary SN~Ia event.
The rise times of the light curves in all bands are very rapid ($<$~10~days).
The explosion nucleosynthesis in these events primarily yields heavy $\alpha$-chain elements (\super{40}Ca
 through \super{56}Ni) and unburned He (up to 80\%) 
\cite{2010ApJ...715..767S,moore_dd13}.
It is interesting that  with the 1~day dynamic-cadence 
experiment on the 
Palomar Transient Factory\epubtkFootnote{\url{http://www.astro.caltech.edu/ptf}}, it 
is expected to annually discover a few such events. 

SN~2010X (PTF~10bhp) with extremely rapid decay ($\tau$~=5~day) was, for 
instance, suggested as
a candidate SN~.Ia~\cite{2010ApJ...723L..98K}. Another candidate may be a faint
type Ib supernova, SN~2005E, in the halo of the nearby isolated galaxy,
NGC~1032. Spectroscopic observations and analysis reveal high velocities of the
ejecta, dominated by helium-burning products including a large amount of
radioactive \super{44}Ti, just as expected for SN~.Ia. However, as noticed by Drout et al. \cite{Drout_SN.Ia_13}, e. g., severely
stripped-envelope SN~Ic explosions may mimic SNe .Ia. 

It is possible that strong flashes of He-burning can be identified with He-novae,
like H-outbursts are identified with classical novae.
A sole known object which may be a prototype of He-novae --- V445~Pup erupted in
2000, and its optical and IR spectra are hydrogen-deficient 
\cite{ashok03a,woudt_steeghs05b}. Provisional \porb\ of V445~Pup is  $\approx0.65$ day 
\cite{2010PZ.....30....4G}. Several other candidate He-novae have been suggested but remain 
still unexplored \cite{rosenbush_he_nov08}. The estimated pre-outburst luminosity of
V445~Pup is   $\log(L/\ls)=4.34\pm0.36$ 
\cite{2009ApJ...706..738W}. Apparently, it is associated with a more massive system
than the typical \am star, but one cannot exclude nova-scale events also among AM CVn's.

\newpage
\section{Gravitational Waves from Compact Binaries with White-Dwarf Components}  
\label{section:waves}

We need to start this section with a caveat that until now virtually 
all studies on detection of low-frequency GWR from compact binaries, with a single exception of~\cite{nissanke_gw12}, have been made having in mind the future launch of LISA mission. At the time of writing 
only several months
elapsed since the decision to cancel 
LISA and the introduction of the eLISA project. Therefore, almost all estimates involving 
sensitivity of the detector are still ``LISA-oriented'' and only in some
cases corrections for the reduced (by about an order of magnitude) eLISA sensitivity are 
available. Nevertheless, as the physics behind GWR emission is unchanged, we 
present in this Section, as examples,  ``LISA-oriented'' results, unless stated otherwise. 

It was suggested initially that contact W~UMa binaries will dominate the galactic gravitational wave background at low
frequencies~\cite{mir65}. However, it was shown in~\cite{ty79a,
  mty_gwr81, eis87, lp87, hbw90} 
that it will be, most probably, totally dominated by detached and
semidetached double white dwarfs.

As soon as it was recognised that the birth rate of the Galactic close
double white dwarfs may be rather high, and even before detection of 
the first close detached 
DD was reported in 1988 \cite{slo88}, in 1987 Evans, Iben, and Smarr~\cite{eis87}
accomplished an analytical study of the detectability of the signal
from the Galactic ensemble of DDs, assuming certain average parameters
for DDs. Their main findings can be formulated as follows. Let us assume
that there exists a certain distribution of DDs over frequency of the
signal $f$ and the strain amplitude $h$: $n(f, h)$. The weakest signal is
$h_\mathrm{w}$. For the time span of observations $\tau_\mathrm{int}$, the frequency resolution bin of the detector is $\Delta f_\mathrm{int}
\approx 1/\tau_\mathrm{int}$. Then, integration of $n(f, h)$ over
amplitude down to a certain limiting $h$ and over $\Delta f$ gives the
mean number of sources per unit frequency resolution bin for a volume defined by $h$. If for a certain $h_n$
\begin{equation}
  \int^\infty_{h_n} \left(\frac{dn}{df\,dh}\right)\,\Delta
  \nu_\mathrm{int} = 1,
  \label{eq:confusion}
\end{equation}
then all sources with $h_n>h>h_\mathrm{w}$ overlap. If in a certain resolution bin $h_n>h_\mathrm{w}$ does not exist, individual sources may be resolved in
this bin for a given integration time (if they are above the detector's
noise level). In the bins where binaries overlap, they produce the 
so-called \textit{``confusion noise''}: an incoherent sum of signals; the
frequency above which the resolution of individual sources becomes
possible was referred to as the ``confusion limit'', $f_\mathrm{c}$. Evans et~al.\
found $f_\mathrm{c} \approx 10 \mathrm{\ mHz}$ and $3 \mathrm{\ mHz}$ for
integration times 10\super{6}~s and 10\super{8}~s,
respectively.

Independently, the effect of confusion of Galactic binaries was
demonstrated by Lipunov, Postnov, and Prokhorov~\cite{lpp87}  
 who used
simple analytical estimates of the GW confusion limited signal  
produced by
unresolved binaries whose evolution is driven by GWs only; in this
approximation, the expected level of the signal depends solely on
the Galactic merger rate of binary WDs (see~\cite{Grishchuk_al01} for
more details). Later, analytic studies of the
GW signal  produced by binary stars at low
frequencies were continued in~\cite{fbhhs89, hbw90, pp98, hb00}.  
A more detailed approach to the estimate of the GW foreground is possible 
using population synthesis 
models~\cite{lp87, wh98, nyp01, nyp04, 2010ApJ...717.1006R, yu_jeff_gwr10, yu_jeff_gwr13}. Convenient analytical expressions which allow to compare results of
different studies as a function of the assumed Galactic model and merger rate of double dwarfs 
are provided by Nissanke et~al.~\cite{nissanke_gw12}.

Note that in the early studies the combined signal of Galactic binaries was dubbed
\emph{``background''} and considered as a \emph{``noise''}. However, Farmer and
Phinney \cite{2003MNRAS.346.1197F} stressed that this signal will be in fact a
\emph{``foreground''} for \emph{``backgrounds''} produced in the early Universe. But only
circa 2008\,--\,2009 the term \emph{``foreground''} became common in discussions of Galactic
binaries. As summarised by Amaro-Seoane et~al.~\cite{2012CQGra..29l4016A}, the
overall level of the foreground is a measure of the total number of
ultra-compact binaries; the spectral shape of the foreground contains
information about the homogeneity of the sample, as models with specific types
of binaries predict a very distinct shape; the geometrical distribution of the sources can be found  by a probe like eLISA. 

Giampieri and Polnarev \cite{1997MNRAS.291..149G}, Farmer and Phinney
\cite{2003MNRAS.346.1197F} and Edlund et~al. \cite{edlund05} showed that, due to
the concentration of sources in the Galactic centre and the inhomogeneity of the
LISA-antenna pattern, the foreground should be strongly modulated
in a year observations (Figure~\ref{fig:gwr_cycle}), with time periods in where the foreground is
by more than a factor two lower than during other periods. This is not the case for
the signal from extra-galactic binaries which should be almost isotropic. The
characteristics of the modulation can provide information on the distribution of
the sources in the Galaxy, as the different Galactic components (thin disk,
thick disk, halo) contribute differently to the modulation. Additionally, during the 
``low signal'' periods, antenna will be able to observe objects that are away
from the Galactic plane.  

\epubtkImage{edlund_cycle.png}{%
  \begin{figure}[htbp]  
    \centerline{\includegraphics[width=0.75\textwidth]{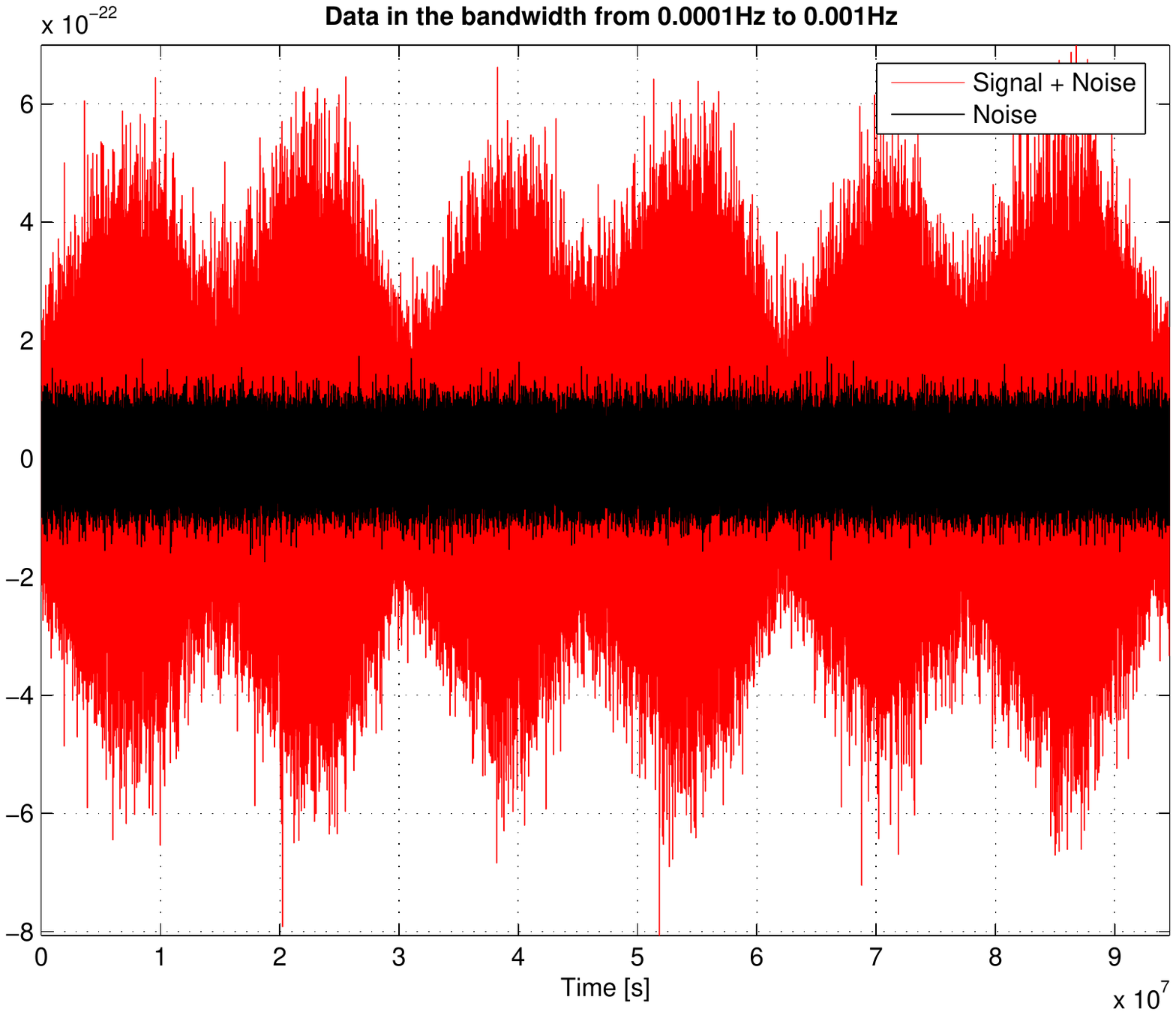}}
    \caption{Simulated time-series of the DD Galactic foreground
    signal of 3 years of data. ``Noise'' is the instrumental LISA
    noise. From \cite{edlund05}, based on computations in \cite{nyp04}.}
    \label{fig:gwr_cycle}
\end{figure}}
\clearpage

Referring the reader for the comprehensive review of eLISA science to 
\cite{2012CQGra..29l4016A}, 
we briefly note that  measurements of individual binaries
provide the following information.
If the system is detached, the evolution of the  signal is dominated by 
gravitational radiation and, in principle, three variables may be measured: 
\begin{equation}
  \label{eq:fders}
  h \propto {\cal M}^{5/3} f^{2/3} D^{-1}, \quad 
  \dot{f} \propto {\cal M}^{5/3} f^{11/3}, \quad
  \ddot{f} = \frac{11}{3} \frac{\dot f}{f},
\end{equation}
where $h$ is the strain amplitude, 
$f$ is the frequency, $\mathcal{M} = (m_1 m_2)^{3/5}/(m_1+m_2)^{1/5}$ is the chirp mass,
and $D$ is  the distance to the source. Thus, the determination of 
$h$, $f$, and $\dot{f}$ 
(which is expected to be possible for 25\% of eLISA sources 
\cite{2012CQGra..29l4016A}) 
provides $\mathcal{M}$ and $D$. 
The value of $\ddot{f}$ (which may be possible for a few high-SNR systems) 
gives information on deviations from the ``pure'' GW signal due to tidal effects 
and  mass-transfer interactions. 
Estimate of the effects of tidal interaction is especially interesting, since it is expected that this effect will be typical for pre-merger systems in which it
may lead to a strong heating, rising the luminosity of each component to 
$\simeq20\,L_{\odot}$, which gives a chance of the 
optical detection on the time scale of a human life; 
moreover,  
it can even affect conditions for thermonuclear burning in the 
inner and outer layers of WD 
\cite{web_iben87a,itf98,2010AIPC.1314..217W,2012ApJ...756L..17F,2013MNRAS.430..274F}.
    
Nelemans et~al.~\cite{nyp01} constructed a population synthesis model of the
gravitational wave signal from the Galactic disk population of binaries
containing two compact objects. The model included detached DDs, semidetached
DDs,%
\epubtkFootnote{The ``optimistic'' model of the population of IDD was
taken \cite{nyp01a}, which assumes effective tidal interaction and
inefficient destruction of hypothetical progenitors of \am stars due
to double-degenerate SN~Ia. Thus, the numbers provided in the table
are upper limits.}
detached systems of NSs and BHs with WD companions, binary NSs and
BHs. For the details of the model we refer the reader to the original
paper and references therein. Table~\ref{table:birthrates} shows the
number of systems with different combinations of components in the
Nelemans et~al.\ model.%
\epubtkFootnote{These numbers, like also for models discussed below,
represent one random realisation of the model and are subject to the
Poissonian noise.}
Note that these numbers strongly depend on the assumptions in the
population synthesis code, especially on the normalisation of stellar
birth rate, star formation history, distributions of binaries over
initial masses of components, their mass-ratios and orbital
separations, treatment of stellar evolution, common envelope
formalism, etc. For binaries with relativistic components (i.e.,
descending from massive stars) an additional uncertainty is brought in
by assumptions on stellar wind mass loss and natal kicks. The factor
of uncertainty in the estimated number of systems of a specific type
may easily exceed ~10 (cf.~\cite{han98, nyp01, ty02, Hurley_al02, nyp04,
2010ApJ...717.1006R,yu_jeff_gwr10}). Thus, these numbers should be
taken with some caution.

Table~\ref{table:birthrates} immediately shows that detached DDs, as expected,
dominate the population of compact binaries, as found by other authors, too.
While Table~\ref{table:birthrates} gives total numbers, examples of
contributions from different formation channels and spatial components of the
Galaxy may be found, e.g., in
\cite{2010ApJ...717.1006R,yu_jeff_gwr10,yu_jeff_gwr13}, with a caveat that
models in these paper are different from the model in \cite{nyp01}. Population
synthesis studies confirmed the expectation that contribution to the signal in
the LISA/eLISA frequency range from thick disk, halo, and bulge binaries should not be
significant, since the binary systems there are predominantly old and either merged or
evolved to too long periods as IDD \cite{2009ApJ...693..383R, yu_jeff_gwr10,
2010ApJ...717.1006R}. It was also shown that the contribution from extra-galactic
binaries to the signal in the LISA/eLISA range will be  $\sim$~1\% of the  foreground
\cite{2003MNRAS.346.1197F}. The list of models computed before 2012 and an 
algorithm allowing to compare results of different calculations to the extent of
their dependence on Galactic model and normalisation may be found
in~\cite{nissanke_gw12}.

Population synthesis computations yield the ensemble of Galactic
binaries at a given epoch with their specific parameters $M_1$, $M_2$,
and $a$ and the Galactic location.
 Figure~\ref{figure:strain} shows examples of the relation
between the frequency of emitted radiation and amplitude of the signals from the 
``typical'' double degenerate system that evolves into contact and
merges, for an initially detached double degenerate system that stably
exchanges matter after the contact, i.e.\ an AM~CVn-type star and its
progenitor, and for an UCXB and its progenitor. For the AM~CVn system
effective spin-orbital coupling is assumed~\cite{nyp01a, mns04}, 
Figure~\ref{f:solheim_stab}. 
The difference in the properties and evolution of DD and IDD 
(merger vs.\ bounce and continuation of evolution with decreasing mass) 
is reflected in the distributions and magnitude of their strains shown in
Figure~\ref{fig:edlund_strain} \cite{edlund05}.  

Note that there is a peculiar difference between WD pairs that
merge and pairs that experience a stable mass exchange. The pairs that
coalesce stop emitting GWs on a relatively short time-scale (of the
order of the period of the last stable orbit, typically a few
minutes)~\cite{loren_gwr05}. Thus, if we would be lucky to
observe a chirping WD and a sudden disappearance of the signal, this
will manifest a merger. However, the chance to detect such an event is small
since the Galactic rate of merging double WDs is
$\sim 10^{-2}\mathrm{\ yr}^{-1}$ only.

For the system with a NS, the mass exchange rate is limited by the
critical Eddington value, and excess of matter is ``re-ejected'' from the
system (see Section~\ref{sec:reemission} and~\cite{ynh02}). Note
that for an AM~CVn-type star it takes only $\sim$~300~Myr
after contact to evolve
to $\log f =-3$ which explains their accumulation at lower $f$. For UCXBs
this time interval is only $\sim$~20~Myr. 

\begin{table}
  \renewcommand{\arraystretch}{1.2}
  \caption[Current birth rates and merger rates per year for
    Galactic disk binaries containing two compact objects and their
    total number in the Galactic disk.]{Current birth rates and merger
  rates per year for Galactic disk binaries containing two compact
  objects and their total number in the Galactic disk~\cite{nyp01}.}
  \label{table:birthrates}
  \centering
  {\small
  \begin{tabular}{lccc}
    \toprule
    Type & Birth rate & Merger rate & Number\\
    \midrule
    Detached DD & 2.5~\texttimes~10\super{-2} & 1.1~\texttimes~10\super{-2} & 1.1~\texttimes~10\super{8} \\
    Semidetached DD & 3.3~\texttimes~10\super{-3} & --- & 4.2~\texttimes~10\super{7} \\
    NS\,+\,WD & 2.4~\texttimes~10\super{-4} & 1.4~\texttimes~10\super{-4} & 2.2~\texttimes~10\super{6} \\
    NS\,+\,NS & 5.7~\texttimes~10\super{-5} & 2.4~\texttimes~10\super{-5} & 7.5~\texttimes~10\super{5} \\
    BH\,+\,WD & 8.2~\texttimes~10\super{-5} & 1.9~\texttimes~10\super{-6} & 1.4~\texttimes~10\super{6} \\
    BH\,+\,NS & 2.6~\texttimes~10\super{-5} & 2.9~\texttimes~10\super{-6} & 4.7~\texttimes~10\super{5} \\
    BH\,+\,BH & 1.6~\texttimes~10\super{-4} & --- & 2.8~\texttimes~10\super{6} \\
    \bottomrule
  \end{tabular}
}
  \renewcommand{\arraystretch}{1.0}
\end{table}

\epubtkImage{figure10.png}{%
  \begin{figure}[htbp]
    \centerline{\includegraphics[width=0.6\textwidth]{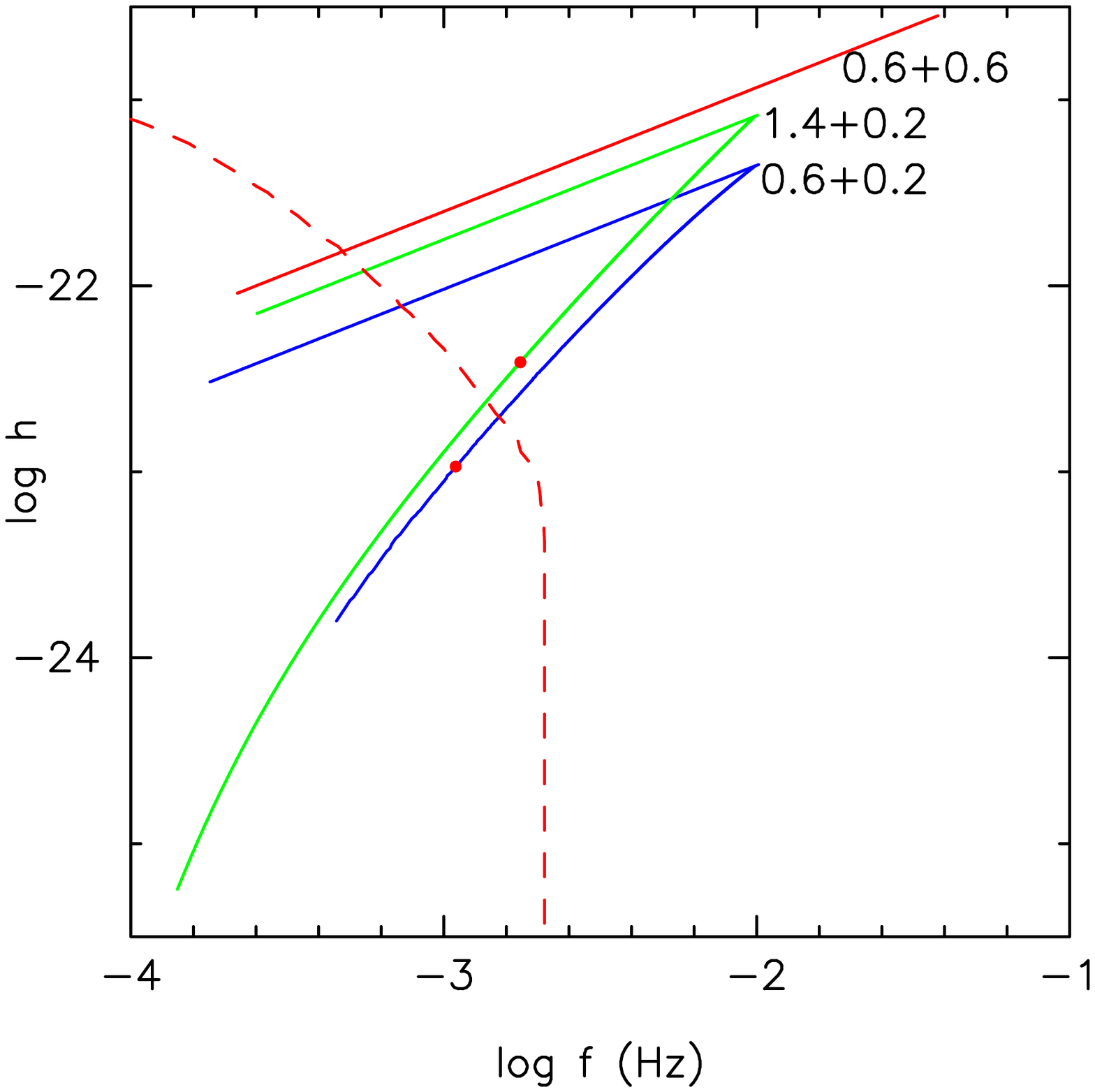}}
    \caption{Dependence of the dimensionless strain amplitude for
      a WD\,+\,WD detached system with initial masses of the
      components of $0.6\,M_\odot + 0.6\,M_\odot$ (red line), a
      WD\,+\,WD system with $0.6\,M_\odot + 0.2\,M_\odot$ (blue line)
      and a NS\,+\,WD system with $1.4\,M_\odot + 0.2\,M_\odot$ (green
      line). All systems have an initial separation of components
      $1\,R_\odot$ and are assumed to be at a distance of
      1~Kpc (i.e.\ the actual strength of the signal has
      to be scaled with factor $1/d$, with $d$ in Kpc). For
      the DD system the line shows an evolution into contact, while
      for the two other systems the upper branches show pre-contact
      evolution and lower branches -- a post-contact evolution with
      mass exchange. The total time-span of evolution covered by the
      tracks is 13.5~Gyr. Red dots mark the positions of
      systems with components mass ratio $q=0.02$ below which the
      conventional picture of evolution with a mass exchange may be
      not valid. The red dashed line marks the position of the
      confusion limit as determined in~\cite{nyp04}.}
    \label{figure:strain}
  \end{figure}
}

\clearpage
\epubtkImage{edlund_strain.png}{%
  \begin{figure}[htbp]
  \centerline{
  \includegraphics[width=0.5\textwidth]{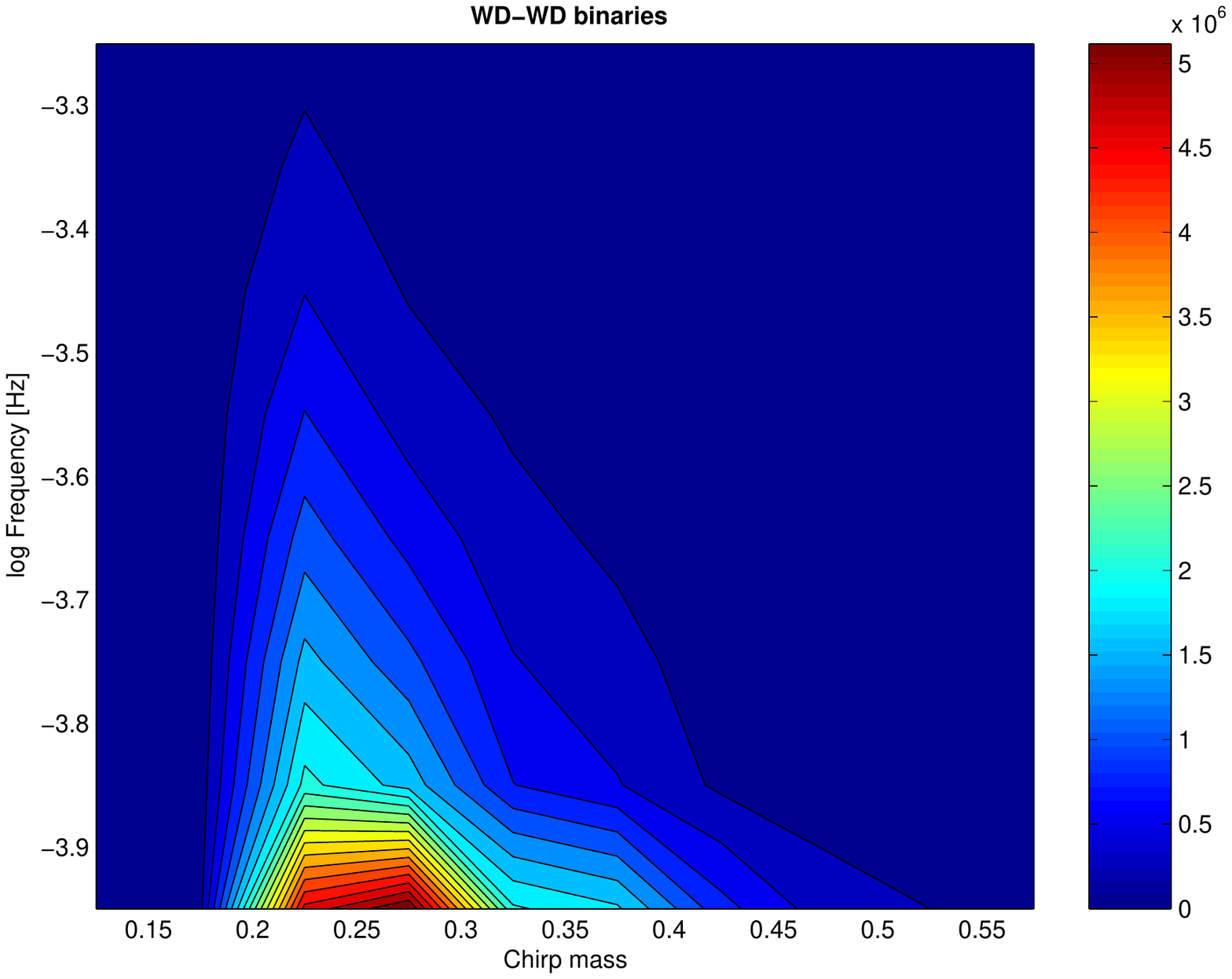}
  \includegraphics[width=0.5\textwidth]{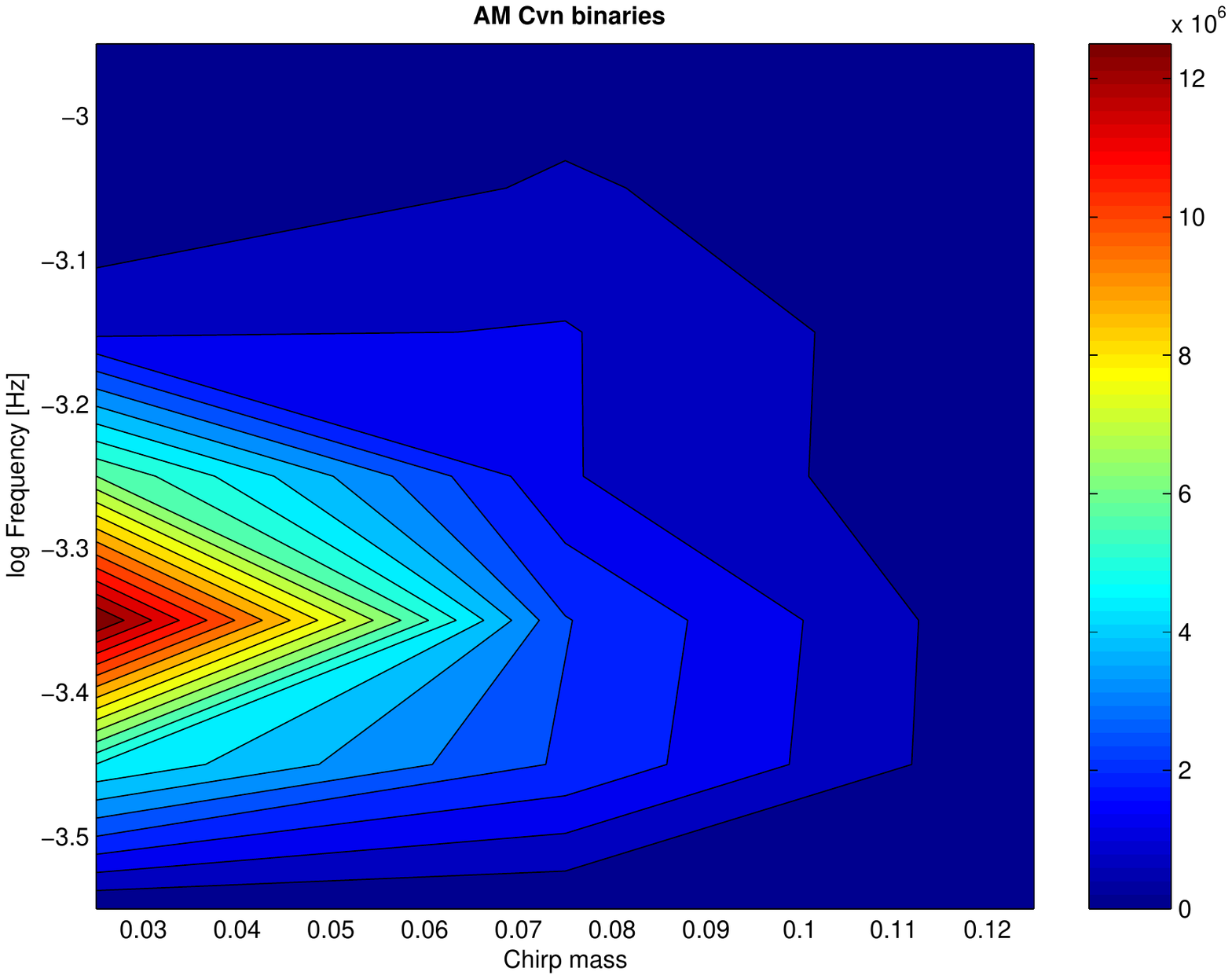}
  }
  \caption{The distribution of Galactic detached WD\,+\,WD binaries
  and interacting WD (\am stars)  as a function of the gravitational
  wave frequency and chirp mass. From~\cite{edlund05}, based on
  computations in~\cite{nyp04}.}
  \label{fig:edlund_strain}
\end{figure}
}
\clearpage
\epubtkImage{figure11.png}{%
  \begin{figure}[htbp]
    \centerline{\includegraphics[width=\textwidth]{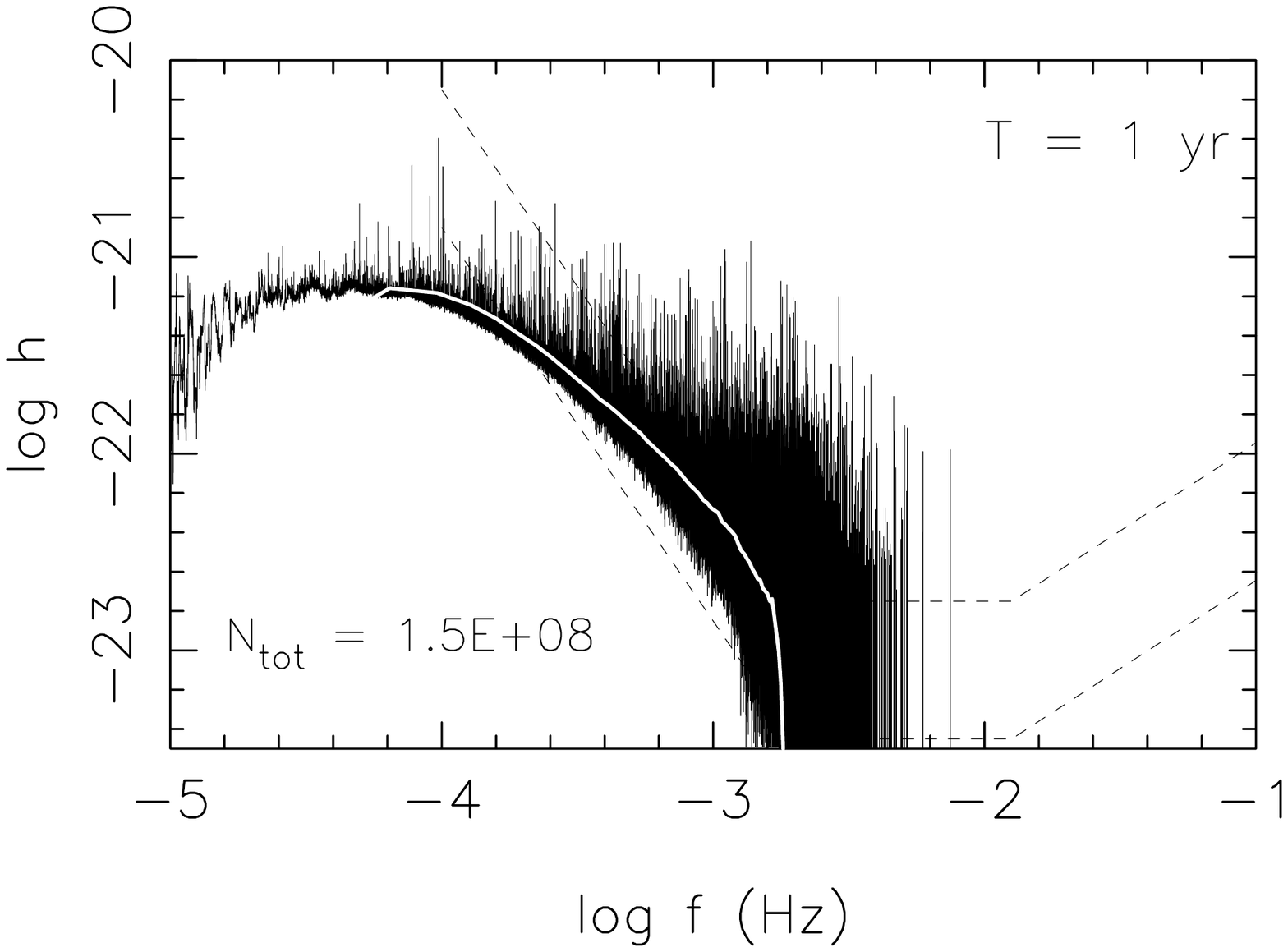}}
    \caption{GWR foreground produced by detached and semidetached
      double white dwarfs as it was expected to be detected by  LISA. The
      assumed integration time is 1~yr. The `noisy' black line gives
      the total power spectrum, the white line shows the average. The dashed
      lines show the expected LISA sensitivity for a $S/N$ of 1 and
      5~\cite{lhh00}. Semidetached  double white dwarfs contribute to
      the peak between $\log f \simeq -3.4$ and $-3.0$. 
      From~\cite{nyp01}.}
    \label{fig:GWR_bg}
\end{figure}
}
\clearpage
\epubtkImage{figure12.png}{%
  \begin{figure}[htbp]
    \centerline{\includegraphics[width=\textwidth]{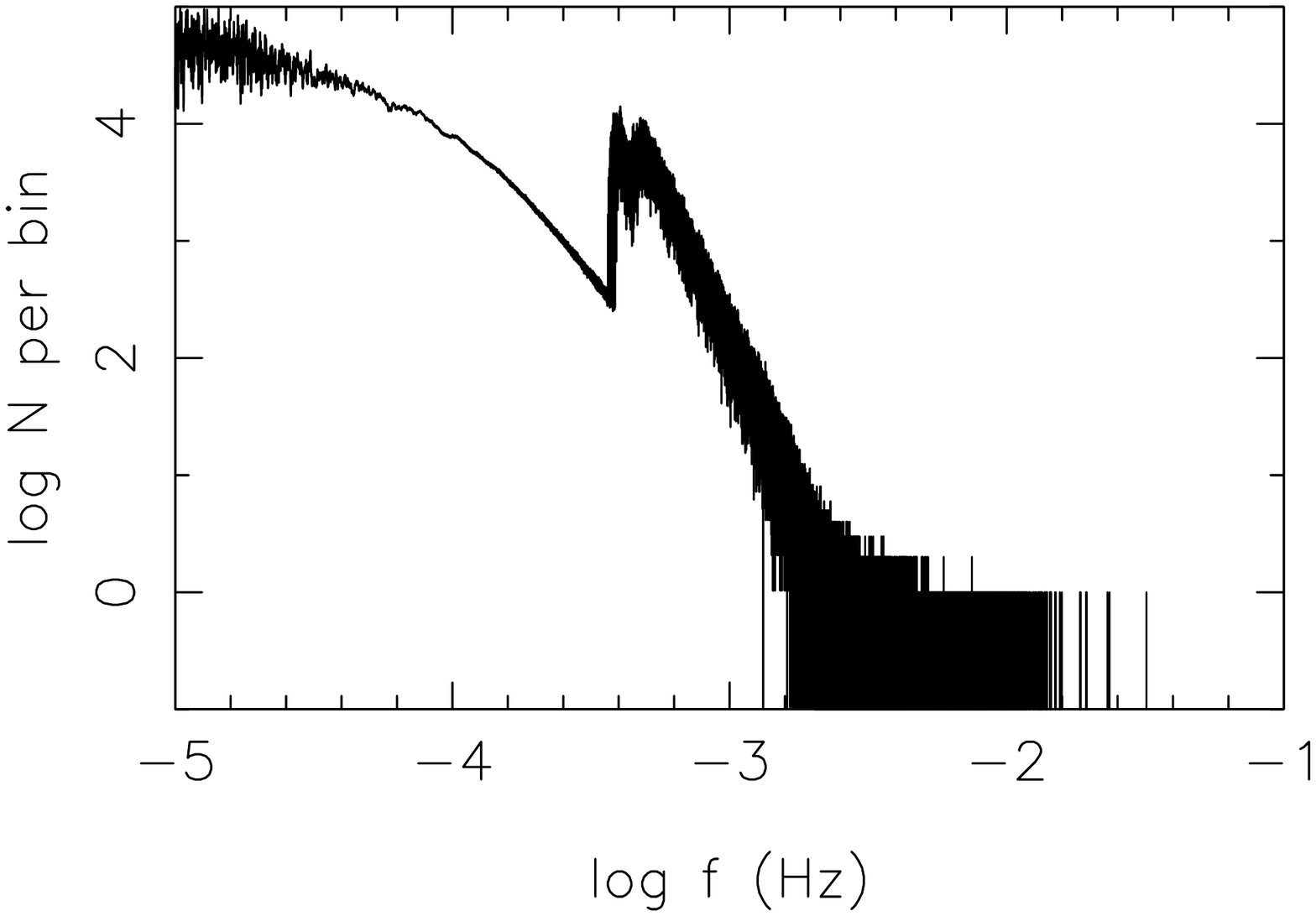}}
    \caption{The number of systems per bin on a logarithmic
      scale. Semidetached double white dwarfs contribute to the peak
      between $\log f \simeq -3.4$ and $-3.0$.
      From~\cite{nyp01}.}
    \label{fig:GWR_bins}
  \end{figure}
}

\epubtkImage{figure13.png}{%
  \begin{figure}[htbp]
\vspace{-0.5cm}
    \centerline{\includegraphics[width=0.6\textwidth,angle=-90]{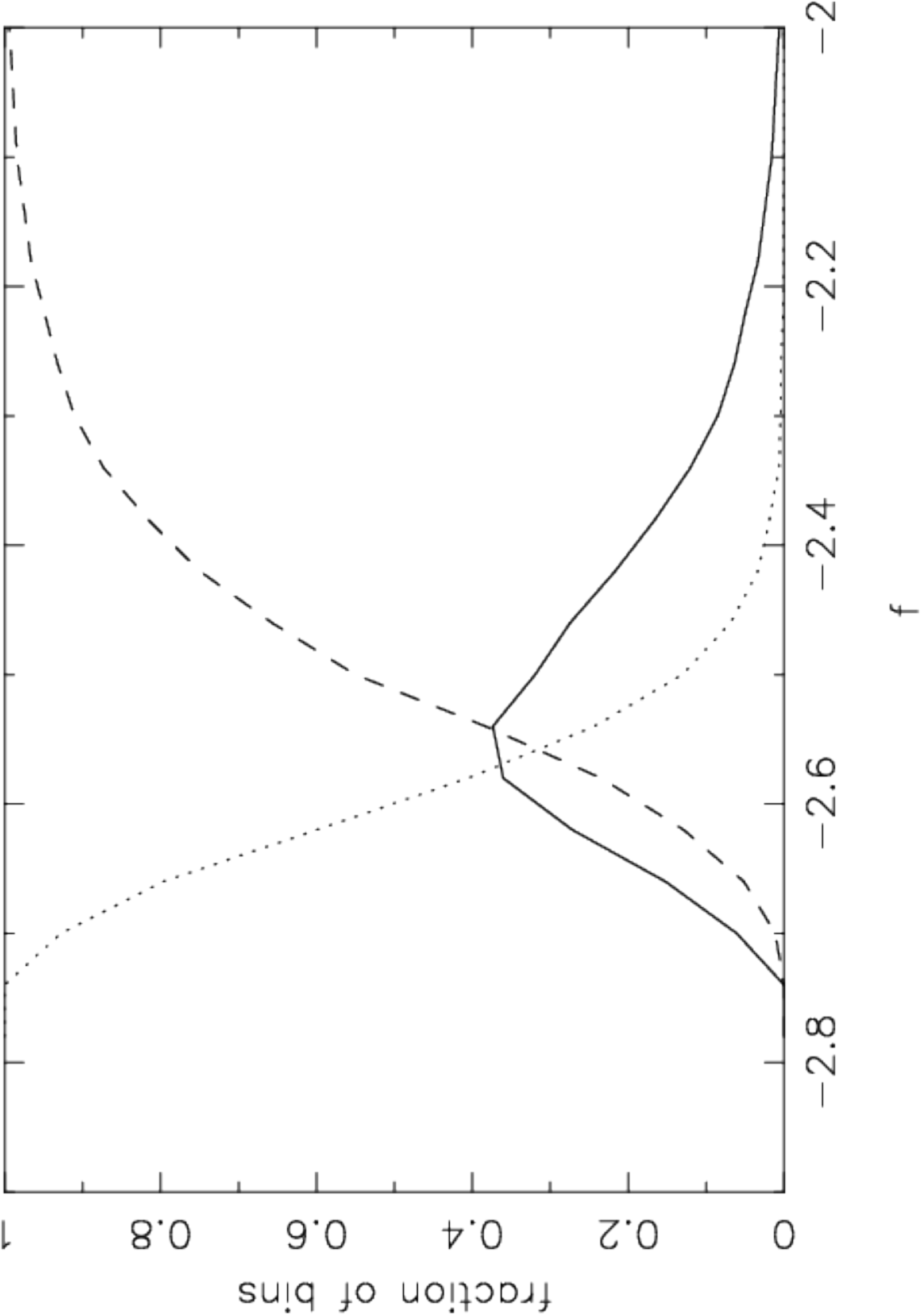}}
    \caption{Fraction of bins that contain exactly one system
      (solid line), empty bins (dashed line), and bins that contain
      more than one system (dotted line) as a function of the signal frequency. 
From~\cite{nyp01}.}
    \label{fig:GWR_fraction}
  \end{figure}
}
\clearpage

In the discussed model, the systems are distributed randomly in the
Galactic disk according to 
\begin{equation}
  \rho(R, z) = \rho_\mathrm{0} \, e^{-R/H} \,
  \mathrm{sech\,}(z/\beta)^2 \mathrm{\ pc}^{-3}\,,
  \label{eq:rho_gal}
\end{equation}
where $H = 2.5 \mathrm{\ kpc}$~\cite{sac97} and $\beta = 200 \mathrm{\
  pc}$. The Sun is located at $R_\odot = 8.5 \mathrm{\ kpc}$ and
$z_\odot = -30 \mathrm{\ pc}$.
Then it is possible to compute the strain amplitude for each system. The
power spectrum of the signal from the population of binaries as it
would be detected by a gravitational wave detector, may be simulated
by computation of the distribution of binaries over $\Delta f = 1/T$ wide
bins, with $T$ being the total integration time. Figure~\ref{fig:GWR_bg}
shows the resulting confusion limited foreground signal. In
Figure~\ref{fig:GWR_bins} the number of systems per bin is
plotted. The assumed integration time is $T=1 \mathrm{\ yr}$. Semidetached double
white dwarfs, which are less numerous than their detached cousins and
have lower strain amplitude, dominate the number of systems per bin in
the frequency interval $-3.4 \lesssim \log f \mathrm{\ (Hz)} \lesssim
-3.0$ producing a peak there, as explained in the comment to
Figure~\ref{figure:strain}.

Results presented below may be, to some extent,  considered as an illustration
of the signal detected by a space-born detector, since they are model-dependent and  ``LISA-oriented'', but they show, at least qualitatively, the features of the signal 
detected by any space-based GW antenna.   

Figure~\ref{fig:GWR_bg} shows that there are many systems with a signal
amplitude much higher than the average in the bins with $f < f_\mathrm{c}$,
suggesting that even in the frequency range seized by the confusion noise
some systems can be detectable above the noise level. The latter will affect 
the ability to detect extra-Galactic massive black-hole binaries 
(Figure~\ref{f:GW_sources}) and to derive their parameters. 

Population synthesis also shows that the notion of a unique ``confusion
limit'' is an artefact of the assumption of a continuous distribution of
systems over their parameters. For a discrete population of sources it
appears that for a given integration time there is actually a range of
frequencies where there are both empty resolution bins and bins containing
more than one system (see Figure~\ref{fig:GWR_fraction}). For this
``statistical'' notion of $f_\mathrm{c}$, Nelemans et~al.~\cite{nyp01} get the
first bin containing exactly one system at $\log f_{res} \mathrm{\ (Hz)} \approx
-2.84$, while up to $\log f \mathrm{\ (Hz)} \approx -2.3$ there are bins
containing more than one system.
Other authors find similar, but slightly more conservative limits on 
$\log f_{res}$: 
$-2.25$ \cite{yu_jeff_gwr10}, 
$-2.44$ \cite{wh98}, 
$-2.52$ \cite{2010ApJ...717.1006R}.

\epubtkImage{verification_elisa.png}{%
  \begin{figure}[htbp]
     \centerline{\includegraphics[width=\textwidth]{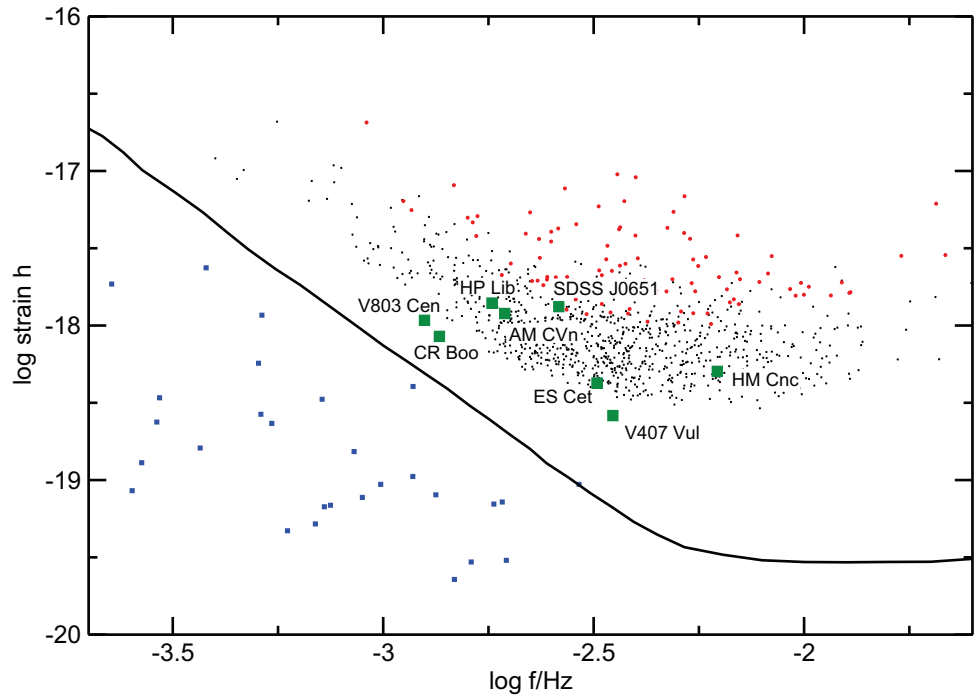}}
    \caption{Strain amplitude spectral density (in Hz$^{-1/2}$) versus frequency 
for the verification binaries and the brightest binaries in the simulated Galactic population of ultra-compact binaries \cite{nyp04}. The solid line shows the sensitivity of
eLISA. The assumed integration time is 2~yrs. 
100 simulated binaries with the largest strain amplitude are shown as red squares.
Observed ultra-compact binaries are shown as blue squares, while the subsample
of them  which can 
serve as verification binaries is marked as green squares. 
From \cite{2013GWN.....6....4A}.
} 
     \label{figure:gwr_back}
  \end{figure}}
\clearpage
Nelemans et al.  \cite{nyp01} also noted the following point. 
Previous studies of GW emission from the AM~CVn systems
 have found that they hardly
contribute to the GW foreground, even despite at
$f \approx(0.3\, - \,1.0)\mathrm{\ mHz}$ they outnumber the detached DDs. This happens
because at these $f$ their chirp mass $\mathcal{M}$ is much smaller than that of a
typical detached system. However, it was overlooked before that at higher frequencies, 
where the number of AM~CVn systems is much smaller, their
$\mathcal{M}$ is similar to that of the detached systems from which
they descend.  This is confirmed also by the latest studies of GW foreground
by Nissanke et~al.,
see Figure~\ref{fig:niss1}.

In \cite{nyp01,nyp04} the following objects were considered as resolved:   single source per frequency bin with
signal-to-instrument noise ratio (SNR)  $\geq 1$ for sources that can be resolved above 
the confusion limit $f_\mathrm{c}$ or SNR $\geq5$ for 
systems with $f < f_\mathrm{c}$ that are detectable above the
noise level. This resulted in approximately equal numbers of resolvable 
detached double degenerates and interacting double degenerates --- about 
11\,000 objects of every kind.

However, the above-mentioned criterion for resolution may be oversimplified. 
A more rigorous approach implies detection using an
iterative identify-and-subtract process, where 
signal-to-noise ratio of a source is evaluated with respect to noise
from both the instrument and the partially subtracted
foreground \cite{timpano06, nissanke_gw12}. 
This procedure was applied by Nissanke et~al. \cite{nissanke_gw12}
to the same sample of binaries as
in \cite{nyp01,nyp04}, for both ``optimistic''(Case~1 below)
 and ``pessimistic'' (Case~2) 
scenarios (see \S~\ref{section:wd_formation} and \ref{section:am-evol}) with different efficiency of the formation of double degenerates. 

\epubtkImage{nissanke1.png}{%
  \begin{figure}[htbp]
  \centerline{\includegraphics[width=\textwidth]{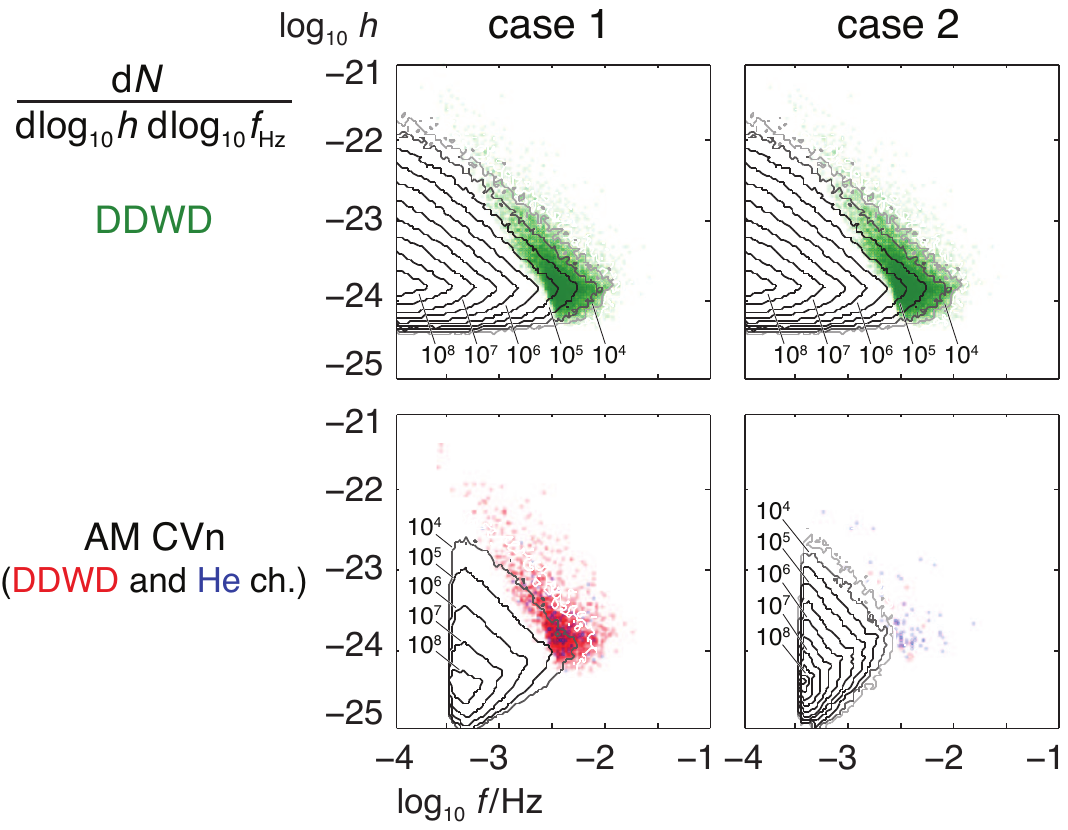}}
    \caption{The number density as a function of GW-strain amplitude
  $h$ and frequency $f$ for close detached WD (top)  and \am stars
  (bottom) in Case 1 and Case 2.  Green dots denote individually
  detected double WD systems (for one year of observation with the
  5-Mkm detector); red dots -- AM CVn systems from the double WD channel
 and blue dots -- AM CVn systems from the He star channel. Courtesy S.~Nissanke.} 
     \label{fig:niss1}
  \end{figure}}
\clearpage
Figure~\ref{fig:niss1} shows the frequency-space density and GW foreground of double WD 
and AM CVn systems for these two scenarios, as computed in \cite{nissanke_gw12}. 
Qualitatively the results are in good agreement 
with \cite{nyp01,nyp04}, despite the predicted number of systems is by a factor of five lower.  

\epubtkImage{nissanke2.png}{%
  \begin{figure}[htbp]
\centerline{\includegraphics[width=\textwidth]{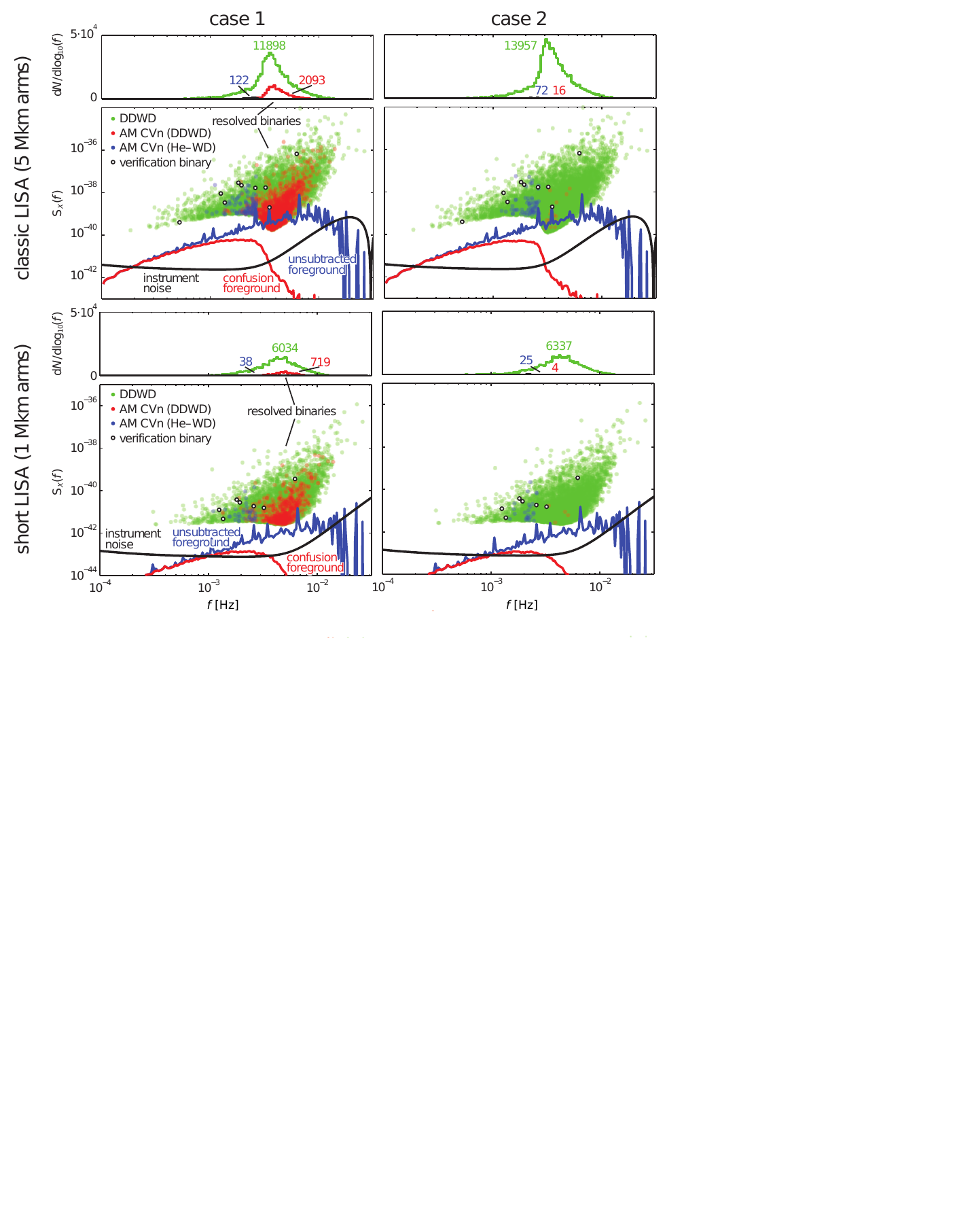}}
  \caption{The frequency-space density and GW foreground of DD and
    AM CVn systems for two scenarios. In each subplot, the bottom
    panel shows the power spectral density of the unsubtracted (blue)
    and partially subtracted (red) foreground, compared to the
    instrumental noise (black). The open white circles indicate the
    frequency and amplitude of the ``verification binaries''.  The
    green dots show the individually detectable DD  systems; the red
    dots show the detectable AM CVn systems formed from DD progenitors
    and the blue dots show the detectable AM CVn systems that arise
    from the He-star--WD channel. The top panel shows histograms of
    the detected sources in frequency space. Courtesy S.~Nissanke.} 
     \label{fig:niss2}
  \end{figure}} 
\clearpage
The frequency-space density and GW foreground of DD and AM CVn systems for two
scenarios are shown in Figure~\ref{fig:niss2} and appropriate numbers
of systems are presented in Table~\ref{tab:niss_table23}.

\begin{table}
\renewcommand{\arraystretch}{1.2}
\caption[Detectable sources (5~Mkm detector -- two upper rows, 1~Mkm
detector -- two lower rows).]{Detectable sources (5~Mkm detector --
  two upper rows, 1~Mkm detector -- two lower rows). Values are shown
  for $\mathrm{SNR}_\mathrm{thr} = 5$, one interferometric observable,
  and one year of observation, with approximate scalings as a function
  of $\rho_\mathrm{eff} = \mathrm{SNR}_\mathrm{thr}
  (T_\mathrm{obs}/\mathrm{yr})^{-1/2}(N_\mathrm{obs})^{-1/2}$ \cite{nissanke_gw12}.}
\label{tab:detLISA}
\centering
\begin{tabular}{lrrr}
\toprule
& \multicolumn{1}{c}{DD} & \multicolumn{1}{c}{AM CVn (DD)} & \multicolumn{1}{c}{AM CVn (He--WD)} \\
\midrule
case 1 &
\rescale{11,898}{1.4} & \rescale{2,093}{1.8} & \rescale{122}{2.8} \\
case 2 &
\rescale{13,957}{1.4} & \rescale{16}{1.7} & \rescale{72}{2.9} \\
\midrule
case 1 &
\rescale{6,034}{1.2} & \rescale{719}{1.4} & \rescale{38}{2.6} \\
case 2 &
\rescale{6,337}{1.2} & \rescale{4}{1.4} & \rescale{25}{2.2} \\
\bottomrule
\renewcommand{\arraystretch}{1.0}
\end{tabular}
\label{tab:niss_table23}
\end{table}

The estimates were made  for two assumed arm-lengths of the detector.
The number of detected sources and the level of residual confusion foreground 
may be scaled by power laws of a single ``effective-SNR
parameter''  $\rho_{\mathrm{eff}} = SNR_{\mathrm{thr}} (T_{\mathrm{obs}}/yr)^{-1/2} (N_{\mathrm{obs}})^{-1/2}$, where  $SNR_{\mathrm{thr}}$ is the detection threshold,
$T_{\mathrm{obs}}$ is the time of observations, and  
$(N_{\mathrm{obs}}$ is the number of Time-Delayed Interferometry observables (TDI --- a specific method to properly time-shift and linearly combine independent Doppler measurements which is necessary in the case of unequal arm-lengths of the 
detector \cite{lrr-2005-4}; 
$N_{\mathrm{obs}}=1$ for two-arms configuration of the detector). 

For the LISA-like configuration,  the number
of detected DD is comparable to earlier estimates for the ``optimistic''
case of \cite{nyp01, nyp04}: $\simeq$~11\,000). The number of 
\am stars which was comparable to the number of DD, now 
strongly drops due to a higher required SNR defined
with respect to both instrument and confusion noise and the usage of
interferometric
observables instead of the strain. For 1~Mkm detector, the effect of changing detection
criterion is much more profound -- by a factor from two to four.
The more rigorous detection criterion changes  the number
of AM CVn stars more than the number of DD, since the former have smaller chirp mass and strain amplitudes. Thus, the conclusion of early studies on the dominance of 
detached DD in the GW foreground formation remains valid. 

\epubtkImage{lisamagn.png}{%
  \begin{figure}[htbp]
\hspace{4cm}\includegraphics[width=\textwidth]{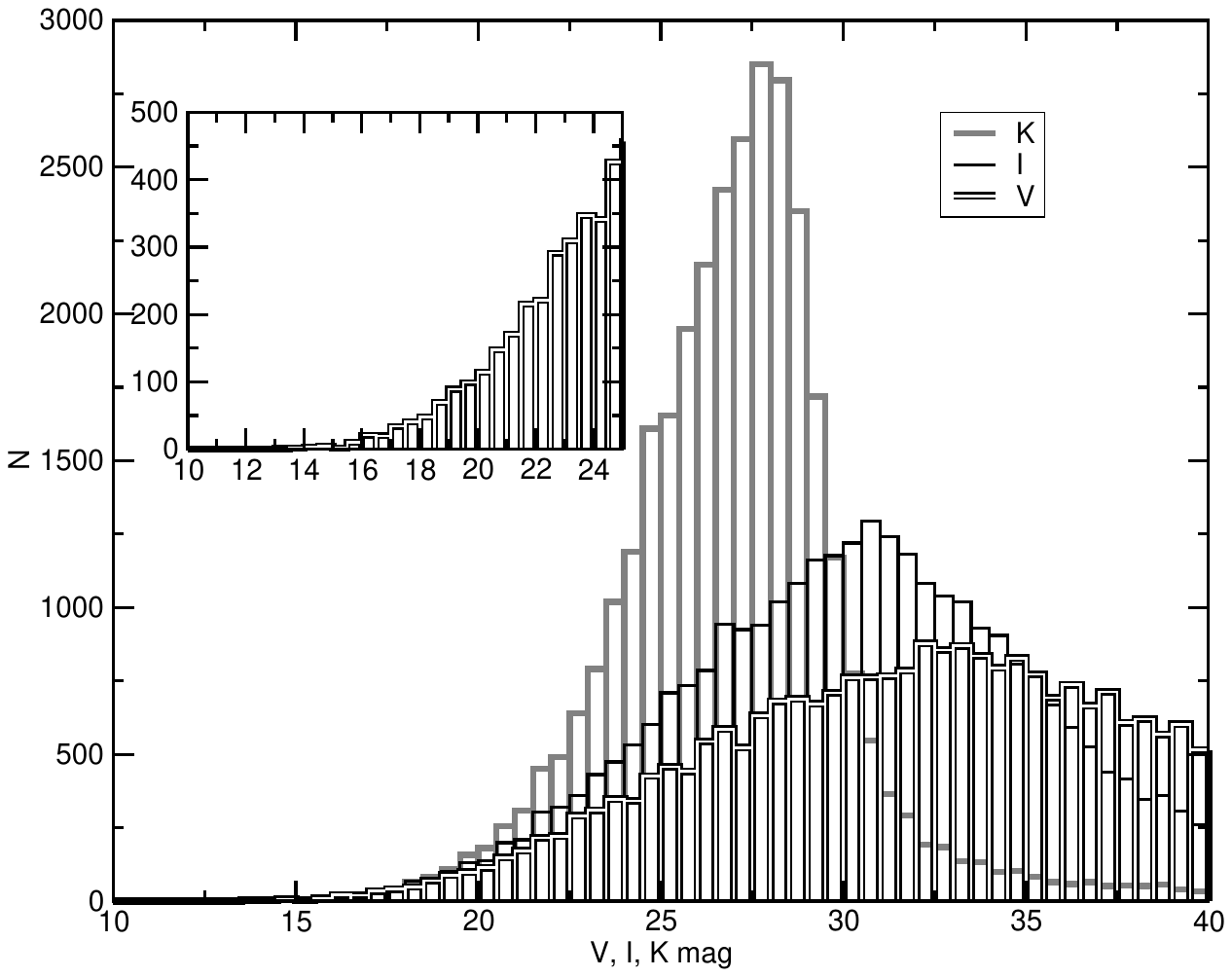}
    \caption{The histogram of apparent magnitudes of the double white
     dwarfs that are estimated to be individually detected by
     LISA. The black-and-white line shows the distribution in the
     $V$-band, the black line the distribution in the $I$-band and the
     grey histogram the distribution in the $K$-band. The Galactic
     absorption is taken into account. The insert shows the bright-end
     tail of the $V$-band distribution. From \cite{2009CQGra..26i4030N}.} 
     \label{fig:magn}
  \end{figure}}
\clearpage 

Peculiarly enough, as 
Figure~\ref{figure:gwr_back} 
shows,
the AM~CVn-type systems appear, in fact, dominant among the so-called
``verification binaries'' for LISA/eLISA: binaries that are well known from
electromagnetic observations and whose radiation is estimated to be
sufficiently strong to be detected; see the list of 30 promising
candidates in~\cite{stroeer06} and references therein.
The perspectives and strategy of discovering additional verification binaries were discussed by Littenberg et~al.~\cite{littenberg_verification13}, who used as an input catalogue the model fom \cite{nyp04}
with minor modifications. It was shown in \cite{littenberg_verification13}
that for the ``optimistic'' scenario of the formation of \am stars, $5\times10^9$\,m arm-length configuration (full LISA), 63\% confidence interval of the sky-location 
posterior distribution area on the celestial
sphere   and limiting stellar magnitude of the sample 
$m\le24$\, the number of candidate double-degenerate verification binaries is 
$\approx$~560. For the ``pessimistic'' scenario, short arm-length 
($1\times10^{9}$\,m, similar to eLISA), $d\Omega \leq1\,\textrm{deg.}^2$,
$m\leq20$ this number reduces to 9. If an additional detection requirement
of being eclipsing is applied, these numbers are reduced by a factor of $\sim$3. 
The conclusion is that the chance for discovering new ``verification binaries''
is associated with the next generation of ground-based telescopes, like the 38-m 
 European Extremely Large Telescope (E-ELT, see \url{http://www.eso.org/public/teles-instr/e-elt/}), 
the 30-m US telescope (TMT, see \url{http://www.tmt.org}),  the 
James Webb Space Telescope (see \url{http://www.jwst.nasa.gov/}). The ESA GAIA mission will also be very 
helpful in localisation of the sources.  Resolved LISA sources are expected to
be most numerous in the $K$-band,  and a  significant fraction of them 
will have $K\aplt29$ (Figure~\ref{fig:magn}), within reach of the
future facilities, whilst current instruments are mostly limited by
$K\aplt(20\mbox{\,--\,}21)$ \cite{2009CQGra..26i4030N}.


The most severe
``astronomical'' problems concerning ``verification binaries'' are
their distances, which for most systems are only estimates, and poorly
constrained component masses. Some features of ``verification binaries'' and prospects of  detection of new objects  in wide-field surveys are also discussed in \cite{2013ASPC..467...47K}.

The issue of ``verification binaries'' is related to the problem of
detection of GWR sources in electromagnetic waves which is discussed
in the next section.

\newpage
\section{AM CVn-Type Stars as Sources of Optical and X-Ray Emission} 
\label{section:opt+x}

Evidently, it is important to study  AM~CVn-stars in all possible
wavebands. The accuracy of GW parameters estimation improves
if the information available from electromagnetic observations is used
 \cite{nyp04,2012AA...544A.153S, 2013AA...553A..82S}. 
GW detectors will measure a combination of 
parameters that determine the GW signal.  If some of these parameters (orbital period, 
position on the celestial sphere)
can be obtained from independent optical or X-ray observations, other
parameters can be determined with higher accuracy.
One of the most interesting features
of the double white dwarfs that can be detected electromagnetically are eclipses, which have large probability for very short \porb\ systems that can be found 
among candidate objects for GW detectors \cite{cooray_ecl04}. Eclipsing systems are especially important since they may provide an information on the absolute parameters of the stars and, possibly, on variation of their orbital periods, as is already shown by the example of a 12.5-min system \epubtkSIMBAD{SDSS~J065133.338+284423.37} 
\cite{2012ApJ...757L..21H}. 
It is estimated that GAIA will be able to detect about 200 eclipsing \am stars
\cite{2013ASPC..467...27N}. 

While the total sample of AM~CVn stars are optical systems, 
those with the shortest orbital periods have been 
expected to be observed with LISA, hopefully, will be observed by eLISA, and 
may be observed both in the 
optical and X-rays thanks to high mass-transfer rates (see
Figure~\ref{figure:dmdtam}). 

A model for electromagnetic emission
properties of the ensemble of the shortest orbital period
($\porb\leq 1500 \mathrm{\ s}$) was constructed in
\cite{nyp04,Nelemans:2007fg,2009CQGra..26i4030N}. 
The ``optimistic'' model
of~\cite{nyp01} was considered.

\paragraph*{Optical emission.}
The luminosity of the accretion disc around a WD can be estimated as 
\begin{equation}
  L_\mathrm{disc} = \frac{1}{2} G M \dot{m}
  \left( \frac{1}{R} - \frac{1}{R_\mathrm{L1}} \right)
  \mathrm{\ erg\ s}^{-1},
  \label{eq:L_acc}
\end{equation}
with $M$ and $R$ being the mass and radius of the accretor, $R_\mathrm{L1}$
being the distance of the first Lagrangian point to the centre of the
accretor, and $\dot{m}$ being the mass transfer rate,
respectively. Optical emission of
the disc was modeled by a single temperature disc extending up 
to 70\% of the Roche lobe of the accretor and radiating as a
black body~\cite{wad84}.

The emission from the donor was treated as the emission of a cooling white
dwarf, using approximations to the cooling WD models of
Hansen~\cite{hansen99}.

The emission from the accretor was treated  as the unperturbed cooling
luminosity of the white dwarf\epubtkFootnote{This is true for
  short-period systems, but becomes an oversimplification at longer
  periods and for $P_{\mathrm{orb}} \gtrsim 40 \mathrm{\ min}$ heating
  of the WD by
  accretion has to be taken into account~\cite{bild_accr06}.}.

The distribution of $V-$, $I-$, $K-$ magnitudes of the sources expected to be resolved 
by LISA was presented in the previous Section in Figure~\ref{fig:magn}
\cite{2009CQGra..26i4030N}.

\paragraph*{X-ray emission.}
For the model of X-ray emission, the ROSAT
0.1\,--\,2.4~keV X-ray band was considered,
taking into account interstellar absorption. The ROSAT band
was chosen because of the discovery of AM~CVn itself~\cite{ull95a} and
\epubtkSIMBAD{RXJ0806.3+1527}~\cite{ipc99} and
\epubtkSIMBAD{V407~Vul}~\cite{mhg_96}) as ROSAT sources and because of the possibility
of comparison with the ROSAT all-sky survey.

Most AM~CVn systems experience a short
(10\super{6}\,--\,10\super{7}~yr) ``direct impact'' stage in the
beginning of mass-transfer~\cite{hb00, nyp01, mns04}. Hence, in the 
modeling the X-ray emission from AM~CVn systems one has to distinguish
two cases: the direct impact and disc accretion.

In the case of the direct impact a small area of the accretor's surface is
heated. One may assume that the total accretion luminosity is emitted
as a black body radiation with a temperature given by
\begin{equation}
  \left( \frac{T_\mathrm{BB}}{T_{\odot}} \right)^4 =
  \frac{1}{s} R^{-2} L_\mathrm{acc},
\end{equation}
where $L_\mathrm{acc}$ and $R$ are in solar units and $L_{\mathrm{acc}}$ is
defined by Eq.~(\ref{eq:L_acc}). The fraction $s$ of the surface that
is radiating depends on the details of the accretion. It was set to
0.001, consistent with expectations for a ballistic stream~\cite{ls75}
and the observed X-ray emission of \epubtkSIMBAD{V407~Vul}, the known
direct-impact system~\cite{mar_ste02}.

In the presence of a disc, the X-ray emission was assumed to be coming from
a boundary layer with temperature~\cite{pringle77}
\begin{equation}
  T_\mathrm{BL} = 5 \times 10^{5} 
  \left( \frac{\dot{m}}{10^{18} \mathrm{\ g\ s}^{-1}} \right)^\frac{2}{9} 
  \left( \frac{M}{M_\odot} \right)^\frac{1}{3}
  \left( \frac{R}{5 \times 10^8 \mathrm{\ cm}} \right)^{-\frac{7}{9}}
  \mathrm{\ K}.
\end{equation}
The systems with an X-ray flux in the ROSAT band higher than
10\super{-13}~erg~s\super{-1}~cm\super{-2} were selected. Then, the intrinsic
flux in this band, the distance and the estimate of the Galactic
hydrogen absorption~\cite{morrison_abs83} can be used to estimate
the detectable flux.

\epubtkImage{figure16.png}{
  \begin{figure}[htbp]
    \centerline{\includegraphics[width=0.5\textwidth]{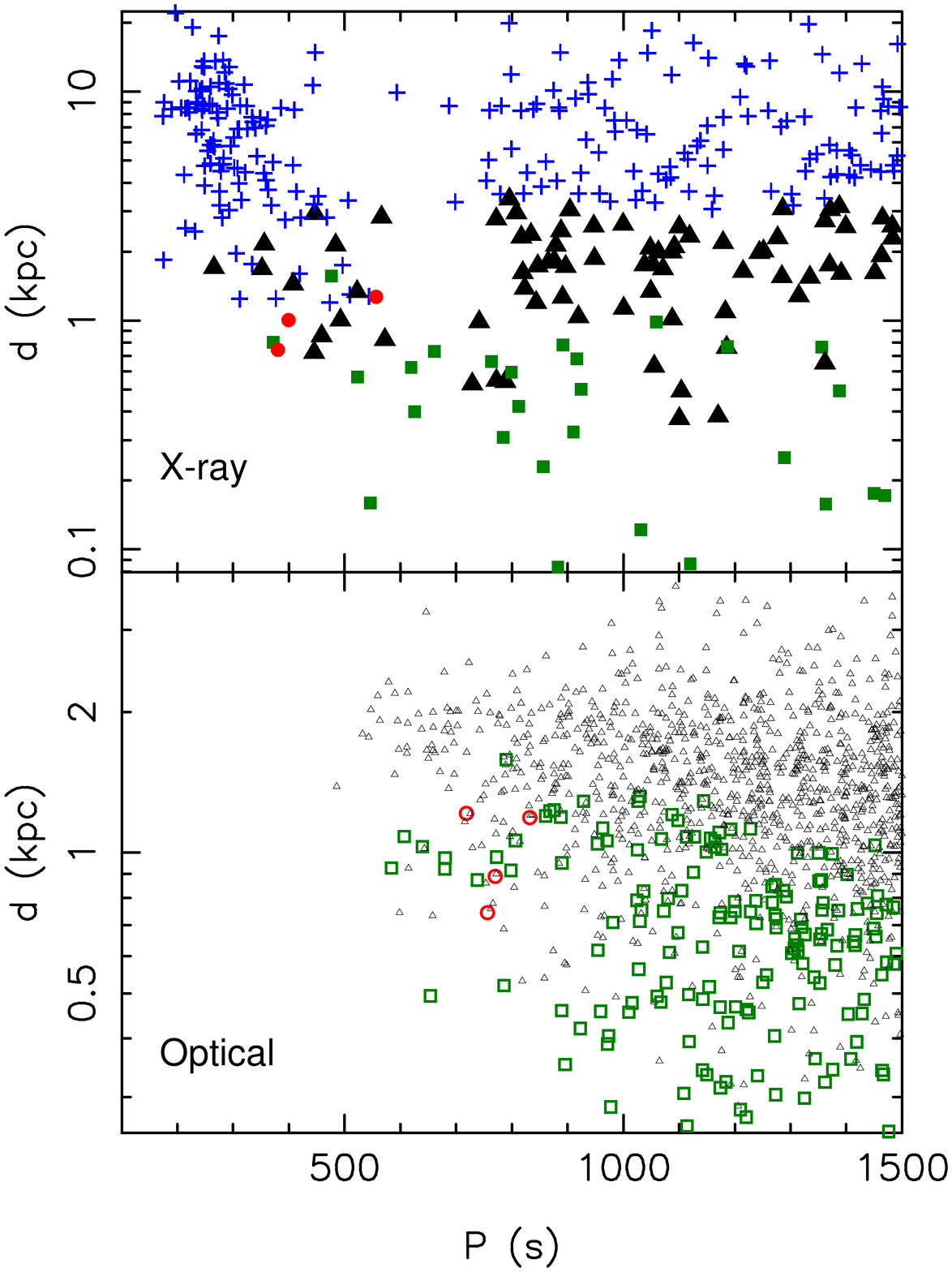}}
    \caption{Distribution of short period AM~CVn-type systems
      detectable in soft X-ray and as optical sources as a function of
      the orbital period and distance. Top panel: systems detectable in
      X-ray only (blue pluses), direct impact systems observable in
      X-ray and \textit{V}-band (red filled circles), systems detectable in
      X-ray with an optically visible donor (green squares), and systems
      detectable in X-ray and with an optically visible disc (large
      filled triangles). Bottom panel: direct impact systems (red open
      circles), systems with a visible donor (green squares), and systems
      with a visible accretion disc (small open triangles). The sample is 
limited by $V=20$.
(Updated figure from~\cite{nyp04}, see also~\cite{url07}.)}
    \label{figure:amcvn_obs}
  \end{figure}}

Figure~\ref{figure:amcvn_obs} presents the resulting model for the sample limited 
by $V=20$\,mag., which is typical for optically detected \am stars.
In the top panel shown  
are 220 systems only detectable in X-ray and 330 systems also detectable in the
\textit{V}-band. One may distinguish two subpopulations in the top panel: In the
shortest period range there are systems with white-dwarf donors with such high
$\dot{M}$ that even sources close to the Galactic centre are detectable.
Spatially, these objects are concentrated in a small area on the sky. At longer
periods the X-rays get weaker (and softer) so only the systems close to the
Earth can be detected. They are more evenly distributed over the sky. Several of
these systems are also detectable in the optical (filled symbols). There are 30
systems that are close enough to the Earth that the donor stars can be seen as
well as the discs (filled squares). Above $P=600 \mathrm{\ s}$ the systems with
helium-star donors show up and have a high enough mass transfer rate to be X-ray
sources, the closer ones of which are also visible in the optical, as these systems
always have a disc. The bottom panel shows the 1230 ``conventional'' AM~CVn
systems, detectable only by optical emission, which for most systems
emanates only from their accretion disc. Of this population 170 objects 
closest to the Earth also have a visible donor. The majority of the optically
detectable systems with orbital periods between 1000 and 1500~s are
expected to show outbursts due to the viscous-thermal disc
instability~\cite{to97,ramsay_am_outbursts12,2012AA...544A..13K} which could 
enhance the chance of their
discovery.

In \cite{nyp04} the estimates of optical and X-ray emission from 
\am stars  were applied to predict their numbers that 
can be detected in gravitational waves and electromagnetic spectrum by LISA. 
These results were also discussed in the previous version of this review.
 However, 
in view of cancellation of the LISA mission and different sensitivity of eLISA,
new insights in the problem of resolution of low-frequency GW sources and the  
recognition that the most promising for detection of new objects 
with future facilities is the IR-band,  these 
results may be considered as outdated and a new study of the problem is necessary.



To conclude this Section, we stress several points that can be important for
the understanding of the formation and evolution of compact WD binaries. 

\begin{itemize}
\item
The major issue concerning compact WD binaries -- DDs, IDDs, UCXBs -- is
their number. Theoretical predictions strongly depend on the assumed
parameters and range within an order of magnitude (see references in
Section~\ref{section:waves}). The treatment of common envelopes and the
distribution of stars over $q$ are, perhaps, the most crucial points.
On the other hand,
observational estimates suffer from numerous selection effects {\bf and resulting 
incompleteness of the samples.
For instance, the local sample of all  WD is considered to be 100\% complete within 
13~pc and  only 85\% complete within 20~pc 
\cite{2009AJ....138.1681S,2014arXiv1401.4989S}}.

\item
The space density of the observed 
AM~CVn-stars is by a factor close to 50 lower than the 
``optimistic'' predictions, but  the situation can be improved:
about half of the known and candidate systems were found within the past
decade, predominantly thanks to systematic searches for candidates in the SDSS and PTF data.
Hopes for future discoveries may be associated with facilities which will become operative
in the next 10 years, like JWST or E-ELT. 

\item
Another key issue in the studies of detached and interacting binary
white dwarfs is the determination of their distances. In this aspect,
the GAIA space probe, which is aimed at constructing a
three-dimensional map of
our Galaxy via measurements of positions and radial velocities of 
about one billion stars (see~\cite{url09}), appears to be the most helpful. 

\end{itemize}

A failure to discover a significant number of detached and interacting
double degenerates or to confirm the current ideas on their structure and
evolution would mean that serious drawbacks exist either in the
implementation of the known stellar evolution physics and observational
statistical data in the population synthesis codes, or in our
understanding of the processes occurring in compact binaries, or in the
treatment of selection effects. Special attention in theoretical
  studies has to be paid to the onset of mass-transfer.

Above we presented some of the current ideas on the formation and evolution of 
compact
binaries that can be interesting for general relativity and cosmology
and on signals that can be expected from them in the 
waveband of space-born GW detectors.
There is another side of the problem -- the analysis of the GW 
signal, would it be detected. This topic is out of the scope of this
brief review. We refer the reader only to several papers discussing
methods of detecting and subtracting individual binary signals from a
data stream with many overlapping signals~\cite{cornish_larson03},
of inferring properties of the distribution of white-dwarf
binaries~\cite{edlund05}, of determining the accuracy of parameter estimation of low-mass
binaries~\cite{takahashi_wd02, rogan_lisa06}, and the discussion of the
wealth
of information that may be provided by 
eLISA \cite{2013GWN.....6....4A}.

\newpage
\section{Conclusions}      
\label{sec:concl}


The current understanding of the evolution of binary stars is
firmly based on observations of many types of binary systems, from
wide non-interacting pairs to very close compact binaries consisting
of stellar remnants -- white dwarfs, neutrons stars, and 
black holes. The largest uncertainties in the
specific parameters of the double compact binary formed at the end
of the evolution of a massive binary system are related to the physical
properties of the pre-supernovae: masses, magnetic fields, equation
of state (for NSs), spins, possible kick velocities, etc. This
situation is due to our limited understanding of both the late stages of
stellar evolution and especially of the supernovae
explosion mechanisms and physics of NS/BH formation. 

The understanding of the origin and
evolution of compact white dwarf binaries also suffers from
incompleteness of our knowledge of white dwarf formation and, in
particular, on the common envelope treatment. The progress in these
fields, both observational and theoretical, will have a major effect
on the understanding of the formation and evolution of compact binary
systems. On the other hand, the phenomenological approach used to
describe these uncertainties proves to be successful in explaining
many observed properties of various binary stars, so the
constraints derived from the studies of binary stars should be taken
into account in modeling stellar evolution and supernovae
explosions.

Of course, specifying and checking the initial distributions of
orbital parameters of binary stars and parameters of binary evolution
(such as evolution in the common envelopes), as well as modelling of 
accretion and merger processes stay in the short-list of
the important actions to be done. Here an essential role belongs to
detailed numerical simulations.

\paragraph*{Further observations of compact binaries.}
Clearly, discoveries of new types of compact binary systems have provided the
largest impetus for studies of binary star evolution. The well
known examples include
the discovery of X-ray binaries, relativistic binary pulsars,
millisecond recycled pulsars and accreting millisecond X-ray pulsars, 
close binary white dwarfs. In the nearest future we expect the
discovery of NS\,+\,BH binaries which are predicted by the massive binary
evolution scenario in the form of binary radio pulsars with
BH companions~\cite{Lipunov_al94, Lipunov_al05, Pfahl_al05}.
Their immediate possible progenitors are observed as well known Galactic 
system Cyg~X-3 and extra-galactic objects like IC10 X-1, NGC300~X-1 harbouring 
Wolf--Rayet stars and NS or BH.

It is
very likely that we already observe the coalescence of double NS/BH
systems as short gamma-ray bursts in other
galaxies~\cite{Gehrels_al05, Nakar07}, and the recent discovery of the 
IR afterglow after short/hard GRB~130603B \cite{Berger_al13, Tanvir_al13}
provided a beautiful confirmation of the expected 
possible electromagnetic phenomena (``kilonova'' or ``macronova'') -- 
the radioactively powered transient, predicted by Li \& Paczy\'{n}ski in 1998
\cite{Li_Paczynski98}. 

It is very likely that 
NS\,+\,BH or BH\,+\,BH binaries can be found first in GW data
analysis~\cite{1993MNRAS.260..675T,Grishchuk_al01,Dominik_al12}. The efforts of the LIGO collaboration
to put constraints on the compact binary coalescences from the
analysis of the existing GW observations are very
important~\cite{Abadie_al12}, as well as the hard work on modeling 
expected signal waveforms  \cite{blanchet_templates14}.

The formation and evolution of compact binaries is a very
interdisciplinary field of modern astrophysics, ranging from
studies of the equation of state for super-dense matter inside neutron
stars and testing effects of strong gravity in relativistic compact
binaries to hydrodynamical simulations of stellar winds, formation and 
evolution of
common envelopes, and stellar explosions.
Therefore, further progress in this rapidly flourishing field of `multi-messenger astronomy', 
which will be made by means of traditional astronomical
observations and new tools, like gravitational wave
and neutrino detectors \cite{2013CQGra..30s3002A}, will undoubtedly have a strong impact on
astronomy and astrophysics as a whole.

\newpage
\section*{Acknowledgements}
\label{section:acknowledgements}

The authors acknowledge the referees for careful reading of the
manuscript and useful comments. 
The authors would like to thank L.P.~Grishchuk, V.M.~Lipunov,
M.E.~Prokhorov, A.G.~Kuranov, N.~Chugai, A.~Tutukov,  A.~Fedorova,
E.~Ergma, M.~Livio, S.F.~Portegies Zwart, J.-P.~Lasota, G.~Dubus,
A.~Renzini, R.~Napiwotzki, E.P.J.~van den Heuvel, G.~Tovmassian,
J.-E.~Solheim, L.~van Haaften, L.~Piersanti, and A.~Tornamb{\'e} for
numerous useful discussions and joint research in the evolution of
binary stars. Writing of this review would have been impossible
without the long-term research cooperation with G.~Nelemans. We
acknowledge him also for useful discussions, help in collecting
information, and updating figures. Useful discussions with
participants of meetings on stellar evolution, binary stars and
\sna are acknowledged. 
An intensive use of the Smithsonian/NASA ADS Astronomy Abstract
Service and arXiv is acknowledged. We thank many of our colleagues for
sending us (p)reprints of their publications. 
Yu. Pachomov is acknowledged for his help in the preparation of the manuscript.

KAP was partially supported by RFBR grants 12-02-00186, 13-02-92119 and  14-02-00657.
LRY was partially supported by the Presidium of the Russian Academy of Sciences
program P-21 and Russian Foundation for Basic Research grant 
14-02-00604.

\newpage
\bibliography{new_lrr_lr_pk2}
\end{document}